\documentclass[11pt]{article}

\PassOptionsToPackage{hidelinks}{hyperref}

\usepackage{jheppub}              
\usepackage{amsmath, amssymb, mathrsfs} 
\usepackage{mathtools}            
\usepackage{physics}              
\usepackage{graphicx}             
\usepackage{pdfpages}             
\usepackage[export]{adjustbox}    
\usepackage{marginnote}           
\usepackage{comment}              
\usepackage{gensymb}              
\usepackage[scr=boondox, cal=esstix]{mathalpha} 
\usepackage[load-configurations=abbreviations]{siunitx} 
\usepackage{todonotes}   
\setuptodonotes{fancyline, color=blue!10, size=\small}

\newlength{\originalVOffset}
 \newlength{\originalHOffset}
\title{Terrestrial Very-Long-Baseline Atom Interferometry: Summary of the Second Workshop}

\abstract{\\
This summary of the second Terrestrial Very-Long-Baseline Atom Interferometry (TVLBAI) Workshop provides a comprehensive overview of our meeting held in London in April 2024~\cite{2ndTVLBAIWorkshop}, building on the initial discussions during the inaugural workshop held at CERN in March 2023~\cite{1stTVLBAIWorkshop}. Like the summary of the first workshop~\cite{TVLBAISummary}, this document records a critical milestone for the international atom interferometry community. It documents our concerted efforts to evaluate progress, address emerging challenges, and refine strategic directions for future large-scale atom interferometry projects. Our commitment to collaboration is manifested by the integration of diverse expertise and the coordination of international resources, all aimed at advancing the frontiers of atom interferometry physics and technology, as set out in a Memorandum of Understanding signed by over 50 institutions~\cite{TVLBAIMOU}.
}

\newpage

\author[1]{Adam~Abdalla,}
\author[2]{Mahiro~Abe,}
\author[3]{Sven~Abend,}
\author[3]{Mouine~Abidi,}
\author[4,5,6,*]{Monika~Aidelsburger,}
\author[3]{Ashkan~Alibabaei,}
\author[7,*]{Baptiste~Allard,}
\author[8]{John~Antoniadis,}
\author[9,*]{Gianluigi~Arduini,}
\author[10]{Nadja~Augst,}
% ...

\affiliation[1]{Technische Universit{\"a}t Darmstadt, Fachbereich Physik, Institut f{\"u}r Angewandte Physik, Schlossgartenstrasse 7, 64289 Darmstadt, Germany}
\affiliation[2]{Department of Physics, Stanford University, Stanford, California 94305, USA}
\affiliation[3]{Leibniz Universit\"at Hannover, Institut f\"ur Quantenoptik, Welfengarten 1, 30167 Hannover, Germany}
\affiliation[4]{Max-Planck-Institut f\"ur Quantenoptik, Hans-Kopfermann-Strasse 1, 85748 Garching, Germany}
\affiliation[5]{Fakult\"at f\"ur Physik, Ludwig-Maximilians-Universit\"at, Schellingstrasse 4, 80799 M\"unchen, Germany}
\affiliation[6]{Munich Center for Quantum Science and Technology (MCQST), Schellingstrasse 4, 80799 M\"unchen, Germany}
\affiliation[7]{Laboratoire Collisions Agr\'{e}gats R\'{e}activit\'{e} UMR5589, University of Toulouse III Paul Sabatier CNRS, 118 Route de Narbonne, Toulouse, FR}
\affiliation[8]{Foundation for Research and Technology-Hellas (FORTH), Heraklion, 70013, Greece}
\affiliation[9]{CERN, CH-1211 Geneva 23, Switzerland}
\affiliation[10]{Institute of Quantum Technologies, German Aerospace Center (DLR), Wilhelm-Runge-Stra\ss{e} 10, 89081 Ulm, Germany}
% ...
\affiliation[*]{Section Editor/Contributor, and/or Workshop Organiser}
\emailAdd{Oliver.Buchmueller@cern.ch} \emailAdd{John.Ellis@cern.ch}

\begin{document}

\maketitle

\section{
Introduction}
\label{Introduction}

This document summarises the discussions and outcomes of the 2nd TVLBAI workshop~\cite{2ndTVLBAIWorkshop}, which gathered international experts to review recent advances in large-scale atom interferometer prototypes and potential future applications of atom interferometry for detecting ultralight dark matter and gravitational waves. The discussions focused on the physical principles and technological advances driving these state-of-the-art systems, and on establishing a structured framework for an international TVLBAI proto-collaboration.

Central to the workshop was the aim to leverage the collective expertise of researchers from various institutions, fostering a dynamic collaborative network to drive strategic discussions and secure funding for future large-scale projects. The sessions were marked by a thorough review of progress since our last meeting, with substantial efforts directed towards formalising the proto-collaboration by defining roles, responsibilities, and strategies for effective communication and coordination.

A primary goal of the TVLBAI activities, under the mandate of the new proto-collaboration, will be the development of a comprehensive roadmap for future kilometre-scale detectors. This roadmap will delineate strategic design choices, technological considerations, and scientific drivers, setting clear timelines and identifying crucial milestones essential for the successful and timely realisation of these ambitious detectors, anticipated to be operational by the mid-2030s. The TVLBAI workshops serve as critical stepping stones towards achieving this goal, ensuring that each phase of the roadmap aligns with the collective vision and capabilities of the international scientific community engaged in this pioneering effort.

Moreover, these workshops play a crucial role in cultivating a sense of community among participants, reinforcing the existing network of experts and advocates committed to pushing the boundaries of atom interferometry. This community spirit is fundamental to sustaining the momentum and ensuring the success of our collective efforts to make pioneering groundbreaking scientific discoveries. Through this summary, we share our vision, highlight the challenges, and explore the exciting potential that lies ahead in the field of atom interferometry.

\subsection{2nd TVLBAI Workshop Participants}
\label{participants}

\begin{figure*}
\centering
\includegraphics[width=0.7\textwidth]{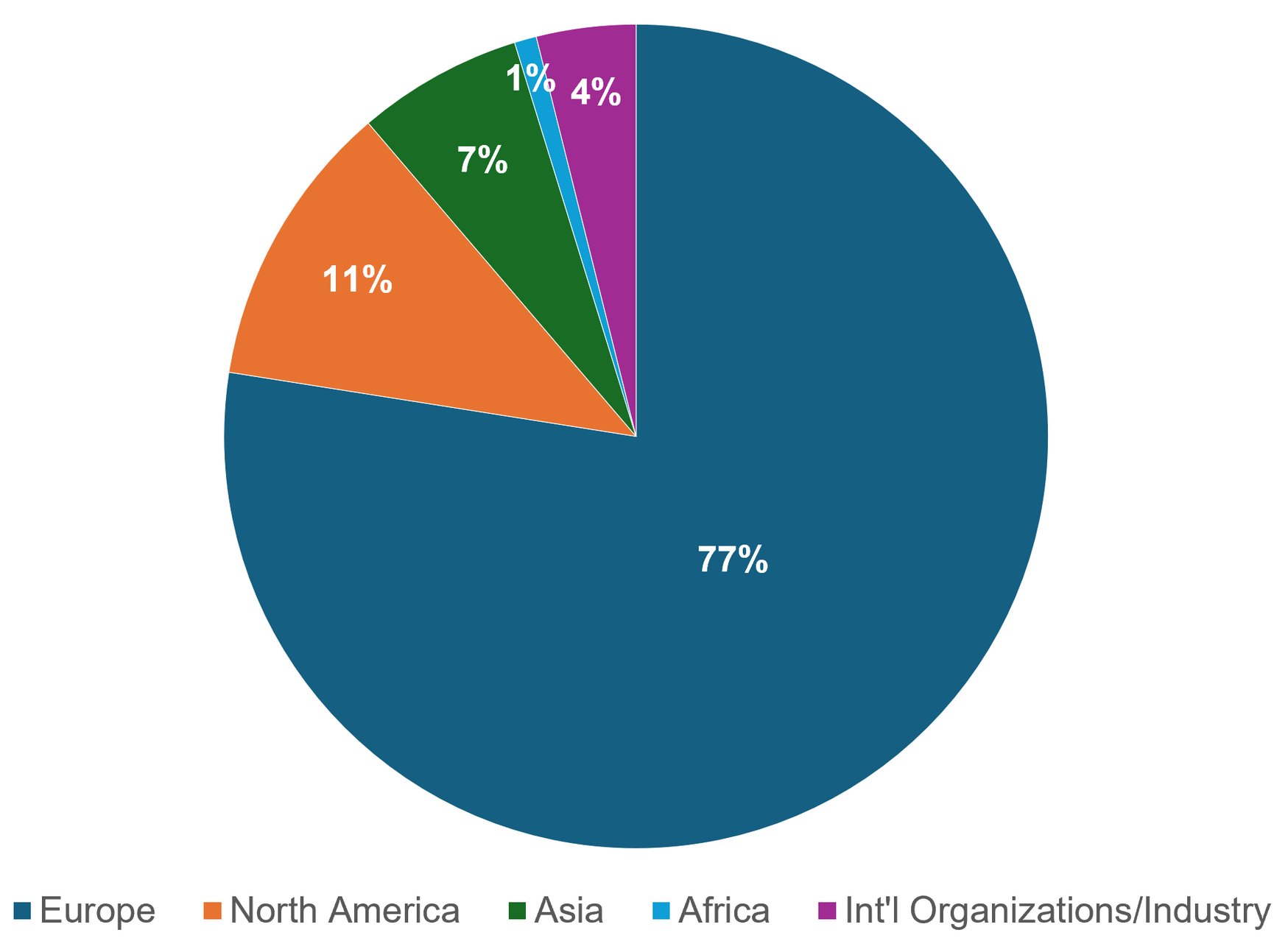}
\caption{Statistics of the geographical distribution of the home institutions of the 276 participants who registered for the Workshop. From~\cite{2ndTVLBAIWorkshop}.}
\label{fig:participants}
\end{figure*}

Figure~\ref{fig:participants} illustrates the diverse geographical distribution of the 276 registered participants in the 2nd TVLBAI Workshop. This breakdown showcases the broad geographical distribution of attendees: 77\% from Europe, 11\% from North America, 7\% from Asia, 4\% from international organisations and industry, and 1\% from Africa. This diversity reflects the geographical scope of the new TVLBAI proto-collaboration currently being formed, which is aimed at fostering global cooperation and strategic dialogue among leading experts in the field.

The participants' diverse geographical distribution aligns with the workshop’s objectives to discuss cutting-edge advancements in large-scale atom interferometer prototypes and their applications in detecting ultralight dark matter and gravitational waves. This gathering not only facilitates the establishment of an international network but also supports the workshop's goal to develop a comprehensive roadmap for a global network of future kilometer-scale detectors. The involvement of such a broad array of experts is crucial for the successful realisation of these ambitious projects with diverse technological approaches, enabling the Workshop to serve as a cornerstone in advancing the initiative in atom interferometry and setting the stage for fundamental scientific discoveries.

\subsection{Setting the Scene: Ultralight Dark Matter \& Gravitational Waves}\label{sec:setting}
~~\\
\noindent
{\bf The Experimental Context}

Participants in this Workshop will be familiar with the basic principle of atom interferometers (summarised here in  Section~\ref{Rasel}) (see also~\cite{Buchmueller2023}), which is similar to that of laser interferometers: clouds of cold atoms are split by lasers into populations of ground and excited states that follow different space-time trajectories before being brought into superposition, where their interference patterns are measured. These patterns could be modified by interactions of coherent waves of ultralight bosonic cold dark matter with the atomic constituents~\cite{Geraci2016,Arvanitaki2018}, or by the passage of gravitational waves~\cite{Dimopoulos:2007cj,Dimopoulos2008a}. {\it Inter alia}, the sensitivity to such effects is enhanced in experiments where the atoms propagate freely over larger distances, hence the drive towards longer baselines on Earth or in space.

There are ongoing 10m projects at Stanford~\cite{Dickerson2013} and in Hannover~\cite{schlippert2020matter}, and another 10m project has been proposed by the AION Collaboration for Oxford in the UK~\cite{Badurina2020}. Several ${\cal O}(100)$m projects are under construction, including MIGA in France~\cite{Canuel_2018}, ZAIGA in China~\cite{Zhan2019} and MAGIS at Fermilab in the US~\cite{Abe2021}, and the AION Collaboration has also proposed a follow-on ${\cal O}(100)$m detector~\cite{Badurina2020}. The focus of this workshop is on a future generation of km-scale detectors, with projects proposed for the Sanford Underground Research facility in the US (see Section~\ref{sites-summary}), Wuhan in China (see Section~\ref{sites-zaiga}), ELGAR in Europe~\cite{Canuel2020}, the Boulby mine in the UK (see Section~\ref{sites-boulby}), and the Gotthard rail tunnel in Switzerland (see Section~\ref{sites-porta_alpina}).\\
~~\\
\noindent
{\bf Searches for Ultralight Dark Matter}

\begin{figure}[h]
\centering 
\includegraphics[width=0.45\textwidth]{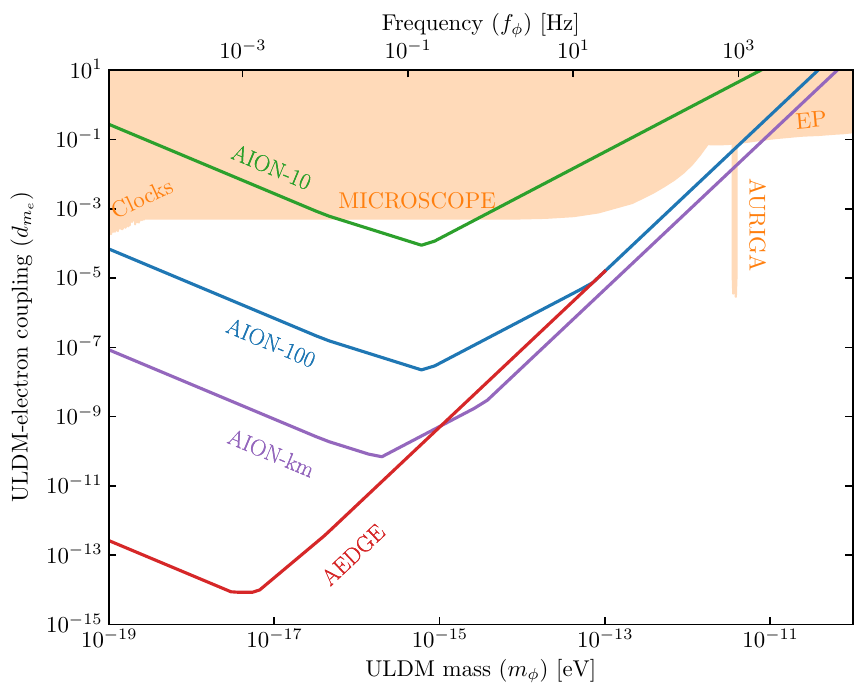}
\includegraphics[width=0.45\textwidth]{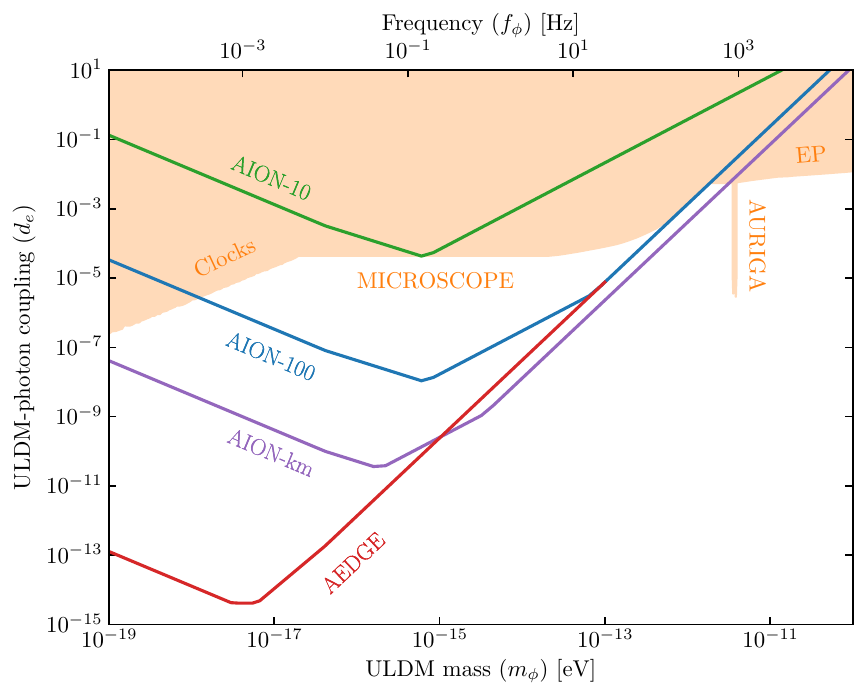}
\caption{Sensitivity projections for linear ULDM couplings to electrons (left panel) and photons (right
panel), neglecting gravity gradient noise. The green (blue) (purple) and red lines are for AION-10 (100) (km)
and AEDGE, respectively. The shaded orange region is excluded by the existing constraints from searches for
violations of the equivalence principle by the MICROSCOPE experiment and with torsion balances, atomic clocks, and the AURIGA experiment, as described in~\cite{Buchmueller2023}, where the assumed experimental specifications can be found. Taken from~\cite{Buchmueller2023}.}
\label{fig:ULDM}
\end{figure}

Figure~\ref{fig:ULDM} displays the prospective sensitivities of long-baseline atom interferometers to possible interactions of cold atoms with ultralight bosonic dark matter, compared with the current sensitivities of atomic clocks at low mass, probes of the universality of free fall (UFF) at intermediate masses, and torsion balances at higher masses. We see that, whereas the sensitivities of 10m atom interferometers begin to be comparable to UFF tests, long-baseline experiments offer orders of magnitude improvements over current sensitivities, e.g., by factors up to $\sim 10^6$ in km-scale experiments and up to $\sim 10^{10}$ in a space-borne experiment such as AEDGE~\cite{ElNeaj2020}.

The main limiting factor for the sensitivity in Fig.~\ref{fig:ULDM} is the atom shot noise, but another potential limiting factor is the Gravity Gradient Noise (GGN) due to seismic vibrations, whose level depends on the site. GGN may be mitigated by locating multiple atomic interferometers in the same vertical shaft, as illustrated in Fig.~\ref{fig:Multigradiometer} below, which can be manipulated with the same laser beam, thereby eliminating laser noise and minimising GGN by making difference measurements. Further mitigation of GGN may be achieved with a network of seismometers around the atom interferometer, which would require further site-specific studies.\\

\noindent
{\bf Searches for Gravitational Waves}

Atom interferometers enable searches for GWs in the deci-Hz frequency range between the sensitivities of terrestrial laser interferometers such as LIGO, Virgo and KAGRA (LVK) (see Section~\ref{LVK}) and space-borne laser interferometers such as LISA, Taiji and TianQin (see Section~\ref{Bayle}). Deci-Hz GW opportunities are  discussed in Section~\ref{Urrutia}, here some examples are introduced briefly.

\begin{figure*}
\centering
\includegraphics[width=0.8\textwidth]{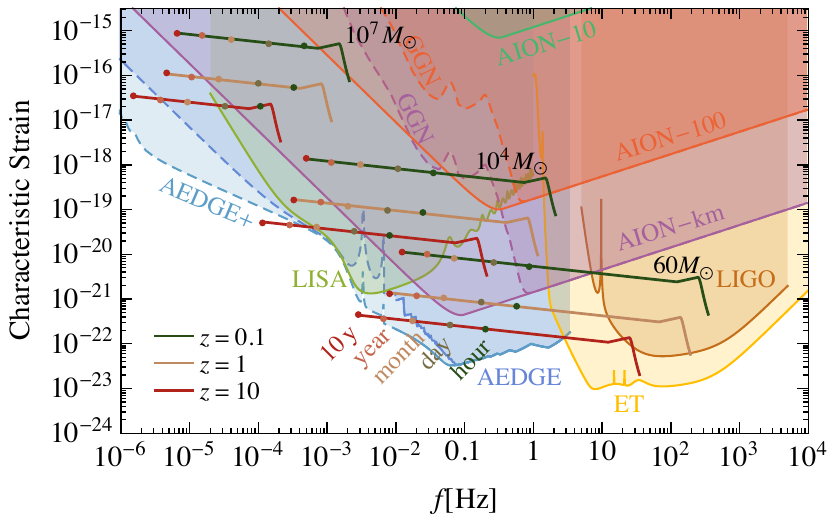}
\caption{Dimensionless strain sensitivities of AION-10, -100 and -km, AEDGE and AEDGE+, compared with those of LIGO, LISA and ET and the signals expected from mergers of equal-mass binaries with combined masses 60, $10^4$ and $10^7$ solar masses. The assumed redshifts are $z = 0.1, 1$ and 10, as indicated. Also shown are the remaining times during inspiral before the final mergers. From~\cite{Badurina:2021rgt}.}
\label{fig:GWgap}
\end{figure*}

As seen in Fig.~\ref{fig:GWgap}, experiments in this frequency range may be sensitive to the final stages of mergers between intermediate-mass black holes weighing $\sim 10^4$ solar masses, as well as the early inspiral stages of lower-mass black-hole binaries whose mergers can subsequently be observed by LVK detectors such as LIGO. These atom interferometer measurements could be used to predict when and in what direction these mergers will take place~\cite{Ellis:2020lxl}, facilitating the preparation of multi-messenger measurements. Likewise, LISA could measure the early inspiral measurements of intermediate-mass mergers to be observed later by atom interferometers. These are examples of the prospective synergies of atom interferometer measurements with laser interferometers, as reviewed in Section~\ref{sec:synergies}.

Measurements of mergers of intermediate black holes are interesting in their own right, as they may help us understand the mechanisms that have formed the supermassive black holes weighing $\gtrsim 10^6$  solar masses that infest galactic nuclei. Were they assembled hierarchically from low-mass seeds provided by the collapses of Population~III stars? Or assembled from intermediate-mass seeds formed in protogalaxies that subsequently merged along with the black holes they contained? Or were the supermassive black holes formed directly in the mergers of protogalaxies?

These questions have become hot topics following the apparent observation by pulsar timing arrays (PTAs) of a stochastic background of nano-Hz GWs commonly thought to have been emitted by supermassive black hole binaries~\cite{NANOGrav:2023hde,EPTA:2023xxk,Zic:2023gta,Xu:2023wog}, and the discovery of a population of high-redshift supermassive black holes in observations using JWST and other telescopes. These two sets of observations are quite consistent and, taken together, may provide important clues to the formation of active galactic nuclei (AGNs) as well as supermassive black holes~\cite{Ellis:2024wdh}.

\begin{figure*}
\centering
\includegraphics[width=0.9\textwidth]{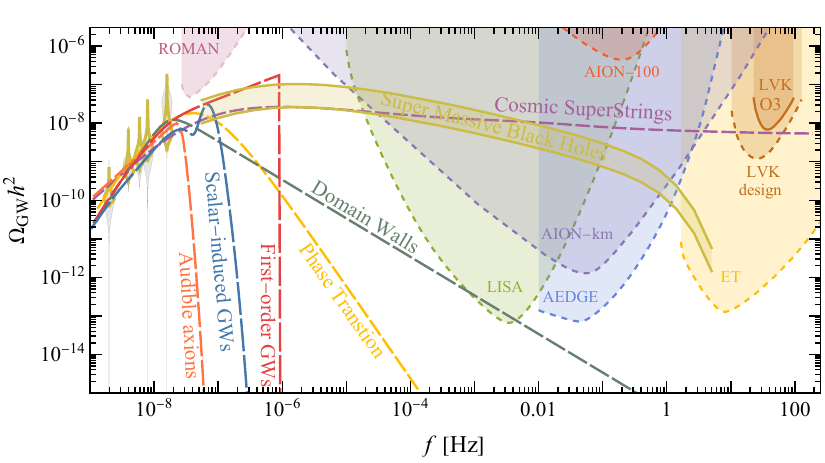}
\caption{Extension of fits to current NANOGrav PTA data~\cite{NANOGrav:2023hde} (grey ``violins") to higher frequencies, indicating the prospective sensitivities to the fractional cosmological energy density of gravitational waves of LVK and planned and proposed future detectors including LISA, AION and AEDGE. For clarity, we include only the first four and the eighth NANOGrav data bins, which have the largest impact on fit quality. 
The green band extends the green supermassive black hole (SMBH) binary ``violins" in the PTA  range to higher frequencies, and shows the mean GW energy density spectrum from SMBH binaries heavier than $10^3 M_\odot$ for $p_{\rm BH} = 0.25 - 1$. Individual SMBH binaries are expected to be measurable in this frequency range. From~\cite{Ellis2023b}.}
\label{fig:MMA}
\end{figure*}

The PTA and JWST data are good news for atom interferometers in at least two ways. They both suggest that the Universe may contain more massive black holes than known previously, and the PTA data suggest that they form binaries relatively easily, a suggestion supported by JWST observations of a population of dual AGNs presumably containing black holes that will subsequently merge. Fig.~\ref{fig:MMA} shows how an extrapolation of the PTA data using the Extended Press-Schechter formalism~\cite{Ellis2023b} to predict rates for higher-frequency gravitational waves from black hole mergers suggests that there may be an observable signal in the deci-Hz range.

Whilst black hole binaries are the default astrophysical interpretation of the PTA signal, many cosmological models invoking physics beyond the Standard Model also fit the data. One example shown in Fig.~\ref{fig:MMA} is provided by cosmic (super)strings, which suggest a broad signal spectrum extending across the LISA, atom interferometer and LVK frequency ranges. Comparisons of the signal strengths in these ranges could provide unique information about the expansion history of the Universe~\cite{Ellis2023b}.

As seen in Fig.~\ref{fig:MMA}, there are alternative cosmological scenarios based on phase transitions, domain walls, etc., that can also fit the PTA data but then would not predict an observable signal at higher frequencies. Alternatively, if such phenomena occur at a higher energy scale, they might abandon the PTA signal to massive black holes while providing a stochastic gravitational-wave background at higher frequencies.

~~\\
\noindent
{\bf Summary}

These examples provide clear evidence that atom interferometers not only have unique reaches for ultralight bosonic dark matter, but also have interesting capabilities for measuring gravitational waves in the deci-Hz range. They open up the possibility of observing the mergers of intermediate mass black holes, and thereby providing information on the possible assembly mechanisms for supermassive black holes. The prospects for such measurements have been enhanced by PTA observations of a stochastic gravitational wave background that might be due to precursors to the biggest bangs since the Big Bang, or perhaps cosmological physics beyond the Standard Model. Very long baseline atom interferometers could help us discover the answer.

In addition to the fundamental physics that is central to this Workshop,~\footnote{See the summary of the inaugural Workshop~\cite{TVLBAISummary} for some other fundamental physics capabilities of long-baseline atom interferometers.} large-scale atom interferometry also offers unique prospects for use in various geosciences. By achieving an extreme high accuracy using a completely independent measurement principle, it can provide a new absolute gravity reference for geodesy in future. All classical gravimeters can then be compared (and even calibrated) against such a novel gravity standard~\cite{Schilling2020, wodey2019towards}.
This high accuracy would enable a higher sensitivity to local and regional mass changes, e.g., caused by ground water variations or geophysical processes. Everything that perturbs the measurements for a certain application can be the signal of interest for another one~\cite{Beaufils2022}. Thus, networks of such large-scale atom interferometers may also be useful for monitoring geophysical and geodynamic processes in the future, e.g., those associated with climate change~\cite{KCLClimateChange}.

\subsection{{Setting the Scene: Cold Atom Technology}}
\label{Rasel}

\begin{figure}[h]
\centering 
\includegraphics[width=7.5cm]{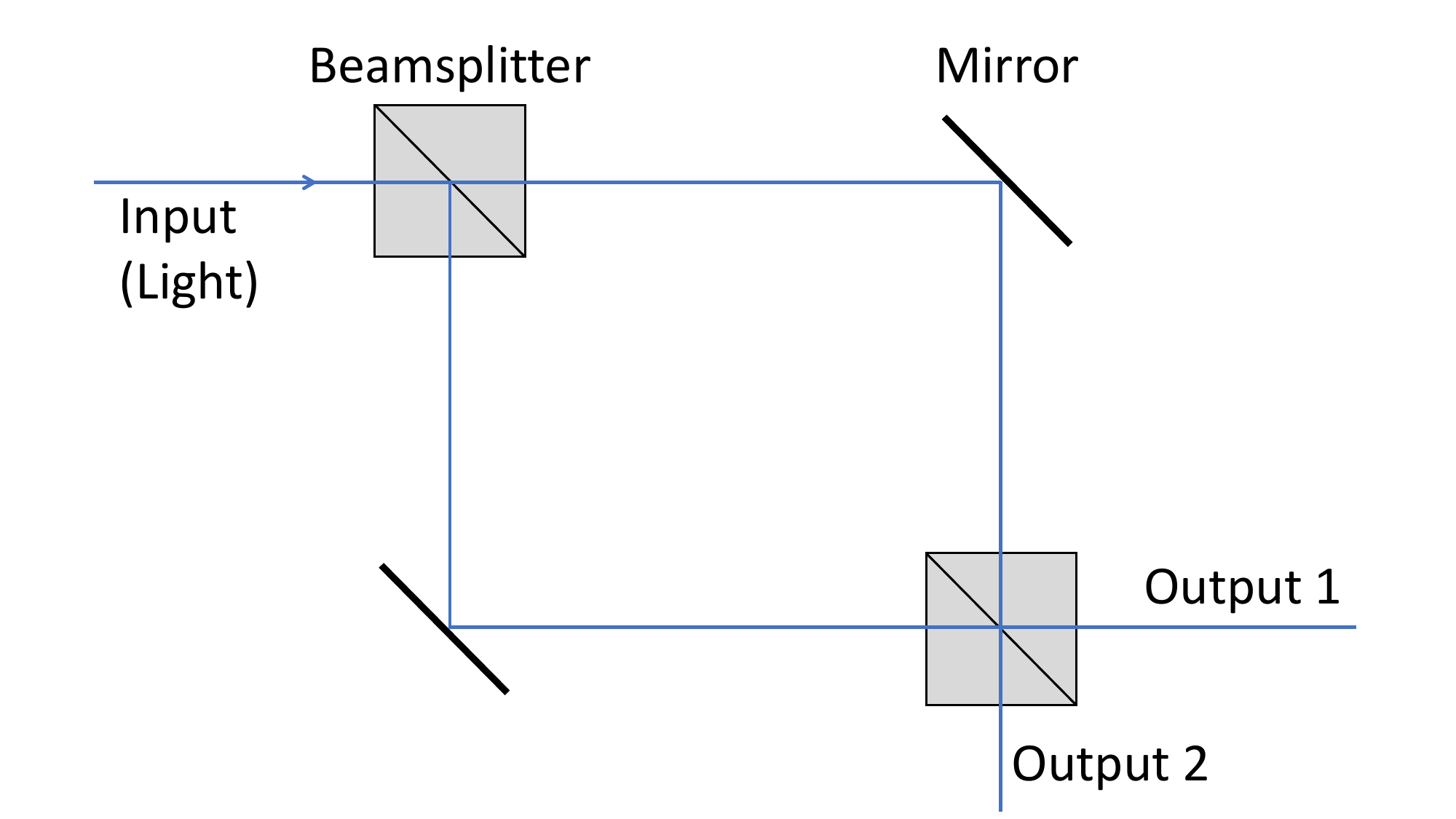}
\includegraphics[width=7.5cm]{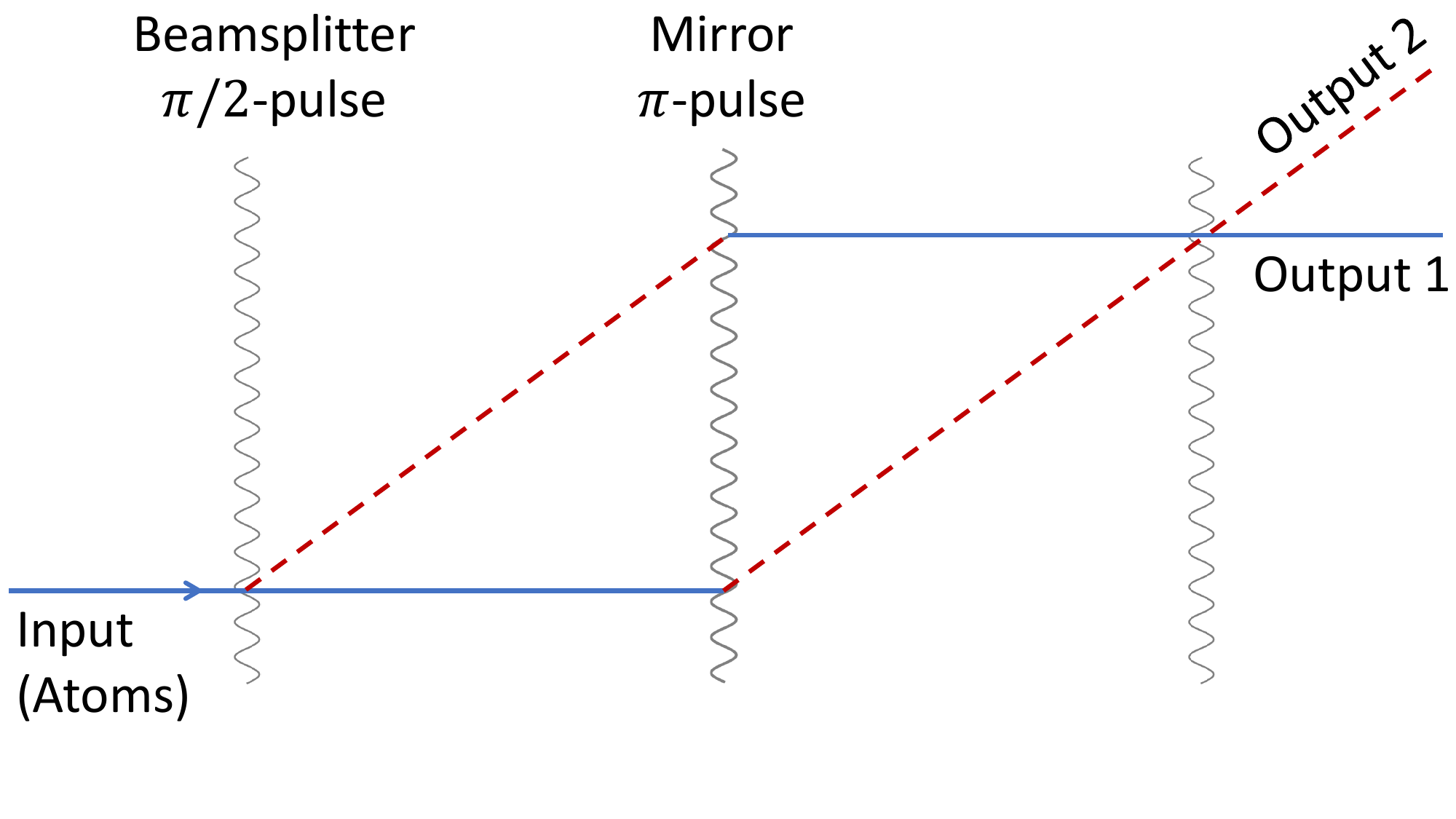}
\caption{ Left: Conceptual outline of a Mach-Zehnder laser interferometer~\cite{zehnder1891neuer,mach1892ueber}.
Right: Conceptual outline of an analogous atom interferometer. Atoms in the ground state, $\ket{g}$, are represented by solid blue lines, the dashed red lines represent atoms in the excited state, $\ket{e}$, and laser pulses are represented by wavy lines. From~\cite{Buchmueller2023}.}
\label{fig:Interferometers}
\end{figure}

Atom interferometers are analogous to Mach-Zehnder optical interferometers, splitting and recombining clouds of cold atoms rather than beams of light. They are capable of detecting extremely small differences in the relative phases of atom clouds that follow different paths, allowing for the measurement of tiny changes in distance, angular velocity, and other physical quantities. Atom interferometers have been made possible by the development of laser technology and quantum optics that enable the beamsplitters and mirrors of optical interferometers to be replaced by atom-light interactions, as illustrated in Fig.~\ref{fig:Interferometers}, taken from~\cite{Buchmueller2023}.

\begin{figure}[h]
\centering 
\includegraphics[width=10cm]{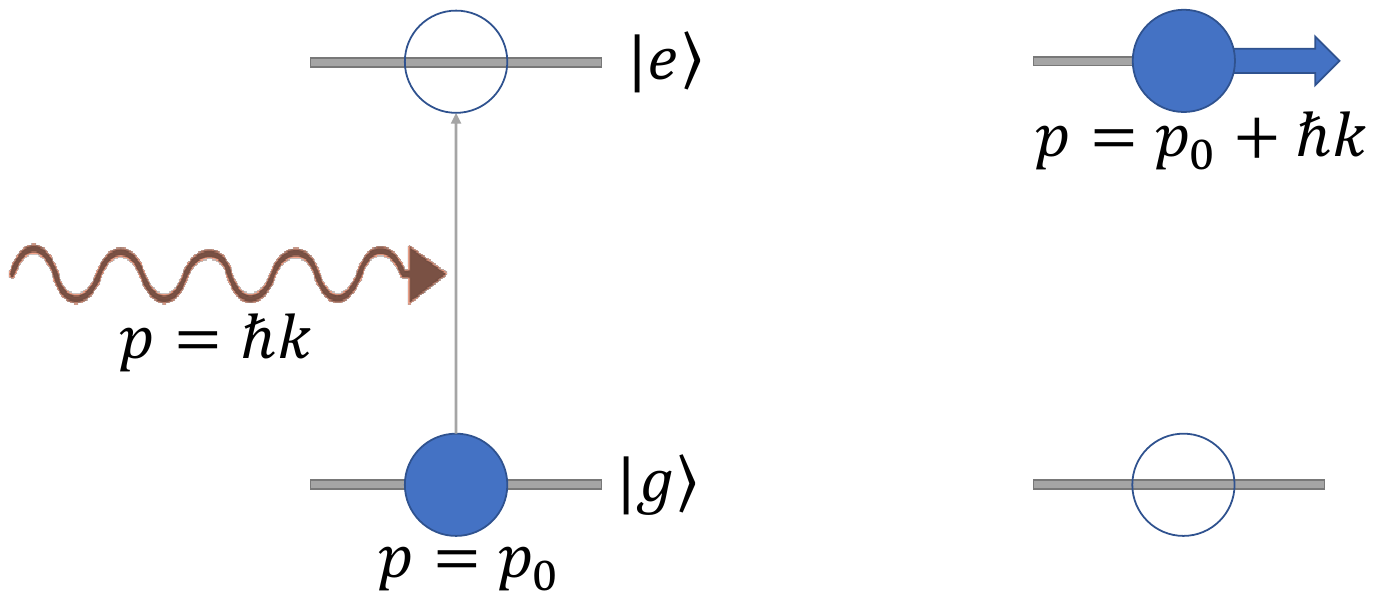}
\vspace{-10mm}
\caption{Every photon carries a momentum $p=\hbar k$ and, upon absorption of the photon, its momentum is transferred to the atom. This forms the basis for manipulating atomic momentum with resonant light. From~\cite{Buchmueller2023}.}
\label{fig:MomentumKick}
\end{figure}

Laser pulses of coherent single-frequency light cause transitions between the atomic ground state and a specific excited state: $|g \rangle \leftrightarrow |e \rangle$, transferring both energy and momentum as illustrated in Fig.~\ref{fig:MomentumKick}, taken from~\cite{Buchmueller2023}. Precise control of the amplitude and duration of the light pulse enables the implementation of a beamsplitter that brings the atom cloud into a superposition of ground and excited states. After this splitting, the two states propagate along different paths and may accumulate different phase shifts due to, for instance, different gravitational fields. After a time $T$, a  second pulse switches the ground and excited states and provides a second momentum kick that acts as a mirror, so that the paths recombine. A final pulse then acts as another beamsplitter before the numbers of cold atoms in the ground and excited states are read out by, e.g., fluorescence imaging.

\begin{figure}
\centering 
\includegraphics[width=12cm]{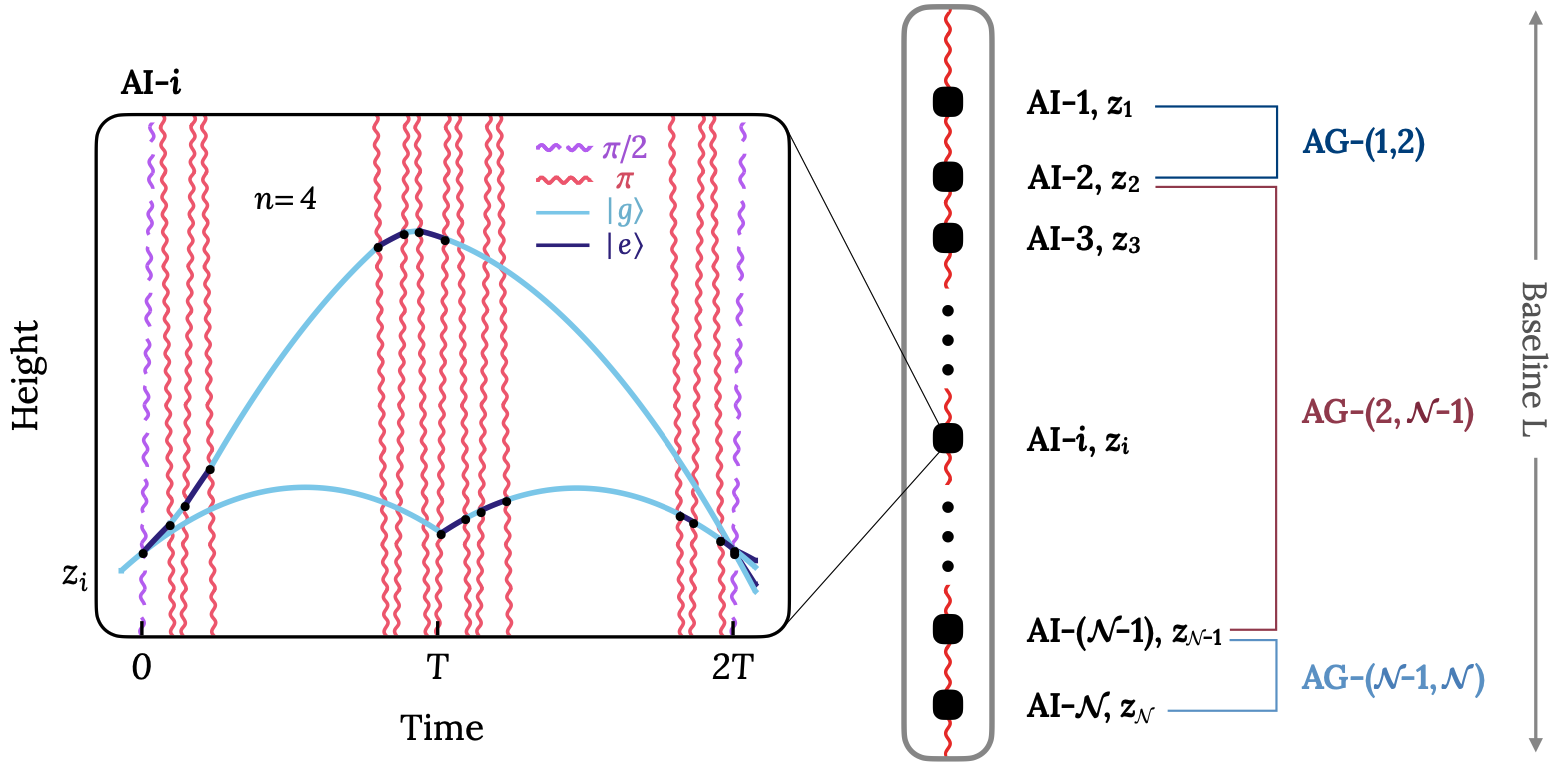}
\vspace{-3mm}
\caption{Schematic representation of a gradiometer with multiple atom interferometers. The left panel illustrates the spacetime diagram of one of the atom interferometers with $n = 4$ LMT kicks. The atoms' excited ($|e\rangle$) and ground ($|g\rangle$) states are shown in dark and light blue, respectively, and $\pi/2-$ and $\pi-$pulses are displayed as wavy purple and red lines. The right panel shows how a series of such atom interferometers may be spaced within a single vertical vacuum tube of length $L$, forming multiple atom gradiometers that combine the atom interferometers $i,j$, labelled as AG$-(i,j)$. From~\cite{Badurina:2022ngn}.}
\label{fig:Multigradiometer}
\end{figure}

The precision with which the final cold atom phase can be read out is limited by shot-noise, which scales $\propto 1/\sqrt{N}$, where $N$ is the number of atoms. Current atom interferometers operate with $\sim 10^6$ atoms per second, and a key area of development in the coming years will be to increase this by orders of magnitude, as discussed in Section~\ref{sources}. Another area of improvement in the coming years will come from squeezing, as discussed in Section~\ref{sec:squeezing}. Another objective will be to achieve large momentum transfers (LMTs) by arranging for the atoms to interact many times with counter-propagating interferometer lasers, as illustrated in the left panel of Fig.~\ref{fig:Multigradiometer}, taken from~\cite{Badurina:2022ngn}, and thereby acquire a large number $n$ of momentum kicks, as discussed in Section~\ref{LMT}. In principle, LMTs can increase the sensitivity of the interferometer $n$-fold and thereby enable high-precision measurements. LMTs with $n \lesssim 400$ have been demonstrated experimentally, and future developments will aim at increasing $n$ further.

The phase noise of the laser used to implement the beamsplitters is a fundamental limitation of atom interferometry. This sensitivity to laser noise is used in atomic clocks to stabilize the laser, enabling high precision for the time-averaged frequency. However, it is a significant limitation for atom interferometers, which search for time-dependent effects such as oscillating dark matter fields or gravitational waves. This limitation can be mitigated by employing gradiometer configurations, where two or more interferometers are interrogated by a common laser, as illustrated in the right panel of Fig.~\ref{fig:Multigradiometer}~\cite{Badurina:2022ngn}. In such a design the laser noise is a common mode for all interferometers and so does not contribute to the differences between the phases measured by the interferometers. The sensitivity scales linearly with the separation, $\Delta r$, between the interferometers. This separation is in practice limited to around the kilometre scale in terrestrial interferometers, which is the longest baseline considered in this Workshop.

\begin{figure}
\centering 
\includegraphics[height=5cm]{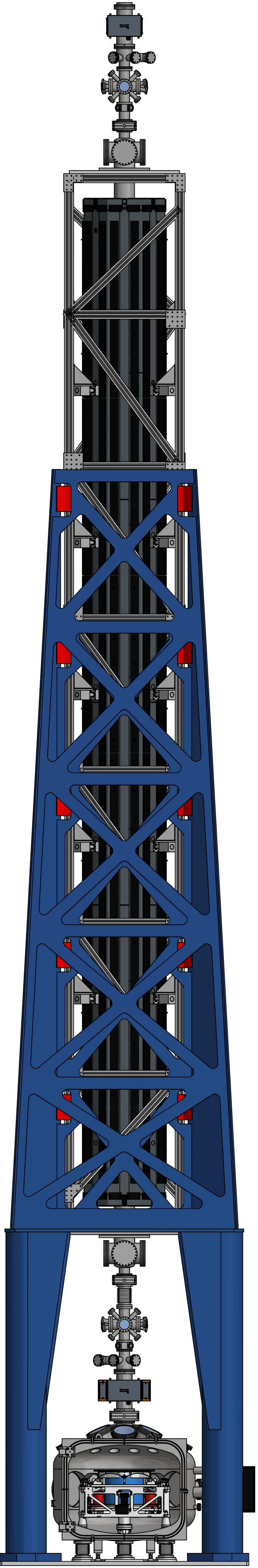}
\hspace{2cm}
\includegraphics[height=4cm]{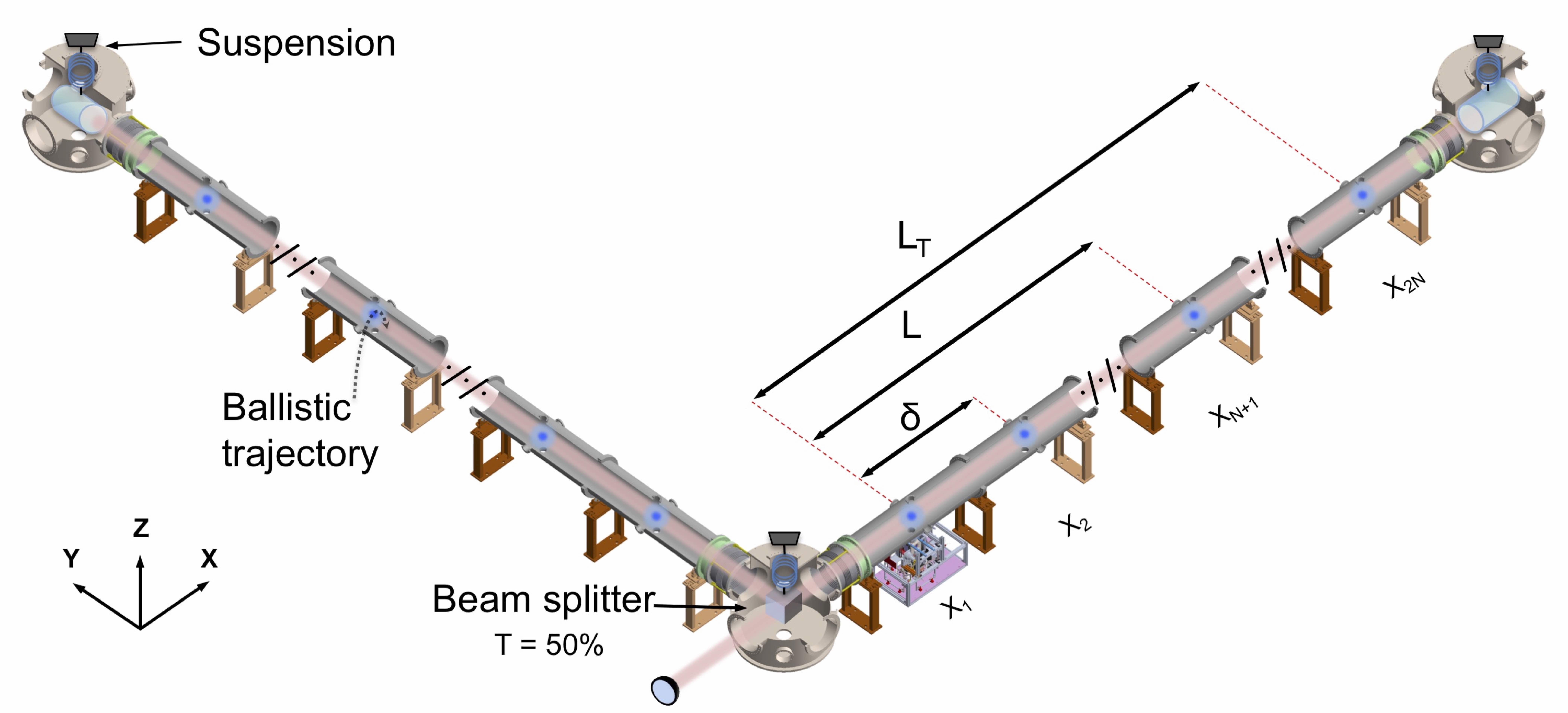}\\
\vspace{-3mm}
\caption{Schematic illustrations of (a) the vertical atom interferometer, VLBAI, located in Hannover (from~\cite{Schilling2020}) and (b) the horizontal geometry of the proposed ELGAR detector (from~\cite{Canuel2020}).}
\label{fig:geometries}
\end{figure}

Broadly speaking, two classes of geometrical design are being considered for long-baseline atom interferometers: vertical and horizontal, as illustrated in Fig.~\ref{fig:geometries}. The vertical design class is illustrated by the VLBAI detector located in Hannover~\cite{schlippert2020matter}, and the horizontal design class by the proposed ELGAR detector~\cite{Canuel_2018}. In vertical designs clouds of cold atoms are launched vertically in a long vacuum pipe where they can be interrogated many times by a vertical laser beam, as illustrated in Fig.~\ref{fig:Multigradiometer}. In horizontal designs a larger number of cold atom sources are launched into horizontal beam pipes where they are interrogated by horizontal laser beams. Several different atomic species are used in the various atom interferometer projects, such as ytterbium (VLBAI), rubidium (VLBAI, MIGA, ZAIGA, ELGAR) and strontium (Stanford, ZAIGA, MAGIS, AION).

In addition to the Hannover VLBAI project, there are vertical atom interferometer projects in the United States (Stanford~\cite{Dickerson2013}, MAGIS~\cite{Abe2021}), in the UK (AION~\cite{Badurina2020}) and in China (ZAIGA~\cite{Zhan2019}). The ZAIGA programme also includes a horizontal detector, and another (MIGA~\cite{Canuel_2018}) is under construction in the the Laboratoire Souterrain {\` a} Bas Bruit (LSBB) in France. Horizontal and vertical detectors are both under consideration for TVLBAI, possibly in combination, operated as a network. We note that, whereas there are coherent plans for developing atom interferometers in the United States (Stanford, MAGIS and potentially SURF) and in China (Wuhan and ZAIGA), long-term plans in Europe are still at the discussion stage. 

\section{Physics}
\label{Physics}

\subsection{Introduction}
\label{PhysicsIntroduction}

The Standard Model (SM) of elementary particles and fields, has been extremely successful in predicting and explaining a plethora of physical phenomena. At the same time, the SM fails to explain extensive observational data on galactic and larger scales based on the observed matter content in the framework of ``normal gravity" and the accelerated rate of the Universe's expansion. It also fails to resolve the hierarchy problem (why the masses of known particles are so much lighter than the fundamental energy scales at which the unification of fundamental forces occurs), the origin of the cosmological matter-antimatter asymmetry, the origin of neutrino masses and other fundamental physics issues. 

The prevailing view is that invisible ``dark matter"  accounts for 84\% of all matter in the Universe~\cite{2022PDG}. 
Despite much effort and advances in detectors, over two decades of direct searches for a very promising candidate, weakly interactive massive particles (WIMPs), 
 have not yielded a discovery~\cite{2024WIMP}. As a result, the past decade has seen unprecedented efforts in dark matter model building at all mass scales as well as the extensive design of numerous new detector types  \cite{2017DM,2023DMQS,Antypas2022}. 
  In particular,  there has been a strong interest in discovering particles that interact very weakly with atomic matter and arise in a number of well-motivated theories \cite{SafBudDem18,Antypas2022}.
Within a broad class of models, dark matter can be composed of bosonic fields associated with ultralight ($\lessapprox 10$~eV) particles that are generally classified  by their spin and intrinsic parity (scalar, pseudoscalar, vector)~ \cite{SafBudDem18,Antypas2022}. 
In this mass range
these particles  are necessarily bosonic and exhibit a large occupation number,
behaving in a ``wave-like’’ manner. Their phenomenology is described by an oscillating classical field.
The coherent oscillations of these ULDM waves would give rise to a diverse range of time-dependent
signals that could be detected using atom interferometers~\cite{Antypas2022,TVLBAISummary}:
\begin{itemize}
\item oscillations of fundamental constants such as the fine-structure constant $\alpha$, and ratio of the electron and proton masses,
\item time-dependent differences in
accelerations between atoms in theories involving vector candidates, and 
\item time-dependent precession of nuclear spins in the case of
pseudoscalar candidates.
\end{itemize}
The search for ULDM in the galactic halo using TVLBAI to probe its energy density near Earth has been discussed in detail in the first workshop and presented in Chapter III.C of \cite{TVLBAISummary}; a summary is given in Section~\ref{sec:setting} with plots of expected limits.  In this workshop, new ideas on using TVLBAI to detect transient ULDM-induced bursts that provide complementary coverage of the parameter space are presented here in Section~\ref{transient}.

  Gravitational waves are another example of physics that interact very weakly with atomic matter but may carry information about fundamental physics targets such as black holes and cosmology. The reaches of the TVLBAI signal targets for astrophysics and new physics detection are summarized in Section~\ref{sec:setting}.
  New physics discovery opportunities for TVLBAI gravitational wave detection were described in the first workshop summary, Chapter III.B \cite{TVLBAISummary}, and gravitational wave opportunities in the deci-Hertz range are discussed here in Section~\ref{grav}. 
  
TVLBAI enables many other new physics searches. Using atom interferometers as freely falling clocks for time-dilation measurements is introduced in Section~\ref{TD}. Proposals for using atom interferometers for tests of quantum mechanics and of atom neutrality were described in the first workshop summary in Chapters III.D and III.F \cite{TVLBAISummary}, respectively.

\subsection{Gravitational wave opportunities in the deci-Hertz range}
\label{grav}

Gravitational waves (GWs) are generated through the acceleration of massive objects. In General Relativity they are transverse waves propagating at the speed of light with two polarizations, typically denoted $+$ and $\times$ to denote their different quadrupolar effects on space-time.  The typical amplitude of a gravitational-wave signal induces a strain of $h \lesssim 10^{-20}$, requiring an instrument capable of measuring a fractional length fluctuation on that scale to observe a signal.  
Gravitational waves from binary mergers were first observed in 2015 \cite{LIGOScientific:2016aoc} by the LIGO Collaboration and, to date, observations of close to 100 mergers from binaries composed of black holes and neutron stars \cite{KAGRA:2021vkt} have been published by the LIGO-Virgo-KAGRA (LVK) network.  These observations have occurred at frequencies from tens of Hz to kilo-Hz.  More recently, groups in the International Pulsar Timing Array have announced the observation of a stochastic background in the nano-Hertz band \cite{Agazie2023, Antoniadis2023a}, which is consistent with the signal expected from a population of supermassive black hole binary mergers \cite{Agazie2023b}.

\begin{figure*}
\centering
\includegraphics[width=0.8\textwidth]{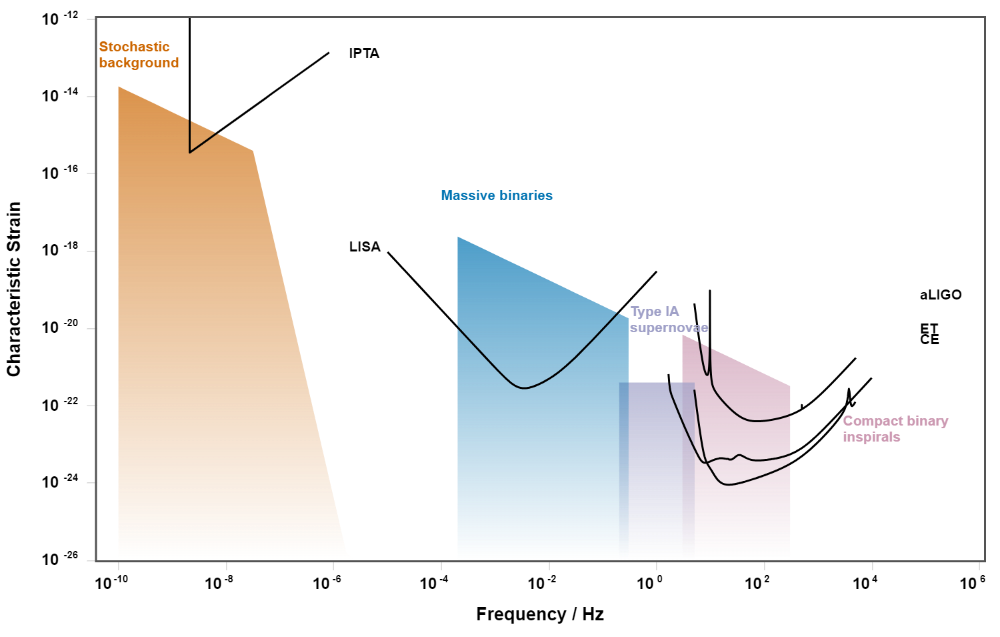}
\caption{Sensitivity of existing and planned GW observatories, and potential GW sources.  The y-axis shows the ``characteristic strain'' which is normalized so that the relative height of the signal above the detector sensitivity provides an estimate of the contribution to the signal to noise ratio. The figure shows a clear gap in instrumental sensitivity in the deci-Hertz range, from $0.1-1\mathrm{Hz}$,  where the existing detectors do not have sensitivity. Figure generated by S.~Fairhurst using {\tt GWPlotter}~\cite{Moore:2014lga}.}
\label{fig:gw_sensitivity}
\end{figure*}

In the coming decade,  ground-based interferometers are expected to increase in sensitivity~\cite{LVKRunPlans}, enabling the observation of binary mergers out to redshifts $z$ of a few, which corresponds to the era of peak star formation.  The proposed terrestrial GW observatories Cosmic Explorer \cite{Evans:2023euw} and Einstein Telescope \cite{Maggiore:2019uih}, will provide sensitivity to binary mergers throughout the universe, at frequencies down to around $5 \, \mathrm{Hz}$.  The LISA mission is scheduled to fly in the mid-2030s \cite{LISA:2017pwj, Colpi:2024xhw}, and will have sensitivities in the milli-Hertz range, enabling the observation of individual supermassive black hole binaries.  The sensitivities of current and future observatories are summarized in Fig.~\ref{fig:gw_sensitivity}, with several GW sources indicated.

Gravitational waves in the deci-Hz range provide an excellent science target for TVLBAI.  As is clear from Fig.~\ref{fig:gw_sensitivity}, other observatories do not have sensitivity in the $0.1-1 \, \mathrm{Hz}$ range.  The potential sensitivity of future observatories is shown for compact binary mergers in Fig.~\ref{fig:GWgap} and for a stochastic background of gravitational waves in Fig.~\ref{fig:MMA}. 
There are several interesting sources of gravitational waves in the deci-Hertz band that will provide unique information that is complementary to the observations from ground-based laser interferometers at higher frequencies and LISA at lower frequencies.  More details of the synergies with other observatories are discussed in Section \ref{sec:synergies}.

\begin{figure*}
\centering
\includegraphics[width=\textwidth]{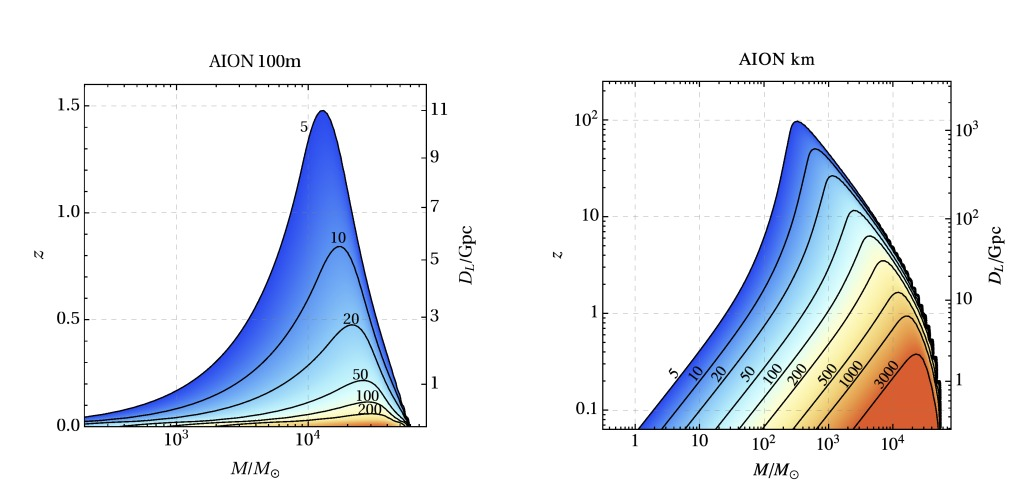}
\caption{Sensitivities of proposed TVLBAI observatories (AION 100m and AION km) to intermediate mass black hole binaries.  The figures show the distance at which the observatory would observe the GW signal from a merging black hole binary of a given total mass as a function of the observed signal to noise ratio.  The sensitivities are generated assuming optimistic noise curves with fully subtracted gravity gradient noise. From~\cite{Badurina2020}.  
}
\label{fig:aion_bbh}
\end{figure*}

The inspiral and merger of intermediate mass black hole binaries is a leading GW source for TVLBAI.  Fig.~\ref{fig:aion_bbh} shows the sensitivity of the proposed AION 100m and km detectors.  Assuming that a signal to noise ratio of 10 is required for confident detection (as is the case for the LVK observations \cite{KAGRA:2021vkt}), then 100~m scale atom interferometers have the potential to observe intermediate mass black hole mergers around $10^{4} M_{\odot}$ in the relatively nearby Universe, while km-scale observatories will have sensitivity to $z\sim 10$, and will enable high-SNR precision measurements for nearby sources~\cite{Torres-Orjuela:2023dle}.  Any observations, particularly in the $10^{3}-10^{4} M_{\odot}$ mass range, will be complementary to LVK or LISA observations which have limited sensitivity to mergers of this mass~\cite{Valiante:2020zhj} (see Fig.~\ref{fig:lisa_et_bbh}).  In addition, even where observations overlap with other detectors, TVLBAI observations will provide complementary information.  For example, they will make measurements at lower frequencies, and hence during an earlier stage of the binary's evolution. Moreover, some formation scenarios lead to binaries formed in eccentric orbits that, through gravitational wave emission, gradually circularize \cite{Peters:1963ux}.  This makes it challenging to observe eccentricity in the LVK frequency range, but observations in the deci-Hertz range, where the eccentricity is larger, provide unique capabilities to differentiate formation models \cite{ArcaSedda2020}.

Type Ia supernovae occur either when a white dwarf accretes matter from a main-sequence companion (single degenerate progenitor) or when two white dwarfs merge (double degenerate progenitor)~\cite{Livio2000, Mandel:2017pzd}.  Observation, or lack of, a gravitational wave signal in the deci-Hertz band provides a unique opportunity to distinguish between these scenarios.  For a double white dwarf system, the gravitational wave emission would continue until $\sim 0.1 \, \mathrm{Hz}$ before the objects merge. However, a white dwarf--main sequence binary would merge at a much lower frequency, due to the size of the main sequence star.  Thus, observation of a deci-Hz GW signal associated with a Type Ia supernova would therefore provide strong evidence of a double degenerate progenitor \cite{Mandel:2017pzd}.  The signals would, ideally, be associated both temporally and spatially and this requires good localization of the GW event.  These events will persist for over a year in the deci-Hz band, and can therefore be localized accurately thanks to the motion of the Earth around the Sun during the observation\cite{Graham2018a}.

\subsection{Transient Targets for Dark Matter Searches using Quantum Sensors}
\label{transient}
ULDM is a promising target for TVLBAI, as outlined in Section~\ref{sec:setting}. In addition to traditional searches for ULDM that probe its energy density near the Earth, searches for transient ULDM-induced bursts are also a viable method that provides complementary coverage of the parameter space.

One widely-studied example is the relativistic burst of scalars emitted in the collapse of a boson star~\cite{Kaup:1968zz,Ruffini:1969qy}, a process called a \emph{bosenova}~\cite{Eby:2016cnq,Levkov:2016rkk}. During collapse, scalars in the non-relativistic boson star infall toward the core, increasing their energy until it approaches their mass. At this point, number-changing annihilation processes~\cite{Eby:2015hyx} of the infalling scalars lead to a rapid conversion of a large fraction (generally between $10-50\%$) of the initial mass of the boson star to more energetic scalars that escape the collapsing star~\cite{Levkov:2016rkk}. 

A search for these relativistic bursts can be characterized by the properties of the emitted scalars, which can be derived from numerical simulations of the boson star collapse~\cite{Levkov:2016rkk}.
Their typical energy, arising from e.g. $3\to 1$ or $4\to 2$ annihilation processes, is of order a few times their mass. As a result, an experiment searching for oscillations of frequency $m_\phi$ may, without modification, be sensitive to this modest shift in frequency, if the amplitude is large enough.
Further, in contrast to cold dark matter (DM), the dispersion in momentum of the emitted scalars is large, implying that a broadband search strategy should be employed. 

In general, the burst propagates an astrophysical distance $\mathcal{R}$ to the Earth before encountering and depositing energy in the detector.
Owing to the large mass density of boson stars being converted rapidly into the burst, even after propagating a distance $\mathcal{R} \sim {\rm pc - kpc}$, the burst density $\rho_\star$ can exceed the local DM density $\rho_{\rm local} \simeq 0.4\,{\rm GeV/cm}^3$ by many orders of magnitude~\cite{Eby:2021ece}. This can be estimated simply as
\begin{equation} \label{eq:transientrhoburst}
    \rho_\star \simeq \frac{\mathcal{E}}{4\pi \mathcal{R}^2 \delta x}\,,
\end{equation}
where $\mathcal{E}$ is the total emitted energy in the bosenova (which will be of order the mass energy of the progenitor boson star) and $\delta x$ is the spatial extent of the burst.

The semi-relativistic nature of the emitted scalars implies that the wave spreads significantly in flight, leading to several important effects~\cite{Eby:2021ece}. First, clearly the spreading dilutes the energy density as $\sim \mathcal{R}^{-2}$, as in Eq.~\eqref{eq:transientrhoburst}. Secondly, for bursts occurring outside the solar system, the length of the burst $\delta x$ will quickly be dominated by the wave spreading rather than the intrinsic duration of the burst at the source, leading to a length that is of order $\mathcal{R}$. Thirdly, the propagation of fast momentum modes away from slow momentum modes in the burst will reduce the amount of destructive interference of the waves by the time they reach the detector, leading to a longer effective timescale of coherent oscillations which is of order $\tau_\star \sim 10^{-2}\mathcal{R}/c$ (see~\cite{Eby:2021ece} for details). 

In summary, one can characterize the sensitivity of an experiment by the ratio of the minimal sensitive coupling in a traditional DM search ($g_{\rm dm}$) to the minimal sensitive coupling in a transient bosenova search ($g_\star$)~\cite{Eby:2021ece}:
\begin{equation} \label{eq:transientsensitivity}
    \frac{g_\star}{g_{\rm dm}} \simeq 
        \left(\frac{\rho_{\rm local}}{\rho_\star}
        \right)^n
        \frac{t_{\rm int}^{1/4}{\rm min}(t_{\rm int}^{1/4},\tau_{\rm dm}^{1/4})}
        {{\rm min}[t_{\rm int}^{1/4},(\delta x/c)^{1/4}]
        {\rm min}[t_{\rm int}^{1/4},\tau_{\star}^{1/4}]}\,,
\end{equation}
where $\tau_{\rm dm}$ is the coherent oscillation timescale for cold DM, $t_{\rm int}$ is the integration time in the experiment, and $n=1/2$ ($n=1$) for ULDM with a linear (quadratic) coupling to the Standard Model. When $g_\star/g_{\rm dm}<1$, a transient search appears to be more sensitive to the coupling $g$ than the corresponding traditional search for cold DM.

The sensitivity of current and future quantum sensing searches to a bosenova arising in various particle physics models was characterized  in~\cite{Arakawa:2023gyq,Arakawa:2024lqr} using Eq.~\eqref{eq:transientsensitivity}. For bursts occurring within pc (kpc) distances, searches for bosonovae were found to be highly sensitive, exceeding the sensitivity of traditional DM searches in the mass range $10^{-21} \lesssim m_\phi\,c^2/{\rm eV} \lesssim 10^{-4}$ ($10^{-21} \lesssim m_\phi\,c^2/{\rm eV} \lesssim 10^{-10}$). Thus a TVLBAI is well-suited for the task of searching for, and discovering, transient signals from bosenovae that occur within our galaxy. Additional work is required to characterize the expected rate of bosenovae, which would allow for constraints to be set also in the absence of a signal; see~\cite{Maseizik:2024qly,Gorghetto:2024vnp,Chang:2024fol} for important recent literature on this subject.

\subsection{Atom interferometers as freely falling clocks for time-dilation measurements}
\label{TD}

As explained above, the two main objectives of TVLBAI facilities
such as MAGIS-100 \cite{Abe2021} and AION \cite{Badurina2020} serve as prototypes for future gravitational antennae in the mid-frequency band and searching for ULDM fields. However, their planned sensitivities may be insufficient for the actual detection of gravitational waves, and the outcome of the ULDM search may simply be a moderate improvement of the bounds on the couplings to the Standard Model fields, especially at the early stages.
It is therefore worth investigating additional applications of such facilities to fundamental physics that can lead to non-vanishing measurements rather than mere null tests.

A possible application of this kind is the local measurement of relativistic time-dilation effects with freely-falling atoms, beyond the reach for state-of-the-art atomic-fountain clocks. In comparison, experiments in TVLBAI facilities can benefit from the much higher energy difference $\Delta E$ for an optical transition between the two clock states (instead of the hyperfine transition in alkali atoms), and from the longer baseline. Nevertheless, conceptual and practical challenges associated with the much larger photon recoil need to be addressed. Following Ref.~\cite{Roura2025}, one can show that a Mach-Zehnder interferometer with laser pulses driving a single-photon transition between the two clock states can act as a freely-falling clock (involving a quantum superposition of the two internal states) with an intermediate state inversion.
As a result of this inversion, the clock runs backwards the second half of the time, leading to a vanishing relative phase between the two clock states in the case of identical times before and after the inversion pulse.
When considering equal durations with respect to a time reference in the laboratory frame, time-dilation effects (due to special relativity and the gravitational redshift) imply a slight imbalance in the proper times for such a freely-falling clock.
Measuring this time dilation requires a suitable measurement scheme that cancels the larger contribution from retardation effects
due to the finite speed of light and the motion of the atoms, which we will refer to as \emph{Doppler effect}. The essential idea can be easily understood in a freely-falling frame, as outlined in the following paragraph.

In the freely-falling frame, vertically propagating light rays follow straight lines with fixed slope in a spacetime diagram. Both time dilation and the Doppler effect simply shift the light rays, with a different shift for each light ray, but keeping the slope unchanged; see Fig.~3 in Ref.~\cite{Roura2025}.
Interestingly, while the shifts caused by time dilation are the same for upward- and downward-propagating light rays, those due to the Doppler effect have opposite sign for upward and downward propagation.
This difference can be exploited to suppress 
the contributions from retardation effects.
Indeed, the shifts due to the Doppler effect can 
be compensated through a suitable frequency chirp of the static source (in the laboratory frame) that emits the laser pulses. In practice, however, one cannot match the gravitational acceleration $\mathbf{g}$ and the initial velocity $\mathbf{v}_0$ of the freely-falling atoms exactly. The remaining contributions due to an imperfect matching can be suppressed by adding up the phase shifts measured for a pair of reversed interferometers involving laser pulses propagating in opposite directions.

As shown in Ref.~\cite{Roura2025}, the phase shift for a Mach-Zehnder interferometer with laser pulses driving the single-photon transition between the two clock states is given by
\begin{equation}
\delta\phi = - 2 \,(\Delta E / \hbar) \left( \bar{\mathbf{v}}_0 \cdot \mathbf{g}\, T^2 + g^2 T^3  \right) / c^2
+ \delta\phi_\text{corr} \,
\label{eq:phase_shift} .
\end{equation}
This coincides with the result for an ideal freely-falling clock with an intermediate inversion pulse following the mid-point trajectory between the two interferometer arms, except for the corrections $\delta\phi_\text{corr}$ that arise due to an imperfect matching of the frequency chirp. These can be suppressed effectively by considering a pair of reversed interferometers, as described above.
Moreover, adding up the phase shift for the pair of reversed interferometers has the added benefit of suppressing the systematic effects associated with light shifts and laser wave-front curvature as well as gravity gradients and rotations.

\begin{figure}[h]
\begin{center}
\includegraphics[width=4.0cm, height=7.0cm]{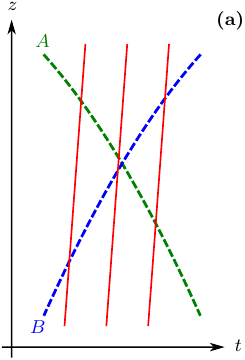}
\hspace{8.0ex}
\includegraphics[width=4.0cm, height=7.0cm]{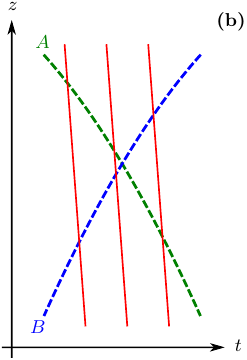}
\end{center}
\caption{Spacetime diagrams in the laboratory frame that depict non-standard gradiometric configurations involving a pair of atom interferometers operated simultaneously with different initial velocities. The two atom clouds are independently launched from the top ($A$) and bottom ($B$) sources and interrogated by three common laser pulses consisting each of two slightly different frequencies so that both interferometers can be resonantly addressed. The mid-point trajectories of the two interferometers are shown as dashed lines, and the central wave fronts of the laser pulses are shown as continuous red lines. The pair of atom interferometers are interrogated by upward propagating pulses in (a), whereas a pair of reversed interferometers are alternatively interrogated by downward propagating ones in (b). From \cite{Roura2025}.}
\label{fig:gradiometric_conf}
\end{figure}

On the other hand, in order to overcome the impact of laser phase noise and vibration noise of the retro-reflection mirror, which would otherwise overwhelm the interferometric signal for sufficiently long times and sensitive interferometers, one needs to consider a suitable differential measurement that does not suppress the signal of interest. This can be achieved with a non-standard gradiometric configuration, depicted in Fig.~\ref{fig:gradiometric_conf}, where a pair of atom interferometers launched independently from two atom sources, located at the top ($A$) and at the bottom ($B$) of the long baseline, are interrogated by common laser pulses. The key aspect is the different velocities of the two simultaneously-interrogated interferometers (requiring two slightly different frequency components in each laser pulse in order to resonantly address both interferometers).
The resulting differential phase shift is then given by
\begin{equation}
\delta\phi_A - \delta\phi_B
= - 2\, (\Delta E / \hbar) \left( \bar{\mathbf{v}}_0^A - \bar{\mathbf{v}}_0^B \right) \cdot \mathbf{g}\, T^2 / c^2 \,
\label{eq:gradiometric} ,
\end{equation}
and can be interpreted as a comparison of the relativistic time-dilation effects
for two freely-falling clocks with different initial velocities and where a precise time reference in the laboratory frame no longer plays an important role.

The experimental implementation with MAGIS-100 should be relatively straightforward, with essentially no requirements in addition to those already planned. For initial velocities $\bar{\mathbf{v}}_0^A = (-20\, \text{m/s})\, \hat{\mathbf{z}}$ and $\bar{\mathbf{v}}_0^B = (40\, \text{m/s})\, \hat{\mathbf{z}}$, a total interferometer time $2T = 2\, \text{s}$ and $N=10^5$ detected atoms, a shot-noise-limited fractional sensitivity of $10^{-5}$ can be reached with just a hundred shots.
Moreover, as discussed in ref.~\cite{Roura2025}, bringing the main systematic effects (including pulse timings, magnetic fields, rotations and gravity gradients) down to that level is feasible with capabilities that have already been demonstrated.
The main challenge will come from the systematic effects associated with temperature gradients \cite{Roura2025}. For example, a temperature gradient of $2\, \text{K/m}$ (a typical value for this kind of facility, see \cite{Arduini2023}) implies systematic corrections at the $10^{-2}$ level. By employing an array of temperature sensors along the baseline, it should be possible to post-correct such systematic effects and reduce the associated uncertainties by at least two orders of magnitude.
It should also be stressed that in addition to the 100-m facilities, preliminary measurements with limited sensitivity will be possible in the smaller-scale prototypes involving 10-m atomic fountains that will become available very soon.

In conclusion, the proposed scheme will enable the local measurement of gravitational time dilation with freely-falling atoms, beyond the reach of state-of-the-art atomic fountain clocks based on microwave transitions, and will offer an excellent opportunity to test and validate atom interferometry experiments spanning baselines up to 100m with guaranteed non-vanishing measurements of relativistic effects.
This measurement scheme differs conceptually from recent proposals to measure gravitational time-dilation effects \cite{Roura2020} in quantum-clock interferometry \cite{Sinha2011,Zych2011,Loriani2019} (or related variants \cite{Roura2021,Ufrecht2020a,DiPumpo2021}). In those, a single clock is prepared in a delocalized quantum superposition of two wave packets at different heights and experiencing a different gravitational redshift, which is then reflected on the interference signal when they are eventually recombined.
In contrast, the interferometric scheme displayed in Fig.~\ref{fig:gradiometric_conf} corresponds to comparing two independent clocks, and rather than being proportional to the arm separation for a single interferometer, the signal depends on the total baseline available, which can be much longer in TVLBAI facilities.

It should also be emphasized that, contrary to the null test involving a differential measurement of two different isotopes described in Ref.~\cite{DiPumpo2023}, where a non-vanishing result would only be obtained in case of violations of the equivalence principle, the interferometric scheme presented here gives a non-trivial result even in the absence of such violations.
This is particularly relevant because to leading order the energy difference for these optical clock transitions is independent of the nuclear mass, and the effect of any violations would be rather suppressed when comparing two isotopes. Moreover, since the comparison involves at least one bosonic isotope, for which the clock transition is forbidden unless a strong magnetic field is applied, an implementation based on single-photon transitions is in practice not viable in a TVLBAI facility.


\section{Synergies of Cold Atom and Laser Interferometry GW Experiments}
\label{sec:synergies}

\subsection{Introduction}
\label{SynergiesIntroduction}

Following the discovery of GWs in the frequency range of 10 to 100~Hz by the LIGO and Virgo experiments, several pulsar timing array (PTA) collaborations have presented evidence for a stochastic background of GWs with frequencies in the nano-Hz range~\cite{NANOGrav:2023hde,EPTA:2023xxk,Zic:2023gta,Xu:2023wog}. The space-borne laser interferometer experiment LISA~\cite{AmaroSeoane2017} has recently been approved by ESA, and there are also proposals in China (Taiji~\cite{Ruan2020} and TianQin~\cite{TianQin:2015yph}) for space-borne laser interferometer experiments targetting similar frequency ranges $\sim 10^{-4}$ to $10^{-2}$~Hz. As discussed during this Workshop, there are interesting opportunities for GW observations in the deci-Hz frequency range intermediate between the ranges where ground- and space-based laser interferometers are optimised: see Figs.~\ref{fig:GWgap} and \ref{fig:MMA}.

Decades of experience with electromagnetic detectors have demonstrated the valuable synergies provided by multi-wavelength observations of astrophysical sources. This Session is devoted to the prospective synergies between GW observations in different frequency ranges. These include synergies between deci-Hz observations and observations in adjacent frequency bands, e.g., the higher frequencies being explored by the LIGO, Virgo and KAGRA (LVK) laser experiments -- and possibly the Einstein Telescope~\cite{Sathyaprakash:2012jk,Maggiore:2019uih} and/or Cosmic Explorer~\cite{Reitze2019,Evans:2023euw} experiments in the future -- and the lower frequencies to be explored by space-borne laser interferometers such as LISA. There are also interesting prospects for deci-Hz observations of the mergers of intermediate-mass black holes that may have synergies with the interpretation of the GW signals reported by the PTA collaborations.

This Session comprises three presentations: one on the status and prospects for science with the LVK experiments, another on the prospects for LISA observations, and a third discussing how long-baseline atom interferometer measurements in the deci-Hz range may be motivated by, and related to, the PTA observations.

\subsection{LIGO, Virgo and KAGRA science and synergies}
\label{LVK}

The ground-based laser interferometric gravitational wave observatories, LIGO, Virgo and KAGRA (LVK) have completed three observing runs \cite{KAGRA:2021vkt} operating as a global network, during which close to one hundred gravitational wave signals from binary mergers have been observed.  Indeed, in less than ten years from the first observation \cite{LIGOScientific:2016aoc}, the current rate of observations has increased to several events per week. To date, the progenitors of all events have been the mergers of two compact objects, either black holes or neutron stars. The vast majority of events have been emitted by binary black holes whose orbits shrink with the emission of gravitational waves until the system finally merges into a single black hole.  In addition a handful of neutron star - black hole and binary neutron star mergers have been detected, most notably the first observation of a binary neutron star merger (GW170817) \cite{LIGOScientific:2017vwq} which was observed also as a Gamma Ray Burst and an electromagnetic transient across a broad range of frequencies \cite{LIGOScientific:2017ync}.

The range of black hole masses observed by the LVK network is determined by the frequency sensitivity of the detectors, with the observing band extending from $\approx 10$Hz to few kHz. The range of black hole mergers to which the detectors are sensitive scales with the frequency as
\begin{equation}\label{eq:f_merge}
    f \approx 100\mathrm{Hz} \frac{100 M_{\odot}}{(1 + z) M}
\end{equation}
where $M$ is the total mass of the binary, in units of solar masses, and $z$ is the redshift of the source.  Thus, binaries comprised of low-mass black holes or neutron stars sweep across the LVK band while mergers of massive stellar origin black holes (with individual masses $\gtrsim 30 M_{\odot}$) merge shortly after entering the LVK band.

LVK observations are beginning to reveal the properties of the black hole population in the nearby universe \cite{KAGRA:2021duu}.  This population shows a preference for approximately equal mass binaries, with the emergence of structure in the mass distribution.  In particular, there are peaks in the primary mass distribution at around $10 M_{\odot}$ and $30 M_{\odot}$. There is an expectation of a mass gap from $70 - 120 M_{\odot}$ where pair-instability supernova would lead to the explosion of massive stars and no black hole remnant. Some events, such as GW190521 \cite{LIGOScientific:2020iuh}, have masses that appear to lie in the putative pair-instability mass gap, although black holes lying in this gap can be formed in other ways such as through hierarchical mergers.

The proposed next generation of terrestrial gravitational wave observatories, the Einstein Telescope (ET) \cite{Sathyaprakash:2012jk,Maggiore:2019uih} and Cosmic Explorer (CE) \cite{Reitze2019,Evans:2023euw}, will be approximately an order of magnitude more sensitive than the current detectors across a broad frequency band. This will enable observations of binary mergers throughout the universe, back to the era of the formation of the first generation of stars. In addition, these detectors will push to lower frequency, with sensitivities down to $5\mathrm{Hz}$ or even $3\mathrm{Hz}$. This will enable CE and ET to observe a broad range of binary mergers, including black hole binaries with masses as large as $\sim 10^4 M_{\odot}$, and enable the observation of heavy stellar origin black hole mergers $\sim 100 M_{\odot}$ at redshift of $z\approx20$, which corresponds to the era of first star formation.  For these early universe systems, the signal is redshifted to lower frequencies as it travels to the Earth, which leads to the $(1 + z)$ factor in the denominator of Equation (\ref{eq:f_merge}).  For even higher-mass or higher-redshift systems, the gravitational wave emission will lie outside the sensitive band of CE and ET.

\begin{figure}
\centering
  \includegraphics[width=10cm]{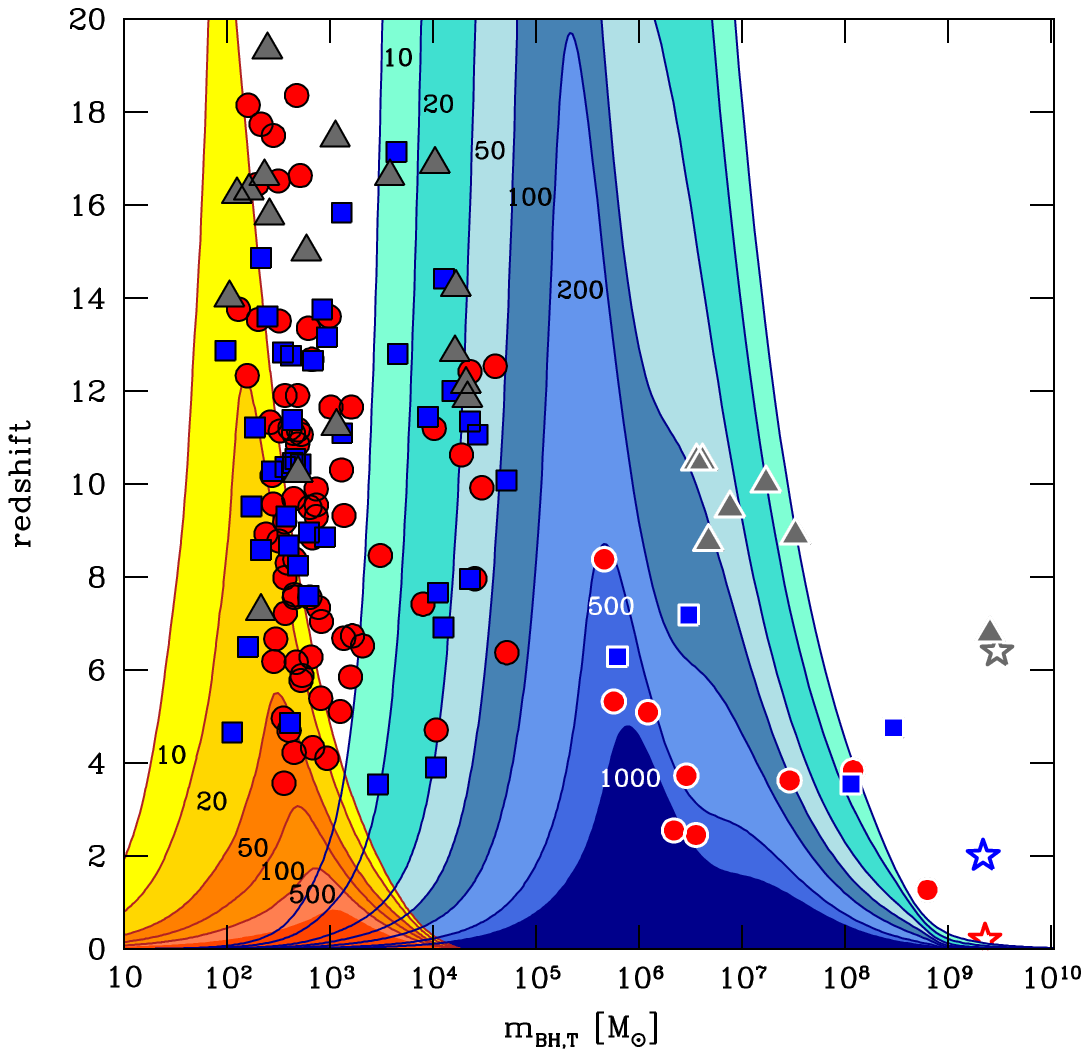}
  \caption{Model predictions for the distribution of binary black hole (BBH) coalescence events in the redshift $z$--$m_{\rm BH,T}$ diagram~\cite{Valiante:2020zhj}, where $m_{\rm BH,T}$ is the total BBH mass. The data points describe cosmologically-driven BH mergers. Grey triangles, blue squares and red circles denote the total masses and redshifts of the coalescences from a simulation forming a $\sim 10^{9} M_{\odot}$ SMBH at $z_{\rm QSO}= 6.4$, $2$ and $0.2$ (represented with stars in the plot). Symbols with white edges indicate mergers involving at least one heavy seed. Color-coded areas represent lines of constant signal-to-noise ratios for ET (yellow/red) and LISA (azure/blue) computed for non-spinning binaries assuming a mass ratio $q=0.5$, which corresponds to the mean value of the merging binaries extracted from our samples. Taken from~\cite{Valiante:2020zhj}.}
  \label{fig:lisa_et_bbh}
\end{figure}

A major question in astronomy is how supermassive black holes (SMBHs), with masses as high as $10^{9} M_{\odot}$ formed by redshift $z \sim 7$, less than 1 billion years after the big bang \cite{Valiante:2020zhj}.  Proposed models include both \textit{light seed} and \textit{heavy seed} black holes, or a combination of the two.  Light seed black holes are expected to form from the collapse of population III stars with a mass of several $100 M_{\odot}$, whereas heavy seeds of mass $\gtrsim 10^{4} M_{\odot}$ form through the direct collapses of gas clouds.  It is expected that through a series of mergers and accretion, these seed black holes will evolve into the supermassive black holes that are observed. Gravitational wave observations provide an ideal venue for unveiling the formation of these SMBHs, as shown in Fig.~\ref{fig:lisa_et_bbh}. The mergers of heavy seeds and the final mergers of SMBHs will be observable by LISA. However, the initial mergers of light seed black holes, with masses $\approx 100 M_{\odot}$ in the early universe will only be visible to the next-generation GW observatories, CE and ET.  Nonetheless, there is a range of black hole mergers which will not be observable by either next-generation ground-based laser interferometers or by LISA.  These are clearly identifiable in Fig.~\ref{fig:lisa_et_bbh} at redshifts above $z=4$ and masses between a few hundred and a few thousand solar masses.  Observation of these signals is vital to ensure that the \textit{light seed} black holes really do evolve all the way from $\sim 100 M_{\odot}$ remnants of the first stars into $\sim 10^{9} M_{\odot}$ super massive black holes. Future atom interferometers with sensitivities in the deci-Hz range provide a unique capability to probe these mergers and therefore complete the picture of light seed evolution to supermassive black hole.

As an additional synergy, we briefly discuss a case where a \textit{non-detection} of a signal by atom interferometers may nonetheless provide significant astrophysical insight.  In \cite{Fairhurst:2023beb}, a detailed investigation of the observability of high-mass stellar BHs at high redshift was performed.  In many cases, only the final few cycles of the gravitational waveform from these events, corresponding to the end of the inspiral then merger, will be observable in CE and ET.  This makes it challenging to extract accurately the parameters of the signal.  One particular challenge is to accurately infer the distance to the signal, as this is largely degenerate with the binary orientation \cite{Mills:2020thr}.  In many cases, this degeneracy can be broken by identifying multiple modes of the gravitational wave signal.  The leading order (2, 2) mode is emitted at twice the orbital frequency while higher modes are typically emitted at higher frequencies with, e.g. the (3, 3) and (4, 4) modes emitted at three and four times the orbital frequency respectively.  However, in some cases where only one mode is identifiable, it was found that the mode could not be unambiguously identified --- the signal could equally well be attributed to the (2, 2) mode of a binary or the (3, 3) mode of a different binary with significantly higher mass and lower redshift.  The former signal would be associated to a merger of light seed black holes in the early universe, while the second would be from intermediate mass black holes (IMBHs).  The amplitude of the IMBH signal would be an order of magnitude larger in the deci-Hz band than the light seed black hole signal.  Thus, the \textit{non-observation} of a signal in an atom interferometer, combined with the observation in CE and ET, would provide clear evidence of the early-universe origin of the event.

\subsection{{LISA science and synergies}}
\label{Bayle}

\begin{figure}[h!]
    \centering
    \includegraphics[width=0.7\textwidth]{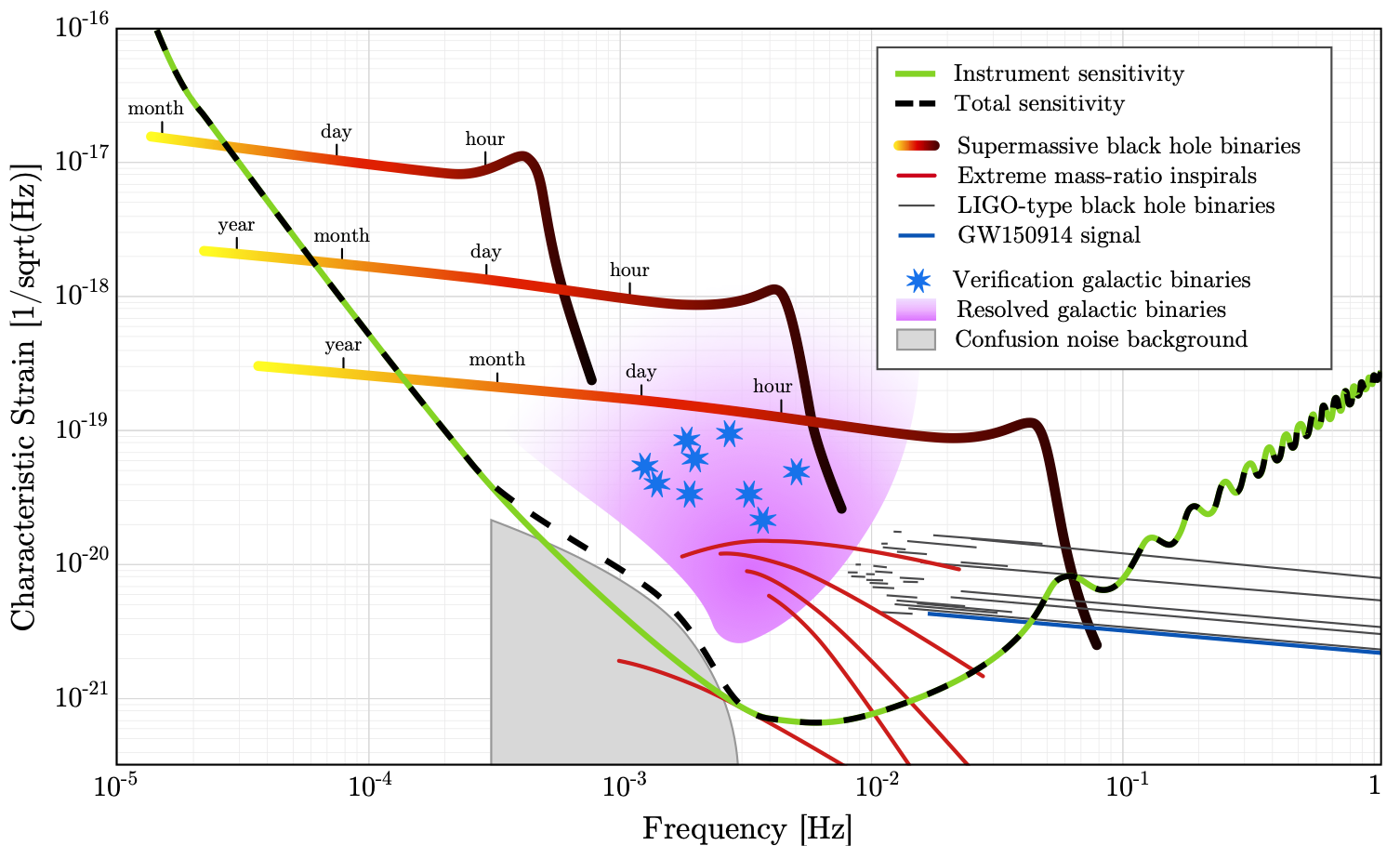}
    \vspace{-3mm}
    \caption{Illustration of the GW science capabilities of LISA, featuring sources including mergers of SMBH binaries of masses $10^7, 10^6$ and $10^5$ solar masses at redshifts $z =3$ (from left to right), extreme mass-ratio infall (EMRI) events, verified and resolved galactic binaries, the confusion noise from unresolved binaries, and early stages of LIGO-type stellar-mass binaries. Adapted from~\cite{AmaroSeoane2017}.}
    \label{fig:Bayle}
\end{figure}

The expected strain sensitivity of LISA as a function of GW frequency is illustrated in Fig.~\ref{fig:Bayle}. The sensitivity is maximised at $f \sim 10^{-2}$~Hz, with an interesting range extending from $\sim 10^{-4}$ to $\sim 10^{-1}$~Hz, as seen in Fig.~\ref{fig:Bayle}. LISA is optimised for the detection of mergers of supermassive black hole (SMBH) binaries and extreme mass-ratio inspirals (EMRIs)~\cite{AmaroSeoane2017}. It will also be able to observe known galactic binaries (blue stars), which will be useful for verification and calibration purposes. Many more galactic binaries are expected to exist, and their GWs signals could be resolved in the region of the frequency/strain plane that is shaded violet. GW signals in the region of the plane that is shaded grey could not be resolved, which reduces the total sensitivity of LISA compared to the ideal instrumental limit over a range of frequencies $\sim 10^{-3}$~Hz.

The frequency of the GW signal from any specific SMBH binary increases during the inspiral phase, with the characteristic strain gradually decreasing until the final stages of infall and merger, which generate a peak in the induced strain that is followed by  a decrease in the GW amplitude during the ringdown phase. Observations of the GW signal during the inspiral stage of binaries with SMBH masses in the range of $10^7$ to $10^5$ solar masses will be possible over long periods of time, as indicated in Fig.~\ref{fig:Bayle}, making possible detailed studies of the evolving GW wave form that can be used to test general relativity, verifying high-precision theoretical predictions. These measurements will also enable detailed properties of the SMBH binary systems to be determined (e.g., masses, redshift, orientation and eccentricity)~\cite{AmaroSeoane2017}. They will also enable the time to merger to be predicted, facilitating multi-messenger observations by astronomical observatories.

In addition to the mergers of SMBHs that could be expected on the basis of the observed cosmological population of SMBHs associated with active galactic nuclei (AGNs), one may anticipate that mergers of intermediate-mass black holes (IMBHs) with masses $< 10^5$ solar masses may also occur at observable rates and distances. Although the GW signals from their early inspiral stages could be observed by LISA, as seen in Fig.~\ref{fig:Bayle}, the IMBH merger and ringdown stages would emit GWs with frequencies in the range $\gtrsim 10^{-1}$~Hz where LISA is less sensitive. In this case there would be interesting synergies between observations by LISA and in the deci-Hz range of frequencies, as could be provided by atom interferometers~\cite{Torres-Orjuela:2024tmu}: see Section~\ref{Urrutia}. Fig.~\ref{fig:GWgap} in Section~\ref{sec:setting} shows the example of a merger of a pair of IMBHs with total mass $10^4$ solar masses~\cite{Badurina2021}. GWs due to inspiral prior to such an event at a redshift $z = 1$ could be measured by LISA over a period of a month, with a TVLBAI (represented here by AION-km) subsequently observing the merger and ringdown stages. The LISA observations would enable the direction, distance and chirp mass of such an event to be estimated, but they could be measured with much greater precision by a TVLBAI, as seen in Fig.~\ref{fig:measurements}~\cite{Ellis:2023iyb}. Such IMBH mergers may play essential roles in the assembly of SMBHs, but the existence and environments in which such events take place are currently unknown.

Another interesting example of possible synergies is provided by the (almost) straight, sloping lines in Fig.~\ref{fig:Bayle} showing GWs at frequencies $\gtrsim 10^{-2}$~Hz due to the early inspiral stages of LIGO-type stellar-mass black hole binaries~\cite{Toubiana2021,Sberna2022}. As shown in Fig.~\ref{fig:GWgap}, their signals would pass across the TVLBAI frequency range before the final merger and ringdown stages were observed by LVK and/or ET/CE. In such a case, TVLBAI observations could sharpen considerably the measurements of the parameters of the binary system made by LISA~\cite{Ellis2020}, facilitating multi-messenger observations of the final merger.

In addition to the astrophysical GW sources exhibited in Fig.~\ref{fig:Bayle}, LISA will also have interesting sensitivities to cosmological sources such as first-order phase transitions~\cite{Caprini2020} and cosmic strings~\cite{Auclair2020}. Their signals extend over a broad frequency range that may extend across the TVLBAI frequency range as well as the LISA range. Joint measurements may help distinguish between possible sources of such stochastic GW backgrounds.

\subsection{Deci-Hz synergies}
\label{Urrutia}

Since the discovery of gravitational waves (GWs) by LIGO~\cite{LIGOScientific:2016aoc}, GWs have proven useful for uncovering a population of stellar-mass black holes (BHs)~\cite{KAGRA:2021vkt,KAGRA:2021duu}, testing Einstein’s theory of gravity in the strong regime~\cite{LIGOScientific:2021sio}, and exploring cosmology~\cite{LIGOScientific:2017ync,LIGOScientific:2021izm}. This represents just the first indication of what future GW observations promise. One of the keys to unlocking the full potential of these observations lies in the exploration of the entire frequency spectrum, ranging from nano- to kilo-Hertz. In this regard, atomic interferometers play a crucial role in bridging the observational gap between LISA~\cite{AmaroSeoane2017} and ground detectors~\cite{LIGOScientific:2014pky,Maggiore:2019uih} that was highlighted in Section~\ref{grav}.
\begin{figure}[h!]
    \centering
    \includegraphics[width=0.7\textwidth]{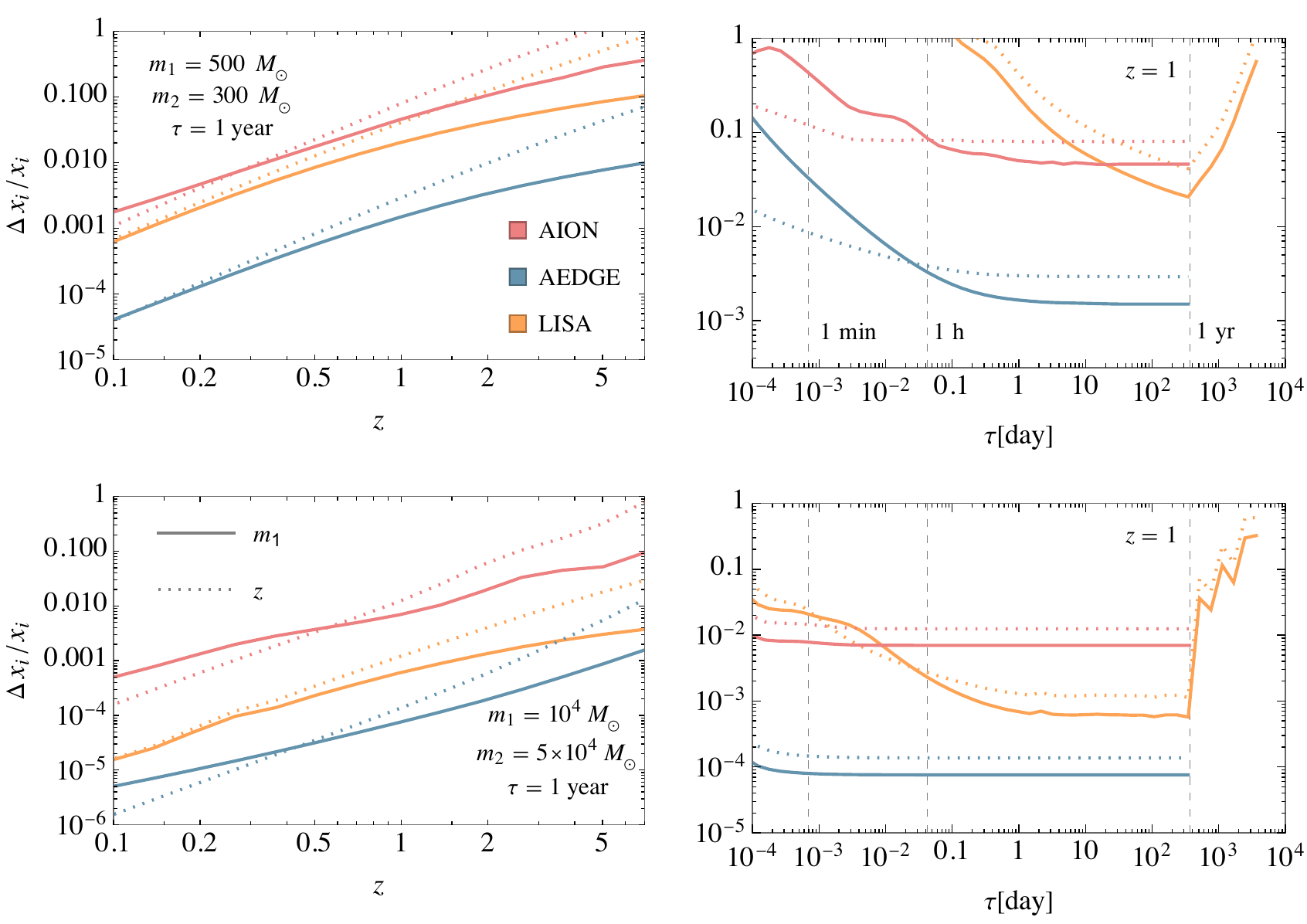}
    \vspace{-3mm}
    \caption{Prospective accuracies for measurements of binary parameters (distance, redshift and chirp mass): The upper and lower panels correspond to two binaries whose component masses are fixed. In the left panels, the binary is observed for the last 1 year and the errors are shown as functions of redshift. In the right panels, $z=1$ and the errors are shown as a function of the binary coalescence time at the beginning of a one-year observation. The errors for the masses of the both BHs are almost the same. From~\cite{Ellis:2023iyb}.}
    \label{fig:measurements}
\end{figure}

In astrophysics, synergies with LISA will emerge in at least two directions. The first involves detecting the same BH binaries as LISA but in different stages of evolution and the second, perhaps more intriguingly, detecting BHs that escape the range of LISA. The frequency range of atomic interferometers is optimal for observing mergers, which are the most violent processes in the evolution of the binary, with chirp masses $\mathcal{M}_c = \left(10^2 - 10^4\right) M_{\odot}$. In this respect, a 1-km TVLBAI will surpass the sensitivity of LISA during the final days of evolution for binaries with $\mathcal{M}_c < 10^3 M_{\odot}$. In the left panels of Fig.~\ref{fig:measurements}, we see the errors in recovering the parameters of the binary as a function of redshift for both LISA, a 1-km TVLBAI and a space-borne atom interferometer (AEDGE). We see how the 1-km TVLBAI can recover the parameters of the binary with $10\%$ precision up to $z\sim 2$ for binaries of $\mathcal{O}(10^2\, M_{\odot})$ and up to $z\sim 5$ for binaries of $\mathcal{O}(10^4\, M_{\odot})$, and how it compares to both LISA and AEDGE. On the right side, we see the precision with which the parameters of the binary can be recovered as a function of the time to merger. We see that for a binary of $\mathcal{O}(10^2\, M_{\odot})$ at $z=1$, a 1-km TVLBAI can still detect the binary in its final stages when it is at less than one day to the merger, whereas LISA can detect it only up to $\tau\sim 1\, {\rm day}$.

\begin{figure}[h!]
    \centering
    \includegraphics[width=0.84\textwidth]{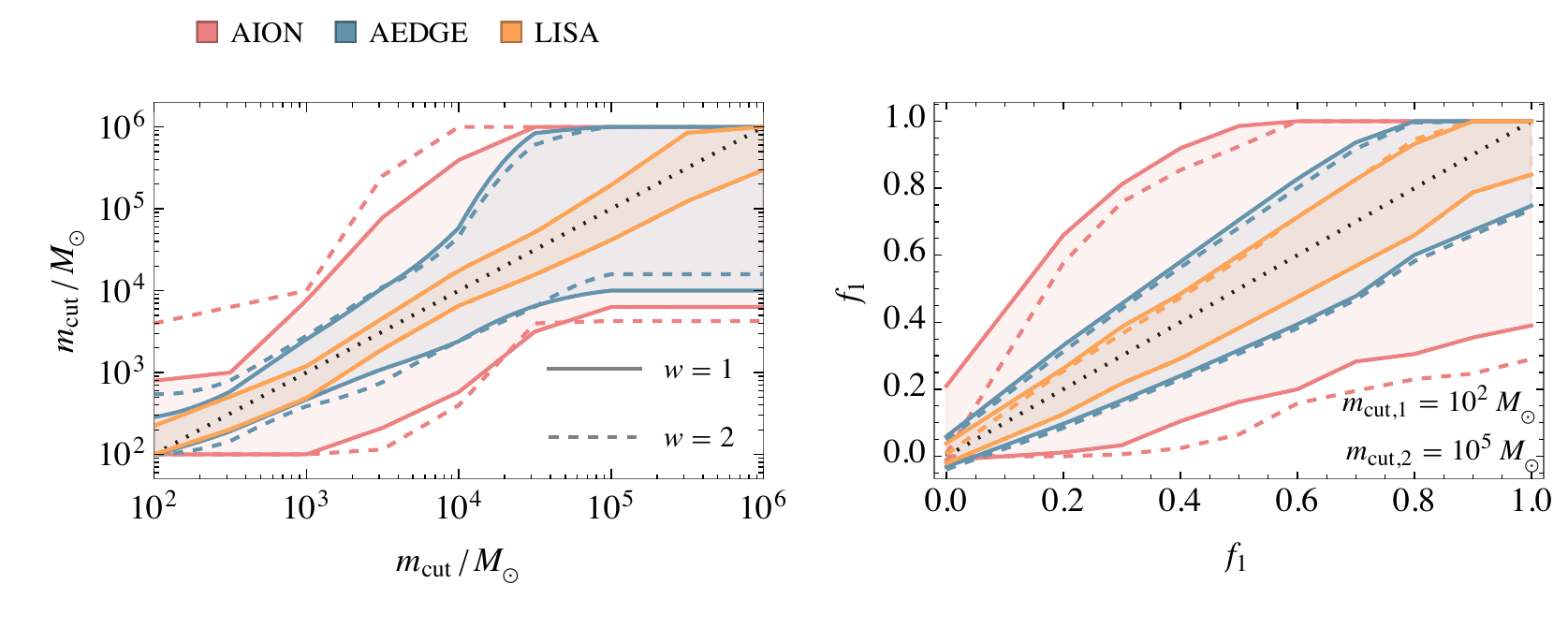}
    \vspace{-3mm}
    \caption{\textit{Left panel:} The 95\% CL accuracy with which LISA, AEDGE and a 1-km TVLBAI could measure the seed mass parameter $m_{\rm cut}$ in the range of $[10^2, 10^6] M_{\odot}$. \textit{Right panel:} The 95\% CL accuracy with which LISA, AEDGE and a 1-km TVLBAI could measure the fraction of light seeds, $f_{\rm 1}$, assuming an input mixture of seeds with masses $10^2$ and $10^5 M_{\odot}$ and $f_2 = 1 - f_1$. The solid and dashed curves in both panels correspond, respectively, to $w=1$ and $w=2$, which is related to the width of the distribution of the seed mass. From~\cite{Ellis:2023iyb}.}
    \label{fig:posteriors}
\end{figure}

BHs in this mass range hold the key to solving the enigma of supermassive black hole (SMBH) formation~\cite{Volonteri:2010wz}. It is believed that these BHs evolved from their ``seeds" by a hierarchy of mergers and through mass accretion. The hierarchical nature of this evolution implies a continuous mass spectrum, ranging from seed masses that have either evolved very little or not at all~\cite{Volonteri:2009vh}, to the SMBHs that populate the centers of the largest galaxies~\cite{Kormendy:2013dxa}. The reason why the origin of SMBHs remains unknown is partly connected to our lack of understanding of the behavior of dark matter on small scales. The lightest halos lack sufficient gas to form stars, making them extremely difficult to detect. These halos and the faintest galaxies, according to cosmological models, host the least evolved BH descendants and hold the key to unravelling the origin of SMBHs~\cite{Chadayammuri:2022bjj}. Atomic interferometers are sensitive to these remnants in the ``light seed" scenario, where the seeds are $\left(10^2-10^3\right)M_{\odot}$, and could detect remnants of these SMBH progenitors, and their non-detection would provide evidence in favor of a ``heavy seed" $(>10^4\, M_{\odot})$ scenario. A 1-km TVLBAI would be sufficient to distinguish between the ``heavy seed" and ``light seed" scenarios. In the left panel of Fig.~\ref{fig:posteriors} we see how well it is possible to recover the seed mass parameter, $m_{\rm cut}$, from binary observations with a 1-km TVLBAI, AEDGE and LISA. We see how a 1-km TVLBAI could distinguish at the 95\% CL a heavy seed scenario with $\mathcal{O}\left(10^5\,M_{\odot}\right)$ from a light seed scenario with $\mathcal{O}\left(10^2\, M_{\odot}\right)$. In the right panel we see that in the case of a mixed formation scenario, where $f_1$ is the fraction of the SMBHs that come from the light seed, it is possible to recover the contribution of each channel with an interesting uncertainty in the value of $f_1$. 

\begin{figure}[h!]
    \centering
    \includegraphics[width=0.9\textwidth]{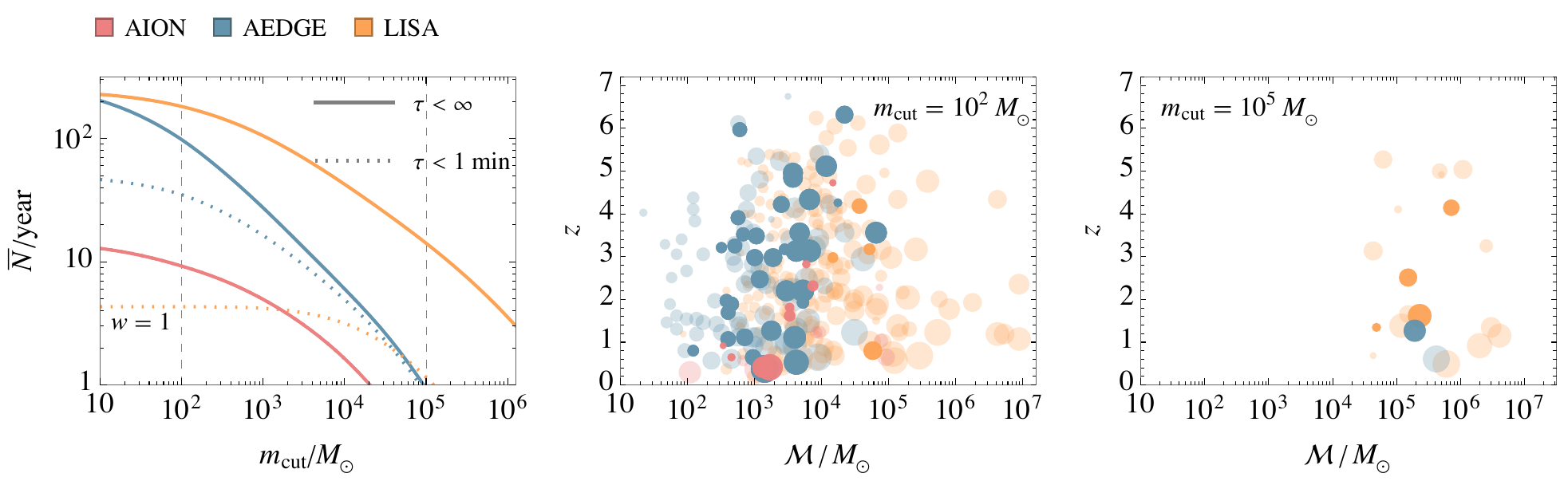}
    \vspace{-3mm}
    \caption{\textit{Left panel:} The expected numbers of binaries detectable by a 1-km TVLBAI (AION), AEDGE and LISA during a year of observation, as functions of $m_{\rm cut}$. The solid curves show all detectable binaries whereas the dotted curves for AEDGE and LISA show only those for which the last 1 minute of the merger is seen: the last minutes of all mergers are seen by AION.
    \textit{Middle and right panels:} Explicit examples of the detectable binaries for a light-seed and a heavy-seed scenario. The sizes of the dots are $\propto \ln[ {\rm SNR}^{-1}]$ with the minimum size corresponding to ${\rm SNR}=10^4$ and the maximal to ${\rm SNR}=10$. The darker dots correspond to binaries for which the last 1 minute of the merger is seen. From~\cite{Ellis:2023iyb}.}
    \label{fig:events}
\end{figure}

The recent evidence from Pulsar Timing Arrays (PTAs) for a stochastic GW background in the nano-Hertz range~\cite{NANOGrav:2023hde,EPTA:2023xxk,Zic:2023gta,Xu:2023wog} provides insight into predictions for LISA and atom interferometers. If the signal is to be interpreted as originating from SMBH binaries, it implies a high efficiency in binary formation and rapid orbital evolution~\cite{NANOGrav:2023hfp,EPTA:2023xxk,Ellis:2023dgf,Raidal:2024odr} during the final stages of hierarchical SMBH evolution. If this mechanism extends to lighter binaries that are part of the same hierarchical process, it predicts that a significant number of binaries should be observable by both LISA and atom interferometers. As can be seen in the left panel of Fig.~\ref{fig:events}, focusing on binaries that can be observed within 1 minute of the merger, the number of events per year for LISA is reduced to only $\sim 5 \,{\rm events}$ for the light-seed case, while a 1-km TVLBAI such as AION is expected to detect around $10$. In the heavy-seed case, LISA will not see the last 1 minute of the binaries if $m_{\rm cut} > 10^5M_{\odot}$. On the other hand, AEDGE would detect most of the binaries until within 1 minute of the merger, except for those lighter than $\mathcal{M}_c\sim 100 M_{\odot}$. A 1-km TVLBAI has a similar range to AEDGE but, because it has less sensitivity, the number of events is decreased to around $10$ if $m_{\rm cut}\sim10^2M_{\odot}$ and further reduced to 1 event per year on average for $2\times 10^4\, M_{\odot}$. 

An explicit realization of the binary population in a light-seed scenario with $(m_{\rm cut}=10^2\, M_{\odot})$ is shown in the middle panel of Fig.~\ref{fig:events}, and in a heavy-seed scenario with $(m_{\rm cut}=10^5\, M_{\odot})$ in the right panel of Fig.~\ref{fig:events}. We see that AEDGE and, to a lesser extent, a 1-km TVLBAI will probe BH binaries over a range of masses that are too light for LISA, covering the mass range covered by terrestrial laser interferometers from $m_{\rm cut}<100 M_{\odot}$ to $\sim 10^4 M_{\odot}$. Considering the final stages within one minute of a merger, AEDGE binaries can explore the range $m_{\rm cut}\in \left(10^2-10^4\right)\, M_{\odot}$ out to redshifts $z\sim 7$. LISA, on the other hand, observes only the last minutes of heavier binaries with $m_{\rm cut}\in\left(10^4-10^6\right)\, M_{\odot}$ and $z<4$. Finally, a 1-km TVLBAI will detect a handful of events per year for $m_{\rm cut}<10^4\, M_{\odot}$ at low redshifts, $z<4$, mainly in the $\left(10^3-10^4 \right)\, M_{\odot}$ mass range and mostly in the last moments before the merger. We refer the reader to~\cite{Ellis:2023iyb} for more details on the computation.

Extrapolating from PTA observations to deci-Hertz frequencies may seem risky, but there is preliminary evidence that the model for SMBH binaries remains valid for lighter BHs. The new observations from JWST are invaluable, as we are witnessing the formation of a SMBH in its most vigorous stage. The interpretation of these results has not yet reached a unanimous consensus: the BHs observed are heavier than expected~\cite{Pacucci:2023oci}, and the density of active galactic nuclei (AGNs), specifically those known as ``little red dots"~\cite{Matthee:2023utn}, as well as the fraction of dual AGNs~\cite{2023arXiv231003067P}, is higher than anticipated. If the merger is interpreted as the final stage of a process that triggers AGNs, which is a reasonable assumption, it allows us to connect the binary model with the JWST observations.
\begin{figure}[h!]
    \centering
    \includegraphics[width=0.4\textwidth]{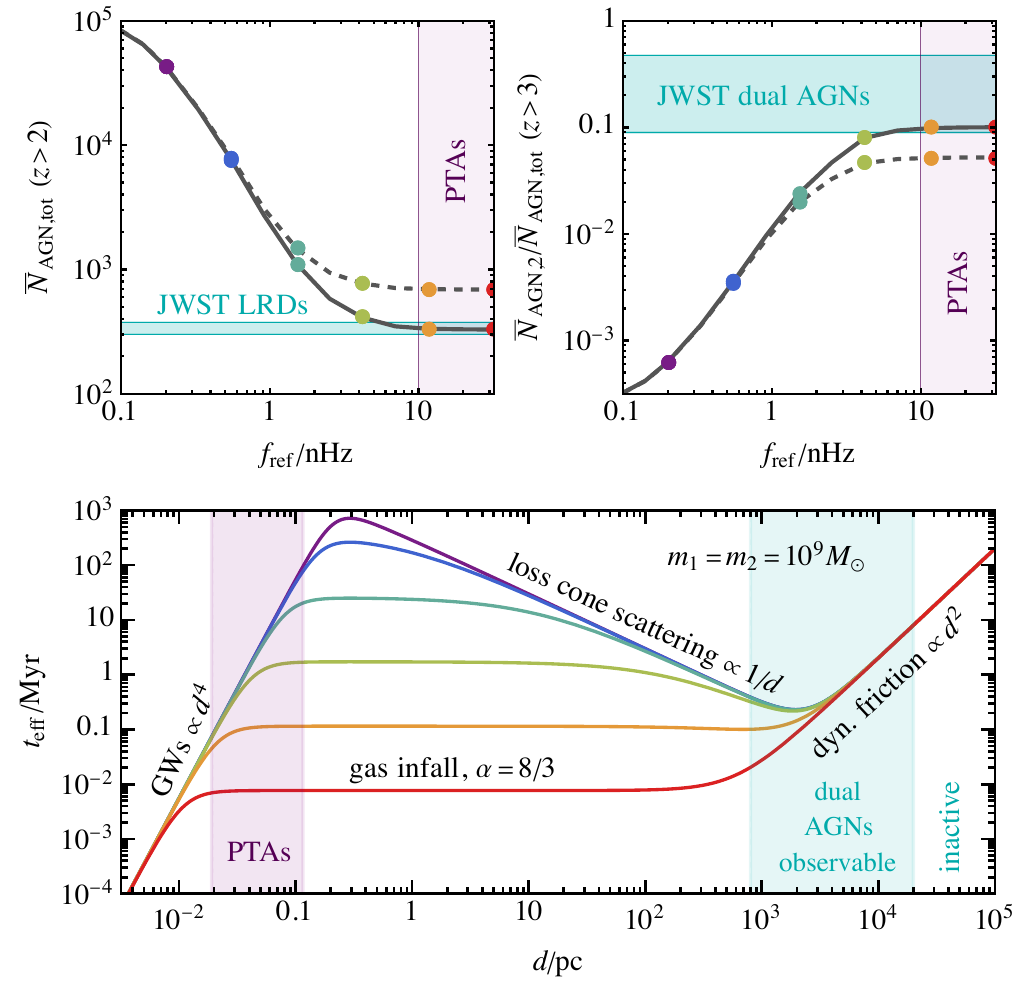}
    \vspace{-3mm}
    \caption{\textit{Upper panels:} The expected total number of AGNs at $z>2$ and the expected dual AGN fraction at $z>3$ as functions of the environmental parameters. \textit{Lower panel:} Illustration of the timescales of the environmental energy loss mechanisms of SMBH binaries as functions of their separations for the same parameter values as in the upper panels. The vertical bands illustrate the separation ranges probed by the PTAs and by the JWST dual AGN observations. The color coding of the curves matches that shown in the upper panels. From~\cite{Ellis:2024wdh}.}
    \label{fig:teff}
\end{figure}
In the upper panels of Fig.~\ref{fig:teff} we show the dependencies of the numbers of AGNs and fractions of dual AGNs as functions of the parameters of the environmental effects. For these parameters, the lower panel of Fig.~\ref{fig:teff} illustrates the timescales of the environmental energy loss mechanisms, including GW emission, gas effects, stellar loss-cone scattering and dynamical friction, as functions of the binary separation. The horizontal bands in the upper panel show the 95\% CL bands of the JWST ``Little red dots" and dual AGN observations calculated assuming the Poisson and binomial distribution, respectively. We can see how, within the 95\% CL region of the NANOGrav fit, it is possible to obtain a coherent description of the evolution of binary systems that explains both the NANOGrav GW signal and the JWST dual AGN and ''little red dots" observations, providing preliminary evidence that the binary evolution picture can be extrapolated to lower masses. For more details on the computation we refer the reader to~\cite{Ellis:2024wdh}. Although the JWST observations are still very recent and their interpretation remains uncertain, there is a synergy between JWST and atom interferometers, as they can probe the same populations of SMBHs at complementary stages in their evolution. 

In conclusion, atom interferometers will enable the detection of the same binaries as LISA when they are at  $z<4$ and with masses of  $\mathcal{O}(10^2\, M_{\odot})$, but at different stages of their evolution. Specifically, they will observe their evolution when less than one day remains before the merger. Heavier binaries with masses $\mathcal{O}(10^4\, M_{\odot})$ will be observed by both LISA and atom interferometers, although atom interferometers will provide better precision in the final minutes before the merger. The BHs to which they are most sensitive, with masses $\mathcal{O}(10^2 - 10^4 M_{\odot})$, hold the key to resolving the mystery of SMBH formation. Although these are closer in mass to the binaries observed by LIGO, they could represent the lighter end of a hierarchical process that culminates in SMBHs at the centers of galaxies. Hence, theoretical models that explain the new PTA observations allow us to extrapolate this hierarchical formation process to higher frequencies. Preliminary evidence that this extrapolation may be correct lies in the compatibility with JWST observations. Under these assumptions, the expected number of binaries for a 1-km TVLBAI in the light-seed scenario is $\mathcal{O}(10)$ per year, in the mass range of  $\mathcal{O}(10^2-10^4) M_{\odot}$  at  $z<4$. In all these binaries, the merger will also be observed. LISA, however, will only observe the merger of the heavier binaries. In the case of the heavy-seed scenario, a 1-km TVLBAI is not expected to observe any binaries. In any case, the observation or non-observation of binaries will be sufficient to determine the origin of SMBHs. As our understanding of the GW background in the nano-Hertz regime improves, or as our interpretation of JWST observations evolves, the predictions will need to be updated, highlighting the strong synergy between atom interferometers and other observatories.

\section{Introduction to cold atom experimental requirements}
\label{sec:compiledintro}
The long-baseline atom interferometers currently in design and development consist of atoms pre-cooled in an \textit{atom source}, and efficiently transferred to an \textit{interferometry vacuum tube} long enough to accommodate the baseline between interferometers. 
The interferometry tube has stringent requirements on vacuum levels and magnetic shielding, with subsequent constraints on optical and mechanical access as discussed in Section~\ref{engineering}. 
This design concept is common across many ultracold atom platforms, where a carefully-designed `science chamber' or glass cell accommodates the vacuum, optical, geometric and magnetic field requirements driven by the experimental science goals. Pre-conditioned atoms are then transferred from a spatially separate atom source that may consist of one or more vacuum chambers\footnotemark. 
\footnotetext{It is important to note that there also exist a large number of single-chamber experiments in which the different stages of the experiment are separated temporally rather than spatially; if the science goals and other experimental requirements allow for this approach, benefits include experimental simplicity, fast repetition rate and a compact design.}

Design and operational requirements on long-baseline interferometers are motivated primarily by their phase sensitivity goals. 
This places a set of interconnected experimental constraints affecting both the atom source and approach to interferometry. This influences apparatus design choices as well as the experimental steps in cooling, manipulating and probing the ultracold atoms. Many of these considerations are shared with cold atom platforms used for quantum simulation, quantum information processing and other forms of quantum sensing, as well as atom interferometry. Sections~\ref{LMT} to \ref{Metrology} outline methods and considerations that primarily apply to long-baseline atom interferometry, but share commonalities with portable or tabletop interferometers or broader ultracold atom platforms, and some that directly link techniques employed in quantum simulation experiments with the requirements we seek here.

State-of-the-art atom interferometric sensors are usually limited in their sensitivity by a fundamental limit that arises whenever such sensors are operated with uncorrelated particles. The phase resolution at this so-called Standard Quantum Limit (SQL) is equal to $\Delta \phi_{\rm SQL}= 1/\sqrt{N}$ for $N$ uncorrelated two-level systems entering the interferometer. The SQL is assumed to be the dominant source of noise in many TVLBAI proposals~\cite{Badurina2020,Abe2021,Canuel2018}: as such, minimizing the SQL, or beating it, will be crucial for allowing TVLBAI detectors to achieve their projected sensitivities. Sensitivity below the SQL is possible by introducing quantum correlations between particles, i.e. quantum entanglement, for example by preparing the atoms in a squeezed state. These entangled states can improve significantly the scaling with atom number~\cite{PhysRevLett.96.010401}, in principle up to the Heisenberg limit where the phase resolution is given by $\Delta \phi_{\rm H}= 1/N$. These beyond-SQL methods are the topic of Section~\ref{sec:squeezing}. 
This also motivates the requirement for a large atom number in each interferometry cycle, hence efficient delivery of a high flux of atoms as discussed in Section~\ref{sources}. A tension exists between these requirements, due to the challenge of achieving the Heisenberg limit in the large ensembles in our target range of $10^3 - 10^6$ atoms~\cite{RMPPezze18}, and a compromise will need to be reached based on technical challenges and achievements in both atom cloud production and squeezing technologies.

The science goals of long-baseline interferometers (e.g., gravitational wave detection and searches for dark matter) place demanding requirements on the acceleration sensitivity that long-baseline interferometers must achieve.  Since the phase response to accelerations scales linearly with the momentum splitting between interferometer arms, one promising route to higher sensitivity is to increase the number of photons exchanged during each atom-light interaction—a technique known as large momentum transfer (LMT) interferometry, discussed in Section~\ref{LMT}.  Previous experiments have demonstrated LMT interferometry with up to $90 \hbar k$ beamsplitters based on sequential two-photon Bragg transitions in an atomic fountain~\cite{Kovachy2015a}, $102 \hbar k$ utilizing multi-photon Bragg transitions~\cite{chio11}, and $40 \hbar k$ using Bloch oscillations in an optical lattice~\cite{mcalpine2020excited}.  Extending the momentum splitting to $1000 \hbar k$, a key technical goal, requires a transfer efficiency of 99.9\% per photon to avoid excessive atom loss.
This high-fidelity LMT requires that the atom cloud experiences an approximately uniform laser intensity across its spatial extent and given technical limitations on laser power (thus reasonable beam size) this motivates a second critical requirement of the atom source: the atom cloud must be as compact as possible while avoiding interaction effects at high densities, and with minimal expansion over the interrogation time of the interferometer. Delivery of atoms from source to interferometry tube must therefore accommodate transport of atoms at temperatures as low as possible, with minimal heating. Methods of minimising the velocity spread of the atom cloud after delivery must also be explored, with options including selection of a narrow velocity class from within an atom cloud, or matter-wave lensing to narrow the overall distribution. Some methods for efficiently transporting and minimising the velocity spread of an atom cloud are discussed in Section~\ref{sources}.

As we explore increasing the scale of atom interferometers we must investigate and understand the variations in our detector systematics and necessary changes to our measurement methods. Metrology is the study of measurement. In our case, this is the determination of the phase shift between precisely controlled quantum states of an ensemble of ultracold atoms. Two components compose this determination: the precision and accuracy with which we can measure the phase of the system, and the detector operation strategy for a given measurement campaign.
For these new scales we are presented with two systematic effects as discussed in Section~\ref{Metrology}: the Coriolis effect, and laser wavefront distortion. These systematics are a subset of new noise sources arising from the increase in detector size and can be severely limiting to our overall instrument sensitivity. Possible mitigation strategies for these effects were discussed. Utilizing the longer baseline of these large-scale detectors raises questions of optimal measurement strategies. A discussion on the best use of the baseline was then presented for targeting maximal phase shift sensitivity configurations for simultaneous dark matter and gravitational wave searches.
These presentations are a preliminary effort showing the change in thought needed for atom interferometry metrology in this new regime. Very-long-baseline atom interferometry metrology is critical for understanding the detector’s limits and is a growing area of investigation with many studies to pursue. 

\section{Atom interferometry: Large momentum transfer techniques}
\label{LMT}
\subsection{Introduction}

TVLBAI interferometers rely on a very large separation between the two arms of the atom interferometer, requiring the use of large momentum transfer (LMT) beam splitters with momentum transfers greater than 1000$\hbar k$. A common approach is to first create a coherent superposition between two momentum states separated by a few photon recoils. This is then followed by an acceleration phase to transfer additional momentum to each interferometer arm \cite{McGuirk2000}. Efficient momentum transfer is critical to prevent atom loss during the process, and the transfers must be fast to fit within the limited interferometer time of flight. In addition, precise control of the laser phases during acceleration is required. Various methods have been proposed to achieve these goals, including single-photon transitions \cite{Wilkason2022atom,Bott2023}, sequences of Raman pulses \cite{McGuirk2000,Kotru2015,Berg2015}, and the acceleration of atoms within an optical lattice using sequences of Bragg pulses \cite{chio11,plot18,Beguin23} or Bloch oscillations \cite{page20,gebb21}. Here we discuss a new approach to developing an optimal quantum acceleration technique based on a Floquet state engineering scheme, showing the application of this approach to lattice-based acceleration, which can be implemented for arbitrary atomic species. We also discuss multi-photon clock atom interferometry and the use of magic-depth optical lattices to manage Bloch Oscillation (BO) phases for LMT interferometry.

\subsection{Optimal Floquet State Engineering for Large Scale Interferometers}
\label{Floquet}

\begin{figure}
  \includegraphics[width=0.9\textwidth]{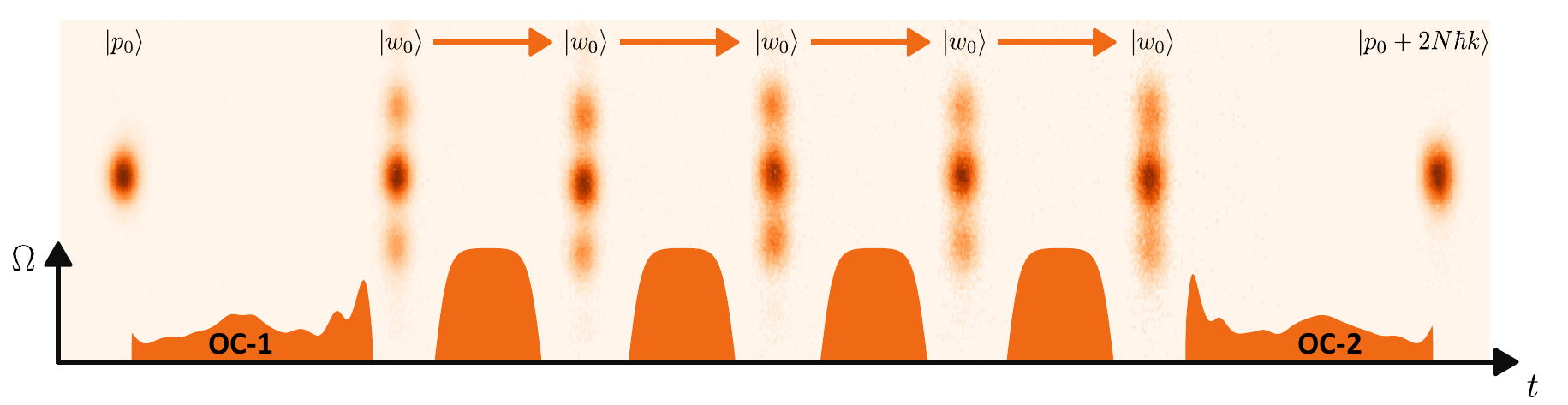}
  \caption{The initial momentum state, $|p_0\rangle$, is prepared in a Floquet state, $|w_0\rangle$, associated with the specific periodic evolution of the optical lattice. The preparation sequence (OC-1) is obtained using optimal control theory (OCT). At the end of the sequence, the state $|w_0\rangle$ is transformed back to the accelerated momentum state (OC-2).}
  \label{fig-floquet}
\end{figure}

The Floquet approach is founded upon the periodic driving of the Hamiltonian in the accelerated frame, which results in dynamics that are well-described by the Floquet formalism \cite{Bukov2015}. Floquet states are eigenstates of the one-period propagator. Consequently, if the system is initiated in a pure Floquet state, $|w_m(t_0)\rangle$, its temporal evolution is periodic. After each pulse with a duration of $\tau$, the atom returns to the same Floquet state, $|w_m\rangle$, leading to a stroboscopic stabilization of the Floquet state in the accelerated frame (see Fig. \ref{fig-floquet}). This results in an almost lossless, coherent acceleration of the atomic wave packet in the laboratory frame.

In order to exploit the stroboscopic stabilization, two possible approaches may be taken. One approach entails modifying the amplitude and phase of the optical lattice with the objective of aligning the Floquet state with the initial state, which is a pure momentum state. An alternative approach employs the resources of quantum optimal control theory (OCT). This method considers any optical lattice profile and introduces a preparation step to transform the initial state into the Floquet state corresponding to the specific lattice profile. This state-to-state preparation step is referred to as “OCT” in Fig.~\ref{fig-floquet}.

This technique has been implemented with a rubidium Bose-Einstein condensate in the specific context of resonant square Bragg pulses. The experimental evidence demonstrates an efficiency of $99.95\%$ per $\hbar k$ and a transfer rate of 2.5 $\mu s$ per $\hbar k$ following a 1200$\hbar k$ transfer. The primary constraint is spontaneous emission, which can be mitigated by increasing the resonant detuning. The use of more powerful laser systems, as in \cite{Kim20}, or the incorporation of optical cavities, could help overcome this limitation. Another significant limitation is the impact of phase and amplitude fluctuations during the acceleration sequence, which can affect the periodicity of the sequence. The method was finally implemented to set up a Mach-Zehnder-type atom interferometer (splitter-mirror-splitter) with a maximum separation of 600$\hbar k$ between the interferometer arms. The visibility was maintained at 18$\%$, and interference patterns were observed with high confidence, indicating that the method does not introduce uncontrollable parasitic phase shifts. However, the separation is limited by the detection volume in the experiment.
This approach effectively integrates previous techniques in LMT atom optics, including independent pulse sequences in quasi-Bragg regimes \cite{chio11,plot18}, sequences that exploit destructive interference between different loss channels \cite{Beguin23}, and continuous Bloch acceleration under both adiabatic and non-adiabatic conditions \cite{page20,gebb21,rahm24}. This scheme enables optimal state-to-state control in large Hilbert spaces, exceeding the capabilities of traditional brute-force numerical methods. By encapsulating the complexity of the problem within the Floquet state, this method enables the engineering of Floquet states or optical lattice shaping to mitigate sensitivities to systematics such as velocity dispersion and lattice amplitude. Moreover, this approach allows for a comprehensive comparison of the relative merits of different methods and an evaluation of their fundamental limits within the context of TVLBAI.

\subsection{Multi-photon clock atom interferometry}
\label{Hogan}

Clock atom interferometry is a new technique for inertial sensing based on optical transitions between long-lived atomic states such as those typically used in atomic clocks.  In contrast to traditional atom interferometry based on two-photon atom optics (e.g., Raman or Bragg transitions), clock atom interferometers make use of single-photon transitions.  Such an approach is possible in alkaline-earth-like atoms (such as strontium) that possess narrow optical transitions with a sufficiently long-lived excited state.  The use of single-photon transitions is an essential ingredient to proposals for atom interferometric gravitational wave detectors \cite{Abe2021} as well as dark matter searches \cite{Arvanitaki2018}.  In these applications, sending light from a single direction at a time for each atom optics pulse allows for better suppression of laser phase noise in long-baseline gradiometers compared to the counter-propagating lasers used for two-photon atom optics \cite{Graham2013}.  Clock atom interferometers are also promising platforms for large momentum transfer (LMT) atom optics, where the sensitivity is enhanced through the application of multiple laser pulses to increase the relative momentum between the interferometer arms.  Given the long lifetime of the excited state, combined with low off-resonant scattering from the nearest transitions many nanometers away, clock atom interferometers enjoy low spontaneous emission losses, allowing in principle for more light pulses than other methods.

Clock atom interferometers may be implemented in several ways, each with different advantages and applications.  Narrowband clock atom interferometry is based on ultranarrow transitions such as ${^1\mathrm{S}_0}\! \rightarrow\! {^3\mathrm{P}_0}$ in $^{87}$Sr, which has an excited state lifetime $>100~\text{s}$ \cite{Graham2013}.  This transition is well-suited to the long-baseline gradiometer applications mentioned above, where the long excited state lifetime allows for light propagation over distances exceeding a kilometre, as required for a TVLBAI detector.  However, the narrow linewidth also implies a weak atom-light interaction strength, resulting in a relatively low Rabi frequency with a narrow Doppler acceptance range.  In contrast, broadband clock atom interferometry makes used of intermediate linewidth transitions such as ${^1\mathrm{S}_0}\! \rightarrow\! {^3\mathrm{P}_1}$ in strontium, which has an excited state lifetime of $22~\textbf{$\mu$s}$.  The stronger coupling strength enabled by this transition supports MHz-level Rabi frequencies, resulting in higher Doppler acceptance and reduced sensitivity to laser frequency noise.  Broadband clock atom interferometry has recently been used to set new records for LMT enhancement using comparatively hot (i.e., laser cooled, not evaporatively cooled) atom ensembles \cite{Rudolph2020,Wilkason2022}.

In atoms like Sr, the ultranarrow ${^1\mathrm{S}_0}\! \rightarrow\! {^3\mathrm{P}_0}$ clock transition is typically only allowed in the fermionic isotope.  However, bosonic isotopes promise considerable advantages, such as their lack of nuclear spin, typically higher natural abundance, as well as simplified laser cooling and state preparation.  To access these desirable properties, coherent multi-photon processes in bosons have been proposed in the context of atomic clocks~\cite{Hong2005, Santra2005, Vitaly2007, Alden2014} to circumvent the lack of direct coupling between the clock states.  Of particular interest is the three-photon excitation ${^1\mathrm{S}_0}\! \rightarrow\! {^3\mathrm{P}_1} \!\rightarrow\! {^3\mathrm{S}_1} \!\rightarrow\! {^3\mathrm{P}_0}$~\cite{Hong2005}, using a set of laser frequencies readily available for laser cooling, repumping, and imaging in Sr.  This excitation approach has recently been used to implement multi-photon clock atom interferometery in bosonic $^{88}$Sr \cite{carman2024collinear}.

\begin{figure}[t]
    \centering
    \vspace*{12pt}\includegraphics[width=\columnwidth]{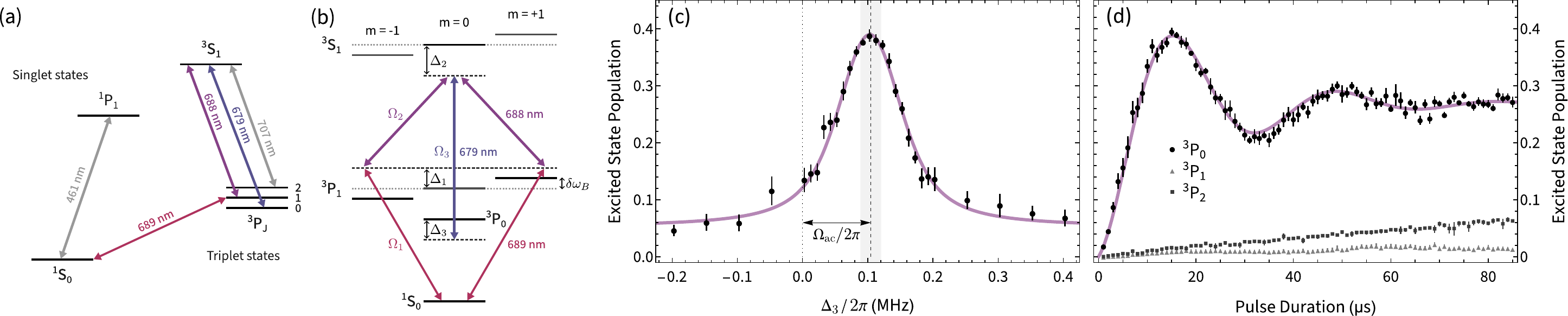}
    \caption{(a) The singlet and triplet states of $^{88}$Sr and the relevant transition wavelength connecting the atomic states, and
    the three-photon transition between $^1\mathrm{S}_0$ and $^3\mathrm{P}_0$ that uses laser light at 689~nm (magenta), 688~nm (purple), and 679~nm (blue).
    (b) Energy levels involved in the three-photon transition with Zeeman sublevels $m$ in the presence of a magnetic field, causing a relative shift $\delta\omega_B$. The polarizations of the optical fields are linear, with 689~nm\,($\Omega_1$) and 688~nm\,($\Omega_2$) normal, and 679~nm\,($\Omega_3$) parallel to the quantization axis. Cumulative frequency detunings of the lasers from the respective $m=0$ states are denoted by $\Delta_1$, $\Delta_2$, and $\Delta_3$. (c) Line scan of the three-photon transition showing the fractional excited state population versus the cumulative laser detuning $\Delta_3$, using $\Delta_{1}/2\pi = 9.95(1)~\text{MHz}$, $\Delta_2/2\pi = -2.54~\text{GHz}$, and $\delta\omega_{B}/2\pi = 21.13(1)~\text{MHz}$ \cite{carman2024collinear}.
    (d) Rabi oscillation at the measured three-photon resonance frequency, showing the fractional excited state populations versus the pulse duration \cite{carman2024collinear}. Circles, triangles, and squares indicate the population in the states ${^3\mathrm{P}_0}$, ${^3\mathrm{P}_1}$, and ${^3\mathrm{P}_2}$, respectively. The fit (solid curve) is an exponentially damped sinusoid with a frequency of $29.9(2)~\text{kHz}$. From~\cite{carman2024collinear}.}
\label{Fig:ThreePhotonSummary}
\end{figure}

The intermediate states and laser wavelengths (689~nm, 688~nm, 679~nm) for this multi-photon process are shown in Fig.~\ref{Fig:ThreePhotonSummary}~(a).  In order to drive the transition coherently, each of the three lasers must have a well-defined relative phase.  To achieve this, the three lasers were phase locked to an optical frequency comb~\cite{carman2024collinear}.  Although such a multi-photon transition requires multiple lasers as opposed to the one laser used for the single-photon transitions described above, the multi-photon clock atom interferometer can still support the same level of laser noise immunity in a long-baseline gradiometer if all laser frequencies are locally phase stabilized and then propagate together collinearly in the same direction to the atoms.

One significant complication is that selection rules naively forbid this three-photon process using collinear light, since this requires all three polarization vectors be coplanar.  As discussed in~\cite{carman2024collinear}, using light at 689~nm with $\mathbf{\hat{x}}$ polarization, followed by 688~nm with $\mathbf{\hat{x}}$ polarization, and then 679~nm with $\mathbf{\hat{z}}$ polarization results in two competing excitation pathways that interfere destructively, leading to zero coupling strength.  To avoid this, a small magnetic field can be used to break the degeneracy between the two paths, as discussed in~\cite{carman2024collinear}.  In this case, the three-photon Rabi frequency is
\begin{equation}\label{eq:ThreePhotonCoupling}
\Omega_{\text{eff}} = \frac{\Omega_1 \Omega_2 \Omega_3}{4\Delta_2} \left(\frac{1}{\Delta_1 - \delta\omega_B} - \frac{1}{\Delta_1 + \delta\omega_B} \right)
\end{equation}
where $\Omega_1$, $\Omega_2$, and $\Omega_3$ are the intermediate single-photon Rabi frequencies and $\Delta_1$ and $\Delta_2$ are the intermediate (cumulative) detunings. The minus sign between the terms exhibits the destructive interference of the two excitation paths (via $m=-1$ and $m=+1$) in the absence of the magnetic field shift $\delta\omega_B$.

Figure~\ref{Fig:ThreePhotonSummary}~(c) and (d) show the first demonstration of the three-photon excitation of the clock transition in bosonic $^{88}$Sr.  The three-photon resonance frequency was determined using the optical frequency comb to set the absolute frequencies of each of the three lasers.  The transition was located experimentally by scanning the frequency of one of the lasers around the predicted resonance (Fig.~\ref{Fig:ThreePhotonSummary}~(c)).  The peak was found at the expected location after accounting for the AC Stark shift $\Omega_{\text{ac}}$.  Figure~\ref{Fig:ThreePhotonSummary}~(d) shows Rabi oscillations in the ${^3\mathrm{P}_0}$ population observed by scanning the pulse duration at the measured three-photon resonance frequency.  The measured Rabi frequency of $29.9(2)\,\text{kHz}$ is in good agreement with the expected three-photon Rabi frequency predicted by Eq.~\ref{eq:ThreePhotonCoupling}.  The noticeable damping of the Rabi oscillation stems mostly from intensity inhomogeneity across the cloud, since in this proof-of-concept experiment the laser beam was tightly focused to increase the peak intensity.  Nevertheless, this transfer efficiency was sufficient to implement a Mach-Zehnder atom interferometer pulse sequence, resulting in an interferometer visibility of $20\%$ \cite{carman2024collinear}.

As a next step, it is straightforward to improve the Rabi oscillation transfer efficiency by using higher-power lasers and bigger beams to reduce inhomogeneous loss.  There is also a wide parameter space to explore in order to optimize the Rabi coupling strength and spontaneous emission loss, including the magnetic field strength and the two intermediate detunings.  In the future, multi-photon clock atom interferometry could provide a possible alternative to the ${^1\mathrm{S}_0}\! \rightarrow\! {^3\mathrm{P}_0}$ transition in $^{87}$Sr in long-baseline quantum sensors such as MAGIS-100 \cite{Abe2021}, offering the potential for improved LMT atom optics performance by engineering an effective transition with strong coupling but long excited-state lifetime.

\subsection{Managing Bloch Oscillation Phases for Large Momentum Transfer Interferometry using Magic-Depth Optical Lattices}
\label{Gupta}

Bragg diffraction and Bloch oscillation (BO) have been two main techniques at the forefront of large-momentum-transfer (LMT) atom optics \cite{more20,chio11,plot18,page20,gebb21,rahm24}. A key systematic effect that challenges the use of LMT within an atom interferometer (AI) is the control of AI phase noise from lattice intensity fluctuations. While the efficiency of BO-LMT is greater than that of Bragg-LMT, BOs are more susceptible to phase noise induced by lattice intensity fluctuations \cite{goch19,mcal20,fitz23,rahm24}. High-efficiency LMT without compromising phase stability is desirable in order to push AI sensitivity forward for fundamental physics in TVLBAI and inertial sensing. Here we discuss the use of excited-band BOs in magic-depth lattices for this purpose.

A linear sweep of the relative frequency of the lattice laser beams induces BOs of atoms in the lattice frame, which converts to an acceleration in the laboratory frame. LMT can be achieved in this way through a sequence of several BOs. In a 2-path Mach-Zehnder AI, one path is accelerated relative to the other using this technique and used to prepare LMT beamplitters, mirrors and recombiners for BO-enhanced AI. During the BO process, the non-accelerated arm is in a higher band of the lattice, undergoing Landau-Zener tunneling to the next higher band with each successive BO \cite{rahm24}.

\begin{figure}
  \includegraphics[width=15cm]{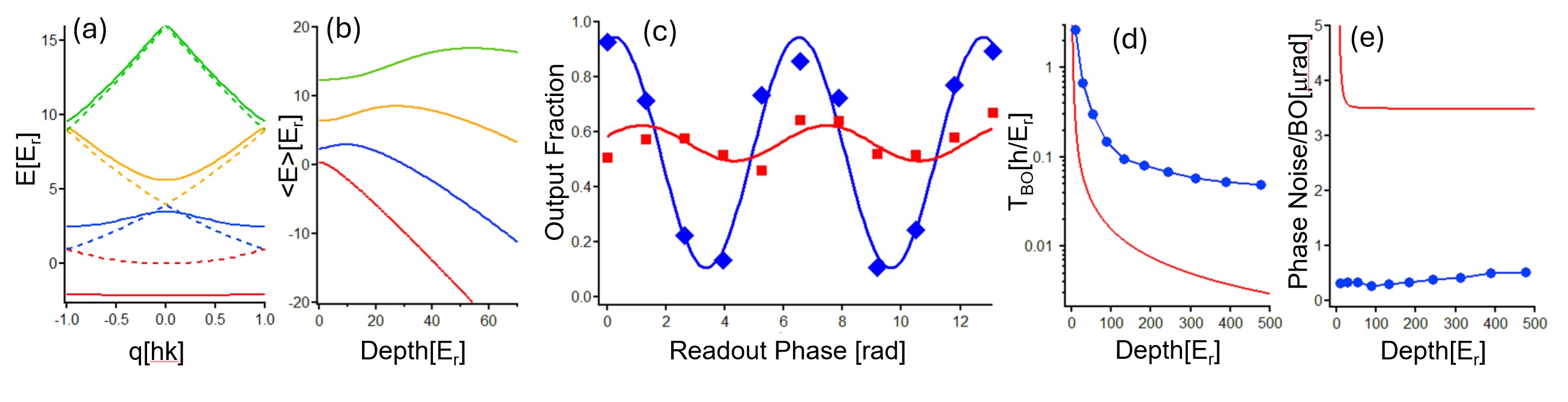}
  \caption{(a) Bloch bands (solid lines) for a sinusoidal optical lattice with a representative depth of $U_0 = 10 E_r$. Dotted lines represent the quadratic free-space dispersion. (b) Average energy over one Brillouin zone of the ground and first two excited bands. The magic depth for each excited band is at its respective local extremum. The ground band does not exhibit any magic depth feature. (c) Observed Mach-Zehnder AI signals with single-BO-acceleration applied at the magic depth for $b=1$ (blue) and for a similar depth for $b=0$ (d) Calculated time per BO ($T_{\rm BO}$) in units of the inverse recoil ($272\,\mu$s for Yb) for operation at 99.9\% efficiency as a function of depth for ground band (red curve) and at the magic depth for excited bands $b$ = 1 to 10 (blue markers joined by lines). (e) Calculated phase noise from 0.5\% intensity noise for the ground band (red curve) and at the magic depth for excited bands $b$ = 1 to 10 (blue markers joined by lines), evaluated at the parameters of (d).}
  \label{gupta_fig}
\end{figure}

The magic depth idea (named in analogy to the magic wavelength for optical clocks) is illustrated in Fig.\,\ref{gupta_fig}(a) and (b), which show the Bloch bands of a sinusoidal lattice and the average band energies $\langle E\rangle$ as a function of lattice depth respectively. Therein $E_r$ is the recoil energy, $q$ is quasi-momentum, $U_0$ is peak lattice depth, and $b$ is band number ($b=0$ is ground, $b=1$ is first-excited etc.). Notably, {\it only} for excited bands does $\langle E\rangle$ feature a local extremum at a particular ($b$-dependent) magic depth $U_{\rm md}$. $\langle E\rangle/\hbar$ corresponds to the rate at which BOs will impart phases on atoms in a particular band. In the presence of unavoidable laser intensity noise (i.e, noise in $U_0$), the BO phase accumulation rate will vary in an uncontrolled way over time, leading to phase noise in a BO-enhanced AI. Unlike for the $b=0$ case, BOs in $b>0$ when performed at $U_{\rm md}$, are immune to intensity fluctuations to first-order. This effect was demonstrated in a small-area horizontal Mach-Zehnder AI in \cite{mcal20} [also see Fig.\,\ref{gupta_fig}(c)]. As can be seen, the AI signal from using BO in $b=1$ at $U_{\rm md}$ (where the subscript indicates the magic depth) is far superior to that from  BO in $b=0$, with 7 times greater visibility.

The residual AI phase noise $\delta \phi_L$ at $U_{\rm md}$ operation stems from weak second-order sensitivity to lattice intensity noise: 
\begin{equation}
{\delta \phi}_L = \frac{1}{2} \frac{\partial^2\langle E\rangle}{\partial U_0^2}\bigg\rvert_{U_{\rm md}} U_{\rm md}^2 \frac{T_{\rm BO}}{\hbar} \epsilon^2
\label{eq:gupta}
\end{equation}
for each BO process. For each magic depth value $U_{\rm md}$ for bands 1 to 10, we plot in Fig.\,\ref{gupta_fig}(d) the calculated value of the period of the Bloch oscillation, $T_{\rm BO}$, for which the per-BO efficiency is 99.9\%. Using these values and the calculated curvature at $U_{\rm md}$ \cite{mcal20} in Eq.\,(\ref{eq:gupta}), we can evaluate the phase noise for a standard 0.5\% level of intensity stability shown in Fig.\,\ref{gupta_fig}(e). We see that 150 mrad phase stability is possible at 1000 recoils (500 BOs). The corresponding analysis for ground-band $b=0$ [solid red lines in Fig.\,\ref{gupta_fig}(d),(e)], show more than an order of magnitude worse phase noise from lattice intensity fluctuations. We note here that we have taken 99.9\% per-BO efficiency as a practical value given other decoherence processes such as spontaneous scattering. We also note that while ground-band BO is clearly faster than excited-band BO [Fig.\,\ref{gupta_fig}(d)], the phase noise disparity is the more important metric [shown in Fig.\,\ref{gupta_fig}(e)], since the interferometer interrogation time for TVLBAI will be much larger than the LMT acceleration and deceleration times. As a practical example, using available laser powers and detuning of $5 \times 10^4$ linewidths from the 556nm intercombination transition in Yb, the use of magic-depth BOs in band $b = 5$ at the corresponding magic depth $U_{\rm md} = 132E_r$ keeps the spontaneous scattering rate below the 0.1\% level per BO.


\section{Atom sources: Scaling atom number and temperature}
\label{sources}

\subsection{Introduction}
\label{SourcesIntroduction}

Long-baseline atom interferometers require \textit{rapid and repeatable delivery} of ultracold atoms from atom source to interferometry tube so as not to limit the repetition rate of what is a fundamentally discrete, rather than continuous, measurement. Besides quasi-continuous modes of operation, rapid sources are likewise required to support interleaved measurement cycles.
In the absence of squeezing to introduce correlations between particles (see Section~\ref{sec:squeezing}) the standard quantum limit associated with quantum projection noise scales as $\Delta \phi_{\rm SQL}= 1/\sqrt{N}$, with $N$ the atom number, while squeezing in principle makes the Heisenberg limit scaling as $\Delta \phi_{\rm H}= 1/N$ accessible. A \textit{large atom number} is required to minimise the noise in measuring the interferometer phase. 
Furthermore, a \textit{narrow thermal momentum distribution} is necessary in order to maintain high-fidelity atom-optics by minimising the expansion of the atom cloud in a Gaussian laser intensity profile and by undercutting velocity selectivity due to Doppler shifts.

In practice, these requirements have complex interdependencies; for example, the phase resolution gains with atom number are in tension with the increased difficulty in implementing squeezing with an increased atom number. A large atom number in a fermionic species will necessitate greater limits on the atom cloud distribution due to the Pauli exclusion principle, thus making matter-wave collimation more challenging. 
However, these are the requirements that motivate the atom source design and share certain commonalities with, and key differences from, ultracold atom platforms for quantum simulation or quantum information processing. 
Techniques vary with atomic species, whose properties determine the handles and limits available, for example collisional properties, ability to tune interactions using a Feshbach resonance, or the contrasting nature of bosons and fermions on reaching quantum degeneracy. 

The four contributions in this Section discuss different aspects of this scaling challenge, covering a range of novel approaches to the solution that are applicable to a variety of experimental platforms. Section~\ref{Bennetts} describes a continuous, high-flux source of strontium atoms, and Section~\ref{Aidelsburger} addresses the challenge of efficient delivery of atoms from atom source to science chamber in the context of a quantum simulation experiment. Section~\ref{Herbst} then considers atom-optics techniques to narrow the velocity distribution of an atom cloud, with control over the shape of the optical potential and taking advantage of the readily-tuneable scattering length of potassium atoms. Finally, Section~\ref{Lan} describes rapid cooling of atoms as a precursor to atom interferometry by electromagnetically-induced transparency in an optical lattice.

\subsection{{Continuous high-flux sources of ultracold strontium atoms}}
\label{Bennetts}

While measurements are a fundamentally discrete process, a continuous source offers arbitrary choice of bandwidth and a detection waveform unconstrained by the repetition rate of the available source. Continuous sources can eliminate sensor dead time and the Dick effect~\cite{Dick1987LOI} that can alias in noise around the measurement frequency. For these reasons, continuous sources of ultracold atoms can offer important advantages for quantum sensors like atom interferometry and continuous optical clocks~\cite{Katori2024_CWBeaminMagicwavelength, Takeuchi_2023, Chen2005ActiveClock, Holland2009PRLMeiser,Cline2022SRL}. Coherent, high brightness matter wave sources based on quantum gases like Bose-Einstein Condensates (BECs) and atom lasers~\cite{Robins2013RevAtomLaser} offer the further advantages of minimising beam expansion, which reduces wavefront aberrations and improves the accuracy of an atom interferometer's pulse sequences. Finally, at the heart of any quantum sensor is the need for high flux to minimise projection noise. Here, we will briefly describe our recent work that demonstrated continuous Bose-Einstein condensation~\cite{Chen2022CWBEC} and review the prospects for demonstrating high-flux continuous atom laser sources in the coming years.

\begin{figure*}[tb]
\centerline{\includegraphics[width=0.95\textwidth]{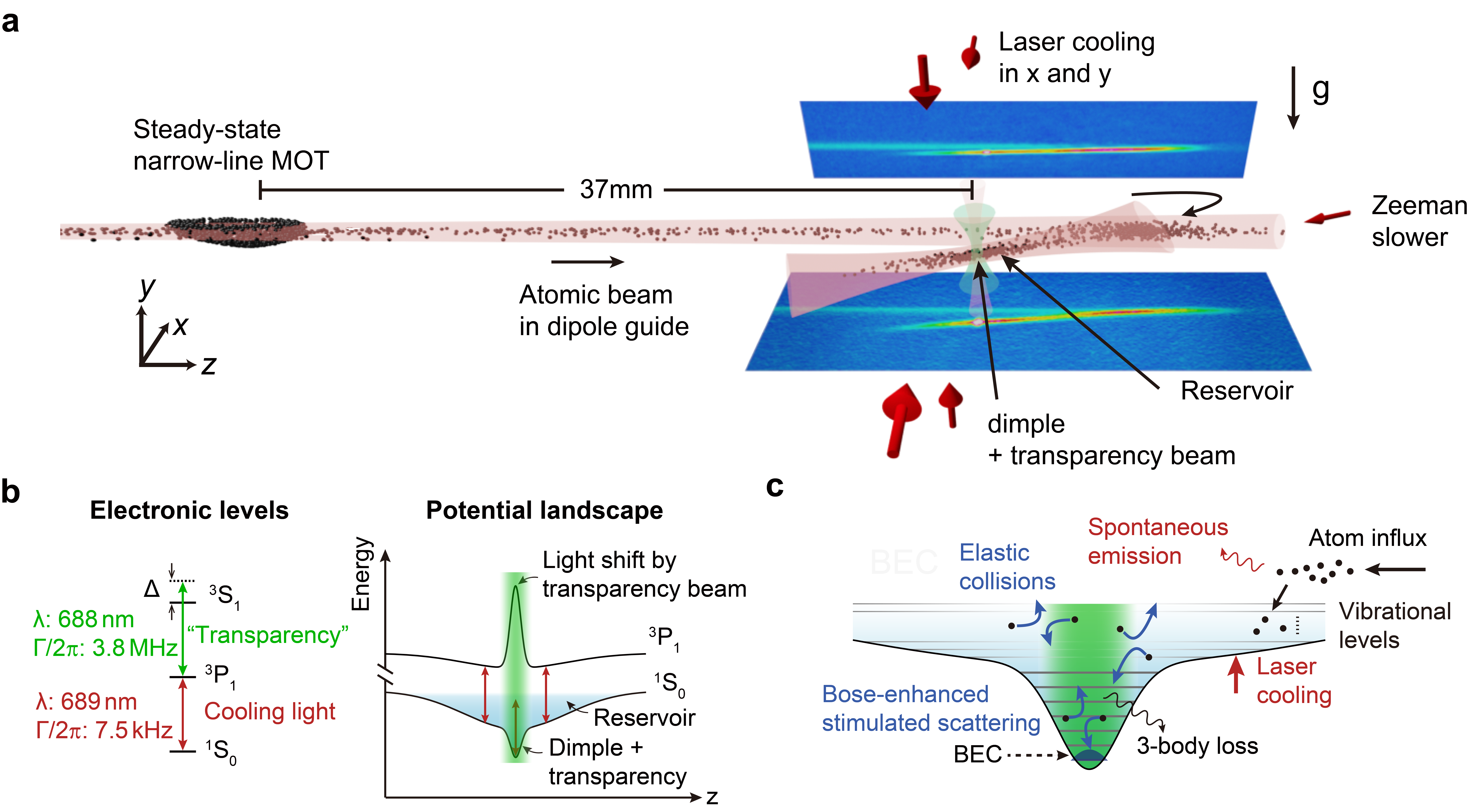}}
\caption{\label{fig:Fig_CWBEC_architecture}\textbar \, Schematic drawing of an apparatus demonstrating continuous Bose-Einstein condensation (a) $^{84}$Sr atoms from a steady-state narrow-line magneto-optical trap (MOT) are continuously out-coupled into a guide and loaded into a crossed-beam dipole trap that forms a large reservoir with a small, deep dimple. Atoms accumulate in the laser-cooled reservoir and densely populate the dimple, where a BEC forms in steady state. (b) By off-resonantly addressing the ${^3\mathrm{P}_1} - {^3\mathrm{S}_1}$~transition using a ``transparency" laser beam, we produce a strong spatially-varying light shift on the $^3\mathrm{P}_1$ electronic state, rendering atoms locally transparent to laser cooling photons addressing the ${^1\mathrm{S}_0} - {^3\mathrm{P}_1}$~transition. This enables condensation in the protected dimple region. (c) Schematic of the potential landscape from both reservoir and dimple trap, and of the dominant mechanisms leading to BEC atom gain and loss. Figures adapted from~\cite{Chen2022CWBEC,Chen2023CWBECRev}.
}
\label{fig:CWBEC}
\end{figure*}

The production of quantum gases has traditionally relied on executing a sequence of cooling steps distributed in time, beginning with laser cooling and concluding with evaporative cooling. This was necessary since the phase-space density achievable by laser cooling alone was limited by effects such as multiple scattering. The result was a practical incompatibility between the laser cooling needed to gather and cool large clouds of atoms and the evaporative cooling needed to reach degeneracy. In recent years, several techniques have been demonstrated (including the electromagnetically-induced transparency (EIT) technique described below) that enable laser cooling to degeneracy~\cite{Stellmer2013LaserCoolingToBEC, Vuletic2017LaserCoolingToBEC, Urvoy2019}. When combined with the distribution of cooling stages over space instead of time, continuous cooling to degeneracy and continuous atom lasers become possible.

In the approach illustrated in Fig.~\ref{fig:CWBEC}, we continuously cool strontium using multiple cooling stages distributed in space. We begin with an oven-based thermal beam source, transverse cooling and a Zeeman slower after which $2.6\times10^9$ $^{88}$Sr atoms/s are gathered by a 2D magneto-optical trap (MOT). These all operate on the 30~MHz broad linewidth ${^1\mathrm{S}_0} - {^1\mathrm{P}_1}$ transition. The beam of atoms from the 2D MOT is then transversally cooled using the 7.5~kHz narrow linewidth ${^1\mathrm{S}_0} -- {^3\mathrm{P}_1}$~transition and falls to a second chamber protected from blue resonant light. Here, $5.1\times10^8$ $^{88}$ Sr atoms/s are loaded into a steady-state red MOT using just the narrow 7.5~kHz $^1\mathrm{S}_0$ - $^3\mathrm{P}_1$ transition~\cite{Bennetts2017SSMOTHighPSD}. Next, atoms from the red MOT are continuously loaded and pushed along a dipole trap guide to create a high phase-space density guided atomic beam with fluxes of $3\times 10^{7}$ $^{88}$Sr atoms/s~\citep{Chen2019Beam} and a phase-space density exceeding 10$^{-4}$. In this way, atoms are continuously transported to a dipole trap where they are collected and further laser cooled. Within this dipole trap a dimple trap accumulates atoms. Here, a light shift of around 4~MHz~\cite{Stellmer2013LaserCoolingToBEC} from a ``transparency'' beam operating 33~GHz blue detuned from the ${^3\mathrm{P}_1} - {^3\mathrm{S}_1}$ transition creates a region protected from scattered resonant light where atoms can condense.

In order to achieve degeneracy we switch to operate using the $^{84}$Sr isotope (natural abundance 0.6\%), which has a scattering length of around 124 a$_0$ that is ideal for thermalisation. Operating at a higher oven temperature, a guided atomic beam flux of $8.6\times 10^{6}$ $^{84}$Sr atoms/s can be achieved. A BEC of $7.4\times 10^3$ atoms is then formed within the dimple trap that can be maintained in steady-state indefinitely. This represents a continuous atom laser, albeit one with no out-coupling. By perturbing the system we can estimate the gain and loss from the BEC giving an average flux of around $2.4\times10^5$ atoms/s being condensed~\cite{Chen2022CWBEC}. In steady-state this gain is balanced by losses dominated by three-body loss. If efficiently out-coupled into a continuous atom laser one might expect a flux of around $10^5$ atoms/s.

Looking ahead, reducing the temperature of the thermal cloud that provides gain to the BEC from the \SI{1.1}{\micro\kelvin} achieved in this system could dramatically improve both the steady-state BEC atom number and its purity. A range of options exist for achieving this including EIT, Raman cooling or Sisyphus cooling~\cite{Chen2024Sisyphus}. Out-coupling a practical continuous atom laser might be achieved using a coherent three-photon transfer of condensed atoms to an untrapped $^3\mathrm{P}_0$ state for which the first steps have already been demonstrated~\cite{He2024Outcoupling, carman2024collinear}. To scale up the flux of not just continuous atom lasers but atom sources in general a number of options present themselves. A new highly efficient Zeeman slower design shows potential for significant improvement~\cite{Feng2023} along with buffer-gas based sources~\cite{Doyle2012BufferGas}. Combining these approaches, not only are continuous atom lasers within reach but also scaling atom lasers to condensed fluxes of $10^7$ atoms/s seems possible. 

\subsection{{Quantum simulation – Engineering \& understanding quantum systems atom-by-atom}}
\label{Aidelsburger}


\begin{figure*}[t]
\centering
\includegraphics{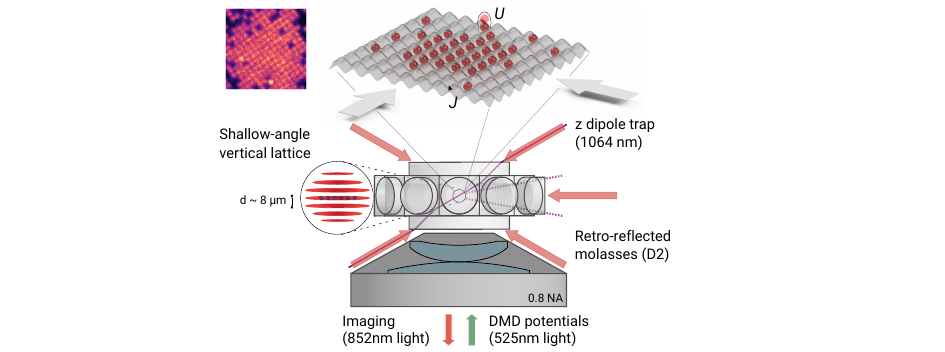}
\caption{Schematic drawing of an illustrative quantum gas microscope setup for $^{133}$Cs atoms. Atoms are confined in a single node of a vertical lattice and a square optical lattice with the Hubbard parameters tunnel coupling $J$ and on-site interaction $U$. For fluorescence imaging near-detuned molasses laser beams at $852\,$nm are employed which simultaneously cool the atoms. The scattered photons are collected with a high numerical aperture (NA) objective. Potential shaping is performed using incoherent light at $525\,$nm and a digital micromirror device (DMD). Upper left: fluorescence image of $^{133}$Cs atoms trapped in a square optical lattice with constant $767\,$nm. Figure adapted from~\cite{impertro_unsupervised_2023}.}
\label{fig:QGM}
\end{figure*}

Increasing repetition rates is crucial for state-of-the-art quantum simulators based on neutral atoms in optical lattices, which directly sample from quantum many-body wavefunctions using high-resolution imaging techniques as employed in quantum gas microscopes~\cite{bakr_probing_2010,sherson_single-atom-resolved_2010,cheuk_quantum-gas_2015,haller_single-atom_2015,parsons_site-resolved_2015,gross_quantum_2021}, as illustrated in Fig.~\ref{fig:QGM}. In these experimental platforms low-entropy initial states are typically generated by employing several different cooling and preparation stages. After laser cooling in a magneto-optical trap, quantum degenerate gases are realized via evaporation in optical dipole traps. In order to generate large arrays of homogeneous filling, the atoms are then adiabatically transferred into an optical lattice while gradually increasing the on-site Hubbard interaction to cross the Mott transition~\cite{greiner_quantum_2002,bakr_probing_2010,sherson_single-atom-resolved_2010}, where fluctuations of the on-site density are strongly suppressed. Using high-resolution imaging techniques the occupation of particles in the lattice can be detected and addressed~\cite{weitenberg_single-spin_2011} after preparing the atoms in a single node of a vertical optical lattice using fluorescence imaging~\cite{gross_quantum_2021}. Moreover, using these imaging optics in reverse allows for high-fidelity shaping of the confining potential on top of the periodic lattice. State-of-the-art quantum gas microscopes are now able to realize high-quality homogeneous lattices with several thousand atoms by projecting box potentials into the atomic plane and by compensating the residual harmonic confinement introduced by the Gaussian profile of the lattice laser beams~\cite{yang_cooling_2020,wei_quantum_2022,impertro_unsupervised_2023,wienand_emergence_2023}. By sampling snapshots from quantum many-body wavefunctions higher-order correlation functions~\cite{kaufman_quantum_2016,lukin_probing_2019,cheneau_light-cone-like_2012,zheng_efficiently_2022} and full counting statistics~\cite{wei_quantum_2022,wienand_emergence_2023} can be accessed which provides invaluable information on properties such as entanglement or quantum transport. More recently, these techniques have been extended to access local orbital operators, such as currents or kinetic energy, by projecting the lattice onto isolated dimers~\cite{impertro_local_2024}.

\begin{figure*}[t]
\centering
\includegraphics{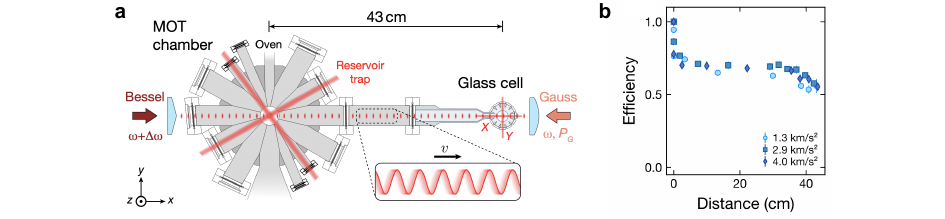}
\caption{(a) Schematic drawing of an illustrative two-chamber vacuum systems: the chamber on the left is used for pre-cooling atoms in a magneto-optical trap (MOT), while the second one (glass cell) is used for quantum simulation experiments, which require large optical access. Atoms are transported between the two sections using a running-wave optical lattice that is generated by interfering a Bessel-type and a Gaussian laser beam. The velocity $v$ and the acceleration are controlled by the relative frequency detuning $\Delta\omega$. 
(b) Round-trip transport efficiency measured as a function of the acceleration, consistent with a one-way transport efficiency of $\sim 75\%$ for the full distance of $43\,$cm.
Figure adapted from~\cite{klostermann_fast_2022}.}
\label{fig:transport}
\end{figure*}

The level of control required for high-fidelity quantum simulation experiments demands excellent optical access and control over the electromagnetic field environment. This often necessitates a two-chamber design, where the science chamber is spatially separated from the oven and pre-cooling stages. Such a two-chamber design requires efficient transport of the atoms over relatively large distances, as illustrated in Fig.~\ref{fig:transport}a. Moreover, it is absolutely crucial to maintain fast cycle times and to develop robust schemes that do not limit the already considerable complexity of optical-lattice based quantum simulators. Optical transport offers many advantages over other schemes that rely on magnetic traps~\cite{greiner_magnetic_2001,lewandowski_observation_2002,pertot_versatile_2009}, since it can be realized independent of the atomic species or molecule used in the apparatus. However, optical traps are typically shallow and transporting atoms in a tigthly-focused optical dipole trap, whose focus position can be controlled dynamically either requires mechanically-movable mounts~\cite{gross_all-optical_2016,couvert_optimal_2008}, which are susceptible to noise, or focus-tunable lenses~\cite{leonard_optical_2014,unnikrishnan_long_2021}. The transport duration is then fundamentally limited by the longitudinal trap frequency. Running-wave optical lattices can offer a convenient solution~\cite{schrader_optical_2001,schmid_long_2006,klostermann_fast_2022}. In these schemes, two counterpropagating laser beams interfere and form a standing wave. Controlling the relative frequency between the two beams enables programmable motion that does not rely on any mechanical parts and can be extremely fast due to the tight confining frequency in the lattice wells along the direction of motion. In Ref.~\cite{klostermann_fast_2022} the running-wave lattice was realized by interfering a Bessel and a Gaussian laser beam (see Fig.~\ref{fig:transport}a) and fast transport of heavy $^{133}$Cs atoms was demonstrated over a distance of $43\,$cm in less than $30\,$ms. In this scheme the Bessel beam acts as a waveguide that holds the atoms against gravity and its interference with the weaker Gaussian beam generates the dynamically-controllable lattice. In this work final velocities of up to $26.6\,$m/s have been achieved and the one-way transport efficiency was evaluated to be $\sim75\%$ (see Fig.~\ref{fig:transport}b). Note, that these efficiencies have been obtained with simple linear frequency ramps leaving room for further improvements by minimizing sudden changes in the acceleration. Moreover, the scheme proved to be extremely robust, which was demonstrated by investigating the stability of a Bose-Einstein condensate over the course of several hours.

\subsection{Matter-wave collimation to picokelvin energies with scattering length and potential shape control}
\label{Herbst}
The instability of Mach-Zehnder atom interferometers~\cite{Kasevich1991,Peters2001} benefits from performing measurements on atomic clouds with large particle numbers at high repetition rates.
Quantum-degenerate gases are ideal candidates for TVLBAI devices, due to their low expansion energy, which minimizes systematic errors related to kinematics and extends pulse separation time without losing contrast~\cite{schlippert2020matter,Hensel2021}. 
However, their preparation through evaporative cooling is time-consuming and current proposals require further collimation to picokelvin energies to meet experimental constraints and requirements~\cite{Hartwig2015,Canuel2020,Badurina2020,Ahlers2022}.
Delta-kick collimation~\cite{Ammann1997} allows access to this regime, but often relies on extended free-fall times before applying the matter-wave lens to minimize atomic interactions that would otherwise drive the expansion~\cite{Kovachy2015, Deppner2021, Gaaloul2022}. 
We present an alternative approach to these challenges using time-averaged optical potentials (TAOPs)~\cite{Roy2016} to control trap frequencies in combination with Feshbach resonances to tailor the atomic scattering length~\cite{Inouye1998}. 
Our method significantly reduces the timescale necessary for evaporative cooling while also achieving state-of-the-art matter-wave collimation with minimal initial free-fall time.  

TAOPs enable the creation of nearly arbitrary potential shapes by applying an rf-modulation to an acousto-optics element.
In a harmonic trap, controlling the potential width allows decoupling trap frequencies and depths, which are linked via the beam waist in standard optical dipole traps (ODTs).
We can create a trap with 65 $\mu$K trap depth and an effective beam waist of 1.4 mm, suitable for loading more than $2\times 10^7$ atoms directly from a D$_1$ grey molasses~\cite{Salomon2013} at 6 $\mu$K.
For our crossed ODT, this configuration requires 16 W optical power per ODT beam at a wavelength of 1064 nm. 
By continuously compressing the trap during evaporative cooling we maintain high trap frequencies, significantly enhancing the evaporation rate~\cite{Roy2016}. 
Furthermore, controlling the scattering length is particularly beneficial for species like $^{39}$K, whose Feshbach resonances can be easily accessed with low magnetic fields~\cite{DErrico2007}, and also offers exciting prospects for species like Sr and Yb with optical resonances using the forbidden intercombination transition~\cite{Enomoto2008, Yan2013}. 
For $^{39}$K we demonstrated the formation of a pure condensate of $6\times10^4$ atoms after only $170$ ms  of evaporative cooling by maintaining a constant trap frequency of $2\pi\times 300$ Hz, while exponentially decreasing the scattering length from $2000\, \text{a}_0$ to $300\, \text{a}_0$, to control three-body losses~\cite{Herbst2024a}.
By optimizing parameters for more efficient evaporation ramps at the expense of the evaporation duration, we achieved BECs of up to $6\times10^5$ particles within $2$ s of evaporative cooling, maintaining the evaporation flux.

To further reduce the expansion energy of the atomic ensemble we use a continuous delta-kick collimation technique directly in the trapping potential.
Similar methods have been successfully implemented with magnetic traps~\cite{Dickerson2013,Kovachy2015,Pandey2021} 
and, in conjunction with a pulsed delta-kick after release, have yielded a record value of $38$ pK in micro-gravity~\cite{Deppner2021}.
In our experiment we induced common mode oscillations of the atomic ensemble's size by lowering the trap frequencies through rapid trap relaxation. 
A well-timed release from the trap at the oscillations' turning point then allows one to obtain an enlarged ensemble with minimal momentum spread along the lensed directions in free fall.
Initially developed with $^{87}$Rb, this method can easily be applied in any energy regime accessible with the ODT, allowing for short-cutting evaporative cooling without atom loss when only certain energy limits must be met~\cite{Albers2022}.
For a BEC, collimation can be further improved by tailoring the scattering length with respect to trap frequency ratio. 
While a minimal scattering length reduces the repulsive force upon release, it also diminishes the excitation amplitude in the trap.
Therefore, the optimal value is a trade-off between these effects and, for higher trap frequency ratios, the optimal scattering length shifts to higher values.
Experimentally, we found the lowest expansion energy in the lensed 2-dimensional plane to be $438 \pm 77$ pK at $10\,\text{a}_0$ in the weak interaction regime. 
This value corresponds to an improvement by more than a factor of two compared to a measurement conducted in the strong interaction regime at $158\,\text{a}_0$~\cite{Herbst2024b}.
For an improved experimental configuration with larger trap frequency ratios and an additional pulsed delta-kick to collimate the vertical axes 25 ms after release, simulations predict a final expansion energy of below 20 pK in three dimensions. 
Finally, we envision energies below $100$ pK, when applying our method to $^{87}$Rb without the use of tuneable interactions.

\subsection{Fast formation of a quantum gas for atom interferometers}
\label{Lan}

Quantum gases, such as Bose-Einstein condensates (BECs), are essential as coherent matter waves in highly sensitive atom interferometers. The traditional method for preparing quantum gases relies on evaporative cooling, which requires specific collisional properties of atoms and typically takes several seconds. This method also suffers from significant atom loss, limiting its efficiency. Enhancing the speed and efficiency of quantum gas preparation would substantially improve the performance of atom interferometers.

A thermal atomic cloud enters the quantum statistical regime when its phase space density, $n\lambda^3$, approaches unity, where $n$ is the atomic density and $\lambda$ is the de Broglie wavelength. Achieving this regime can be ideally realized by unit filling of atoms in a three-dimensional (3D) optical lattice, where the atomic wavefunction matches the size of each lattice site. A recent study by Xin et al.~\cite{xin2024fast} demonstrates that combining a 3D optical lattice with EIT cooling enables the gas to reach the quantum regime within just 10 milliseconds after sub-Doppler cooling.

The process of unit filling in the lattice is achieved by sequentially switching off lattice beams in different directions, forming one-dimensional (1D) tube traps or two-dimensional (2D) pancake traps. Atoms oscillate within these traps, and the full 3D lattice is reactivated when the atoms converge towards the trap center. This results in a fivefold increase in atomic density to approximately $10^{13}$ cm$^{-3}$, matching the density of the lattice sites. Following each aggregation, a 1 ms EIT cooling step is applied, with 62\% of the atoms achieving the 3D ground state. The EIT cooling process involves two coherent Raman beams—control and cooling—coupled to a three-level atomic system. The control beam tailors the absorption spectrum of the cooling beam, strongly suppressing heating transitions in the sideband cooling \cite{Mor00, Hua21}.

In the final stage, the atomic wave function is expanded by gradually reducing the lattice potential, achieving an effective temperature of 137 nK. With lattice beam waists of 60 µm, 60 µm, and 140 µm, approximately $6\times10^5$ atoms are in the optical lattice, retaining 95\% of the initial atom number that was loaded. This method is versatile, allowing for the production of large atom number condensates or achieving even lower temperatures. For example, with an atomic density of $10^{12}$ cm$^{-3}$ after sub-Doppler cooling, a BEC containing $10^9$ atoms can be achieved using lattice beams with 1 mm waists. Larger lattice spacings can yield even lower temperatures. Loading the atoms into a tailored optical potential can also enable the realization of a trapped atomic gyroscope. Additionally, after adiabatic expansion, atoms can be loaded into a single dipole trap and further cooled using delta-kick cooling, producing ultracold temperatures suitable for free-space atom interferometers. Notably, the entire cooling process can take less than 10 ms, allowing for the use of near-detuned lattice beams with reduced power consumption.

\section{Squeezing and multipartite entanglement for atom interferometry}\label{sec:squeezing}

 
 

\subsection{Introduction}
\label{SqueezingIntroduction}


Section~\ref{sec:compiledintro} introduced the science motivations for exploring squeezed states in order to overcome the Standard Quantum Limit (SQL) corresponding to $\Delta \phi_{\rm SQL}= 1/\sqrt{N}$ for $N$ uncorrelated atoms entering the interferometer and exploiting quantum entanglement to improve this scaling with atom number~\cite{PhysRevLett.96.010401}, possibly as far as the Heisenberg limit $\Delta \phi_{\rm H}= 1/N$. While attaining the Heisenberg limit in ensembles of $10^3-10^6$ atoms is technically challenging~\cite{RMPPezze18}, the so-called spin-squeezed states~\cite{Kitagawa93} offer significant metrological gain even for large $N$ while being relatively robust against losses and decoherence. Reduced quantum uncertainties of up to a factor 100 (20 dB) in variance relative to the SQL have been observed and various methods exist for their deterministic or conditional preparation~\cite{Hosten2016,Cox2016}. A gain of 100 in a complete interferometer would be of great practical significance since this would mean that either the sensor would reach the required precision 100 times faster or that 100 times fewer atoms (i.e., smaller densities) are needed. The demonstrated possibility of using squeezed states for interrogation times up to 1 second~\cite{huang2023} has further strengthened the technological relevance for real-life sensors.

While the preparation of spin-squeezed states has been very successful in recent years, little metrological gain has been attained in complete interferometric sequences. In atomic clocks, for example, local oscillator noise puts stringent limits on the achievable gain. Other effects such as non-uniform entanglement or environmental fluctuations (e.g., magnetic fields) have also been identified as problematic in this context. 

In the following contributions we explore some of the significant progress towards addressing these issues in atom interferometers with separated arms. This setting certainly requires the development of special techniques since, for example, inhomogeneities play a more important role here than for, e.g., optically-trapped atoms in an atomic clock.

The following Sections describe different methods of squeezing that could be compatible with atom interferometry in a freely-falling cloud.

\subsection{Progress towards a squeezed-state atom interferometer in a ring cavity}
\label{Hosten}

There has been growing interest in the last few years in translating the achieved spin-squeezing results into interferometers with spatial arm separations to allow for similar improvements for inertial sensing ~\cite{szigeti2021}, e.g., for navigation, gravimetry, or for bridging the frequency gap between planned and existing gravitational wave detectors. Initial experiments have already begun to yield results~\cite{malia2022,Greve2022,Cassens24}, with reference~\cite{Cassens24}  sensing a gravitational acceleration. Nevertheless, efforts are still at an early phase, with all results showing less than 2 dB of metrological gain. It remains to be seen if technologically exciting levels of entanglement enhancement can also be obtained for inertial atomic sensors.

To this end, we have been developing a squeezed-state atom interferometer inside an optical ring cavity, where the atomic wavepacket splitting and combining take place along the optical axis in a travelling-wave dipole trap established by a cavity mode. This cavity also enables the generation of squeezing and precision sensing of the atomic states. In addition to the established high levels of squeezing that can be achieved with optical cavities, an important motivation for the specific setup is the intended mitigation of squeezing degradation that follows if a different `atomic mode' than the one that is squeezed is probed during the interferometer sequence~\cite{wu2020}.

\begin{figure*}
\centering
\includegraphics[width=\textwidth]{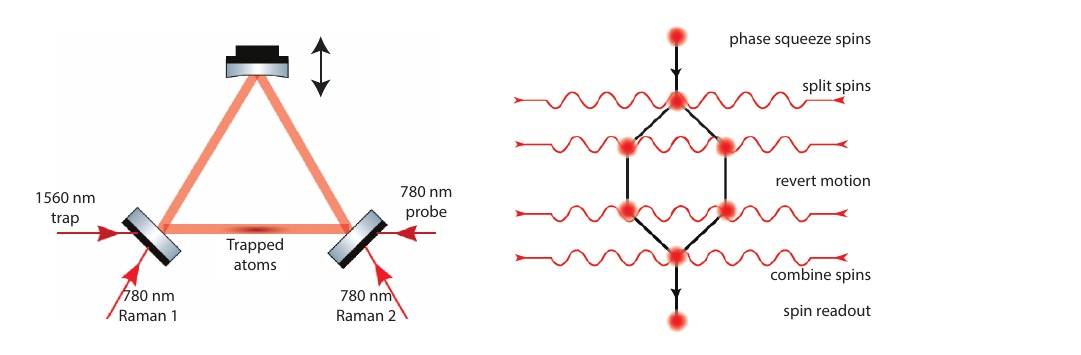}
\caption{The travelling wave cavity and the simplified experimental protocol used in~\cite{Hosten2016}.}
\label{fig:hosten_protocol}
\end{figure*}

The key aspects of our experiment are illustrated in Fig.~\ref{fig:hosten_protocol}. An ensemble of up to 1 million $^{87}$Rb atoms are trapped in a 1560 nm mode of a $\sim$100,000 finesse cavity, which is oriented in the horizontal plane in the lab. The cavity can resonantly support two counter-propagating optical tones at 780 nm, thanks to a mm-range tunable cavity length, allowing real-time cavity free-spectral-range tuning. The two optical tones serve as Raman beams to facilitate mapping of the internal spin states to momentum states. Initially, spin-squeezed states will be prepared utilizing the cavity as in reference~\cite{Hosten2016}. Raman-$\pi$ pulses will then follow to transfer the opposite spins to opposite momentum states separated by 4$\hbar$k, mapping the quantum correlations that were established between the spins to the two arms of an interferometer. Four Raman beam operations split the spin states spatially, and then bring them back, mapping the phase shifts incurred in the two arms to spin degrees of freedom. The phase readout is then performed with established cavity-based spin measurements~\cite{Hosten2016}.

Currently we have $\sim$10 $\mu$K atoms trapped in the cavity mode. An atom interferometric sequence inside this traveling-wave trap and the generation of squeezed states are currently under experimental investigation. Following a successful implementation of each, we aim to demonstrate squeezing-enhanced sensing of an acceleration signal, e.g., due to a tilt of the optical table. Although the specific experimental system has not been built for direct integration into long-baseline interferometers, lessons learned from this system can inform the design principle for effective utilization of squeezed states in such interferometers. Alternatively, it might be possible to utilize the method of releasing the atoms from the cavity after squeezing~\cite{Malia2020} for utilization in long-baseline interferometers.

\subsection{Progress towards a squeezed-state atom interferometer in a linear cavity}
\label{Hobson}

\begin{figure*}
\centering
\includegraphics[width=0.9\textwidth]{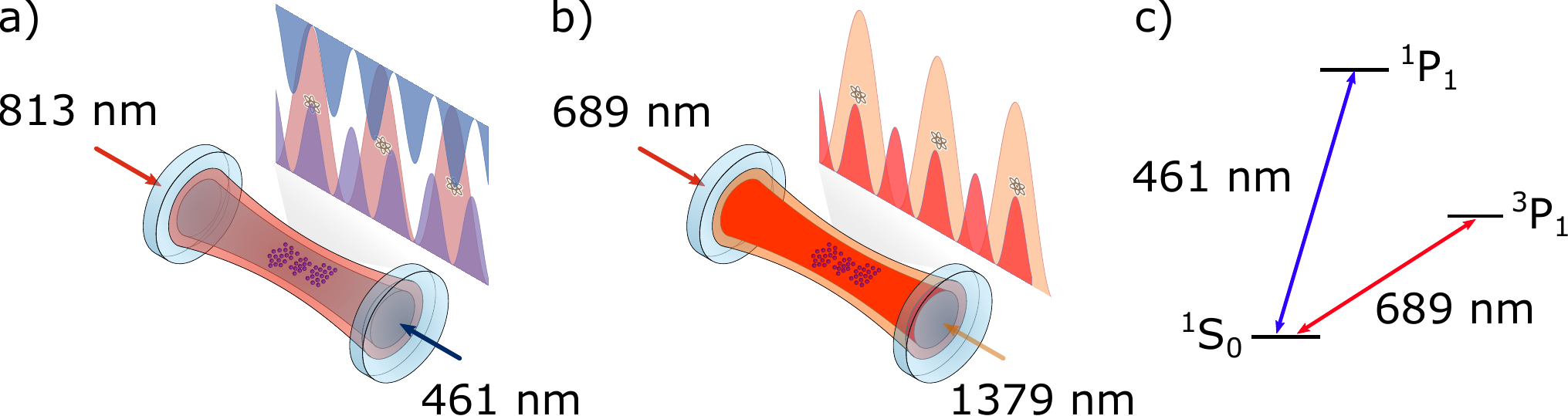}
\caption{Linear cavity configurations for the preparation of squeezed states of Sr. a) In an atomic clock, we used the broad 461~nm transition to carry out quantum nondemolition measurement of Sr trapped in an 813~nm lattice, with the intention to create measurement-based squeezing~\cite{hobson_cavity-enhanced_2019}. Coherence-preserving measurements were achieved~\cite{bowden_improving_2020}, but metrologically useful squeezing was impeded by the probe modes being incommensurate with the lattice sites. b) In planned work, we will use the narrow-line cooling transition at 689~nm to probe Sr atoms in a 1379~nm lattice. Importantly, in the new setup, the atoms are trapped at intensity peaks of the probe, providing near-uniform atom-cavity coupling. 
c) Relevant transitions in Sr.}
\label{fig:Squeezing_linear_cavity}
\end{figure*}

A linear cavity is a relatively simple geometry which has been used in a series of squeezed atomic clocks and atom interferometers, using various atomic species including Rb~\cite{leroux_implementation_2010,Hosten2016}, Sr~\cite{robinson_direct_2022} and Yb~\cite{Braverman2019,Pedrozo-penafiel2020}. However, the simplicity of the linear cavity comes with a compromise: the probe laser field forms a standing wave inside the cavity, typically resulting in non-uniform coupling between the cavity field and the atoms - see Fig.~\ref{fig:Squeezing_linear_cavity}. The non-uniform coupling reduces the effective number of atoms participating in the squeezing process, and can introduce excess anti-squeezing of the atomic state---an important obstacle to metrologically useful states~\cite{braverman_impact_2018}.

A neat trick, which can achieve near-uniform atom-cavity coupling despite the standing-wave cavity field, is to trap the atoms in a series of lattice sites aligned with antinodes of the probe field. An example of such a lattice/probe configuration is depicted in Fig.~\ref{fig:Squeezing_linear_cavity}b. This trick was first applied to Rb using a 1560~nm lattice in combination with a 780~nm probe~\cite{Hosten2016}, in an experiment that established a long-standing record of 20~dB squeezing on the microwave clock states of \textsuperscript{87}Rb.

In the work proposed here, we build on our previous work in cavity-enhanced Sr atomic clocks~\cite{hobson_cavity-enhanced_2019,bowden_improving_2020}, integrating the key innovation of a commensurate 1379~nm lattice and 689~nm probe. In the latest setup, we have so far trapped cold atoms in a 1379~nm lattice, inside a cavity with \num{2e5} finesse at 689~nm, and observed cavity length noise compatible with high-fidelity quantum nondemolition measurements of the atom population in the \textsuperscript{1}S\textsubscript{0} M\textsubscript{F} = 9/2 state in \textsuperscript{87}Sr. In-vacuum coils are installed to apply state rotations between the M\textsubscript{F} states, which should allow the preparation of a squeezed state between a pair of M\textsubscript{F} states before mapping to separate arms of the atom interferometer~\cite{Pedrozo-penafiel2020,malia_distributed_2022}.

In previous experiments, the narrow 689~nm transition in Sr has facilitated very precise cavity-based measurements of atom number in a linear cavity~\cite{norcia_strong_2016}, an important step towards squeezing. However, it remains an open question whether the technical complications of linear cavities can be overcome sufficiently to deploy squeezing in a TVLBAI. Key challenges include the mitigation of non-unitary evolution during the squeezing sequence~\cite{braverman_impact_2018}, the implementation of high-fidelity state rotations, the mapping of squeezing onto momentum-split states, the realisation of fluorescence readout with noise below the standard quantum limit, and the control of the atom-interferometer phase to within the narrow region in which squeezed states perform better than coherent states~\cite{pezze_heisenberg-limited_2020}.

\subsection{Quantum-enhanced BEC interferometry}
\label{Corgier}

Bose Einstein condensates (BECs) have been  pinpointed as ideal candidates to realize entanglement-enhanced free-fall atom interferometry measurements~\cite{Szigeti20, Corgier21b}. Recently, a first proof of principle quantum-enhanced gravimeter has been experimentally demonstrated paving the way for future developments~\cite{Cassens24}. 
For inertial sensors, the major constraints are the production of entangled states, the delocalisation of the entanglement into a superposition of two different well-defined and well-separated external states at the input port~\cite{Anders2021} and the detection of the quantum state at the output port of the interferometer. To fulfill these conditions, the idea proposed in~\cite{Szigeti20, Corgier21a, Corgier21b} involves state preparation based on an interferometer sequence itself to generate a spin-squeezed state through One-Axis Twisting dynamic~\cite{Kitagawa93}.

On the one hand, in free-fall configurations the fast BEC expansion drastically limits the atom-atom interaction time and current theoretical studies predict only modest sub-SQL sensitivity gain~\cite{Szigeti20}. On the other hand, even though trapped configurations allowed for long interaction times and therefore large squeezing strength during the state preparation, the presence of interactions during the interferometer sequence again led to modest sub-SQL sensitivity gain~\cite{Corgier21a}. 
In both cases, the splitting efficiency of the BEC in different external states suffers from atom number fluctuation and therefore density change~\cite{Hartmann2020} shot-to-shot, as well as mode mismatch~\cite{Poulsen02} due to BEC shape deformation~\cite{Burchianti20}. 

\begin{figure*}
\centering
\includegraphics[width=0.9\textwidth]{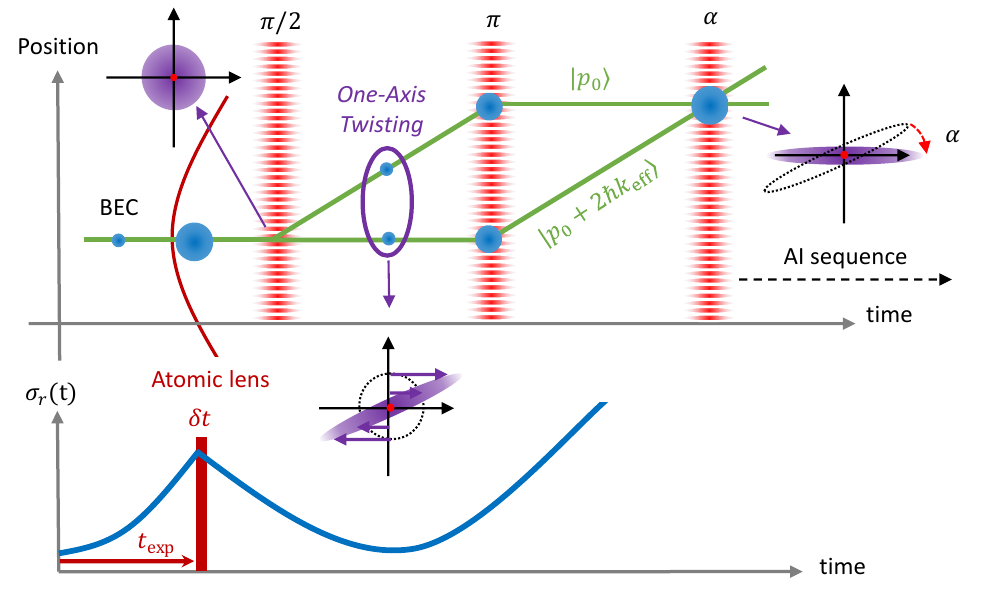}
\caption{Principle of the delta-kick squeezing protocol proposed in Ref.~\cite{Corgier21b}.
Top: step preparation of a spin-squeezed state thanks to the combination of an atomic lens aiming to focus the BEC and a Mach-Zehnder interferometer. 
Bottom: Evolution of the size of the BEC.}
\label{fig:Rcorgier_DKS}
\end{figure*}

Fig.~\ref{fig:Rcorgier_DKS} summarizes the concept behind the method called ``delta-kick squeezing"~\cite{Corgier21b}. 
The idea is to take advantage of the control of the atom-atom interactions in free fall thanks to an external trap, briefly switch on after a preliminary expansion of $t_{\rm exp}$ and switch off after a time $\delta t$, acting on the overall as a lens for the matter-wave. 
This method allows one to focus the BEC at a given time after an initial pre-expansion step. 
Combined with a beam-splitter pulse, it is therefore possible to split a BEC into well-defined and well-separated external states while the BEC is diluted, allowing for optimal splitting efficiency, e.g., without BEC shape deformation and minimizing the impact of shot-to-shot atom number fluctuation. 
The squeezing strength depends on the atomic lens parameters at the time when the lens is switched on, $t_{\rm exp}$, and its duration, $\delta t$. 
The two arms of the interferometer are then deflected and the orientation of the quantum state can be manipulated with a third atom-light interaction used to initialize the interferometer sequence (not shown). 

While with Raman diffraction the squeezing strength can only be positive or null, Bragg diffraction enables one to tune the sign from positive to negative values. 
As such, Bragg diffraction opens the possibility to use a non-linear detection scheme, known to be extremely robust against atom number detection uncertainty, and extend the work of~\cite{Davis16, Hosten2016Science} to BEC interferometry. 

In practice, absolute measurements are often limited by the onset of systematic effects and/or phase noise inherent to the interferometer itself or to the environment. 
As a direct consequence, it can be difficult to apply the opportune rotation of the spin-squeezed state at the input port of the interferometer and to operate the interferometer at the optimal working point, e.g., mid-fringe.
Experiments aiming at precision measurements often benefit from a differential
configuration where common phase noise can be rejected.   
This work directly extends to robust differential measurement schemes where the combination of spin-squeezing with a mode-swapping method allows one to enhance the sensitivity of a differential phase measurement at the expense of the generation of a single two-mode spin-squeezed state\,\cite{Corgier23}.  

\subsection{Experimental atom interferometry with entanglement-enhanced resolution}
\label{Klempt}

The improvement of measurement resolution by entanglement was demonstrated for internal degrees of freedom~\cite{RMPPezze18} and for laser-cooled ensembles~\cite{malia2022,Greve2022}, but an inertial signal was not retrieved.
In a recent work, we have now enhanced the measurement of gravitational accelaration by squeezing for the first time~\cite{Cassens24}.

\begin{figure*}
\centering
\includegraphics[width=\textwidth]{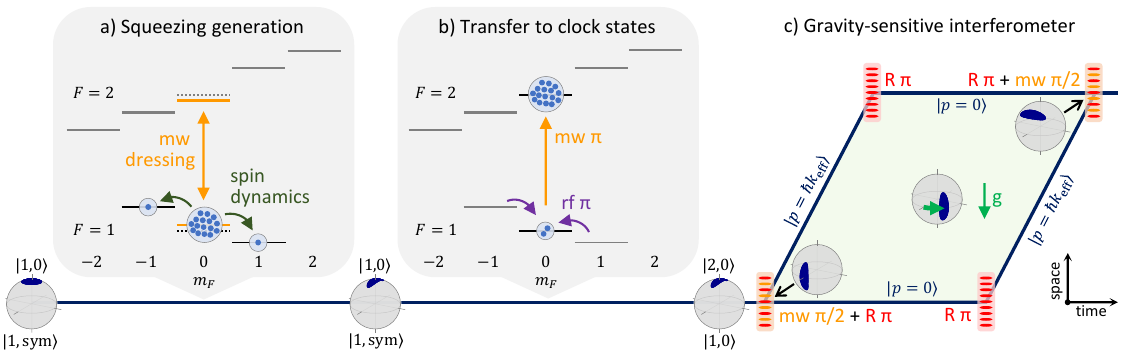}
\caption{Operation of the entanglement-enhanced gravimeter.
(a) A two-mode squeezed vacuum is generated in the $|1,\text{sym}\rangle=\frac{1}{\sqrt{2}}\left( |1,+1\rangle + |1,-1\rangle \right)$ state by spin-mixing dynamics (dark green). This process is activated by a microwave (mw; orange) dressing field that counteracts the quadratic Zeeman shift. (b) Employing circularly-polarized radiofrequency (rf; purple) and microwave pulses, single-mode squeezing is transferred to the magnetically-insensitive clock states. Steps (a) and (b) happen in internal states of the atoms. (c) An interferometric sequence is created by microwave $\pi/2$ pulses. Raman-laser (R; red) $\pi$ pulses transfer $\hbar k_\text{eff}$ momentum and render the interferometer sensitive to gravity.}
\label{fig:EntGrav}
\end{figure*}

In our experiments entanglement is generated in the internal degrees of freedom of $^{87}\text{Rb}$ Bose-Einstein condensates by spin-changing collisions that create pairs of atoms in the $|F=1,m_F=\pm1\rangle$ state from a reservoir of atoms in the $|1,0\rangle$ state (Fig.~\ref{fig:EntGrav}a).
This process is activated by microwave dressing that counteracts the quadratic Zeeman shift.
In this way, a two-mode squeezed vacuum state can be generated that shows reduced fluctuations in a certain observable.
A combination of microwave and circularly-polarized radiofrequency pulses then transforms the two-mode squeezed state into single-mode squeezing in the magnetically insensitive clock states $|F,m_F=0\rangle$ (Fig.~\ref{fig:EntGrav}b).
Subsequently, microwave and Raman-laser pulses form a light-pulse interferometer that enables the measurement of the absolute gravitational acceleration (Fig.~\ref{fig:EntGrav}c).
Using this approach, we achieve a sensitivity of $-3.9^{+0.6}_{-0.7}\,$dB below a coherent-state reference and $-1.7^{+0.4}_
{-0.5}\,$dB below the theoretical standard quantum limit~\cite{Cassens24}.

We have demonstrated that entanglement generation is compatible with delta-kick collimation to reduce the expansion of the atomic cloud.
Furthermore, quantum density fluctuations can be counteracted by appropriate adjustment of the squeezing angle~\cite{feldmann2023}.
The concept of squeezing generation in the internal degrees of freedom and subsequent transfer to inertially-sensitive momentum modes presented here is therefore scalable to large atom numbers and a well-suited improvement for large-scale atom interferometers aiming at the highest precision.
The method is in particular convenient for differential measurements such as gravity gradiometry or gravitational wave detection because in such scenarios common-mode noise contributions like vibrations are cancelled.

A direct extension of this concept would be its demonstration at much longer interferometry times. 
For this purpose, the application in a large-baseline apparatus such as TVLBAI~\cite{lezeik2022} could be planned.
Another scenario of interest is the evaluation of the entanglement-enhancement under microgravity to prepare spaceborne applications. 
This is envisioned in the INTENTAS project that aims at performing entanglement-enhanced interferometry in the Einstein Elevator facility~\cite{lotz2017} under micro-gravity conditions, resulting in seconds of free-fall time and therefore an improved sensitivity by multiple orders of magnitude.

\section{Atom interferometry: Metrology \& Systematics}
\label{Metrology}

\subsection{{Introduction}}
\label{MetrologyIntroduction}

Large-scale interferometers proposed for future gravitational wave (GW) detectors and dark matter (DM) searches demand unprecedented sensitivity and accuracy, surpassing the current capabilities of atom interferometers. As a result, extensive studies are underway to evaluate the limitations imposed by spurious effects and to determine the optimal measurement strategies for these detectors.

In this Section, we examine two systematic effects relevant to these new, larger scales: laser wavefront distortions and the Coriolis effect. These effects are parts of a broader set of systematics that arise with the increase in baseline length and could significantly limit the overall sensitivity of the instrument. Potential mitigation strategies are discussed to address these limitations. Additionally, the extended baseline of these large-scale detectors necessitates a re-evaluation of optimal measurement strategies.

The studies presented here represent preliminary investigations into adapting atom interferometry metrology to this novel regime of very long baseline detectors. It is crucial to understand the limitations of these interferometers in order to refine their design and define the optimal measurement techniques for these new infrastructures.

\subsection{Wave distortion and other systematic effects in high precision atom interferometry}
\label{Distortion}

Atomic interferometry is a tool that enables to measure some fundamental constants with the utmost precision. In order to measure precisely the recoil of an atom absorbing a photon, one can use the Bloch oscillation technique to transfer many recoils to atoms, and an atom interferometer to accurately measure this recoil. From this measurement, one can determine the ratio between Planck's constant and the mass of the atom under study, and this ratio provides a determination of the fine structure constant $\alpha$. Using this technique, it was possible to measure $\alpha$ with a relative precision of 80 ppt \cite{morel2020}.

One of the main limitations of this measurement is the need for precise knowledge of the wavefronts. When lasers interact with atoms, the phase of the laser is added to that of the atom, inducing a recoil proportional to the phase gradient. This gradient is well known in the case of a plane wave and is $k=\omega / c$, with $\omega$ the laser pulsation, but in the case of a real beam, a correction must be made. This effect is common to most atomic interferometers. For a Gaussian beam, the correction is $\delta k/k = 1/k^2w^2$ at the center of the beam, where $w$ is the waist of the Gaussian beam \cite{weiss1994}. In the case of a beam in the paraxial approximation along the $z$ axis, which can be described by its real amplitude A(x, y) and phase $\phi(x, y)$, one can show that the recoil is given by 
\begin{equation}
\frac{\delta k}{k} =
		{-\frac{1}{2}\Big|\Big|
		\frac{\overrightarrow{\nabla}_{\perp}\phi}{k}
		\Big|\Big|^2}
		+ \frac{1}{2k^2}\frac{\Delta_{\perp}A}{A}
\end{equation}

In this formula, the first term in $\overrightarrow{\nabla}_{\perp}\phi$, corresponds to the angle the wavefront makes with the $xy$ plane and therefore to a reduction in the recoil component along the $z$ axis. The second term, involving the Laplacian, corresponds to the Gouy phase term already identified in a Gaussian beam. This term can be positive or negative. When there are amplitude fluctuations, one might think that the effect averages out to zero. However, there is a systematic negative effect in the measurement of $h/m$, linked to a survivor bias: atoms are more likely to survive when the amplitude is high, where on average the Laplacian will be negative~\cite{bade2018}. This effect can be modelled using Monte Carlo simulation. 

The Laplacian effect is all the more important the smaller the characteristic size of the fluctuations. The small-scale quality of the wave is important, and intensity fluctuations must be limited. A simple technique is to let the beam propagate freely. 
We would like to emphasize that while the effect of aberrations has long been studied, it is also necessary to control the quality of the beam on a small scale.

In order to better compare simulations with experiment and measure wavefront defects more accurately, one needs to reduce the size of the cloud at the moment of interaction with the lasers, which requires an initially smaller and cooler cloud. To achieve this, one may use a Bose-Einstein condensate. A preliminary study has shown that this method can be used to see spatial fluctuations in recoil as a function of measurement position \cite{carrez2023}. 

In conclusion, it is important to control small-scale fluctuations of the laser beam intensity. A Bose-Einstein condensate can be used to probe these fluctuations locally and characterize. In this way, one may hope to improve the $h/m$ measurement. It should also be noted that a recoil velocity measurement for which many recoils are transferred gives very good precision on the in situ measurement of the laser wave front. This study therefore offers possibilities beyond the measurement of $h/m$.

\subsection{Coriolis Force Compensation for Long Baseline Atom Interferometry}
\label{Coriolis}

Terrestrial atom interferometers can be highly sensitive to Coriolis forces, which induce phase shifts that scale with the product of the Earth's rotation rate and the initial velocity of the atom. Schemes for suppressing Coriolis-induced phase shifts in interferometers with baseline lengths $\leq 10$m, including counter-rotating a retro-reflecting mirror against Earth's rotation \cite{Dickerson2013, Lan2012} and operating the interferometer in a multi-loop configuration \cite{Dubetsky2006, wang2024robust}, have been demonstrated. However, these schemes break down in longer baseline interferometers owing to atom-laser misalignment. In extreme cases, the interferometer laser beam can miss the atom cloud entirely, but lesser misalignments are also detrimental in part because they lead to reduced Rabi frequencies for the atom-laser interactions. Larger Rabi frequencies increase the efficiency of atom-optics operations and enable shorter pulse durations, which facilitates performing more LMT enhancement pulses in an interferometer sequence. Here we outline a new method for achieving Coriolis force suppression in long baseline atom interferometers which keeps the interferometer beam aligned with the atom cloud (further details can be found in \cite{glick2024coriolis}).

Traditional methods for suppressing Coriolis-induced phase shifts, such as counter-rotating a retro-reflecting mirror \cite{Dickerson2013}, can lead to atom-laser misalignment in long baseline interferometers. In this method, the position of the mirror sets the point about which the reflected beam pivots. Atom-laser misalignment increases with the lever arm between this pivot point and the atom cloud launch point. In longer baseline interferometers, this lever arm may need to be large to accommodate magnetic shielding or lattice launching systems. In a gradiometer configuration, this distance can be much greater than engineering constraints require, resulting in even greater atom-laser misalignment.

\begin{figure}
\centering
\includegraphics[width=6in]{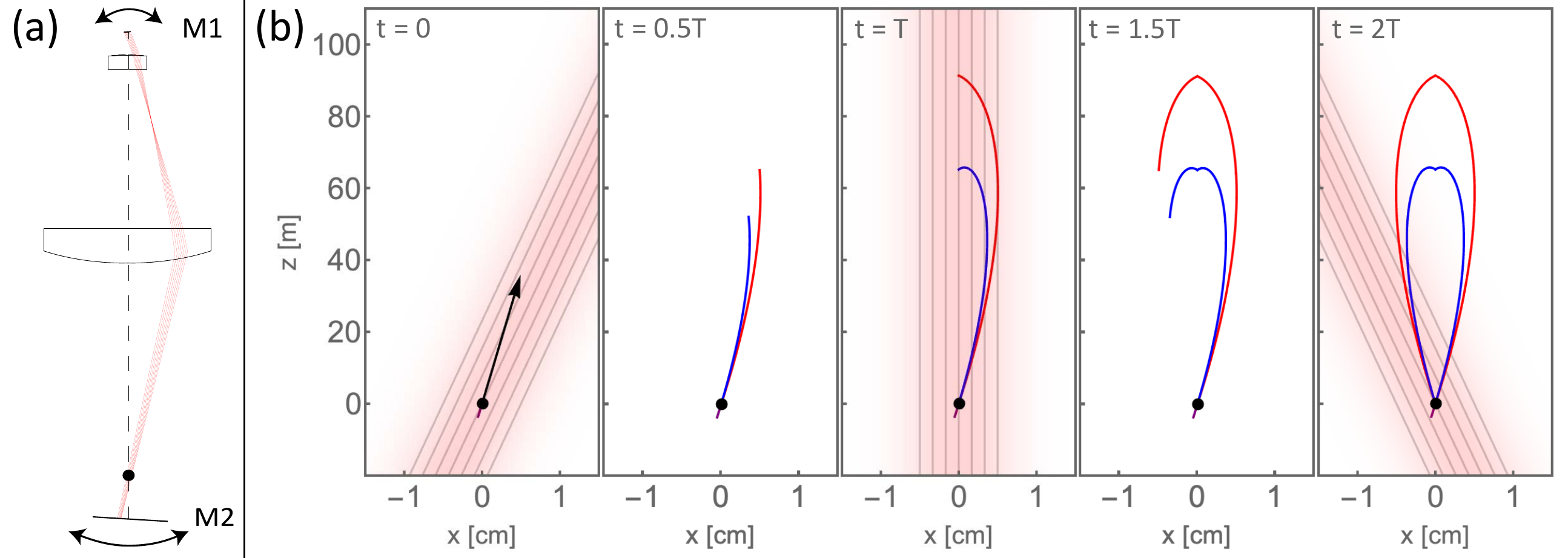}
\caption{\label{fig:coriolis_compensation_overview}
Overview of Coriolis force compensation for long baseline atom interferometry.
(a) A schematic (not to scale) of the optical setup associated with the Coriolis force compensation method. A mirror prior to the beam expanding telescope (M1) rotates simultaneously with a mirror at the other end of the interferometer baseline (M2) during an interferometer sequence. The pivot point of the interferometer beam is indicated by a black dot.
(b) A Mach-Zehnder interferometer sequence with a $T=4$ s interrogation time, and $1000\hbar k$ momentum separation, where the two arms of the interferometer (red and blue solid lines) span an $\approx 80$m distance. The rotation vector from the rotating earth is taken to be along the $y$-axis (in and out of the page) and the interferometer axis is taken to be along $z$. The black arrow indicates the direction of the initial atom launch.
}
\end{figure}

In an alternate method, the interferometer beam can be delivered such that it is centered on the atom cloud during each atom-optics operation while it undergoes Coriolis-force suppressing counter-rotation, even with a large lever arm between the atom launch point and the retro-reflecting mirror. The scheme involves adding a second piezo-actuated mirror before a beam-expanding telescope and rotating both this mirror and the retro-reflecting mirror simultaneously during an experiment cycle (see Fig. \ref{fig:coriolis_compensation_overview}a). This optical configuration allows for the upward and downward propagating components of the interferometer beam to be co-linear, and allows for adjustment of the pivot point of the interferometer beam by adjusting the distance between the pre-telescope mirror and the first telescope lens. Optical proof-of-concept tests for this system are currently underway. Fig.~\ref{fig:coriolis_compensation_overview}b illustrates the scheme for a Mach-Zehnder sequence in a long baseline interferometer. The two interferometer arms are centered on the interferometer beam for each of the three atom-optics operations as the beam rotates. This scheme also works in a gradiometer configuration, so long as the initial kinematics of the different atoms clouds can be adjusted independently. The impact of errors in the counter-rotation rate for this scheme has been studied. In a gradiometer configuration, a cancellation of the Coriolis and centrifugal contributions provides an important suppression of the susceptibility of the differential phase shift to these errors (see \cite{glick2024coriolis} for further details).

Scaling this scheme to km-scale baselines poses additional challenges, including increased cost. The initial angle of the interferometer beam is proportional to the product of Earth's rotation rate and the interferometer duration. Longer baselines allow for longer durations, which require larger initial angles. For a km-scale detector, in an operating mode where interferometer trajectories span the full baseline, and the pivot point is set to be $\approx 1$km from the beam-expanding telescope, the combination of these large initial angles and extended lever arm results in meter-scale beam deflections by the telescope. This requires optics and vacuum tubes with correspondingly large diameters, which can be costly.

\subsection{Baseline optimization for large-scale detectors}
\label{Fabio}

\begin{figure}[h]
    \centering
	\includegraphics{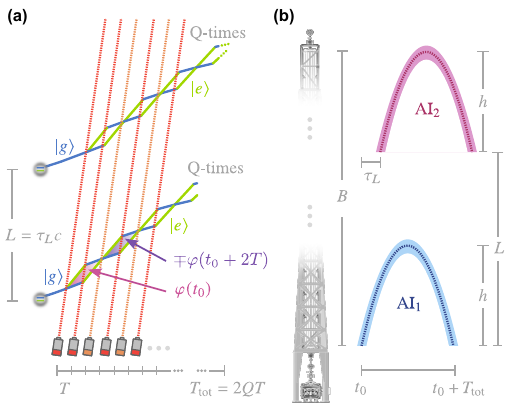}
    \caption{{\sffamily\bfseries (a)} 
    Spacetime diagram of a gradiometer consisting of two atom interferomters separated by a distance $L=\tau_L c$.
    Each atom interferomter consists of $Q$ subsequent basic Mach-Zehnder interferomters (MZIs) generated by single-photon pulses (dotted, red), where each subsequent basic MZI is created at multiples of the interrogation time $T$. 
    The first scheme ($-\varphi$) is generated by the pulses indicted in red, while the additional $\pi$ pulses shown in yellow are only present in the second ($+\varphi$) scheme.
    The overall interrogation time $T_\text{tot}=2QT$ scales with the number of basic MZIs $Q$.
    We indicate the ground state $\ket{g}$ of the atom in blue and its excited state $\ket{e}$ in green. 
    The phase difference from one basic MZI $\varphi(t_0)$ depends on the initial time $t_0$ and is identical but shifted in time in the $+\varphi$ scheme.
    On the contrary, in the first $-\varphi$ scheme it alternates its sign in subsequent MZIs, since the roles of both arms are interchanged.
    {\sffamily\bfseries (b)}
    The spatial extension $h$ and the midpoint trajectory (dashed) of the two atomic fountains that are used as generalized MZIs $\text{AI}_1$ (blue) and $\text{AI}_2$ (red) are shown.
    They are separated by a distance $L$ distributed along the baseline $B$ of the detector, while their start and end is delayed by a time $\tau_L=L/c$, stemming from the finite propagation time of the light between the two atomic ensembles.
    Their actual finite spatial extension originating from the atomic recoil and subsequent wave-packet propagation is illustrated by the shaded area surrounding the respective midpoint trajectories.
    This Figure was taken from Ref.~\cite{DiPumpo2024}; licensed under a Creative Commons Attribution (CC BY) license.
    }
    \label{fig:setup}
\end{figure}

Atom interferometers can be used to detect dark matter (DM)~\cite{Geraci2016,Arvanitaki2018} and gravitational waves (GW)~\cite{Graham2013,Norcia2017}.
As discussed before in this work, one typically uses gradiometric setups~\cite{Chaibi2016,Canuel2018,Badurina2020,Abe2021} to isolate small signals, such as DM and GWs.
Such a gradiometer with two spatially separated atom interferometers with separation $L$ is depicted in Fig.~\ref{fig:setup}a.

The detection of scalar ultralight DM with atom interferometers is possible due to its coupling to the mass-energy of the atom~\cite{Arvanitaki2015,Graham2016,Safronova2018,Derr2023}.
As such, it modulates the atomic transition frequency $\Omega(t)=\Omega_0+\bar{\varepsilon}\delta\Omega\cos{\left(\omega t+\phi\right)}$, where $\Omega_0$ is the unperturbed atomic transition frequency, $\bar{\varepsilon}$ is the mean coupling of the source mass to the DM field, $\omega$ is the DM frequency, and we have neglected any spatial dependence of the DM field.
On the contrary, scalar DM does not influence the laser phase~\cite{DiPumpo2022} to leading order.
As a consequence, the atom's time-varying transition frequency between light pulses delivers the dominant signature of scalar DM in atom interferometers.
For a single Mach-Zehnder interferometer (MZI), we find~\cite{Arvanitaki2018} the phase
\begin{align}
\begin{split}
    \varphi(t_0)=-\int\limits_{\mathrlap{t_0}}^{\mathrlap{t_0+T}}{\mathrm{d} t\,\Omega(t)}+\int\limits_{\mathrlap{t_0+T}}^{\mathrlap{t_0+2T}}{\mathrm{d} t\,\Omega(t)},
\end{split}
\end{align}
with initial time $t_0$, the middle light pulse acting at $T$, and the final pulse at $2T$, see Fig.~\ref{fig:setup}a for $Q=1$.

Contrarily to DM, the coupling of GWs is mediated through the diffracting light field~\cite{Dimopoulos2009,Graham2016b}.
Conversely, GWs have no direct effect on the atoms at a Newtonian gravitational level~\cite{Misner2017}.
As such, a GW with strain $\mathcal{h}$ induces an oscillating laser phase $\delta\Psi(t)=-c k_\ell \mathcal{h}\sin{\left(\omega t+\phi\right)}/(2\omega)$, with GW frequency $\omega$, where $k_\ell$ is the light's wave number, and where we assumed the GW to propagate orthogonal to the atom interferometer's baseline.
This oscillating laser phase is imprinted on the atoms during the diffracting light pulses.
Hence, for a single MZI we obtain~\cite{Graham2016b} the phase 
\begin{align}
\begin{split}
   \varphi(t_0)=&-\int\limits_{\mathrlap{t_0}}^{\mathrlap{t_0+2T}}{\mathrm{d} t\,\delta\Psi(t)} \left[\delta_{t_0}-2\delta_{T+t_0}+\delta_{2T+t_0}\right] 
\end{split}
\end{align}
with delta functions $\delta_{t'}=\delta\left(t-t'\right)$.

We now generalize the MZI sequence to $Q$ subsequent butterfly-like schemes, see Fig.~\ref{fig:setup}a.
There, two choices for the respective consecutive interferometers arise: Either one interchanges the roles of the arms after each MZI (neglecting the yellow mirror pulses in Fig.~\ref{fig:setup}a) or one retains them (including the yellow mirror pulses).
The total signal both for DM as well as GWs is then found~\cite{DiPumpo2024} to be
\begin{align}
\begin{split}
   \Phi(t_0)=\sum\limits_{q=1}^{Q}\left(\mp 1\right)^{q-1}\varphi\left(t_0+2(q-1) T\right),
\end{split}
\end{align}
where the minus sign represents interchanging roles and the plus sign represents retaining roles.
The interchanging scheme would cancel leading-order DC gravitational effects~\cite{Kleinert2015,DiPumpo2023} and, thus, it could potentially suppress gravity-gradient~\cite{DAmico2017,Roura2017} noise~\cite{Junca2019,Mitchell2022}.
From this signal, one finds the differential phase $\delta \Phi = \Phi(t_0+\tau_L)-\Phi(t_0)$ for two generalized multiloop MZIs separated by the initial time delay $\tau_L=L/c$, see Fig.~\ref{fig:setup}a.
Since $\phi$ is unknown, one measures the amplitude $\Phi_\text{S}=\big[2\int_0^{2\pi}{\mathrm{d}\phi \,\delta \Phi^2/(2\pi)}\big]^{1/2}$, where the subscript S denotes the signal type, either DM or a GW, instead of the mere differential phase.
Moreover, we replace every light pulse by many large-momentum-transfer (LMT) pulses~\cite{Graham2013,Graham2016,Schubert2019,Rudolph2020}, enhancing the number of interaction points.
Thus, we find for $\omega \tau_L \ll1$ the amplitudes
\begin{align}
\begin{split}
   &\Phi_\text{DM}=  \bar{\varepsilon} 4 \delta \Omega  \tau_L N \left|\mathcal{Q}_\mp(\omega T,Q)\right|, \\
   &\Phi_\text{GW}=  \mathcal{h} 2 k_\ell L N \left| \mathcal{Q}_\mp(\omega T,Q)\right|,
\end{split}
\end{align}
where $N$ is the number of LMT pulses, with $\mathcal{Q}_+(\omega T, Q) =  \frac{1}{2} \sin (Q \omega T) \tan \frac{\omega T}{2}$ for retaining roles of arms and 
\begin{equation}
    \mathcal{Q}_-(\omega T, Q) =   
    \begin{cases}
     \sin^2 \frac{\omega T}{2} \cos \left(Q\omega T\right) / \cos \omega T & \text{for } Q\text{ odd} \\
     \sin^2 \frac{\omega T}{2} \sin \left(Q\omega T\right) / \cos \omega T & \text{for } Q\text{ even}
    \end{cases}
\end{equation}
for interchanging roles.
In a resonant-mode operation, we find~\cite{DiPumpo2024} $\left|\mathcal{Q}_+(\pi, Q) =  Q\right|$ for $\omega T = \pi$ and $\left|\mathcal{Q}_-(\pi/2, Q) =  Q/2\right|$ for $\omega T = \pi/2$, where the latter resonance condition is only approximately valid, in particular for small $Q$.

If we consider the parameter uncertainties
$$\Delta \bar \varepsilon = \Delta \Phi_\text{DM}/( 4 \delta \Omega  \tau_L N \left| \mathcal{Q}_\mp\right|), \; \Delta\mathcal{h} = \Delta \Phi_\text{GW}/( 2 k_\ell L N \left| \mathcal{Q}_\mp\right|) \, ,$$
and assume a weak time dependence $| Q \mathrm{d} \Delta \Phi_S /\mathrm{d} T|_{\omega T=\text{res}}\ll 1$, we find~\cite{DiPumpo2024} with $T_\text{tot}=2QT$ in resonant-mode operation the uncertainties
\begin{align}
\begin{split}
   &\Delta \bar \varepsilon = \frac{\pi}{2}\frac{\Delta \Phi_\text{DM}  }{ N \delta \Omega \omega  \tau_L T_\text{tot}}, \\
   &\Delta\mathcal{h} = \pi\frac{\Delta \Phi_\text{GW}  }{ N k_\ell \omega  L T_\text{tot}},
\end{split}
\end{align}
where we observe that they are equal for both the interchanging scheme and the retaining scheme.
In the following, we consider $\tau_L = (B-h)/c$, where $B$ is the baseline of the experiment and $h$ the height of a single fountain, and take into account a parabola flight of the atoms, see Fig.~\ref{fig:setup}b, with $T_\text{tot}\cong \sqrt{ 8 h /g}$.
Making the restrictive assumption that the uncertainties $\Delta\bar\varepsilon$ and $\Delta\mathcal{h}$ are moreover also independent of $T_\text{tot}$, we can minimize the inverse of $\left(B-h\right)h^{1/2}$, and find $h=B/3$, i.\,e. 30\%~\cite{DiPumpo2024} of the baseline, for an optimal baseline exploitation both for GW and DM detection.

In order to generalize our treatment, we make less restrictive assumptions and assume shot-noise limitation $\Delta \Phi_S = \sqrt{2/(\nu n_\text{at})}$, with integration time $T_\text{int}= \nu T_\text{tot}$, for both uncertainties.
In this case, we have to minimize the inverse of $\left(B-h\right)h^{1/4}$, obtaining $h=B/5$, which is $20\%$~\cite{DiPumpo2024} of the total baseline, for an optimal exploitation both for GW and DM detection.

In summary, we have presented a method~\cite{DiPumpo2024} for exploiting the baseline optimally both for DM detection as well as for GW detection with atom interferomters.
Our treatment could in the future be refined by including recoil effects in the (interrupted) parabola flights and by connecting the number of subsequent MZIs $Q$ with the number of LMT pulses $N$.

\subsection{Quantum sensing with ultracold atoms in phase-modulated optical lattices}
\label{Weidner}

The concept of atom inertial sensing using ultracold bosons in phase-modulated (or ``shaken'') optical lattices came about in 2017~\cite{anderson2017a}, with an experimental demonstration of a 1D shaken lattice accelerometer in 2018~\cite{anderson2018}. This so-called shaken lattice interferometry (SLI) has a number of potential advantages in that the system can be tuned to the signal of interest, including AC~\cite{anderson2017a} and DC~\cite{anderson2018} biases, and the atoms remain trapped throughout the interferometry sequence, which increases robustness. However, like with most new technologies, a number of questions remain, including the ultimate sensitivity of the devices and their practicality as a useful and deployable sensor.

SLI works by loading atoms into the ground state of a shallow lattice, leading to atoms trapped in the lowest Bloch band, as the shallow lattice lends itself to atom delocalization in position space (and thus localization in momentum space). The atoms' momentum population is then modified via phase modulation, leading to the overall Hamiltonian (when considering the 1D case)
\begin{equation}
    \label{eq:SLI_H}
    H(x,t) = \frac{p^2}{2m} - \frac{V_0}{2}\cos{\big(2k_\mathrm{L}x + \phi(t)\big)} + U_\mathrm{a}(x,t)
\end{equation}
where $m$ is the mass of the atom, $V_0$ is the lattice depth, usually given in units of the recoil energy $E_\mathrm{R} = \hbar^2k_\mathrm{L}^2/2m$ for $k_\mathrm{L} = 2\pi/\lambda_\mathrm{L}$ and $\lambda_\mathrm{L}$ the lattice wavelength. The modulation function is given by $\phi(t)$ and the applied potential $U_\mathrm{a}(x,t)$ is the signal that we wish to sense, e.g., $U_\mathrm{a}(x,t) = max$ for a DC acceleration $a$.

\begin{figure}
    \centering
    \includegraphics[scale=0.6]{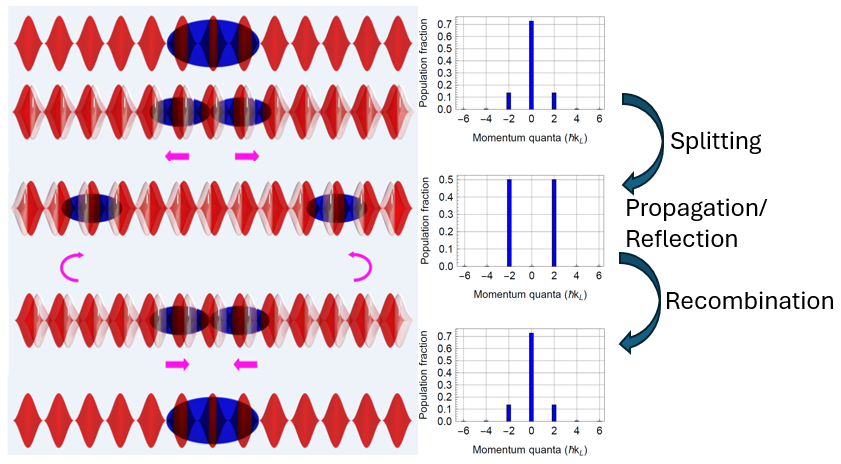}
    \caption{A diagram of a one-dimensional shaken lattice interferometry protocol designed to measure accelerations along the lattice axis. The atoms start in the ground Bloch state of the lattice (as indicated by the momentum state on the right). They then undergo splitting into the $\pm2\hbar k_\mathrm{L}$ state for some time before the protocol is run in reverse, reflecting and then recombining the atoms. In the absence of an applied acceleration potential $U_\mathrm{a} = max, a = 0$, the atoms will be recombined into their original state, but if $a\neq 0$, the atoms' final state will change, and it is this signal that we detect, e.g., through time-of-flight measurements. Parts of this figure adapted with permission from Ref.~\cite{weidner2018}.}
    \label{fig:SLI}
\end{figure}

The first implementations of SLI mimicked a typical atom interferometer by implementing a splitting protocol, where the atoms equally populated the $\pm2n\hbar k_\mathrm{L}$ states for some integer $n$, then allowing the atoms to propagate in the modulated lattice before the protocol is reversed, effectively reflecting the atoms and recombining them into their initial ground state, for the case where $U_\mathrm{a}(x,t) = 0$. A diagram of this is shown in Fig.~\ref{fig:SLI}. If $U_\mathrm{a}(x,t) \neq 0$, then the final momentum state population of the atoms changes as a function of the applied signal. 

The desired shaking function is typically found via some learning algorithm, and it has been shown that the underlying physics is, unsuprisingly, driven by transitions between different Bloch bands~\cite{anderson2018a}. Excellent results regarding the generation of momentum state populations of atoms in the lattice have been shown in Ref.~\cite{gueryodelin}. Machine learning methods have also been applied specifically to SLI both theoretically, to obtain a rotation sensor~\cite{Holland2021}, and experimentally~\cite{ledesma}. A similar sensor has also recently~\cite{ledesma2} been used to demonstrate a two-axis accelerometer.

The future of SLI relies on a verification of its utility as an inertial sensor, and while its applications to VLBAI are currently unclear, such interferometers could be used, e.g., to measure systematics along the arms of a VLBAI, and/or the optimal control methods used to find SLI protocols could be applied in a proposed TVBLAI. 

\section{Possible site options for a TVLBAI}
\label{Sites}


\subsection{Introduction}
\label{SitesIntroduction}

The main infrastructure required to deploy a TVLBAI is a tunnel or shaft with length or depth equal to the experiment baseline of $\gtrsim 100$m, respectively. There are already several facilities around the world that feature tunnels or shafts with those characteristics, and are therefore potential candidates to host TVLBAI experiments. An exhaustive list of site options was discussed in~\cite{TVLBAISummary}. For completeness, Section~\ref{sites-summary} summarises this information. Then, Sections~\ref{sites-boulby}, \ref{sites-porta_alpina} and \ref{sites-canfranc} provide updates from the Boulby Underground Laboratory, Porta Alpina and Laboratorio Subterr\'aneo de Canfranc, respectively. Finally, Section~\ref{sites-zaiga} describes the current status and plans of ZAIGA, including details of the Wuhan 10~m and proposed 240~m atom interferometers.

\subsection{Summary of site options}\label{sites-summary}

Potential site options have been selected based on two types of criteria:
\begin{itemize}
\item Infrastructure requirements, i.e., the ability to accommodate a fully operating experiment. These requirements are summarised in Table \ref{table-sites-requirements}.
\item Environmental requirements, i.e., the sources of noise that could degrade the sensitivity of such experiment. The most important effect is gravity gradient noise (GGN), namely stochastic matter perturbations in the environment that propagate to the atom beam through gravitational coupling and therefore cannot be shielded. GGN cannot be measured by any existing instrument other than atom interferometers, but correlates with vibrational noise. For this reason, sites with very low vibrational noise levels are required. Besides GGN, sites with low electromagnetic noise are also preferred.
\end{itemize}

\begin{table}
\centering
\begin{tabular}{|l|l|}
\hline
{\bf Component}       & \multicolumn{1}{c|}{{\bf Requirements}}                    \\
\hline
Laser laboratory      & At least 50 m$^2$ area                                     \\
                      & 35 kW electrical power                                     \\
                      & Air cooling: 30 kV heat load, 1 $^\circ$C stability        \\
                      & Maximum distance to interferometry region: 50 m            \\
\hline
Interferometry region & Tunnel or shaft with required baseline length              \\
                      & Full access to entire tunnel or shaft                      \\
\hline
Atom sources          & 2 to 10 units, equally spaced over the tunnel or shaft     \\
                      & 1$\times$1$\times$2 m$^3$ volume, 200 kg weight (per unit) \\
                      & 10 kW power consumption (per unit)                         \\
\hline
\end{tabular}
\caption{Infrastructure requirements for TVLBAI experiments.}
\label{table-sites-requirements}
\end{table}

\begin{figure}
\centering
\includegraphics[width=0.9\linewidth]{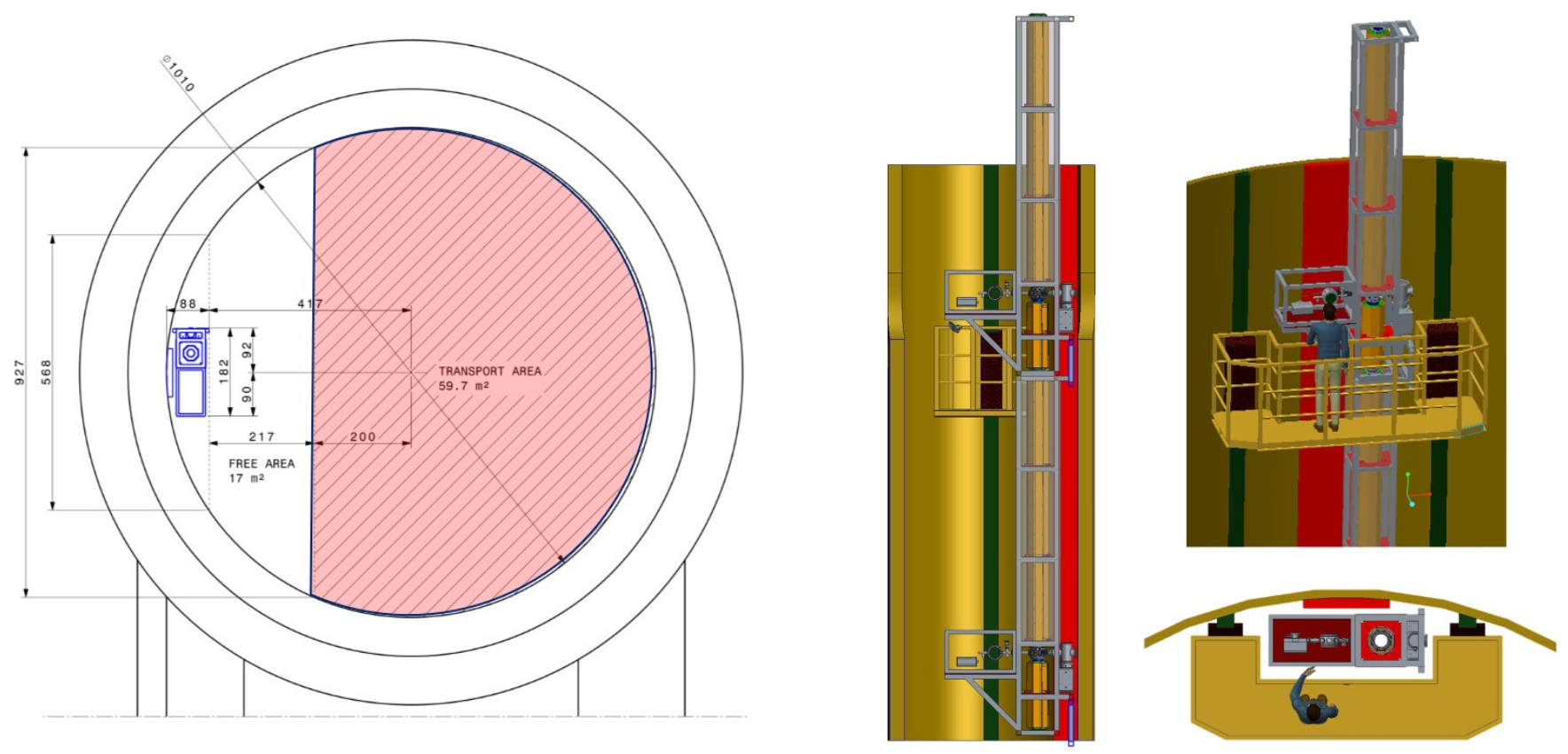}
\caption{Proposed design for a vertical 100 m TVLBAI experiment in the CERN PX46 shaft. Left: top view showing the free area and the proposed layout for the atom interferometer (blue lines). Right: drawings showing the atom interferometer and the access elevator. From~\cite{Arduini2023}.}
\label{fig-sites-cern}
\end{figure}

Note that these definitions are similar, but not identical, to those used in~\cite{TVLBAISummary}.
\newline
\newline
{\bf CERN}: A feasibility study of hosting a vertical 100m atom interferometer at CERN was published in 2023~\cite{Arduini2023}. The proposed infrastructure is the PX46 shaft at Point 4 (143m depth, 10.1m diameter). The top of this shaft is enclosed by a large building that could host the laser laboratory. This study provides a tentative design that includes the atom interferometer and an elevator within the available shaft area, see Fig. \ref{fig-sites-cern}, and a radiation shield at the bottom corridor that connects to the LHC tunnel. The estimated cost to prepare the site for a TVLBAI experiment is $\sim$1.5M CHF. 
The vibrational and electromagnetic noise have been measured, finding levels within the acceptable ranges. Moreover, magnetic field variations due to LHC operations ($\sim$50 nT) follow a predictable pattern, and therefore could be subtracted.

The LHC Long Shutdowns in 2026-29 and 2034-35~\cite{LHCshutdowndates} have been identified as major opportunities for installation works.
\newline
\newline
{\bf Sanford Underground Research Facility (SURF)}~\cite{Lesko:2015sma}: This facility, opened in 2007, is the home of leading experiments in dark matter (LZ) and neutrino physics (Majorana Demonstrator, LBNF/DUNE). It features two experimental areas at 1500m depth (Davis Campus, Ross Campus), each one accessed by a shaft. There is strong community support endorsing more space for science at SURF, with a plan to excavate two new caverns (100$\times$20$\times$24 m$^3$ each) near the Ross Campus for next-generation experiments by $\sim$2030. In addition, there is interest in developing a vertical facility at SURF, that could host a TVLBAI experiment. An initial evaluation study was completed in 2022, identifying four infrastructures for a medium-sized vertical facility ($\sim$100 m) and two infrastructures for a large-sized vertical facility ($\sim$1 km), see Table \ref{table-sites-surf}.
\newline
\newline
{\bf Laboratoire Souterrain \`a Bas Bruit (LSBB)}~\cite{Gaffet2009}: Located in a decommissioned military facility near Rustrel (France), LSBB has $\sim$4 km of nearly horizontal underground galleries, featuring very low seismic and electromagnetic noise. Two new perpendicular galleries of 150 m each have been excavated at $\sim$1 km from the laboratory entrance, that currently host the MIGA experiment.
\newline
\newline
{\bf Callio Lab}~\cite{Joutsenvaara2021}: This laboratory is located in the Pyh\"asalmi mine (Finland). It started in 2000 as the Centre for Underground Physics in Pyh\"asalmi (CUPP), and has continued as the Callio Lab since 2015. Callio Lab features four laboratories at different depths, down to 1430 m, and the area of the deepest laboratory is 120 m$^2$. All laboratories can be accessed by an elevator or an inclined tunnel (11 km to lowest laboratory), enabling shipments by truck. The estimated cost of developing a 100 m tunnel at Callio Lab is 350k EUR, including excavation, ventilation, electrical installation and water lines.

\begin{table}
\centering
\begin{tabular}{|l|c|l|}
\hline
{\bf Infrastructure} & {\bf Available depth} & \multicolumn{1}{c|}{{\bf Comments}} \\
\hline
\#6 Winze            &  140 m                &                                     \\
\hline
Milliken Winze       &  460 m                &                                     \\
\hline
Milnarich Shaft      &  240 m                & Need to remove concrete plug        \\
\hline
\#31 Exhaust Raise   &  230 m                & Need to verify dimensions           \\
\hline
\#5 Shaft            & 1500 m                &                                     \\
\hline
Ellison Shaft        & 1000 m                & Need to remove concrete plug        \\
\hline
\end{tabular}
\caption{Potential infrastructures to host a vertical facility at SURF. Vertical infrastructures dug from an underground level are referred to as winzes. In all cases, diameters range between 2.4 and 5 m.}
\label{table-sites-surf}
\end{table}

\subsection{Boulby Underground Laboratory}\label{sites-boulby}

The Boulby Underground Laboratory~\cite{Murphy:2012zz}, located in the Boulby Mine in the North East of England, is the deepest underground science facility in the UK. The laboratory features 4000 m$^3$ of clean space (ISO 6 and 7) at a depth of 1.1 km, and hosts more than ten collaborative projects in three main areas, namely 1) low-background particle physics, 2) Earth and environmental sciences, and 3) astrobiology and planetary exploration. 
The Boulby Development Project is a plan to expand the facility on a medium-to-long time scale, in two stages. Stage 1, starting in 2024 and finisingh around 2028, will develop a clean manufacturing and multi-science laboratory at a depth of 1.1 km ($\sim$30,000 m$^3$). Among other purposes, this extension aims to become a world-leading facility for the underground study of quantum technologies applied to fundamental science and seismic monitoring, and will be suitable for the installation of an AION-10 or AION-20 experiment. Stage 2, expected to begin around 2030, will build a full science laboratory at a depth of 1.3 km ($\sim$90,000 m$^2$).

\begin{figure}
\centering
\includegraphics[width=0.9\linewidth]{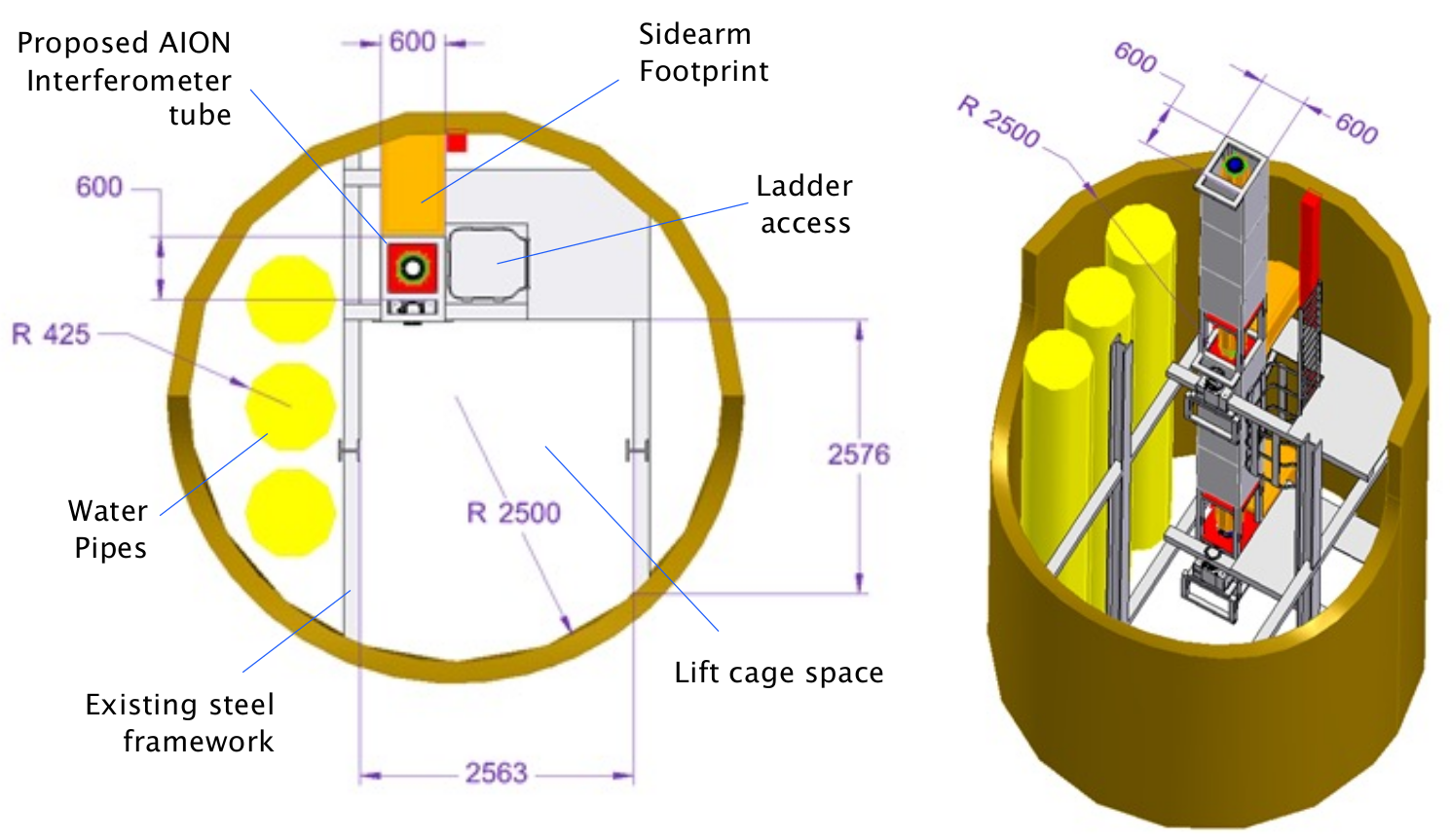}
\caption{Proposed design for a vertical 100 m TVLBAI experiment in the Boulby Mine tailings shaft. Existing water pipes are indicated in yellow.}
\label{fig-sites-boulby}
\end{figure}

In addition, the Boulby Underground Laboratory features three shafts that could be considered for TVLBAI experiments. The main mining site has two shafts with a depth of 1.1 km each, for transportation of personnel and rock respectively. A third shaft for tailings is located at $\sim$500 m from the main mining site, with 180 m depth and 5 m diameter. The top of this shaft is enclosed by a building, equipped with a 3 T crane with a lift cage. This shaft has been considered as an option to deploy a vertical 100 m atom interferometer. A study has provided a tentative design that includes such an atom interferometer and a ladder access within the available shaft area, keeping the existing lift cage, see Fig.~\ref{fig-sites-boulby}. In addition, the top building fulfills the requirements for hosting the laser laboratory. Currently, there is ongoing work to assess the seismic, magnetic and thermal conditions of this site option.

\subsection{Porta Alpina}\label{sites-porta_alpina}

The concept underlying this proposal is to use the infrastructures of the Gotthard Base Tunnel under the Swiss Alps to deploy a vertical TVLBAI experiment. Opened in 2016, the Gotthard Base Tunnel is the longest railway tunnel (57 km) and the deepest traffic tunnel in the world.

\begin{figure}
\centering
\includegraphics[width=0.9\linewidth]{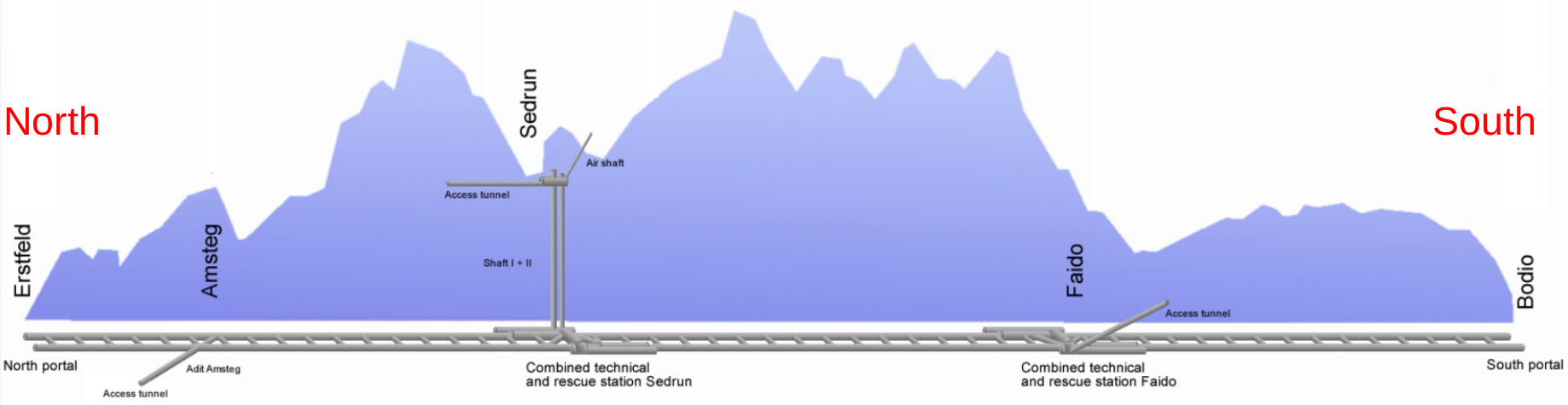}
\caption{Longitudinal section of the Gotthard Base Tunnel, showing the location of the two 800 m shafts at the Sedrun access site.}
\label{fig-sites-porta_alpina}
\end{figure}

In order to reduce the construction time, the excavation of the Gotthard Base Tunnel proceeded not only from their ends (Erstfeld, Bodio), but also from three additional intermediate locations (Amsteg, Sedrun, Faido), accessed by auxiliary tunnels and shafts, see Fig. \ref{fig-sites-porta_alpina}. In particular, the access site at Sedrun features two shafts with a depth of 800 m each, currently used for maintenance and ventilation, whose diameters are 8.6 and 7 m diameter respectively. Both shafts can be accessed from Sedrun through a 1 km horizontal tunnel, and one of them is equipped with a hoist.
The bottom ends of the shafts are connected to the Gotthard Base Tunnel, and four large halls (38$\times$10$\times$5.5 m$^3$ each) were built in the context of the Porta Alpina project to construct an underground passenger station in the Gotthard Base Tunnel connected to Sedrun via an elevator. However, this project was placed on hold in 2007.

Due to its characteristics, the Sedrun access site has been also proposed as an option to host a vertical TVLBAI experiment. 
A site visit to assess the feasibility of such a project took place in November 2024, and it is planned to follow this up with exploratory studies of seismological and electromagnetic noise, in collaboration with Swiss Federal Railways and the local authorities.

\subsection{Laboratorio Subterr\'aneo de Canfranc}\label{sites-canfranc}

The Laboratorio Subterr\'aneo de Canfranc (LSC)~\cite{Ianni:2016fjt} is an underground research facility located on the Southern side of the Pyrenees, in Spain. It is accessed by road through the Somport tunnel. The current laboratory has been operating since June 2010, and features two halls to host experiments (40$\times$15$\times$12 m$^3$ and 15$\times$10$\times$7 m$^3$ respectively). Services available at LSC include an ISO 7 clean room (35.5 m$^2$, that can be partly upgraded to ISO 6 upon demand), a mechanical workshop and offices. In addition, an external building near the Somport tunnel entrance hosts the LSC headquarters, and provides another mechanical workshop, laboratories and offices.

\begin{figure}
\centering
\includegraphics[width=0.9\linewidth]{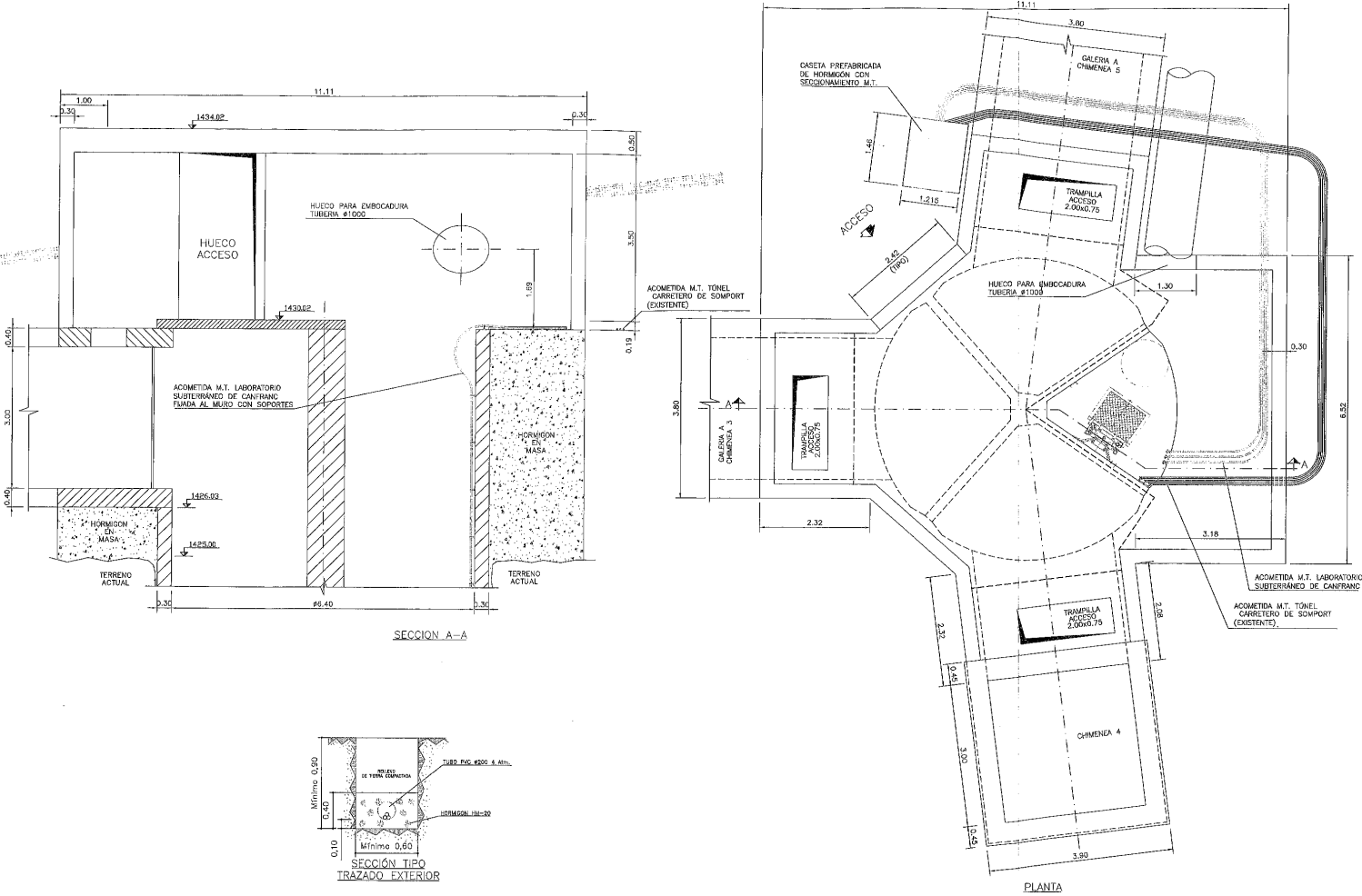}
\caption{Drawings of the top part of the Rioseta ventilation shaft at LSC. Left: side view. Right: top view. The maintenance sector includes the drawing of the elevator structure. The remaining three sectors are covered by a concrete layer, and provide free space for a potential laser laboratory.}
\label{fig-sites-canfranc}
\end{figure}

The laboratory has two infrastructures that could be used for both vertical and horizontal TVLBAI experiments:
\begin{itemize}
\item The Rioseta ventilation shaft (220 m depth, 6.4 m diameter), see Fig. \ref{fig-sites-canfranc}, whose cross section is divided into four sectors. One of those four sectors is used for maintenance, and is equipped with an elevator. While the usage of the other three sectors is restricted to ventilation. The maintenance sector is available to deploy a TVLBAI experiment. The top of the shaft is enclosed by a building with a free area of $\sim$50 m$^2$, accessible by vehicle (500 m from the main road).
\item The old railway tunnel (7874 m length, featuring an orthogonal arm of 200 m), that runs parallel to the Somport road tunnel at a nearly constant distance of $\sim$100 m. It is connected to LSC through a 20 m corridor. This tunnel is currently used as an evacuation route, and could be used to deploy a horizontal TVLBAI experiment.
\end{itemize}
The vibrational noise in the old railway tunnel has already been measured, and was found to have the lowest vibrational noise (above 2 Hz) among 15 international facilities~\cite{Beker_2012}. The midnight-to-midday noise variations are negligible, due to the low anthropogenic contributions. Similar noise conditions are expected at the Rioseta shaft (50 m distant from the Somport road tunnel).

\subsection{ZAIGA}\label{sites-zaiga}
\begin{figure}
\centering
\includegraphics[width=0.6\linewidth]{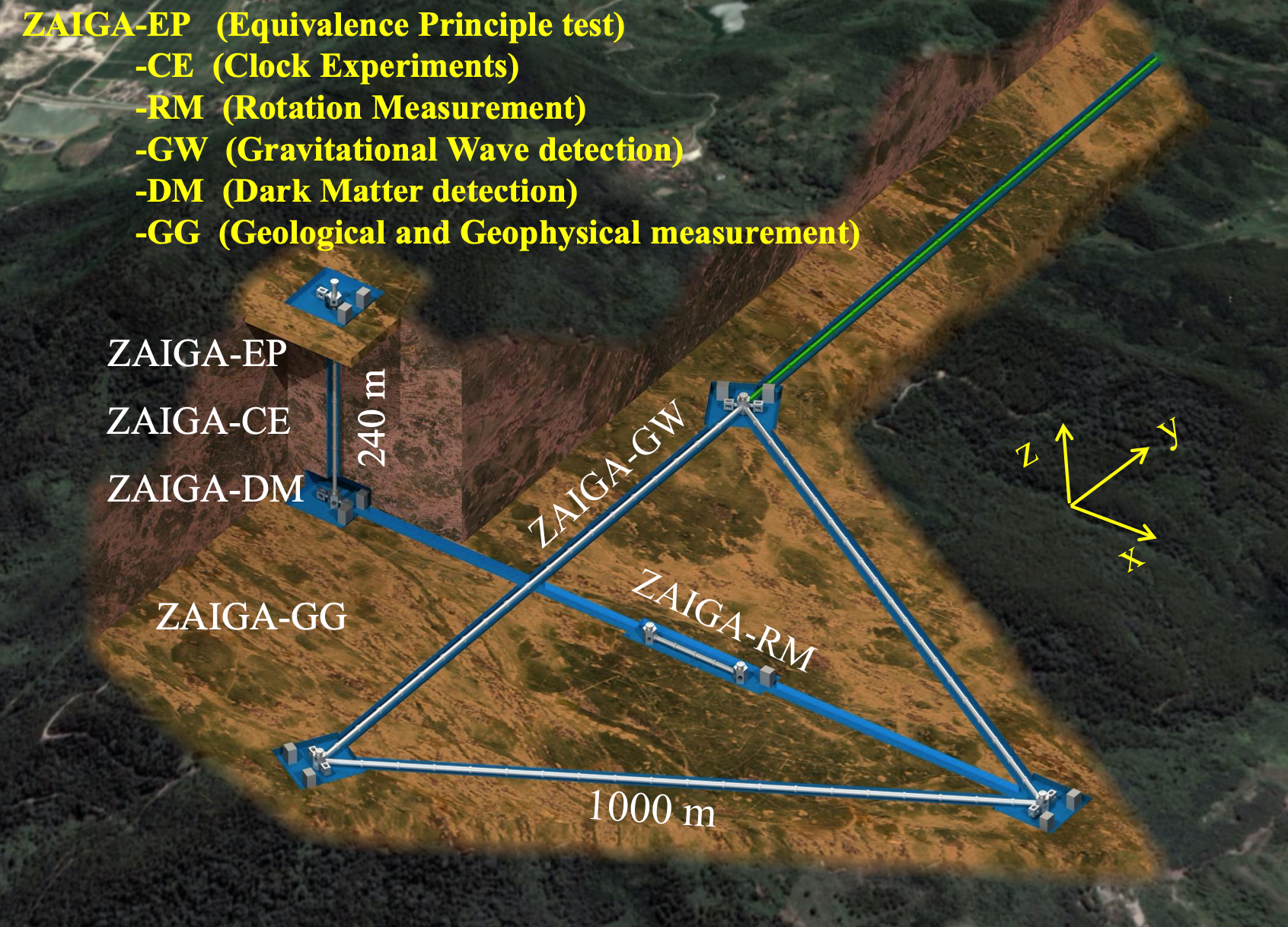}
\caption{Layout of the proposed TVLBAI array at ZAIGA~\cite{Zhan2019}.}
\label{fig-sites-zaiga}
\end{figure}

The ZAIGA programme~\cite{Zhan2019} aims to develop a series of atom interferometers in Zhaoshan (Wuhan, China) for fundamental physics and geoscience. It consists of three phases, see Fig. \ref{fig-sites-zaiga}:
\begin{itemize}
\item Phase I (now to 2027) is already funded, and preliminary design is ongoing. The infrastructure will consist of a 1.4 km tunnel and a 240 m shaft to host a vertical atom interferometer of the same length. This phase will also feature a 20m atomic gyroscope, and a 10m dual Rb/Sr interferometer.
\item Phase II (2027 - 2035) is a plan to develop three atom interferometers of 1 km each, arranged as an equilateral triangle. The science objectives include Weak Equivalence Principle (WEP) tests, measurement of the Lense-Thirring effect, GW detection, and ULDM search.
\item Phase III (after 2035) is a possible extension of one atom interferometer from 1 km to 3 km, in order to improve the sensitivity to GWs and ULDM.
\end{itemize}
Currently, the Wuhan 10m atom interferometer is demonstrating the ability to test the WEP, achieving several improvements in sensitivity over the last years~\cite{Zhan2019}.

\section{Site \& Engineering Challenges for Large-Scale Atom Interferometers}
\label{engineering}

\subsection{{Introduction}}
\label{EngineeringIntroduction}
This Section contains reports on the technical challenges that are arising in the design and installation of the present generation of multi-metre atom interferometers on sites outside the confines of conventional laboratories. These illustrate some of the issues that will also arise in the future construction of km-scale TVLBAI detectors.

\subsection{{Structural Stability and Instrument Installation of AION-10 in the Oxford Beecroft Building}}
\label{Oxford}

\begin{figure}
\centering
\includegraphics[width=0.5\linewidth]{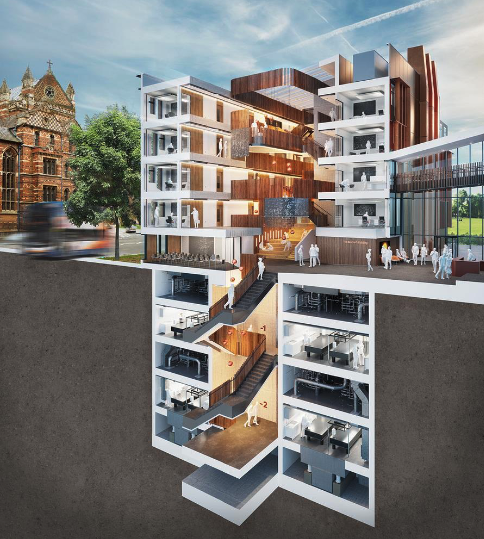}
\caption{Illustration of the Beecroft Building of the Oxford Physics Department where AION-10 is to be installed in the basement stairwell~\cite{HawkinsBrown2019}.}
\label{fig:Beecroft}
\end{figure}

The AION Collaboration~\cite{Badurina2020} plans to install its first-stage 10~m experiment (AION-10) in the basement of the Beecroft Building of the Oxford Physics Department, which has been designed to provide a stable environment and contains a suitable location for the required laser system.  The building is illustrated in Fig.~\ref{fig:Beecroft}, where the basement stairwell to be occupied by AION-10 is clearly visible.

\begin{figure}
\centering
\includegraphics[width=0.2\linewidth]{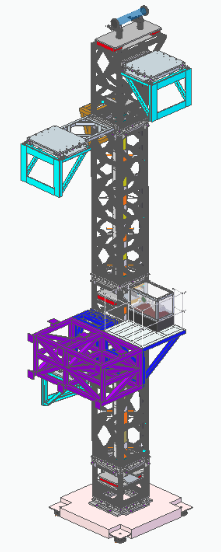}
\hspace{1cm}
\includegraphics[width=0.37\linewidth]{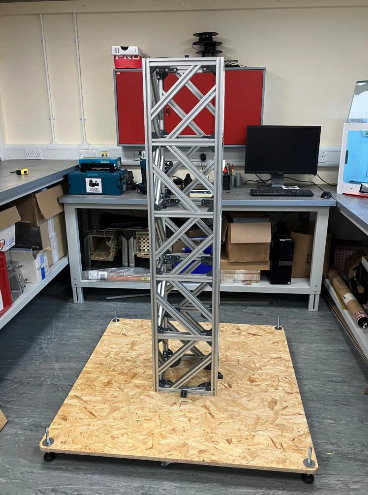}
\caption{Left: The proposed conceptual design of the support structure for AION-10. Right: 1/3-scale prototype of the frame for one of the main modules, used to validate the modal and response analysis model.}
\label{fig:ProposedDesign}
\end{figure}

In preparation for the construction of AION-10, engineering studies have been made of issues related to structural stability and the installation of the detector.~\footnote{There have also been studies of anthropogenic and synanthropic noise sources in such a built environment and their possible mitigation~\cite{Carlton2023}.} The design of the support structure has been developed with careful attention to stability requirements, notably the requirement that the two camera assemblies, which are to be located $\sim 5$~m apart, must be fixed within 100 nm relative to each other during data-taking. With this requirement in mind, various options for the structure design have been considered, and the preferred option is shown in the left panel of Fig.~\ref{fig:ProposedDesign}. 

The subjects of modal and response analysis of the structure have been explored, as well as strategies for vibration control. Options considered for supporting the instrument have included tensioned cables and rigid structures, with the latter being favoured because of its higher lowest-mode frequency $\sim 30$~Hz, well separated from the frequency range of interest for physics measurements. Among the challenges in the analysis have been uncertainties associated with materials, manufacturing and assembling, the selection of the most appropriate damping factor, the choice of single vibration input setting for a structure with multiple support points, and the impact of the building structure. An additional vibration survey will be needed before finalization of the design, including vibration measurements for multiple points in the stairwell with synchronised acquisition of data across the points to preserve the correct relationships between the vibrations at different points. These will then be incorporated into a multi-input vibration model to analyse the structure behaviour in a more realistic setting.

For ease of transport and installation, the AION-10 instrument will consist of two 5~m main modules and smaller modules such as a 2~m telescope section. A 1/3-scale prototype of the frame for one of the main modules has been constructed: see the right panel of Fig.~\ref{fig:ProposedDesign}. Its main purposes are to validate the analysis model on a smaller scale, study the damping factor and test its response.

The modules of the AION-10 instrument will be installed in the Beecroft Building encased in their aluminium support frames, which will provide space for attachments such as cold atom sources as well as stability support, and each high-vacuum chamber will have a magnetic shield that needs to be kept as a single unit in order to meet the shielding requirements. The frames will also protect the modules during transport between the module assembly area and the Beecroft Building and during installation. The modules will be lifted into the building and lowered into place using a dedicated crane system.

These engineering studies have established a baseline design for AION-10 that appears able to meet the physics requirements and can be installed in the basement stairwell of the Oxford Beecroft building. 

\subsection{{Progress and Challenges in MAGIS-100 Construction at Fermilab}}
\label{Fermilab}

All large-scale experiments require considerable effort to create a suitable space for installation and operation. Basic infrastructure, experiment equipment delivery, and personnel access are the main logistical considerations. There are also environmental challenges that include addressing vibrations, thermal gradients, magnetic fields, and ground water. Costs are largest when these spaces are purpose-built, so finding existing spaces that can be modified to become suitable is often a cost-effective solution. 

\begin{figure}
\centering
\includegraphics[width=1\linewidth]{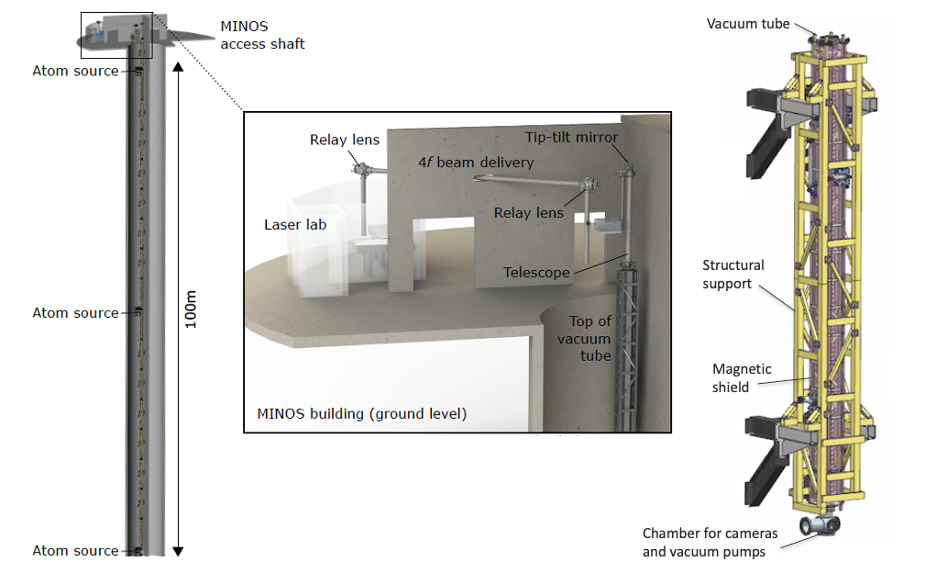}
\caption{Conceptual sketches depict the overall layout of MAGIS-100 (left), a larger view of the ground-level region (middle), and a modular section of the interferometry region (right) \cite{Abe2021}.}
\label{fig:MAGIS 100 layout}
\end{figure}

The MAGIS-100 project at Fermilab~\cite{Abe2021} is one example of adapting to an existing space; the MINOS access shaft for the underground experiment region will be modified to house the equipment for MAGIS-100. There are many benefits of being inside a building which already has crane coverage, adequate work space, and other standard infrastructure established. Unique site challenges include working in a narrow vertical space with a curved wall beyond the reach of the existing crane, accommodating other established uses of the shaft, and engineering delivery systems within these restrictions. These delivery systems must allow precise installation and alignment of the experiment components and positioning personnel at critical locations to safely reach the equipment for installation, tuning, and maintenance. An overview of the MAGIS-100 conceptual layout is shown in Fig.~\ref{fig:MAGIS 100 layout}.

\begin{figure}
\centering
\includegraphics[width=1\linewidth]{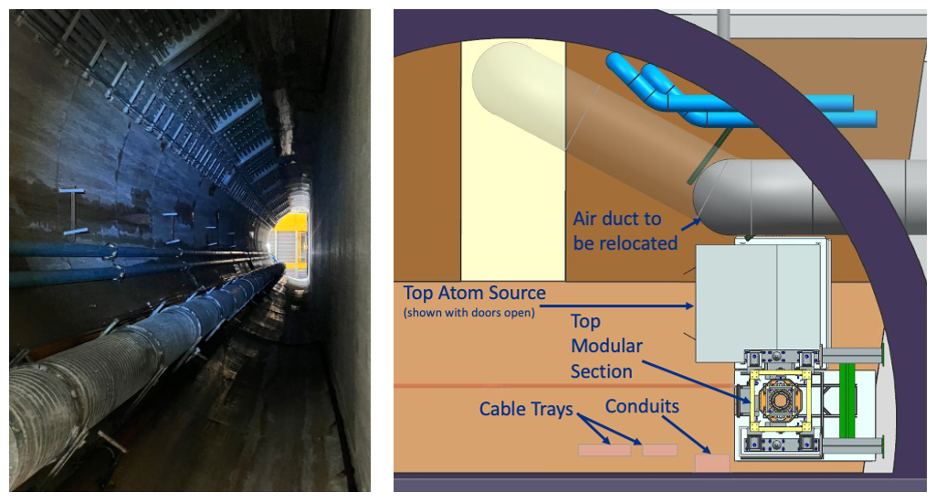}
\caption{The MAGIS-100 site: photograph of the shaft viewed from below (left) and conceptual plan view of equipment which will be added in that space (right).}

\label{fig:MAGIS 100 shaft}
\end{figure}

The MAGIS-100 site characteristics have significantly influenced designs for the overall experiment layout and all sub-systems. The interdependence of the atom source, laser, optical, camera, magnetic, vacuum, structural, alignment, network, and control systems mandates close collaboration between these sub-system designs, including coordination with planning for installation and access. For example, the interferometry tube region will be split into 17 modular sections to allow intricate assembly and qualification prior to installation in the shaft. Each modular section contains a vacuum tube, redundant bake system apparatus, environmental monitoring, magnetic shields and coils, and a mechanical frame for structural stability and alignment features. Connection nodes between these sections will be the main points of access from the shaft and are where the experiment sub-systems have the highest amount of integration. Similarly, atom sources are designed for rigorous qualification prior to installation, and component layout is driven by limited access from specific locations once in the shaft. A photograph of the shaft and a conceptual plan view of the equipment that will be added in the available space in the shaft is shown in Fig.~\ref{fig:MAGIS 100 shaft}. 

Each sub-system has significant technical challenges set by experimental requirements, and installation and personnel access systems have raised a considerable number of difficult engineering issues to address due to specific site features. Coordinating all the design requirements entails careful attention to interfaces, and frequent communication across the collaboration is essential.


\def\RB {Ramsey-Bord\'e }
\section{Additional Topics}
\label{Additional}

    
   


\subsection{A single-photon large-momentum-transfer atom interferometry scheme for strontium with application to determining the fine-structure constant}
\label{Foot}

It has been known for many years that large-momentum-transfer (LMT) techniques in atom interferometry can enhance the recoil-phase sensitivity quadratically with the number of LMT pulses \cite{Borde1993}. Following the results in Ref. \cite{schelfhout2024single} we present 1) the calculation of differential phase between a pair of single-photon \RB atom interferometers using neutral optical clock atoms and 2) the possible precision to which the fine structure constant could be measured using this scheme with current atom interferometry technology.

The fine structure constant $(\alpha)$ is a empirical parameter of the standard model characterizing the strength of the electromagnetic interaction between elementary charged particles. It arises in the definition of the Rydberg constant $R_\infty = \frac{1}{2hc}\alpha^2 m_e c^2$ where $h$ is the Planck constant, $c$ is the speed of light in vacuum, and $m_e$ is the electron rest mass. The leading experimental determinations of $\alpha$, currently rely on atomic photon-recoil measurements from \RB atom interferometry with LMT used to provide an increase in sensitivity.

\begin{figure}
\centering
\includegraphics[width=6in]{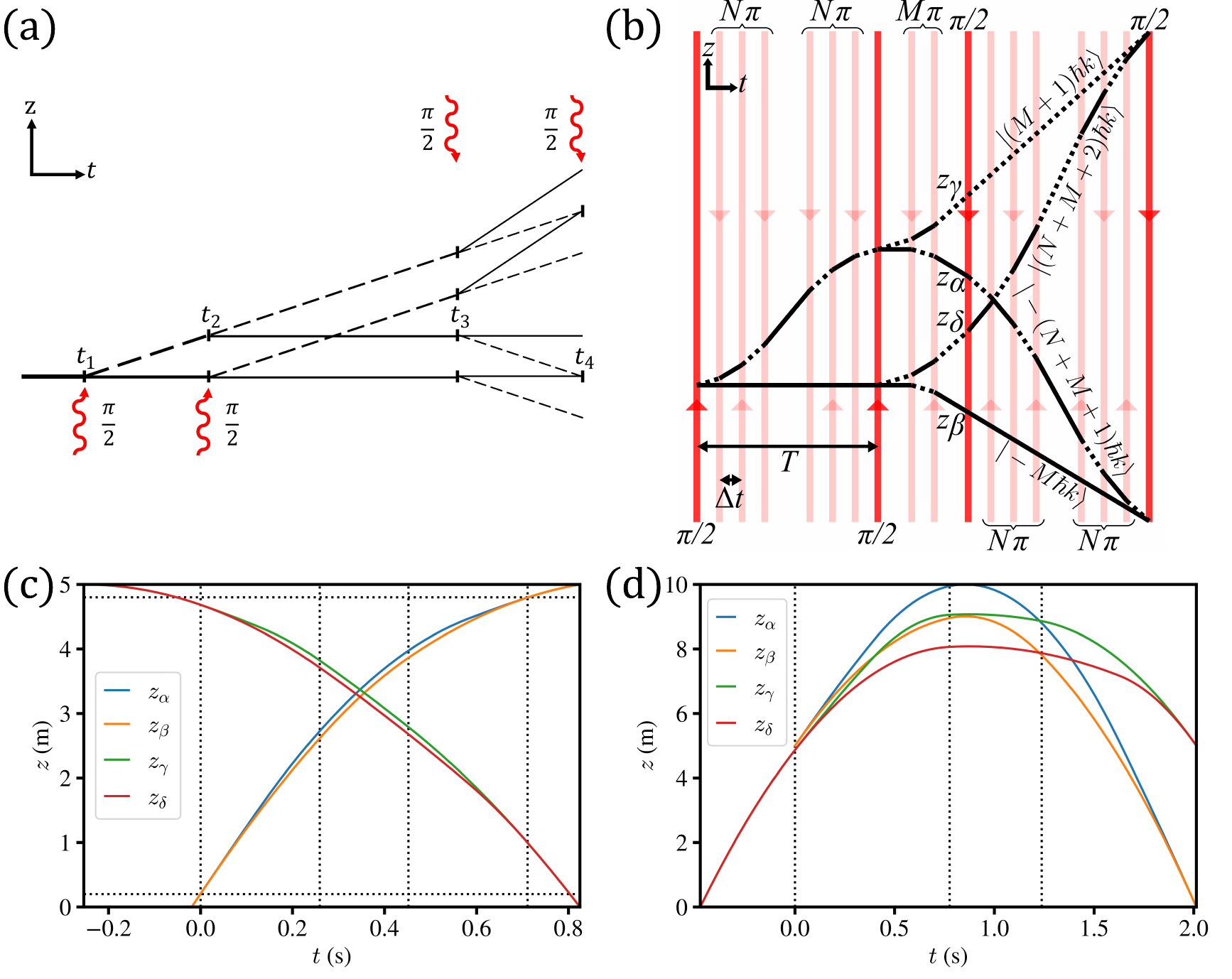}
\caption{ a) Schematic spacetime diagram of a \RB scheme. The ground (excited) state is shown with a solid (dashed) lines.  b) Schematic spacetime diagram of an enhanced \RB atom interferometer with $N=3$ and $M=2$ pulses shown. c) Optimal trajectories for the `X' configuration for a 5m atom interferometer. The vertical dotted lines represent the $\pi/2$-pulses and the horizontal lines bound the interferometry region [0.2m, 4.8m]. d) Optimal trajectories for the Fountain configuration with atomic sources at 0 \& 5m and vertical dotted lines represent the $\pi/2$-pulses, with the final pulse at $t=2$s. Figures from \cite{schelfhout2024single}. 
}
\label{fig:RB_simple_finite_c}
\end{figure}

The simplest form of a \RB atom interferometer consists of four $\pi/2$ pulses at times $t_i$, where pulses 3 and 4 counterpropagate relative to the first two \cite{Borde1984,Borde1989}. Regardless of the duration between the second and third pulses, provided that the duration $T$ between the first two pulses $T=(t_2-t_1)$ is equal to the duration between the last two pulses $(t_4-t_3)$, the two interferometers will close, as demonstrated in Fig.~\ref{fig:RB_simple_finite_c} (a).

The differential phase between these two interferometers, accounting for the time taken for light to propagate between them to order $\mathcal{O}(1/c)$, is given by
\begin{equation}\label{eq:differential_phase_simple_finite_c_main}
\Delta\Phi = \Delta\phi_{\mathrm{top}} - \Delta\phi_{\mathrm{bottom}} = \frac{2 \hbar k^2 T}{m} \left[1 - \frac{1}{c} \left(\frac{2 \hbar k}{m} - g (t_4 + t_3 - t_2) \right) \right] \, , 
\end{equation}
where $g$ is the gravitational field.

To leading order, this phase difference is proportional to the atom's photon-recoil frequency $\omega_{rec}=\frac{\hbar k^2}{2m}$ where LMT pulses can be used to enhance the sensitivity. This has been shown with Bragg diffraction and Bloch oscillations schemes used to improve the fine structure determination as in \cite{Parker2018}. Here we consider the single-photon analogy whereby we apply $\pi$ pulses in alternating directions to increase (and then close) the wavepacket separation. The LMT pulses within a \RB interferometer can take two types which differ in their effect depending on the zone in which they are applied. The pulses can occur between pulses $1-2$ \& $3-4$, or they can occur between pulses 2 and 3, as shown in Fig.~\ref{fig:RB_simple_finite_c} (b). The first set ($N$ pulses) acts to increase (and then cancel) the recoil frequency separation. The cancellation is required for the interferometer to close. The second set ($M$ pulses), acting between pulses 2 and 3 allows additional enhancement of the recoil-frequency measurement whilst also being able to nullify the phase arising from first-order spatial variation in gravitational acceleration. From Ref. \cite{schelfhout2024single}, the differential phase is given by

\begin{equation}
    \Delta\Phi = \frac{(N+1) (N + 2 M + 2) \hbar k^2 (T - N \Delta t_{\mathrm{LMT}})}{m} - \frac{N(N+1) (N + 2) \hbar k^2 \Delta t_{\mathrm{LMT}}}{3 m} + O\left(\frac{1}{c}\right),
\end{equation}
where $\Delta t_{\mathrm{LMT}}$ is the time separation between LMT pulses. This ultimately results in a quadratic scaling of the \RB\ phase with LMT order and offers an advantage compared to the linear scaling with Bloch oscillation order in Rb and Cs experiments.

Gravity gradients are an important systematic effect to consider in ground-based atom interferometry experiments \cite{Roura2017,DAmico2017} and are a limiting factor in photon recoil measurements. Controlling the launch from two separate sources allows the gravity gradient phase to be exactly cancelled when using offset simultaneous conjugate \RB\ atom interferometry \cite{Zhong2020}. This cancellation can be achieved in either an ``X" configuration or a fountain configuration as shown in Fig.~\ref{fig:RB_simple_finite_c} with the fountain configuration offering higher sensitivity but with the trajectories crossing the position of the higher atomic source.

We can consider these (optimal) trajectories in an "X" configuration that maximise sensitivity whilst cancelling gravity gradient noise for an instrument of length $L=3\,\mathrm{m}$ using Sr and Yb, and assume an experimental resolution of $1\,\mathrm{mrad}$. This theorised interferometer offers a twofold improvement in the precision of measuring the fine-structure constant compared to current standards. Additionally, it represents the highest precision in absolute atomic mass measurement, capable of resolving the mass difference between strontium's ground and excited states. Further precision improvements to $\alpha$ could be possible with better relative mass measurements of the electron and atomic isotopes, necessary for future Standard Model tests involving the electron magnetic moment.

Improvements of an order of magnitude in mass measurement precision are achievable with instruments of $L=10\,\mathrm{m}$, particularly in a fountain configuration. The sensitivities for any interferometer size could be improved with shorter LMT pulses, making use of higher Rabi frequencies. Whilst the calculations here are independent of any particular instrument, there are in-progress experiments with Sr or Yb on scales near $10\,\mathrm{m}$, like AION-10 at Oxford~\cite{Badurina2020}, the Sr prototype at Stanford~\cite{Dickerson2013}, and the VLBAI-Teststand~\cite{schlippert2020matter} in Hannover.

It is important to note that the phase shift contributions from laser pulse propagation delays between interferometers already exceed $1\,\mathrm{mrad}$ for the given trajectories and increase with longer baselines. This is therefore extremely relevant to proposed long-baseline interferometric gravitational-wave observatories such as AION~\cite{Badurina2020}, MAGIS~\cite{Abe2021}, MIGA~\cite{Canuel_2018}, ELGAR~\cite{Canuel2020}, and ZAIGA~\cite{Zhan2019}. Detailed analysis of these effects through intermediate-scale prototypes is crucial for advancing toward very long baseline atom interferometry.

\subsection{Atom interferometer using a spatially-localized beamsplitter}
\label{Clade}

Atom interferometry using light pulses has made it possible to measure inertial quantities such as rotation, acceleration or gravity gradient, as well as constants such as the gravitational constant or atomic recoil. All these experiments use continuous-wave (cw) laser sources to manipulate atomic wave packets. 

We have demonstrated that it is possible to implement coherent atomic beam splitters based on stimulated Raman transitions using pico second lasers \cite{solaro2022}. There are two main reasons for exploring this new technique. As with high-resolution spectroscopy, a first motivation for using a pulsed laser (or frequency comb) rather than a cw laser is to extend matter-wave interferometry to a wider spectral range and more atomic species. The second motivation is linked to the fundamental difference between using a continuous-wave laser and a pulsed laser. In the former case, laser-atom interaction takes place at the location of the atoms, whereas in the latter it is determined by the overlap area of counter-propagating ultrashort pulse pairs. 

\begin{figure}[b]
    \includegraphics[width=.9\linewidth]{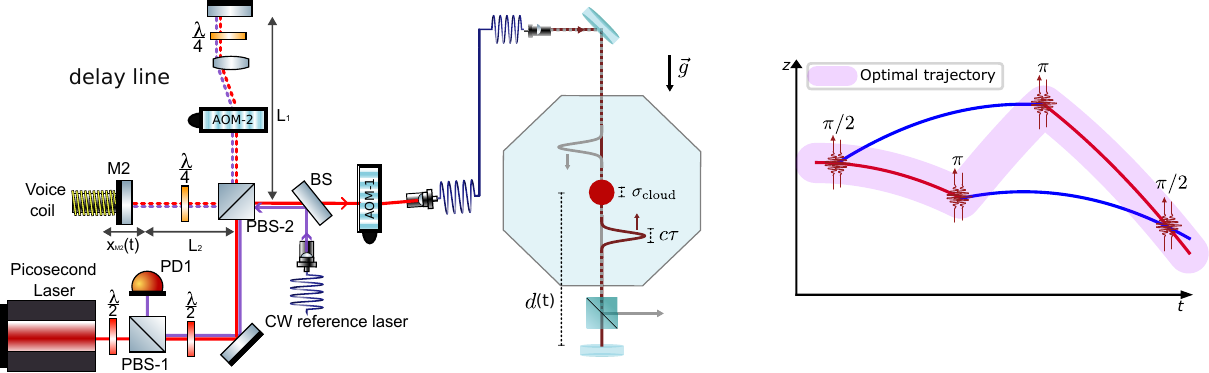}
    \caption{\label{fig:clade_figure_FCAI}
    Schematic of the experimental setup: the pico-second laser is split in two, with one part passing through an acousto-optic modulator and a delay line. This device controls both the delay and the phase between the two pulse trains. The diagram on the right shows the trajectory of the overlap zone in an interferometer in a gravimeter configuration. Figure taken from~\cite{debavelaere2024}.
    }
\end{figure}

Our experimental setup uses a cloud of cold rubidium atoms. Fig.~\ref{fig:clade_figure_FCAI} shows the device used to control precisely the delay and phase between the pulse trains that will overlap at the atom's position. Thanks to a moving mirror, we can now modify this zone in real time to individually address each arm of the interferometer, while controlling the phase using acousto-optic modulators. In this way, we've been able to follow the trajectory of the atoms for around 50 ms \cite{debavelaere2024}. With such a duration, the separation of the two arms of the interferometer is such that two $\pi$ pulses at the middle of the interferometer have to be made individually by moving the stage. This technique therefore allows each arm of the interferometer to be individually interrogated.
As the interaction between laser pulses and atoms is localized, laser beams induce two-photon light shifts only on the atoms being interrogated. This is not the case when using a continuous laser, where light shifts inevitably affect both interfering wave packets, leading to spurious phase shifts, loss of contrast and systematic biases in the measured quantity.


\section{Towards a Proto-Collaboration}
\label{Proto}


\subsection{Experience with Proto-Collaborations}
\label{Experience}

The assembly and constitution of a large international collaboration of scientists and engineers with diverse areas of expertise typically passes through several stages. 
Workshops such as this one are extremely useful to develop and exchange ideas and consider options for large scientific projects such as one or more atom interferometers with lengths in the range 100m to 1km.
In order to move towards a possible realization of such an ambitious programme, it will be important to bring the vision to the next level of organization.
The formation of a proto-collaboration is proposed as the next step in the evolution of international Terrestrial Very Long Baseline Atom Interferometer (TVLBAI) studies.

The concept of proto-collaboration is frequently adopted in big sciences such as high-energy physics (HEP) as a step in the evolution of a project, as a tool to help the participating groups focus and organize coherently towards achieving an agreed scientific goal. 
It is based on a Memorandum of Understanding (MoU), which is not a legally binding agreement, that interested participants can sign in order to join the proto-collaboration. 
The MoU defines the common scientific goals of the project and includes an organigram with a structure that is agreed between the participants. 
TVLBAI studies are now entering a phase where such an MoU will potentially be very beneficial.

Recent examples of proto-collaborations in HEP include that for T2HK, the Tokai to HyperKamiokande Neutrino Experiment.
This was formed in 2015 with the signatures of 73 institutes from 15 countries with about 200 members.
Its formation sent the clear message to the Japanese authorities and national funding agencies that a substantial international community was interested in this experiment. 
In addition to fostering coordinated efforts towards a common goal, it served to attract other potential collaborators, and by 2019 this proto-collaboration had doubled in size and turned into a full collaboration, and is now a fully-fledged and -funded experiment.

Another neutrino project that seeks to follow a similar trajectory is THEIA, which aims to construct a hybrid water Cerenkov/liquid scintillator detector. 
It was established as a proto-collaboration in 2022, with 31 participating institutes and about 100 scientists and engineers. 
Currently an R\&D project, THEIA might become a stand-alone project or become part of the DUNE neutrino experiment.

CORE (COmpact detectoR for the EIC - an electron--ion collider under construction at BNL) is a proto-collaboration that was formed in 2022 with 25 institutes and about 60 scientists and engineers. 
The EIC is a flagship nuclear physics project in the US, and it is confidently expected that CORE will become a fully-fledged collaboration and be approved as an experiment at BNL.

The following are earlier examples of HEP proto-collaborations. 
Mu3e is an experiment searching for flavour-violating muon decays that made a successful transition to an approved experiment.
Likewise, the long-baseline neutrino facility (LBNF) made a successful transition to an approved accelerator project. 
On the other hand, the ILD project is a proto-collaboration that is  awaiting approval of the ILC where it would be a detector.
There is also a proto-collaboration for R\&D for a calorimeter detector that has started recently.

The ongoing TVLBAI  studies already bring together a healthy and much-needed combination of expertise from different fields. 
A proto-collaboration will provide a framework for interested parties to focus on a common future vision and roadmap, with goals such as preparing a Conceptual Design Report.
The proto-collaboration will allow us to speak with a common voice to the outside world and provide a framework for organizing workshops and meetings. 
It should have a minimal formal structure including an institutional board (IB) with representatives of all participating institutes, able to make project decisions and guided by an elected chairperson. 
This will facilitate preparing coherent requests for resources in the future. 
Once significant funding is secured  or approval obtained at some level, we may proceed to a full collaboration agreement. 

\subsection{The TVLBAI Proto-Collaboration}
\label{MoU}

The development of future atom interferometers beyond the 10m scale will require the collaboration of several Laboratories and Institutions. Furthermore, there would be a clear advantage if several interferometers are eventually built to operate them in a network. To facilitate working together and defining clear collective goals, which could also help addressing funding agencies for support in these activities, it was proposed to invite all interested parties (Institutes, Laboratories, Universities) to sign a Memorandum of Understanding (MoU). This expresses the wish of the scientific community to draft a framework for the development and realization of Terrestrial Very-Long-Baseline Atom Interferometry (TVLBAI) experiments, with a view to executing a Conceptual Design Study for a TVLBAI (the “TVLBAI Study”). The MoU establishes a common basis among the Participants for the collaborative effort required for the TVLBAI Study: see~\cite{TVLBAIMOU} for the full text of the MoU, which has been signed by over 50 institutions. 

We summarise here the scope of the MoU: 
\begin{itemize}
    \item{The main focus of the TVLBAI Study Group shall be to prepare a full science and technology Roadmap accompanied by a Conceptual Design Report for a potential TVLBAI project that may consist of one or more demonstrators  in various locations using a combination of design options, e.g., vertical and horizontal geometries and different cold atom species, operated as a network.}
    \item{As part of the Roadmap, the TVLBAI Study shall provide a baseline concept for such a TVLBAI project, performance expectations, and assess the associated key risks, as well as the cost drivers. It shall also identify an R\&D path to demonstrate the feasibility of a TVLBAI project and support its performance claims.}
    \item{Depending on the results of the Roadmap, there could be one or several conceptual design studies for specific experiments that may be undertaken, under the aegis of the TVLBAI, and that may be developed independently by the participants possibly together with other external partners. In addition to scientific and technical aspects, these should take into account environmental impacts, particularly in cases where extensive civil engineering would be required.}
    \item{The TVLBAI Study supports sharing of ideas and encourages free and open exchange of scientific and technical knowledge, expertise, engineering designs, and equipment, within provisions further detailed in the MoU.}
    \item{Potential synergies with other projects shall be explored and used where beneficial to the TVLBAI Study.}
\end{itemize}
To execute these activities, an International Collaboration Board (ICB) is to be formed, whose mandate is to oversee the TVLBAI Study, define the strategies and channel contributions from the Participants. The Board elects a Study Chair who may be supported by a Coordination Group in executing its mandate.
This MoU is an agreement \textit{inter pares} and is not centred nor focused on any specific laboratory or project. Nevertheless, CERN has offered its support in establishing the MoU through its legal service, and will act as central repository for collecting all documents signed by the individual parties and the Study Chair as representative of the ICB.

As stated in the MoU, participants in the TVLBAI Study intend to organise, conduct and disseminate their collaborative research work with due regard for equity, diversity and inclusion.

\appendix
\section*{Appendix: Poster Session}
\addcontentsline{toc}{section}{Appendix: Poster Session}
\hspace{-5mm}

\subsection*{A.1 Introduction}
\addcontentsline{toc}{subsection}{A.1 Introduction}

\begin{figure*}
\centering
\includegraphics[width=0.6\textwidth]{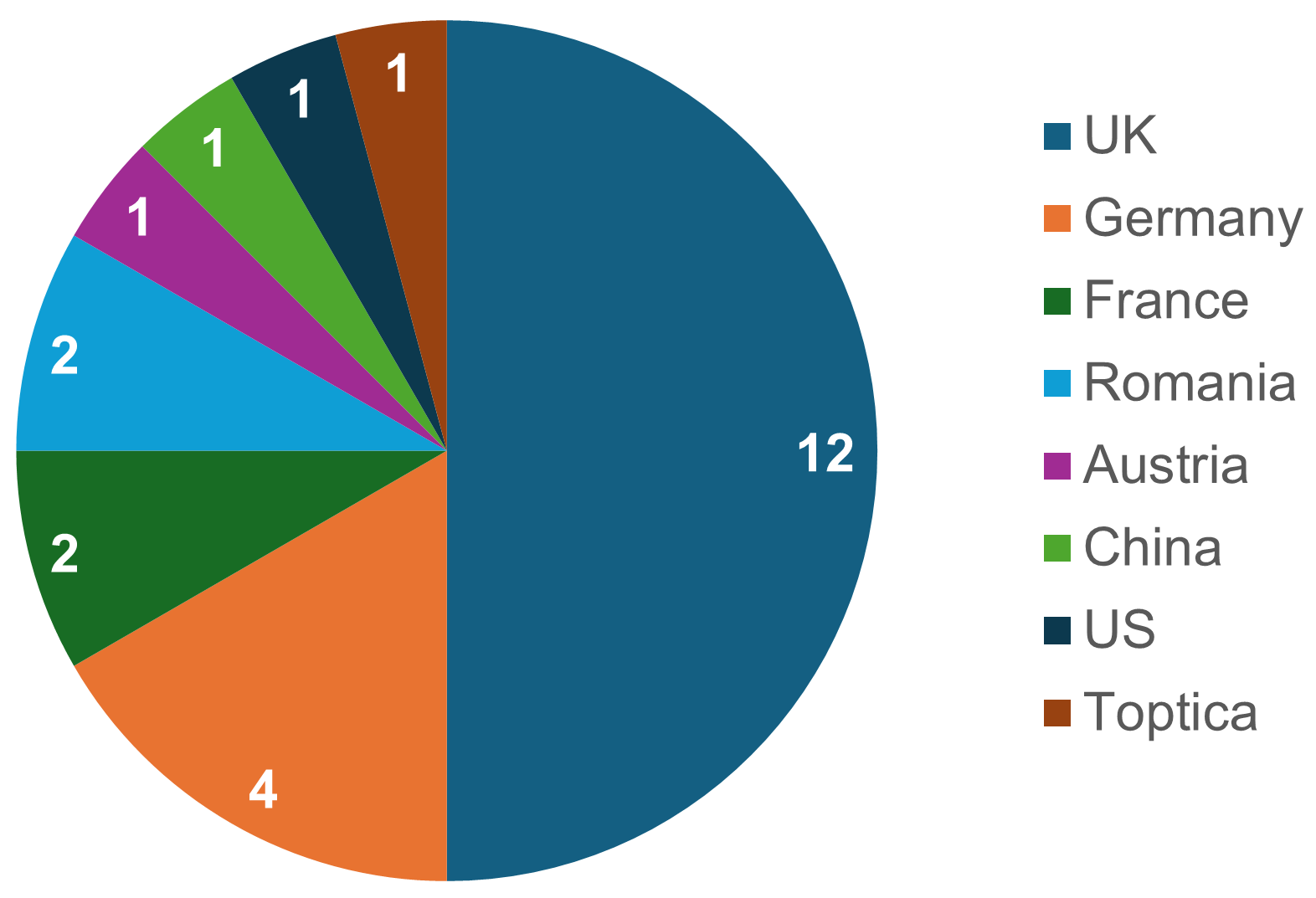}
\caption{Statistics of the geographical distribution of the poster presenters' home institutions.}
\label{fig:poster_participants}
\end{figure*}

A 3-hour poster session during the workshop provided 23 participants with an opportunity to showcase their research. Presenters consisted of PhD students and other early career researchers (ECRs), offering them a platform to discuss their work and build connections with fellow academics. The session featured representation from 12 institutes across the globe, including the US, China, Europe and the UK, as shown in Fig.~\ref{fig:poster_participants}. As seen there, industry  was also represented, with participation and sponsorship from TOPTICA Photonics, a manufacturer of lasers for quantum technologies, biophotonics and material inspection.

Although the talks centered on long-baseline atom interferometry, the poster session encompassed a wide range of topics, facilitating discussions across diverse subject areas. Additional contributions covered fields such as shorter baseline experiments, nuclear interferometry and optical clocks, promoting the exchange of ideas and techniques that are relevant across different applications. The variety of posters, representing both theoretical and experimental approaches, enabled valuable interdisciplinary discussions. The session successfully encouraged networking and collaboration, fostering the development of connections among ECRs.

\subsection*{A.2 Abstract Titles}
\addcontentsline{toc}{subsection}{A.2 Abstract Titles}

Below is a list of the titles of the abstracts for the posters presented. The full abstracts can be found on the workshop Indico page. \footnote{\url{https://indico.cern.ch/event/1369392/timetable/}}

\begin{itemize}
    \item Alice Josset: ``The Atom Interferometer Observatory and Network (AION) detector"
    \item Ashkan Alibabaei: ``Investigating the fundamental limits of Large Momentum Transfer (LMT) Atom Interferometry"
    \item Chung Chuan Hsu: ``Atom Interferometry Observatory and Network (AION) for dark matter and gravitational waves detection"
    \item Daniel Derr, Enno Giese: ``Internal structure of atoms for dark matter detection and tests of the Einstein equivalence principle"
    \item Elizabeth Pasatembou: ``Testing the boundaries of fundamental physics with atomic clocks"
    \item Florentina Pislan: ``Catalogues of potential gravitational wave sources for low to mid-frequency detectors"
    \item Gedminas Elertas: ``Developing a phase-shear detection platform for MAGIS and AION projects"
    \item Hannah Banks: ``The nuclear interferometer as a detector for ultra-light dark matter"
    \item Jiajun Chen, Yijun Tang: ``Efficient cooling and transport of strontium atoms for the AION detector"
    \item John Carlton: ``Characterising noise in long-baseline terrestrial atom interferometer experiments"
    \item Jonathan Ramwell (Toptica): ``Small table-top display for Toptica"
    \item Jordan Gué: ``Expected experimental signals in atom interferometers from scalar dark matter non universally coupled to standard matter"
    \item Junjie Jiang: ``Testing the equivalence principle with atom interferometry"
    \item Kamran Hussain: ``AION and MAGIS: Probing gravitational waves and searching for ultra-light dark matter"
    \item Leonardo Badurina: ``Extending the physics case of atom gradiometers to ultra-heavy dark matter"
    \item Leonie Hawkins: ``Upgrading a frequency standard fountain for atom interferometry"
    \item Ludovico Iannizzotto Venezze: ``Using continuously-operating optical clocks for ultra-light dark matter search"
    \item Maria Isfan: ``Developing a quantum neural network based low latency pipeline for gravitational wave data analysis"
    \item Michael Werner: ``A novel interferometer scheme for measuring spacetime curvature and observing the gravitational Aharonov-Bohm effect"
    \item Oliver Ennis: ``Enhancing large momentum transfer in the AION project"
    \item Sebastian Wald: ``Entangled Mach-Zehnder type Atom Interferometer in an optical, propagating-wave cavity"
    \item Selyan Beldjoudi: ``Achieving Large Momentum Transfer through stroboscopic stabilization of a Floquet state"
    \item Simon Hack: ``Setting up a lattice atom interferometer for precision measurements and searches for new physics"
    \item Thomas Walker: ``Designs for an ultracold strontium source for atom interferometry and optical atomic clock experiments"
    \item Vishu Gupta: ``The Very Long Baseline Atom Interferometry (VLBAI) facility for highly precise inertial measurements"
\end{itemize}

\subsection*{A.3 ECR Engagement and Fostering Collaboration}
\addcontentsline{toc}{subsection}{A.3 ECR Engagement and Fostering Collaboration}

The poster session provided an excellent platform for all participants to engage with the research of PhD students and other ECRs. It offered an opportunity for in-depth discussions on specific research topics related to the development of atom interferometers and the exploration of fundamental physics using quantum technology. 

The diversity in university representation and research topics demonstrated the keen interest of ECRs in joining the planned long-baseline atom interferometry proto-collaboration. The formation of this proto-collaboration can significantly benefit PhD students and ECRs by providing an organizational framework for easier exchange of ideas and fostering collaboration. Additionally, it will enhance their learning opportunities through interactions with more experienced researchers and help them diversify their skills. This, in turn, will accelerate the research and development of technology and increase research output. It will also foster innovation and creativity as PhDs and ECRs from diverse backgrounds and with various skills collaborate. 

The proto-collaboration will facilitate the establishment of a database of PhDs and ECRs trained in quantum technology-related skills, providing easier access to talent by academic institutions within the collaboration, the broader academic community, and industry.  

Involving PhD students and ECRs in the proto-collaboration can enhance its impact both within the academic setting and beyond. This was evidenced already by the poster session at this workshop, which facilitated a significant exchange of ideas and knowledge between the presenters and participants.


\bibliographystyle{JHEP}
\bibliography{main}

\providecommand{\href}[2]{#2}\begingroup\raggedright\begin{thebibliography}{100}

\bibitem{2ndTVLBAIWorkshop}
``Second {T}errestrial {V}ery-{L}ong-{B}aseline {A}tom {I}nterferometry
  {W}orkshop, {Imperial College}, {April} 2024.''
  \url{https://indico.cern.ch/event/1369392/}, 2024.

\bibitem{1stTVLBAIWorkshop}
``{First Terrestrial Very-Long-Baseline Atom Interferometry Workshop, CERN,
  March 2023}.'' \url{https://indico.cern.ch/event/1208783/}, 2023.

\bibitem{TVLBAISummary}
S.~Abend et~al., \emph{Terrestrial {V}ery-{L}ong-{B}aseline {A}tom
  {I}nterferometry: Workshop {S}ummary},
  \href{https://doi.org/10.1116/5.0185291}{\emph{AVS Quantum Science}
  {\bfseries 6} (2024) 024701}
  [\href{https://arxiv.org/abs/2310.08183}{{\ttfamily 2310.08183}}].

\bibitem{TVLBAIMOU}
``Memorandum of {U}nderstanding for the {T}errestrial {V}ery {L}ong {B}aseline
  {A}tom {I}nterferometer {S}tudy.''
  \url{https://indico.cern.ch/event/1369392/attachments/2789312/5096609/TVLBAI%20Study%20MOU%20%20Final%20.pdf}.

\bibitem{Buchmueller2023}
O.~Buchmueller, J.~Ellis and U.~Schneider, \emph{{Large-scale atom
  interferometry for fundamental physics}},
  \href{https://doi.org/10.1080/00107514.2023.2239008}{\emph{Contemp. Phys.}
  {\bfseries 64} (2023) 93} [\href{https://arxiv.org/abs/2306.17726}{{\ttfamily
  2306.17726}}].

\bibitem{Geraci2016}
A.A.~Geraci and A.~Derevianko, \emph{{Sensitivity of atom interferometry to
  ultralight scalar field dark matter}},
  \href{https://doi.org/10.1103/PhysRevLett.117.261301}{\emph{Phys. Rev. Lett.}
  {\bfseries 117} (2016) 261301}
  [\href{https://arxiv.org/abs/1605.04048}{{\ttfamily 1605.04048}}].

\bibitem{Arvanitaki2018}
A.~Arvanitaki, P.W.~Graham, J.M.~Hogan, S.~Rajendran and K.~Van~Tilburg,
  \emph{Search for light scalar dark matter with atomic gravitational wave
  detectors}, \href{https://doi.org/10.1103/PhysRevD.97.075020}{\emph{Phys.
  Rev. D} {\bfseries 97} (2018) 075020}.

\bibitem{Dimopoulos:2007cj}
S.~Dimopoulos, P.W.~Graham, J.M.~Hogan, M.A.~Kasevich and S.~Rajendran,
  \emph{{Gravitational Wave Detection with Atom Interferometry}},
  \href{https://doi.org/10.1016/j.physletb.2009.06.011}{\emph{Phys. Lett. B}
  {\bfseries 678} (2009) 37} [\href{https://arxiv.org/abs/0712.1250}{{\ttfamily
  0712.1250}}].

\bibitem{Dimopoulos2008a}
S.~Dimopoulos, P.W.~Graham, J.M.~Hogan, M.A.~Kasevich and S.~Rajendran,
  \emph{Atomic gravitational wave interferometric sensor},
  \href{https://doi.org/10.1103/PhysRevD.78.122002}{\emph{Phys. Rev. D}
  {\bfseries 78} (2008) 1753}.

\bibitem{Dickerson2013}
S.M.~Dickerson, J.M.~Hogan, A.~Sugarbaker, D.M.S.~Johnson and M.A.~Kasevich,
  \emph{Multiaxis inertial sensing with long-time point source atom
  interferometry},
  \href{https://doi.org/10.1103/PhysRevLett.111.083001}{\emph{Phys. Rev. Lett.}
  {\bfseries 111} (2013) 083001}.

\bibitem{schlippert2020matter}
D.~Schlippert, C.~Meiners, R.~Rengelink, C.~Schubert, D.~Tell, {\'E}.~Wodey
  et~al., \emph{Matter-wave interferometry for inertial sensing and tests of
  fundamental physics},  in \emph{Proceedings of the {E}ighth {M}eeting on
  {CPT} and {L}orentz {S}ymmetry}, pp.~37--40, World Scientific, 2020,
  \href{https://doi.org/10.1142/9789811213984_0010}{DOI}.

\bibitem{Badurina2020}
L.~Badurina, E.~Bentine, D.~Blas, K.~Bongs, D.~Bortoletto, T.~Bowcock et~al.,
  \emph{{AION}: an {A}tom {I}nterferometer {O}bservatory and {N}etwork},
  \href{https://doi.org/10.1088/1475-7516/2020/05/011}{\emph{J. Cosmol.
  Astropart. Phys.} {\bfseries 2020} (2020) 011}.

\bibitem{Canuel_2018}
B.~Canuel et~al., \emph{{Exploring gravity with the MIGA large scale atom
  interferometer}},
  \href{https://doi.org/10.1038/s41598-018-32165-z}{\emph{Sci. Rep.} {\bfseries
  8} (2018) 14064} [\href{https://arxiv.org/abs/1703.02490}{{\ttfamily
  1703.02490}}].

\bibitem{Zhan2019}
M.-S.~Zhan, J.~Wang, W.-T.~Ni, D.-F.~Gao, G.~Wang, L.-X.~He et~al.,
  \emph{{ZAIGA}: {Z}haoshan long-baseline {A}tom {I}nterferometer {G}ravitation
  {A}ntenna}, \href{https://doi.org/10.1142/S0218271819400054}{\emph{Int. J.
  Mod. Phys. D} {\bfseries 29} (2020) 1940005}.

\bibitem{Abe2021}
M.~Abe, P.~Adamson, M.~Borcean, D.~Bortoletto, K.~Bridges, S.P.~Carman et~al.,
  \emph{{Matter-wave Atomic Gradiometer Interferometric Sensor (MAGIS-100)}},
  \href{https://doi.org/10.1088/2058-9565/abf719}{\emph{Quantum Sci. Technol.}
  {\bfseries 6} (2021) 044003}.

\bibitem{Canuel2020}
B.~Canuel, S.~Abend, P.~Amaro-Seoane, F.~Badaracco, Q.~Beaufils, A.~Bertoldi
  et~al., \emph{{ELGAR}---a {E}uropean {L}aboratory for {G}ravitation and
  {A}tom-interferometric {R}esearch},
  \href{https://doi.org/10.1088/1361-6382/aba80e}{\emph{Class. Quantum Grav.}
  {\bfseries 37} (2020) 225017}.

\bibitem{ElNeaj2020}
Y.A.~El-Neaj, C.~Alpigiani, S.~Amairi-Pyka, H.~Ara\'ujo, A.~Balaž, A.~Bassi
  et~al., \emph{{AEDGE}: {A}tomic {E}xperiment for {D}ark matter and {G}ravity
  {E}xploration in space},
  \href{https://doi.org/10.1140/epjqt/s40507-020-0080-0}{\emph{EPJ Quantum
  Technol.} {\bfseries 7} (2020) 127}.

\bibitem{Ellis:2020lxl}
J.~Ellis and V.~Vaskonen, \emph{{Probes of gravitational waves with atom
  interferometers}},
  \href{https://doi.org/10.1103/PhysRevD.101.124013}{\emph{Phys. Rev. D}
  {\bfseries 101} (2020) 124013}
  [\href{https://arxiv.org/abs/2003.13480}{{\ttfamily 2003.13480}}].

\bibitem{Badurina:2021rgt}
L.~Badurina, O.~Buchmueller, J.~Ellis, M.~Lewicki, C.~McCabe and V.~Vaskonen,
  \emph{{Prospective sensitivities of atom interferometers to gravitational
  waves and ultralight dark matter}},
  \href{https://doi.org/10.1098/rsta.2021.0060}{\emph{Phil. Trans. A. Math.
  Phys. Eng. Sci.} {\bfseries 380} (2021) 20210060}
  [\href{https://arxiv.org/abs/2108.02468}{{\ttfamily 2108.02468}}].

\bibitem{NANOGrav:2023hde}
{\scshape NANOGrav} collaboration, \emph{{The NANOGrav 15 yr Data Set:
  Observations and Timing of 68 Millisecond Pulsars}},
  \href{https://doi.org/10.3847/2041-8213/acda9a}{\emph{Astrophys. J. Lett.}
  {\bfseries 951} (2023) L9}
  [\href{https://arxiv.org/abs/2306.16217}{{\ttfamily 2306.16217}}].

\bibitem{EPTA:2023xxk}
{\scshape EPTA \& InPTA} collaboration, \emph{{The second data release from the
  European Pulsar Timing Array - IV. Implications for massive black holes, dark
  matter, and the early Universe}},
  \href{https://doi.org/10.1051/0004-6361/202347433}{\emph{Astron. Astrophys.}
  {\bfseries 685} (2024) A94}
  [\href{https://arxiv.org/abs/2306.16227}{{\ttfamily 2306.16227}}].

\bibitem{Zic:2023gta}
{\scshape Parkes Pulsar Timing Array} collaboration, \emph{{The Parkes Pulsar
  Timing Array Third Data Release}},
  \href{https://arxiv.org/abs/2306.16230}{{\ttfamily 2306.16230}}.

\bibitem{Xu:2023wog}
{\scshape Chinese Pulsar Timing Array} collaboration, \emph{{Searching for the
  nano-Hertz stochastic gravitational wave background with the Chinese Pulsar
  Timing Array Data Release I}},
  \href{https://arxiv.org/abs/2306.16216}{{\ttfamily 2306.16216}}.

\bibitem{Ellis:2024wdh}
J.~Ellis, M.~Fairbairn, G.~H\"utsi, J.~Urrutia, V.~Vaskonen and H.~Veerm\"ae,
  \emph{{Consistency of JWST Black Hole Observations with NANOGrav
  Gravitational Wave Measurements}},
  \href{https://arxiv.org/abs/2403.19650}{{\ttfamily 2403.19650}}.

\bibitem{Ellis2023b}
J.~Ellis, M.~Fairbairn, G.~Franciolini, G.~H\"utsi, A.~Iovino, M.~Lewicki
  et~al., \emph{{What is the source of the {PTA} {GW} signal?}},
  [\href{https://arxiv.org/abs/arXiv:2308.08546}{{\ttfamily
  arXiv:2308.08546}}].

\bibitem{Schilling2020}
M.~Schilling, E.~Wodey, L.~Timmen, D.~Tell, K.H.~Zipfel, D.~Schlippert et~al.,
  \emph{Gravity field modelling for the {H}annover 10 m atom interferometer},
  \href{https://doi.org/10.1007/s00190-020-01451-y}{\emph{J. Geod.} {\bfseries
  94} (2020) 203003}.

\bibitem{wodey2019towards}
{\'E}.~Wodey, M.~Schilling, D.~Tell, C.~Schubert, D.~Schlippert, W.~Ertmer
  et~al., \emph{Towards gravity reference stations with very long baseline atom
  interferometry.},  in \emph{Geophysical Research Abstracts}, vol.~21, 2019.

\bibitem{Beaufils2022}
Q.~Beaufils, L.A.~Sidorenkov, P.~Lebegue, B.~Venon, D.~Holleville, L.~Volodimer
  et~al., \emph{Cold-atom sources for the {M}atter-wave laser {I}nterferometric
  {G}ravitation {A}ntenna ({MIGA})},
  \href{https://doi.org/10.1038/s41598-022-23468-3}{\emph{Sci. Rep.} {\bfseries
  12} (2022) 231102}.

\bibitem{KCLClimateChange}
``{Cold Atoms and Climate Change (Inter-disciplinary Workshop on Environmental
  Applications of Quantum Sensors), March 2022}.''
  \url{https://indico.kcl.ac.uk/event/268/}, (2022).

\bibitem{zehnder1891neuer}
L.~Zehnder, \emph{Ein neuer {I}nterferenzrefraktor}, {\emph{Zeitschrift f{\"u}r
  Instrumentenkunde} {\bfseries 11} (1891) 275}.

\bibitem{mach1892ueber}
L.~Mach, \emph{Ueber einen {I}nterferenzrefraktor}, {\emph{Zeitschrift f{\"u}r
  Instrumentenkunde} {\bfseries 12} (1892) 89}.

\bibitem{Badurina:2022ngn}
L.~Badurina, V.~Gibson, C.~McCabe and J.~Mitchell, \emph{{Ultralight dark
  matter searches at the sub-Hz frontier with atom multigradiometry}},
  \href{https://doi.org/10.1103/PhysRevD.107.055002}{\emph{Phys. Rev. D}
  {\bfseries 107} (2023) 055002}
  [\href{https://arxiv.org/abs/2211.01854}{{\ttfamily 2211.01854}}].

\bibitem{2022PDG}
R.L.~{Workman} et~al., \emph{{Review of Particle Physics}},
  \href{https://doi.org/10.1093/ptep/ptac097}{\emph{Progress of Theoretical and
  Experimental Physics} {\bfseries 2022} (2022) 083C01}.

\bibitem{2024WIMP}
G.~{Arcadi}, D.~{Cabo-Almeida}, M.~{Dutra}, P.~{Ghosh}, M.~{Lindner},
  Y.~{Mambrini} et~al., \emph{{The Waning of the WIMP: Endgame?}},
  \href{https://doi.org/10.48550/arXiv.2403.15860}{\emph{arXiv e-prints} (2024)
  arXiv:2403.15860} [\href{https://arxiv.org/abs/2403.15860}{{\ttfamily
  2403.15860}}].

\bibitem{2017DM}
M.~{Battaglieri} et~al., \emph{{US Cosmic Visions: New Ideas in Dark Matter
  2017: Community Report}},
  \href{https://doi.org/10.48550/arXiv.1707.04591}{\emph{arXiv e-prints} (2017)
  arXiv:1707.04591} [\href{https://arxiv.org/abs/1707.04591}{{\ttfamily
  1707.04591}}].

\bibitem{2023DMQS}
A.~Chou, K.~Irwin, R.H.~Maruyama, O.K.~Baker, C.~Bartram, K.K.~Berggren et~al.,
  \emph{Quantum sensors for high energy physics},
  \textcolor{blue}{arXiv:2311.01930}, 2023.

\bibitem{Antypas2022}
D.~Antypas et~al., \emph{New horizons: Scalar and vector ultralight dark
  matter},  \textcolor{blue}{arXiv:2203.14915}, 2022.

\bibitem{SafBudDem18}
M.S.~{Safronova}, D.~{Budker}, D.~{DeMille}, D.F.J.~{Kimball}, A.~{Derevianko}
  and C.W.~{Clark}, \emph{{Search for New Physics with Atoms and Molecules}},
  {\emph{Rev. Mod. Phys} {\bfseries 90} (2018) 025008}.

\bibitem{LIGOScientific:2016aoc}
{\scshape LIGO Scientific, Virgo} collaboration, \emph{{Observation of
  Gravitational Waves from a Binary Black Hole Merger}},
  \href{https://doi.org/10.1103/PhysRevLett.116.061102}{\emph{Phys. Rev. Lett.}
  {\bfseries 116} (2016) 061102}
  [\href{https://arxiv.org/abs/1602.03837}{{\ttfamily 1602.03837}}].

\bibitem{KAGRA:2021vkt}
{\scshape KAGRA, VIRGO, LIGO Scientific} collaboration, \emph{{GWTC-3: Compact
  Binary Coalescences Observed by LIGO and Virgo during the Second Part of the
  Third Observing Run}},
  \href{https://doi.org/10.1103/PhysRevX.13.041039}{\emph{Phys. Rev. X}
  {\bfseries 13} (2023) 041039}
  [\href{https://arxiv.org/abs/2111.03606}{{\ttfamily 2111.03606}}].

\bibitem{Agazie2023}
{\scshape {NANOG}rav} collaboration, \emph{The {NANOG}rav 15 yr data set:
  Evidence for a gravitational-wave background},
  \href{https://doi.org/10.3847/2041-8213/acdac6}{\emph{ApJL} {\bfseries 951}
  (2023) L8}.

\bibitem{Antoniadis2023a}
{\scshape {{EPTA}}} collaboration, \emph{{The second data release from the
  {E}uropean {P}ulsar {T}iming {A}rray {IV}. {S}earch for continuous
  gravitational wave signals}},
  [\href{https://arxiv.org/abs/arXiv:2306.16226}{{\ttfamily
  arXiv:2306.16226}}].

\bibitem{Agazie2023b}
{\scshape {NANOG}rav} collaboration, \emph{The {NANOG}rav 15 yr data set:
  Constraints on supermassive black hole binaries from the gravitational-wave
  background}, \href{https://doi.org/10.3847/2041-8213/ace18b}{\emph{ApJL}
  {\bfseries 952} (2023) L37}.

\bibitem{Moore:2014lga}
C.J.~Moore, R.H.~Cole and C.P.L.~Berry, \emph{{Gravitational-wave sensitivity
  curves}}, \href{https://doi.org/10.1088/0264-9381/32/1/015014}{\emph{Class.
  Quant. Grav.} {\bfseries 32} (2015) 015014}
  [\href{https://arxiv.org/abs/1408.0740}{{\ttfamily 1408.0740}}].

\bibitem{LVKRunPlans}
``{LIGO, Virgo and KAGRA Observing Run Plans}.''
  \url{https://observing.docs.ligo.org/plan/#}, 2024.

\bibitem{Evans:2023euw}
M.~Evans et~al., \emph{{Cosmic Explorer: A Submission to the NSF MPSAC ngGW
  Subcommittee}},  \href{https://arxiv.org/abs/2306.13745}{{\ttfamily
  2306.13745}}.

\bibitem{Maggiore:2019uih}
M.~Maggiore et~al., \emph{{Science Case for the Einstein Telescope}},
  \href{https://doi.org/10.1088/1475-7516/2020/03/050}{\emph{JCAP} {\bfseries
  03} (2020) 050} [\href{https://arxiv.org/abs/1912.02622}{{\ttfamily
  1912.02622}}].

\bibitem{LISA:2017pwj}
{\scshape LISA} collaboration, \emph{{Laser Interferometer Space Antenna}},
  \href{https://arxiv.org/abs/arXiv:1702.00786}{{\ttfamily arXiv:1702.00786}}.

\bibitem{Colpi:2024xhw}
M.~Colpi et~al., \emph{{LISA Definition Study Report}},
  \href{https://arxiv.org/abs/arXiv:2402.07571}{{\ttfamily arXiv:2402.07571}}.

\bibitem{Torres-Orjuela:2023dle}
A.~Torres-Orjuela, \emph{{Detecting intermediate-mass black hole binaries with
  atom interferometer observatories: Using the resonant mode for the merger
  phase}}, \href{https://doi.org/10.1116/5.0162505}{\emph{AVS Quantum Sci.}
  {\bfseries 5} (2023) 045002}
  [\href{https://arxiv.org/abs/2306.08898}{{\ttfamily 2306.08898}}].

\bibitem{Valiante:2020zhj}
R.~Valiante, M.~Colpi, R.~Schneider, A.~Mangiagli, M.~Bonetti, G.~Cerini
  et~al., \emph{{Unveiling early black hole growth with multifrequency
  gravitational wave observations}},
  \href{https://doi.org/10.1093/mnras/staa3395}{\emph{Mon. Not. Roy. Astron.
  Soc.} {\bfseries 500} (2020) 4095}
  [\href{https://arxiv.org/abs/2010.15096}{{\ttfamily 2010.15096}}].

\bibitem{Peters:1963ux}
P.C.~Peters and J.~Mathews, \emph{{Gravitational radiation from point masses in
  a Keplerian orbit}},
  \href{https://doi.org/10.1103/PhysRev.131.435}{\emph{Phys. Rev.} {\bfseries
  131} (1963) 435}.

\bibitem{ArcaSedda2020}
M.A.~Sedda, C.P.L.~Berry, K.~Jani, P.~Amaro-Seoane, P.~Auclair, J.~Baird
  et~al., \emph{The missing link in gravitational-wave astronomy: discoveries
  waiting in the decihertz range},
  \href{https://doi.org/10.1088/1361-6382/abb5c1}{\emph{Class. Quantum Grav.}
  {\bfseries 37} (2020) 215011}.

\bibitem{Livio2000}
M.~{Livio}, \emph{{The Progenitors of Type Ia Supernovae}},  in \emph{Type Ia
  Supernovae, Theory and Cosmology}, J.C.~{Niemeyer} and J.W.~{Truran}, eds.,
  p.~33, Jan., 2000,
  \href{https://doi.org/10.48550/arXiv.astro-ph/9903264}{DOI}
  [\href{https://arxiv.org/abs/astro-ph/9903264}{{\ttfamily
  astro-ph/9903264}}].

\bibitem{Mandel:2017pzd}
I.~Mandel, A.~Sesana and A.~Vecchio, \emph{{The astrophysical science case for
  a decihertz gravitational-wave detector}},
  \href{https://doi.org/10.1088/1361-6382/aaa7e0}{\emph{Class. Quant. Grav.}
  {\bfseries 35} (2018) 054004}
  [\href{https://arxiv.org/abs/1710.11187}{{\ttfamily 1710.11187}}].

\bibitem{Graham2018a}
P.W.~Graham and S.~Jung, \emph{Localizing gravitational wave sources with
  single-baseline atom interferometers},
  \href{https://doi.org/10.1103/PhysRevD.97.024052}{\emph{Phys. Rev. D}
  {\bfseries 97} (2018) 024052}.

\bibitem{Kaup:1968zz}
D.J.~Kaup, \emph{{Klein-Gordon Geon}},
  \href{https://doi.org/10.1103/PhysRev.172.1331}{\emph{Phys. Rev.} {\bfseries
  172} (1968) 1331}.

\bibitem{Ruffini:1969qy}
R.~Ruffini and S.~Bonazzola, \emph{{Systems of selfgravitating particles in
  general relativity and the concept of an equation of state}},
  \href{https://doi.org/10.1103/PhysRev.187.1767}{\emph{Phys. Rev.} {\bfseries
  187} (1969) 1767}.

\bibitem{Eby:2016cnq}
J.~Eby, M.~Leembruggen, P.~Suranyi and L.C.R.~Wijewardhana, \emph{{Collapse of
  Axion Stars}}, \href{https://doi.org/10.1007/JHEP12(2016)066}{\emph{JHEP}
  {\bfseries 12} (2016) 066}
  [\href{https://arxiv.org/abs/1608.06911}{{\ttfamily 1608.06911}}].

\bibitem{Levkov:2016rkk}
D.G.~Levkov, A.G.~Panin and I.I.~Tkachev, \emph{{Relativistic axions from
  collapsing Bose stars}},
  \href{https://doi.org/10.1103/PhysRevLett.118.011301}{\emph{Phys. Rev. Lett.}
  {\bfseries 118} (2017) 011301}
  [\href{https://arxiv.org/abs/1609.03611}{{\ttfamily 1609.03611}}].

\bibitem{Eby:2015hyx}
J.~Eby, P.~Suranyi and L.C.R.~Wijewardhana, \emph{{The Lifetime of Axion
  Stars}}, \href{https://doi.org/10.1142/S0217732316500905}{\emph{Mod. Phys.
  Lett. A} {\bfseries 31} (2016) 1650090}
  [\href{https://arxiv.org/abs/1512.01709}{{\ttfamily 1512.01709}}].

\bibitem{Eby:2021ece}
J.~Eby, S.~Shirai, Y.V.~Stadnik and V.~Takhistov, \emph{{Probing relativistic
  axions from transient astrophysical sources}},
  \href{https://doi.org/10.1016/j.physletb.2021.136858}{\emph{Phys. Lett. B}
  {\bfseries 825} (2022) 136858}
  [\href{https://arxiv.org/abs/2106.14893}{{\ttfamily 2106.14893}}].

\bibitem{Arakawa:2023gyq}
J.~Arakawa, J.~Eby, M.S.~Safronova, V.~Takhistov and M.H.~Zaheer,
  \emph{{Detection of Bosenovae with Quantum Sensors on Earth and in Space}},
  \href{https://arxiv.org/abs/2306.16468}{{\ttfamily 2306.16468}}.

\bibitem{Arakawa:2024lqr}
J.~Arakawa, M.H.~Zaheer, J.~Eby, V.~Takhistov and M.S.~Safronova,
  \emph{{Bosenovae with Quadratically-Coupled Scalars in Quantum Sensing
  Experiments}},  \href{https://arxiv.org/abs/2402.06736}{{\ttfamily
  2402.06736}}.

\bibitem{Maseizik:2024qly}
D.~Maseizik and G.~Sigl, \emph{{Distributions and Collision Rates of ALP Stars
  in the Milky Way}},  \href{https://arxiv.org/abs/2404.07908}{{\ttfamily
  2404.07908}}.

\bibitem{Gorghetto:2024vnp}
M.~Gorghetto, E.~Hardy and G.~Villadoro, \emph{{More Axion Stars from
  Strings}},  \href{https://arxiv.org/abs/2405.19389}{{\ttfamily 2405.19389}}.

\bibitem{Chang:2024fol}
J.H.~Chang, P.J.~Fox and H.~Xiao, \emph{{Axion Stars: Mass Functions and
  Constraints}},  \href{https://arxiv.org/abs/2406.09499}{{\ttfamily
  2406.09499}}.

\bibitem{Roura2025}
A.~Roura, \emph{Atom interferometer as a freely falling clock for time-dilation
  measurements}, \href{https://doi.org/10.1088/2058-9565/ad9e2e}{\emph{Quantum
  Sci. Technol.} (2025) in press}
  [\href{https://arxiv.org/abs/2402.11065}{{\ttfamily 2402.11065}}].

\bibitem{Arduini2023}
G.~Arduini et~al., \emph{{A Long-Baseline Atom Interferometer at {CERN}:
  {C}onceptual Feasibility Study}},
  [\href{https://arxiv.org/abs/arXiv:2304.00614}{{\ttfamily
  arXiv:2304.00614}}].

\bibitem{Roura2020}
A.~Roura, \emph{Gravitational redshift in quantum-clock interferometry},
  \href{https://doi.org/10.1103/PhysRevX.10.021014}{\emph{Phys. Rev. X}
  {\bfseries 10} (2020) 021014}.

\bibitem{Sinha2011}
S.~Sinha and J.~Samuel, \emph{Atom interferometry and the gravitational
  redshift}, \href{https://doi.org/10.1088/0264-9381/28/14/145018}{\emph{Class.
  Quantum Grav.} {\bfseries 28} (2011) 145018}.

\bibitem{Zych2011}
M.~Zych, F.~Costa, I.~Pikovski and u.~Brukner, \emph{Quantum interferometric
  visibility as a witness of general relativistic proper time},
  \href{https://doi.org/10.1038/ncomms1498}{\emph{Nat. Commun.} {\bfseries 2}
  (2011) 166}.

\bibitem{Loriani2019}
S.~Loriani, D.~Schlippert, C.~Schubert, S.~Abend, H.~Ahlers, W.~Ertmer et~al.,
  \emph{Atomic source selection in space-borne gravitational wave detection},
  \href{https://doi.org/10.1088/1367-2630/ab22d0}{\emph{New J. Phys.}
  {\bfseries 21} (2019) 063030}.

\bibitem{Roura2021}
A.~Roura, C.~Schubert, D.~Schlippert and E.M.~Rasel, \emph{Measuring
  gravitational time dilation with delocalized quantum superpositions},
  \href{https://doi.org/10.1103/PhysRevD.104.084001}{\emph{Phys. Rev. D}
  {\bfseries 104} (2021) 084001}.

\bibitem{Ufrecht2020a}
C.~Ufrecht, F.~Di~Pumpo, A.~Friedrich, A.~Roura, C.~Schubert, D.~Schlippert
  et~al., \emph{Atom-interferometric test of the universality of gravitational
  redshift and free fall},
  \href{https://doi.org/10.1103/PhysRevResearch.2.043240}{\emph{Phys. Rev.
  Research} {\bfseries 2} (2020) 043240}.

\bibitem{DiPumpo2021}
F.~Di~Pumpo, C.~Ufrecht, A.~Friedrich, E.~Giese, W.P.~Schleich and W.G.~Unruh,
  \emph{Gravitational redshift tests with atomic clocks and atom
  interferometers},
  \href{https://doi.org/10.1103/PRXQuantum.2.040333}{\emph{PRX Quantum}
  {\bfseries 2} (2021) 411}.

\bibitem{DiPumpo2023}
F.~Di~Pumpo, A.~Friedrich, C.~Ufrecht and E.~Giese,
  \emph{Universality-of-clock-rates test using atom interferometry with {T}$^3$
  scaling}, \href{https://doi.org/10.1103/PhysRevD.107.064007}{\emph{Phys. Rev.
  D} {\bfseries 107} (2023) 411}.

\bibitem{AmaroSeoane2017}
{LISA Collaboration}, \emph{{{L}aser {I}nterferometer {S}pace {A}ntenna}},
  [\href{https://arxiv.org/abs/arXiv:1702.00786}{{\ttfamily
  arXiv:1702.00786}}].

\bibitem{Ruan2020}
W.-H.~Ruan, Z.-K.~Guo, R.-G.~Cai and Y.-Z.~Zhang, \emph{Taiji program:
  {G}ravitational-wave sources},
  \href{https://doi.org/10.1142/S0217751X2050075X}{\emph{Int. J. Mod. Phys. A}
  {\bfseries 35} (2020) 2050075}.

\bibitem{TianQin:2015yph}
{\scshape TianQin} collaboration, \emph{{TianQin: a space-borne gravitational
  wave detector}},
  \href{https://doi.org/10.1088/0264-9381/33/3/035010}{\emph{Class. Quant.
  Grav.} {\bfseries 33} (2016) 035010}
  [\href{https://arxiv.org/abs/1512.02076}{{\ttfamily 1512.02076}}].

\bibitem{Sathyaprakash:2012jk}
B.~Sathyaprakash et~al., \emph{{Scientific Objectives of Einstein Telescope}},
  \href{https://doi.org/10.1088/0264-9381/29/12/124013}{\emph{Class. Quant.
  Grav.} {\bfseries 29} (2012) 124013}
  [\href{https://arxiv.org/abs/1206.0331}{{\ttfamily 1206.0331}}].

\bibitem{Reitze2019}
D.~Reitze, R.X.~Adhikari, S.~Ballmer, B.~Barish, L.~Barsotti, G.~Billingsley
  et~al., \emph{Cosmic {Explorer}: The {U}.{S}. {Contribution} to
  {Gravitational}-{Wave} {Astronomy} beyond {LIGO}}, {\emph{Bull. Am. Astron.
  Soc.} {\bfseries 51} (2019) 1}.

\bibitem{LIGOScientific:2017vwq}
{\scshape LIGO Scientific, Virgo} collaboration, \emph{{GW170817: Observation
  of Gravitational Waves from a Binary Neutron Star Inspiral}},
  \href{https://doi.org/10.1103/PhysRevLett.119.161101}{\emph{Phys. Rev. Lett.}
  {\bfseries 119} (2017) 161101}
  [\href{https://arxiv.org/abs/1710.05832}{{\ttfamily 1710.05832}}].

\bibitem{LIGOScientific:2017ync}
{\scshape LIGO Scientific, Virgo, Fermi GBM, INTEGRAL, IceCube, AstroSat
  Cadmium Zinc Telluride Imager Team, IPN, Insight-Hxmt, ANTARES, Swift, AGILE
  Team, 1M2H Team, Dark Energy Camera GW-EM, DES, DLT40, GRAWITA, Fermi-LAT,
  ATCA, ASKAP, Las Cumbres Observatory Group, OzGrav, DWF (Deeper Wider Faster
  Program), AST3, CAASTRO, VINROUGE, MASTER, J-GEM, GROWTH, JAGWAR,
  CaltechNRAO, TTU-NRAO, NuSTAR, Pan-STARRS, MAXI Team, TZAC Consortium, KU,
  Nordic Optical Telescope, ePESSTO, GROND, Texas Tech University, SALT Group,
  TOROS, BOOTES, MWA, CALET, IKI-GW Follow-up, H.E.S.S., LOFAR, LWA, HAWC,
  Pierre Auger, ALMA, Euro VLBI Team, Pi of Sky, Chandra Team at McGill
  University, DFN, ATLAS Telescopes, High Time Resolution Universe Survey,
  RIMAS, RATIR, SKA South Africa/MeerKAT} collaboration, \emph{{Multi-messenger
  Observations of a Binary Neutron Star Merger}},
  \href{https://doi.org/10.3847/2041-8213/aa91c9}{\emph{Astrophys. J. Lett.}
  {\bfseries 848} (2017) L12}
  [\href{https://arxiv.org/abs/1710.05833}{{\ttfamily 1710.05833}}].

\bibitem{KAGRA:2021duu}
{\scshape KAGRA, VIRGO, LIGO Scientific} collaboration, \emph{{Population of
  Merging Compact Binaries Inferred Using Gravitational Waves through GWTC-3}},
  \href{https://doi.org/10.1103/PhysRevX.13.011048}{\emph{Phys. Rev. X}
  {\bfseries 13} (2023) 011048}
  [\href{https://arxiv.org/abs/2111.03634}{{\ttfamily 2111.03634}}].

\bibitem{LIGOScientific:2020iuh}
{\scshape LIGO Scientific, Virgo} collaboration, \emph{{GW190521: A Binary
  Black Hole Merger with a Total Mass of $150 M_{\odot}$}},
  \href{https://doi.org/10.1103/PhysRevLett.125.101102}{\emph{Phys. Rev. Lett.}
  {\bfseries 125} (2020) 101102}
  [\href{https://arxiv.org/abs/2009.01075}{{\ttfamily 2009.01075}}].

\bibitem{Fairhurst:2023beb}
S.~Fairhurst, C.~Mills, M.~Colpi, R.~Schneider, A.~Sesana, A.~Trinca et~al.,
  \emph{{Identifying heavy stellar black holes at cosmological distances with
  next-generation gravitational-wave observatories}},
  \href{https://doi.org/10.1093/mnras/stae443}{\emph{Mon. Not. Roy. Astron.
  Soc.} {\bfseries 529} (2024) 2116}
  [\href{https://arxiv.org/abs/2310.18158}{{\ttfamily 2310.18158}}].

\bibitem{Mills:2020thr}
C.~Mills and S.~Fairhurst, \emph{{Measuring gravitational-wave higher-order
  multipoles}}, \href{https://doi.org/10.1103/PhysRevD.103.024042}{\emph{Phys.
  Rev. D} {\bfseries 103} (2021) 024042}
  [\href{https://arxiv.org/abs/2007.04313}{{\ttfamily 2007.04313}}].

\bibitem{Torres-Orjuela:2024tmu}
A.~Torres-Orjuela, \emph{{Black hole spectroscopy with ground-based atom
  interferometer and space-based laser interferometer gravitational wave
  detectors}},  \href{https://arxiv.org/abs/arXiv:2405.10551}{{\ttfamily
  arXiv:2405.10551}}.

\bibitem{Badurina2021}
L.~Badurina, O.~Buchmueller, J.~Ellis, M.~Lewicki, C.~McCabe and V.~Vaskonen,
  \emph{Prospective sensitivities of atom interferometers to gravitational
  waves and ultralight dark matter},
  \href{https://doi.org/10.1098/rsta.2021.0060}{\emph{Phil. Trans. R. Soc. A.}
  {\bfseries 380} (2022) 035}.

\bibitem{Ellis:2023iyb}
J.~Ellis, M.~Fairbairn, J.~Urrutia and V.~Vaskonen, \emph{{Probing Supermassive
  Black Hole Seed Scenarios with Gravitational-wave Measurements}},
  \href{https://doi.org/10.3847/1538-4357/ad27d5}{\emph{Astrophys. J.}
  {\bfseries 964} (2024) 11}
  [\href{https://arxiv.org/abs/2312.02983}{{\ttfamily 2312.02983}}].

\bibitem{Toubiana2021}
A.~Toubiana, L.~Sberna, A.~Caputo, G.~Cusin, S.~Marsat, K.~Jani et~al.,
  \emph{Detectable environmental effects in {GW}190521-like black-hole binaries
  with {LISA}},
  \href{https://doi.org/10.1103/PhysRevLett.126.101105}{\emph{Phys. Rev. Lett.}
  {\bfseries 126} (2021) 101105}.

\bibitem{Sberna2022}
L.~Sberna, S.~Babak, S.~Marsat, A.~Caputo, G.~Cusin, A.~Toubiana et~al.,
  \emph{Observing {GW}190521-like binary black holes and their environment with
  {LISA}}, \href{https://doi.org/10.1103/PhysRevD.106.064056}{\emph{Phys. Rev.
  D} {\bfseries 106} (2022) 064056}.

\bibitem{Ellis2020}
J.~Ellis and V.~Vaskonen, \emph{Probes of gravitational waves with atom
  interferometers},
  \href{https://doi.org/10.1103/PhysRevD.101.124013}{\emph{Phys. Rev. D}
  {\bfseries 101} (2020) 124013}.

\bibitem{Caprini2020}
C.~Caprini, M.~Chala, G.C.~Dorsch, M.~Hindmarsh, S.J.~Huber, T.~Konstandin
  et~al., \emph{Detecting gravitational waves from cosmological phase
  transitions with {LISA}: an update},
  \href{https://doi.org/10.1088/1475-7516/2020/03/024}{\emph{J. Cosmol.
  Astropart. Phys.} {\bfseries 2020} (2020) 024}.

\bibitem{Auclair2020}
P.~Auclair, J.J.~Blanco-Pillado, D.G.~Figueroa, A.C.~Jenkins, M.~Lewicki,
  M.~Sakellariadou et~al., \emph{Probing the gravitational wave background from
  cosmic strings with {LISA}},
  \href{https://doi.org/10.1088/1475-7516/2020/04/034}{\emph{J. Cosmol.
  Astropart. Phys.} {\bfseries 2020} (2020) 034}.

\bibitem{LIGOScientific:2021sio}
{\scshape LIGO Scientific, VIRGO, KAGRA} collaboration, \emph{{Tests of General
  Relativity with GWTC-3}},  \href{https://arxiv.org/abs/2112.06861}{{\ttfamily
  2112.06861}}.

\bibitem{LIGOScientific:2021izm}
{\scshape LIGO Scientific, VIRGO} collaboration, \emph{{Search for Lensing
  Signatures in the Gravitational-Wave Observations from the First Half of
  LIGO\textendash{}Virgo\textquoteright{}s Third Observing Run}},
  \href{https://doi.org/10.3847/1538-4357/ac23db}{\emph{Astrophys. J.}
  {\bfseries 923} (2021) 14}
  [\href{https://arxiv.org/abs/2105.06384}{{\ttfamily 2105.06384}}].

\bibitem{LIGOScientific:2014pky}
{\scshape LIGO Scientific} collaboration, \emph{{Advanced LIGO}},
  \href{https://doi.org/10.1088/0264-9381/32/7/074001}{\emph{Class. Quant.
  Grav.} {\bfseries 32} (2015) 074001}
  [\href{https://arxiv.org/abs/1411.4547}{{\ttfamily 1411.4547}}].

\bibitem{Volonteri:2010wz}
M.~Volonteri, \emph{{Formation of Supermassive Black Holes}},
  \href{https://doi.org/10.1007/s00159-010-0029-x}{\emph{Astron. Astrophys.
  Rev.} {\bfseries 18} (2010) 279}
  [\href{https://arxiv.org/abs/1003.4404}{{\ttfamily 1003.4404}}].

\bibitem{Volonteri:2009vh}
M.~Volonteri and P.~Natarajan, \emph{{Journey to the $M_{\rm BH} - \sigma$
  relation: the fate of low mass black holes in the Universe}},
  \href{https://doi.org/10.1111/j.1365-2966.2009.15577.x}{\emph{Mon. Not. Roy.
  Astron. Soc.} {\bfseries 400} (2009) 1911}
  [\href{https://arxiv.org/abs/0903.2262}{{\ttfamily 0903.2262}}].

\bibitem{Kormendy:2013dxa}
J.~Kormendy and L.C.~Ho, \emph{{Coevolution (Or Not) of Supermassive Black
  Holes and Host Galaxies}},
  \href{https://doi.org/10.1146/annurev-astro-082708-101811}{\emph{Ann. Rev.
  Astron. Astrophys.} {\bfseries 51} (2013) 511}
  [\href{https://arxiv.org/abs/1304.7762}{{\ttfamily 1304.7762}}].

\bibitem{Chadayammuri:2022bjj}
U.~Chadayammuri, A.~Bogdan, A.~Ricarte and P.~Natarajan, \emph{{Constraints
  From Dwarf Galaxies on Black Hole Seeding and Growth Models With Current and
  Future Surveys}},
  \href{https://doi.org/10.3847/1538-4357/acbea6}{\emph{Astrophys. J.}
  {\bfseries 946} (2023) 51}
  [\href{https://arxiv.org/abs/2212.04693}{{\ttfamily 2212.04693}}].

\bibitem{NANOGrav:2023hfp}
{\scshape NANOGrav} collaboration, \emph{{The NANOGrav 15 yr Data Set:
  Constraints on Supermassive Black Hole Binaries from the Gravitational-wave
  Background}},
  \href{https://doi.org/10.3847/2041-8213/ace18b}{\emph{Astrophys. J. Lett.}
  {\bfseries 952} (2023) L37}
  [\href{https://arxiv.org/abs/2306.16220}{{\ttfamily 2306.16220}}].

\bibitem{Ellis:2023dgf}
J.~Ellis, M.~Fairbairn, G.~H\"utsi, J.~Raidal, J.~Urrutia, V.~Vaskonen et~al.,
  \emph{{Gravitational waves from supermassive black hole binaries in light of
  the NANOGrav 15-year data}},
  \href{https://doi.org/10.1103/PhysRevD.109.L021302}{\emph{Phys. Rev. D}
  {\bfseries 109} (2024) L021302}
  [\href{https://arxiv.org/abs/2306.17021}{{\ttfamily 2306.17021}}].

\bibitem{Raidal:2024odr}
J.~Raidal, J.~Urrutia, V.~Vaskonen and H.~Veerm\"ae, \emph{{Eccentricity
  effects on the SMBH GW background}},
  \href{https://arxiv.org/abs/2406.05125}{{\ttfamily 2406.05125}}.

\bibitem{Pacucci:2023oci}
F.~Pacucci, B.~Nguyen, S.~Carniani, R.~Maiolino and X.~Fan, \emph{{{JWST CEERS}
  and {JADES} Active Galaxies at $z = 4-7$ Violate the Local
  $M_{\bullet}-M_{\star}$ Relation at $>3\sigma$: Implications for Low-mass
  Black Holes and Seeding Models}},
  \href{https://doi.org/10.3847/2041-8213/ad0158}{\emph{Astrophys. J. Lett.}
  {\bfseries 957} (2023) L3}
  [\href{https://arxiv.org/abs/2308.12331}{{\ttfamily 2308.12331}}].

\bibitem{Matthee:2023utn}
J.~Matthee et~al., \emph{{Little Red Dots: An Abundant Population of Faint
  Active Galactic Nuclei at z \ensuremath{\sim} 5 Revealed by the EIGER and
  FRESCO JWST Surveys}},
  \href{https://doi.org/10.3847/1538-4357/ad2345}{\emph{Astrophys. J.}
  {\bfseries 963} (2024) 129}
  [\href{https://arxiv.org/abs/2306.05448}{{\ttfamily 2306.05448}}].

\bibitem{2023arXiv231003067P}
M.~{Perna}, S.~{Arribas}, I.~{Lamperti}, C.~{Circosta}, E.~{Bertola},
  P.G.~{P{\'e}rez-Gonz{\'a}lez} et~al., \emph{{A surprisingly high number of
  dual active galactic nuclei in the early Universe}},
  \href{https://arxiv.org/abs/arXiv:2310.03067}{{\ttfamily arXiv:2310.03067}}.

\bibitem{Canuel2018}
B.~Canuel, A.~Bertoldi, L.~Amand, E.~Pozzo~di Borgo, T.~Chantrait, C.~Danquigny
  et~al., \emph{Exploring gravity with the {MIGA} large scale atom
  interferometer}, \href{https://doi.org/10.1038/s41598-018-32165-z}{\emph{Sci.
  Rep.} {\bfseries 8} (2018) 2689}.

\bibitem{PhysRevLett.96.010401}
V.~Giovannetti, S.~Lloyd and L.~Maccone, \emph{Quantum metrology},
  \href{https://doi.org/10.1103/PhysRevLett.96.010401}{\emph{Phys. Rev. Lett.}
  {\bfseries 96} (2006) 010401}.

\bibitem{RMPPezze18}
L.~Pezz\`e, A.~Smerzi, M.K.~Oberthaler, R.~Schmied and P.~Treutlein,
  \emph{Quantum metrology with nonclassical states of atomic ensembles},
  \href{https://doi.org/10.1103/RevModPhys.90.035005}{\emph{Rev. Mod. Phys.}
  {\bfseries 90} (2018) 035005}.

\bibitem{Kovachy2015a}
T.~Kovachy, P.~Asenbaum, C.~Overstreet, C.A.~Donnelly, S.M.~Dickerson,
  A.~Sugarbaker et~al., \emph{Quantum superposition at the half-metre scale},
  \href{https://doi.org/10.1038/nature16155}{\emph{Nature} {\bfseries 528}
  (2015) 530}.

\bibitem{chio11}
S.~Chiow, T.~Kovachy, H.~Chien and M.A.~Kasevich, \emph{102 $\hbar k$ large
  area atom interferometers},
  \href{https://doi.org/10.1103/PhysRevLett.107.130403}{\emph{Phys. Rev. Lett.}
  {\bfseries 107} (2011) 130403}.

\bibitem{mcalpine2020excited}
K.E.~McAlpine, D.~Gochnauer and S.~Gupta, \emph{Excited-band {B}loch
  oscillations for precision atom interferometry},
  \href{https://doi.org/10.1103/PhysRevA.101.023614}{\emph{Physical Review A}
  {\bfseries 101} (2020) 023614}.

\bibitem{McGuirk2000}
J.M.~McGuirk, M.J.~Snadden and M.A.~Kasevich, \emph{Large area light-pulse atom
  interferometry},
  \href{https://doi.org/10.1103/PhysRevLett.85.4498}{\emph{Phys. Rev. Lett.}
  {\bfseries 85} (2000) 4498}.

\bibitem{Wilkason2022atom}
T.~Wilkason, M.~Nantel, J.~Rudolph, Y.~Jiang, B.E.~Garber, H.~Swan et~al.,
  \emph{Atom interferometry with {F}loquet atom optics},
  \href{https://doi.org/10.1103/PhysRevLett.129.183202}{\emph{Phys. Rev. Lett.}
  {\bfseries 129} (2022) 183202}.

\bibitem{Bott2023}
A.~Bott, F.~Di~Pumpo and E.~Giese, \emph{Atomic diffraction from single-photon
  transitions in gravity and {Standard-Model} extensions},
  \href{https://doi.org/10.1116/5.0174258}{\emph{AVS Quantum Sci.} {\bfseries
  5} (2023) 044402}.

\bibitem{Kotru2015}
K.~Kotru, D.L.~Butts, J.M.~Kinast and R.E.~Stoner, \emph{Large-area atom
  interferometry with frequency-swept {Raman} adiabatic passage},
  \href{https://doi.org/10.1103/PhysRevLett.115.103001}{\emph{Phys. Rev. Lett.}
  {\bfseries 115} (2015) 103001}.

\bibitem{Berg2015}
P.~Berg, S.~Abend, G.~Tackmann, C.~Schubert, E.~Giese, W.~Schleich et~al.,
  \emph{Composite-light-pulse technique for high-precision atom
  interferometry},
  \href{https://doi.org/10.1103/PhysRevLett.114.063002}{\emph{Phys. Rev. Lett.}
  {\bfseries 114} (2015) 063002}.

\bibitem{plot18}
B.~Plotkin-Swing, D.~Gochnauer, K.~McAlpine, E.~Cooper, A.~Jamison and
  S.~Gupta, \emph{{Three-Path Atom Interferometry with Large Momentum
  Separation}},
  \href{https://doi.org/10.1103/PhysRevLett.121.133201}{\emph{Phys. Rev. Lett.}
  {\bfseries 121} (2018) 133201}.

\bibitem{Beguin23}
A.~B\'eguin, T.~Rodzinka, L.~Calmels, B.~Allard and A.~Gauguet, \emph{Atom
  interferometry with coherent enhancement of bragg pulse sequences},
  \href{https://doi.org/10.1103/PhysRevLett.131.143401}{\emph{Phys. Rev. Lett.}
  {\bfseries 131} (2023) 143401}.

\bibitem{page20}
Z.~Pagel, W.~Zhong, R.H.~Parker, C.T.~Olund, N.Y.~Yao, C.~Yu et~al.,
  \emph{{Symmetric Bloch oscillations of matter waves}},
  \href{https://doi.org/10.1103/PhysRevA.102.053312}{\emph{Phys. Rev. A}
  {\bfseries 102} (2020) 053312}.

\bibitem{gebb21}
M.~Gebbe, S.~Abend, S.~Jan-Niclas, M.~Gersemann, H.~Ahlers, H.~Müntinga
  et~al., \emph{Twin-lattice atom interferometry},
  \href{https://doi.org/doi.org/10.1038/s41467-021-22823-8}{\emph{Nature
  Communications} {\bfseries 12} (2021) 2544}.

\bibitem{Bukov2015}
M.~Bukov, L.~D'Alessio and A.~Polkovnikov, \emph{{Universal high-frequency
  behavior of periodically driven systems: from dynamical stabilization to
  Floquet engineering}},
  \href{https://doi.org/10.1080/00018732.2015.1055918}{\emph{Advances in
  Physics} {\bfseries 64} (2015) 139}.

\bibitem{Kim20}
M.~Kim, R.~Notermans, C.~Overstreet, J.~Curti, P.~Asenbaum and M.A.~Kasevich,
  \emph{{40 W, 780 nm laser system with compensated dual beam splitters for
  atom interferometry}}, \href{https://doi.org/10.1364/OL.404430}{\emph{Opt.
  Lett.} {\bfseries 45} (2020) 6555}.

\bibitem{rahm24}
T.~Rahman, A.~Wirth-Singh, A.~Ivanov, D.~Gochnauer, E.~Hough and S.~Gupta,
  \emph{{Bloch Oscillation Phases investigated by Multi-path Stuckelberg Atom
  Interferometry}},
  \href{https://doi.org/10.1103/PhysRevResearch.6.L022012}{\emph{Phys. Rev.
  Res.} {\bfseries 6} (2024) L022012}.

\bibitem{Graham2013}
P.W.~Graham, J.M.~Hogan, M.A.~Kasevich and S.~Rajendran, \emph{{New Method for
  Gravitational Wave Detection with Atomic Sensors}},
  \href{https://doi.org/10.1103/PhysRevLett.110.171102}{\emph{Phys. Rev. Lett.}
  {\bfseries 110} (2013) 171102}.

\bibitem{Rudolph2020}
J.~Rudolph, T.~Wilkason, M.~Nantel, H.~Swan, C.M.~Holland, Y.~Jiang et~al.,
  \emph{{Large Momentum Transfer Clock Atom Interferometry on the
  $\mathrm{689~nm}$ Intercombination Line of Strontium}},
  \href{https://doi.org/10.1103/PhysRevLett.124.083604}{\emph{Phys. Rev. Lett.}
  {\bfseries 124} (2020) 083604}.

\bibitem{Wilkason2022}
T.~Wilkason, M.~Nantel, J.~Rudolph, Y.~Jiang, B.E.~Garber, H.~Swan et~al.,
  \emph{{Atom Interferometry with Floquet Atom Optics}},
  \href{https://doi.org/10.1103/PhysRevLett.129.183202}{\emph{Phys. Rev. Lett.}
  {\bfseries 129} (2022) 183202}.

\bibitem{Hong2005}
T.~Hong, C.~Cramer, W.~Nagourney and E.N.~Fortson, \emph{{Optical Clocks Based
  on Ultranarrow Three-Photon Resonances in Alkaline Earth Atoms}},
  \href{https://doi.org/10.1103/PhysRevLett.94.050801}{\emph{Phys. Rev. Lett.}
  {\bfseries 94} (2005) 050801}.

\bibitem{Santra2005}
R.~Santra, E.~Arimondo, T.~Ido, C.H.~Greene and J.~Ye, \emph{{High-Accuracy
  Optical Clock via Three-Level Coherence in Neutral Bosonic
  $^{88}\mathrm{Sr}$}},
  \href{https://doi.org/10.1103/PhysRevLett.94.173002}{\emph{Phys. Rev. Lett.}
  {\bfseries 94} (2005) 173002}.

\bibitem{Vitaly2007}
V.D.~Ovsiannikov, V.G.~Pal'chikov, A.V.~Taichenachev, V.I.~Yudin, H.~Katori and
  M.~Takamoto, \emph{{Magic-wave-induced
  $^{1}\mathrm{S}_{0}-^{3}\mathrm{P}_{0}$ transition in even isotopes of
  alkaline-earth-metal-like atoms}},
  \href{https://doi.org/10.1103/PhysRevA.75.020501}{\emph{Phys. Rev. A}
  {\bfseries 75} (2007) 020501(R)}.

\bibitem{Alden2014}
E.A.~Alden, K.R.~Moore and A.E.~Leanhardt, \emph{{Two-photon $E1$-$M1$ optical
  clock}}, \href{https://doi.org/10.1103/PhysRevA.90.012523}{\emph{Phys. Rev.
  A} {\bfseries 90} (2014) 012523}.

\bibitem{carman2024collinear}
S.P.~Carman, J.~Rudolph, B.E.~Garber, M.J.~Van~de Graaff, H.~Swan, Y.~Jiang
  et~al., \emph{Collinear three-photon excitation of a strongly forbidden
  optical clock transition}, {\emph{\textcolor{blue}{arXiv:2406.07902}} (2024)
  }.

\bibitem{more20}
L.~Morel, Y.~Zhao, P.~Clade and S.~Guellati-Khelifa, \emph{{Determination of
  the fine-structure constant with an accuracy of 81 parts per trillion}},
  \href{https://doi.org/10.1038/s41586-020-2964-7}{\emph{Nature} {\bfseries
  588} (2020) 61}.

\bibitem{goch19}
D.~Gochnauer, K.~McAlpine, B.~Plotkin-Swing, A.~Jamison and S.~Gupta,
  \emph{{Bloch-band picture for light-pulse atom diffraction and
  interferometry}},
  \href{https://doi.org/10.1103/PhysRevA.100.04361}{\emph{Phys. Rev. A.}
  {\bfseries 100} (2019) 043611}.

\bibitem{mcal20}
K.~McAlpine, D.~Gochnauer and S.~Gupta, \emph{{Excited-band Bloch oscillations
  for precision atom interferometry}},
  \href{https://doi.org/10.1103/PhysRevA.101.023614}{\emph{Phys. Rev. A.}
  {\bfseries 101} (2020) 023614}.

\bibitem{fitz23}
F.~Fitzek, J.-N.~Kirsten-Siemß, E.~Rasel, N.~Gaaloul and K.~Hammerer,
  \emph{Accurate and efficient bloch-oscillation-enhanced atom interferometry},
  {\emph{\textcolor{blue}{arXiv:2306.09399}} (2023) }.

\bibitem{Dick1987LOI}
R.~Sydnow, ed., \emph{Local oscillator induced instabilities in trapped ion
  frequency standards}, U.S. Naval Observatory, Washington, DC, 1987.

\bibitem{Katori2024_CWBeaminMagicwavelength}
S.~Okaba, R.~Takeuchi, S.~Tsuji and H.~Katori, \emph{Continuous generation of
  an ultracold atomic beam using crossed moving optical lattices},
  \href{https://doi.org/10.1103/PhysRevApplied.21.034006}{\emph{Phys. Rev.
  Appl.} {\bfseries 21} (2024) 034006}.

\bibitem{Takeuchi_2023}
R.~Takeuchi, H.~Chiba, S.~Okaba, M.~Takamoto, S.~Tsuji and H.~Katori,
  \emph{Continuous outcoupling of ultracold strontium atoms combining three
  different traps},
  \href{https://doi.org/10.35848/1882-0786/accb3c}{\emph{Applied Physics
  Express} {\bfseries 16} (2023) 042003}.

\bibitem{Chen2005ActiveClock}
J.~Chen, \emph{Active optical clock},
  \href{https://doi.org/10.1007/s11434-009-0073-y}{\emph{Chinese Science
  Bulletin} {\bfseries 54} (2009) 348}.

\bibitem{Holland2009PRLMeiser}
D.~Meiser, J.~Ye, D.R.~Carlson and M.J.~Holland, \emph{Prospects for a
  millihertz-linewidth laser},
  \href{https://doi.org/10.1103/PhysRevLett.102.163601}{\emph{Phys. Rev. Lett.}
  {\bfseries 102} (2009) 163601}.

\bibitem{Cline2022SRL}
J.R.K.~Cline, V.M.~Schäfer, Z.~Niu, D.J.~Young, T.H.~Yoon and J.K.~Thompson,
  \emph{{Continuous collective strong coupling between atoms and a high finesse
  optical cavity}},  2022,
  \href{https://doi.org/10.48550/arXiv.2211.00158}{DOI}
  [\href{https://arxiv.org/abs/2211.00158}{{\ttfamily 2211.00158}}].

\bibitem{Robins2013RevAtomLaser}
N.~Robins, P.~Altin, J.~Debs and J.~Close, \emph{Atom lasers: {P}roduction,
  properties and prospects for precision inertial measurement},
  \href{https://doi.org/https://doi.org/10.1016/j.physrep.2013.03.006}{\emph{Physics
  Reports} {\bfseries 529} (2013) 265}.

\bibitem{Chen2022CWBEC}
C.-C.~Chen, R.~Gonz{\'a}lez~Escudero, J.~Min{\'a}{\v{r}}, B.~Pasquiou,
  S.~Bennetts and F.~Schreck, \emph{Continuous {B}ose--{E}instein
  condensation},
  \href{https://doi.org/10.1038/s41586-022-04731-z}{\emph{Nature} {\bfseries
  606} (2022) 683}.

\bibitem{Chen2023CWBECRev}
C.-C.~Chen, S.~Bennetts and F.~Schreck, \emph{Chapter six - the path to
  continuous bose-einstein condensation},  in \emph{Advances in Atomic,
  Molecular, and Optical Physics}, L.F.~DiMauro, H.~Perrin and S.F.~Yelin,
  eds., vol.~72 of \emph{Advances In Atomic, Molecular, and Optical Physics},
  pp.~361--430, Academic Press (2023),
  \href{https://doi.org/https://doi.org/10.1016/bs.aamop.2023.04.004}{DOI}.

\bibitem{Stellmer2013LaserCoolingToBEC}
S.~Stellmer, B.~Pasquiou, R.~Grimm and F.~Schreck, \emph{Laser cooling to
  quantum degeneracy},
  \href{https://doi.org/10.1103/PhysRevLett.110.263003}{\emph{Phys. Rev. Lett.}
  {\bfseries 110} (2013) 263003}.

\bibitem{Vuletic2017LaserCoolingToBEC}
J.~Hu, A.~Urvoy, Z.~Vendeiro, V.~Cr{\'e}pel, W.~Chen and V.~Vuleti{\"{u}},
  \emph{Creation of a {B}ose-condensed gas of {$^{87}$Rb} by laser cooling},
  \href{https://doi.org/10.1126/science.aan5614}{\emph{Science} {\bfseries 358}
  (2017) 1078}
  [\href{https://arxiv.org/abs/https://www.science.org/doi/pdf/10.1126/science.aan5614}{{\ttfamily
  https://www.science.org/doi/pdf/10.1126/science.aan5614}}].

\bibitem{Urvoy2019}
A.~Urvoy, Z.~Vendeiro, J.~Ramette, A.~Adiyatullin and V.~Vuleti\'c,
  \emph{Direct laser cooling to {B}ose-{E}instein {C}ondensation in a dipole
  trap}, \href{https://doi.org/10.1103/PhysRevLett.122.203202}{\emph{Phys. Rev.
  Lett.} {\bfseries 122} (2019) 203202}.

\bibitem{Bennetts2017SSMOTHighPSD}
S.~Bennetts, C.-C.~Chen, B.~Pasquiou and F.~Schreck, \emph{Steady-state
  magneto-optical trap with 100-fold improved phase-space density},
  \href{https://doi.org/10.1103/PhysRevLett.119.223202}{\emph{Phys. Rev. Lett.}
  {\bfseries 119} (2017) 223202}.

\bibitem{Chen2019Beam}
C.-C.~Chen, S.~Bennetts, R.G.~Escudero, B.~Pasquiou and F.~Schreck,
  \emph{Continuous guided strontium beam with high phase-space density},
  \href{https://doi.org/10.1103/PhysRevApplied.12.044014}{\emph{Phys. Rev.
  Appl.} {\bfseries 12} (2019) 044014}.

\bibitem{Chen2024Sisyphus}
C.-C.~Chen, J.L.~Siegel, B.D.~Hunt, T.~Grogan, Y.S.~Hassan, K.~Beloy et~al.,
  \emph{Clock-line-mediated sisyphus cooling},
  \href{https://doi.org/10.1103/PhysRevLett.133.053401}{\emph{Phys. Rev. Lett.}
  {\bfseries 133} (2024) 053401}.

\bibitem{He2024Outcoupling}
J.~He, B.~Pasquiou, R.G.~Escudero, S.~Zhou, M.~Borkowski and F.~Schreck,
  \emph{{Coherent Three-Photon Excitation of the Strontium Clock Transition}},
  2024, \href{https://doi.org/10.48550/arXiv.2406.07530}{DOI}
  [\href{https://arxiv.org/abs/2406.07530}{{\ttfamily 2406.07530}}].

\bibitem{Feng2023}
C.-H.~Feng, P.~Robert, P.~Bouyer, B.~Canuel, J.~Li, S.~Das et~al., \emph{High
  flux strontium atom source},
  \href{https://doi.org/10.1088/2058-9565/ad310b}{\emph{Quantum Science and
  Technology} {\bfseries 9} (2024) 025017}.

\bibitem{Doyle2012BufferGas}
N.R.~Hutzler, H.-I.~Lu and J.M.~Doyle, \emph{The buffer gas beam: An intense,
  cold, and slow source for atoms and molecules},
  \href{https://doi.org/10.1021/cr200362u}{\emph{Chemical Reviews} {\bfseries
  112} (2012) 4803}
  [\href{https://arxiv.org/abs/https://doi.org/10.1021/cr200362u}{{\ttfamily
  https://doi.org/10.1021/cr200362u}}].

\bibitem{impertro_unsupervised_2023}
A.~Impertro, J.F.~Wienand, S.~Häfele, H.~von Raven, S.~Hubele, T.~Klostermann
  et~al., \emph{An unsupervised deep learning algorithm for single-site
  reconstruction in quantum gas microscopes},
  \href{https://doi.org/10.1038/s42005-023-01287-w}{\emph{Communications
  Physics} {\bfseries 6} (2023) 1}.

\bibitem{bakr_probing_2010}
W.S.~Bakr, A.~Peng, M.E.~Tai, R.~Ma, J.~Simon, J.I.~Gillen et~al.,
  \emph{Probing the {Superfluid}–to–{Mott} {Insulator} {Transition} at the
  {Single}-{Atom} {Level}},
  \href{https://doi.org/10.1126/science.1192368}{\emph{Science} {\bfseries 329}
  (2010) 547}.

\bibitem{sherson_single-atom-resolved_2010}
J.F.~Sherson, C.~Weitenberg, M.~Endres, M.~Cheneau, I.~Bloch and S.~Kuhr,
  \emph{Single-atom-resolved fluorescence imaging of an atomic {Mott}
  insulator}, \href{https://doi.org/10.1038/nature09378}{\emph{Nature}
  {\bfseries 467} (2010) 68}.

\bibitem{cheuk_quantum-gas_2015}
L.W.~Cheuk, M.A.~Nichols, M.~Okan, T.~Gersdorf, V.V.~Ramasesh, W.S.~Bakr
  et~al., \emph{Quantum-{Gas} {Microscope} for {Fermionic} {Atoms}},
  \href{https://doi.org/10.1103/PhysRevLett.114.193001}{\emph{Physical Review
  Letters} {\bfseries 114} (2015) 193001}.

\bibitem{haller_single-atom_2015}
E.~Haller, J.~Hudson, A.~Kelly, D.A.~Cotta, B.~Peaudecerf, G.D.~Bruce et~al.,
  \emph{Single-atom imaging of fermions in a quantum-gas microscope},
  \href{https://doi.org/10.1038/nphys3403}{\emph{Nature Physics} {\bfseries 11}
  (2015) 738}.

\bibitem{parsons_site-resolved_2015}
M.F.~Parsons, F.~Huber, A.~Mazurenko, C.S.~Chiu, W.~Setiawan, K.~Wooley-Brown
  et~al., \emph{Site-{Resolved} {Imaging} of {Fermionic} $^6${Li} in an
  {Optical} {Lattice}},
  \href{https://doi.org/10.1103/PhysRevLett.114.213002}{\emph{Physical Review
  Letters} {\bfseries 114} (2015) 213002}.

\bibitem{gross_quantum_2021}
C.~Gross and W.S.~Bakr, \emph{Quantum gas microscopy for single atom and spin
  detection}, \href{https://doi.org/10.1038/s41567-021-01370-5}{\emph{Nature
  Physics} {\bfseries 17} (2021) 1316}.

\bibitem{greiner_quantum_2002}
M.~Greiner, O.~Mandel, T.~Esslinger, T.W.~Hänsch and I.~Bloch, \emph{Quantum
  phase transition from a superfluid to a {Mott} insulator in a gas of
  ultracold atoms}, \href{https://doi.org/10.1038/415039a}{\emph{Nature}
  {\bfseries 415} (2002) 39}.

\bibitem{weitenberg_single-spin_2011}
C.~Weitenberg, M.~Endres, J.F.~Sherson, M.~Cheneau, P.~Schauß, T.~Fukuhara
  et~al., \emph{Single-spin addressing in an atomic {Mott} insulator},
  \href{https://doi.org/10.1038/nature09827}{\emph{Nature} {\bfseries 471}
  (2011) 319}.

\bibitem{yang_cooling_2020}
B.~Yang, H.~Sun, C.-J.~Huang, H.-Y.~Wang, Y.~Deng, H.-N.~Dai et~al.,
  \emph{Cooling and entangling ultracold atoms in optical lattices},
  \href{https://doi.org/10.1126/science.aaz6801}{\emph{Science} {\bfseries 369}
  (2020) 550}.

\bibitem{wei_quantum_2022}
D.~Wei, A.~Rubio-Abadal, B.~Ye, F.~Machado, J.~Kemp, K.~Srakaew et~al.,
  \emph{Quantum gas microscopy of {Kardar}-{Parisi}-{Zhang} superdiffusion},
  \href{https://doi.org/10.1126/science.abk2397}{\emph{Science} {\bfseries 376}
  (2022) 716}.

\bibitem{wienand_emergence_2023}
J.F.~Wienand, S.~Karch, A.~Impertro, C.~Schweizer, E.~McCulloch, R.~Vasseur
  et~al., \emph{Emergence of fluctuating hydrodynamics in chaotic quantum
  systems}, \href{https://doi.org/10.1038/s41567-024-02611-z}{\emph{Nature
  Physics} (2024) }.

\bibitem{kaufman_quantum_2016}
A.M.~Kaufman, M.E.~Tai, A.~Lukin, M.~Rispoli, R.~Schittko, P.M.~Preiss et~al.,
  \emph{Quantum thermalization through entanglement in an isolated many-body
  system}, \href{https://doi.org/10.1126/science.aaf6725}{\emph{Science}
  {\bfseries 353} (2016) 794}.

\bibitem{lukin_probing_2019}
A.~Lukin, M.~Rispoli, R.~Schittko, M.E.~Tai, A.M.~Kaufman, S.~Choi et~al.,
  \emph{Probing entanglement in a many-body–localized system},
  \href{https://doi.org/10.1126/science.aau0818}{\emph{Science} {\bfseries 364}
  (2019) 256}.

\bibitem{cheneau_light-cone-like_2012}
M.~Cheneau, P.~Barmettler, D.~Poletti, M.~Endres, P.~Schauß, T.~Fukuhara
  et~al., \emph{Light-cone-like spreading of correlations in a quantum
  many-body system}, \href{https://doi.org/10.1038/nature10748}{\emph{Nature}
  {\bfseries 481} (2012) 484}.

\bibitem{zheng_efficiently_2022}
Y.-G.~Zheng, W.-Y.~Zhang, Y.-C.~Shen, A.~Luo, Y.~Liu, M.-G.~He et~al.,
  \emph{Efficiently {Extracting} {Multi}-{Point} {Correlations} of a {Floquet}
  {Thermalized} {System}},
  \href{https://doi.org/10.48550/arXiv.2210.08556}{\emph{arXiv:2210.08556}
  (2022) }.

\bibitem{impertro_local_2024}
A.~Impertro, S.~Karch, J.F.~Wienand, S.~Huh, C.~Schweizer, I.~Bloch et~al.,
  \emph{Local {Readout} and {Control} of {Current} and {Kinetic} {Energy}
  {Operators} in {Optical} {Lattices}},
  \href{https://doi.org/10.1103/PhysRevLett.133.063401}{\emph{Physical Review
  Letters} {\bfseries 133} (2024) 063401}.

\bibitem{greiner_magnetic_2001}
M.~Greiner, I.~Bloch, T.W.~Hänsch and T.~Esslinger, \emph{Magnetic transport
  of trapped cold atoms over a large distance},
  \href{https://doi.org/10.1103/PhysRevA.63.031401}{\emph{Physical Review A}
  {\bfseries 63} (2001) 031401}.

\bibitem{lewandowski_observation_2002}
H.J.~Lewandowski, D.M.~Harber, D.L.~Whitaker and E.A.~Cornell,
  \emph{Observation of {Anomalous} {Spin}-{State} {Segregation} in a {Trapped}
  {Ultracold} {Vapor}},
  \href{https://doi.org/10.1103/PhysRevLett.88.070403}{\emph{Physical Review
  Letters} {\bfseries 88} (2002) 070403}.

\bibitem{pertot_versatile_2009}
D.~Pertot, D.~Greif, S.~Albert, B.~Gadway and D.~Schneble, \emph{Versatile
  transporter apparatus for experiments with optically trapped
  {Bose}–{Einstein} condensates},
  \href{https://doi.org/10.1088/0953-4075/42/21/215305}{\emph{Journal of
  Physics B: Atomic, Molecular and Optical Physics} {\bfseries 42} (2009)
  215305}.

\bibitem{gross_all-optical_2016}
C.~Gross, H.C.J.~Gan and K.~Dieckmann, \emph{All-optical production and
  transport of a large $^6${Li} quantum gas in a crossed optical dipole trap},
  \href{https://doi.org/10.1103/PhysRevA.93.053424}{\emph{Physical Review A}
  {\bfseries 93} (2016) 053424}.

\bibitem{couvert_optimal_2008}
A.~Couvert, T.~Kawalec, G.~Reinaudi and D.~Guéry-Odelin, \emph{Optimal
  transport of ultracold atoms in the non-adiabatic regime},
  \href{https://doi.org/10.1209/0295-5075/83/13001}{\emph{EPL} {\bfseries 83}
  (2008) 13001}.

\bibitem{leonard_optical_2014}
J.~Léonard, M.~Lee, A.~Morales, T.M.~Karg, T.~Esslinger and T.~Donner,
  \emph{Optical transport and manipulation of an ultracold atomic cloud using
  focus-tunable lenses},
  \href{https://doi.org/10.1088/1367-2630/16/9/093028}{\emph{New Journal of
  Physics} {\bfseries 16} (2014) 093028}.

\bibitem{unnikrishnan_long_2021}
G.~Unnikrishnan, C.~Beulenkamp, D.~Zhang, K.P.~Zamarski, M.~Landini and
  H.-C.~Nägerl, \emph{Long distance optical transport of ultracold atoms: {A}
  compact setup using a {Moiré} lens},
  \href{https://doi.org/10.1063/5.0049320}{\emph{Review of Scientific
  Instruments} {\bfseries 92} (2021) 063205}.

\bibitem{schrader_optical_2001}
D.~Schrader, S.~Kuhr, W.~Alt, M.~Müller, V.~Gomer and D.~Meschede, \emph{An
  optical conveyor belt for single neutral atoms},
  \href{https://doi.org/10.1007/s003400100722}{\emph{Applied Physics B}
  {\bfseries 73} (2001) 819}.

\bibitem{schmid_long_2006}
S.~Schmid, G.~Thalhammer, K.~Winkler, F.~Lang and J.H.~Denschlag, \emph{Long
  distance transport of ultracold atoms using a {1D} optical lattice},
  \href{https://doi.org/10.1088/1367-2630/8/8/159}{\emph{New Journal of
  Physics} {\bfseries 8} (2006) 159}.

\bibitem{klostermann_fast_2022}
T.~Klostermann, C.R.~Cabrera, H.~von Raven, J.F.~Wienand, C.~Schweizer,
  I.~Bloch et~al., \emph{Fast long-distance transport of cold cesium atoms},
  \href{https://doi.org/10.1103/PhysRevA.105.043319}{\emph{Physical Review A}
  {\bfseries 105} (2022) 043319}.

\bibitem{Kasevich1991}
M.~Kasevich and S.~Chu, \emph{Atomic interferometry using stimulated {R}aman
  transitions}, \href{https://doi.org/10.1103/PhysRevLett.67.181}{\emph{Phys.
  Rev. Lett.} {\bfseries 67} (1991) 181}.

\bibitem{Peters2001}
A.~Peters, K.Y.~Chung and S.~Chu, \emph{High-precision gravity measurements
  using atom interferometry},
  \href{https://doi.org/10.1088/0026-1394/38/1/4}{\emph{Metrologia} {\bfseries
  38} (2001) 25}.

\bibitem{Hensel2021}
T.~Hensel, S.~Loriani, C.~Schubert, F.~Fitzek, S.~Abend, H.~Ahlers et~al.,
  \emph{Inertial sensing with quantum gases: a comparative performance study of
  condensed versus thermal sources for atom interferometry},
  \href{https://doi.org/10.1140/epjd/s10053-021-00069-9}{\emph{The European
  Physical Journal D} {\bfseries 75} (2021) }.

\bibitem{Hartwig2015}
J.~Hartwig, S.~Abend, C.~Schubert, D.~Schlippert, H.~Ahlers, K.~Posso-Trujillo
  et~al., \emph{Testing the universality of free fall with rubidium and
  ytterbium in a very large baseline atom interferometer},
  \href{https://doi.org/10.1088/1367-2630/17/3/035011}{\emph{New J. Phys.}
  {\bfseries 17} (2015) 035011}.

\bibitem{Ahlers2022}
{\scshape STE-QUEST} collaboration, \emph{{STE-QUEST: Space Time Explorer and
  QUantum Equivalence principle Space Test}},
  \href{https://arxiv.org/abs/arXiv:2211.15412}{{\ttfamily arXiv:2211.15412}}.

\bibitem{Ammann1997}
H.~Ammann and N.~Christensen, \emph{Delta kick cooling: A new method for
  cooling atoms},
  \href{https://doi.org/10.1103/PhysRevLett.78.2088}{\emph{Phys. Rev. Lett.}
  {\bfseries 78} (1997) 2088}.

\bibitem{Kovachy2015}
T.~Kovachy, J.M.~Hogan, A.~Sugarbaker, S.M.~Dickerson, C.A.~Donnelly,
  C.~Overstreet et~al., \emph{Matter wave lensing to picokelvin temperatures},
  \href{https://doi.org/10.1103/PhysRevLett.114.143004}{\emph{Phys. Rev. Lett.}
  {\bfseries 114} (2015) 143004}.

\bibitem{Deppner2021}
C.~Deppner, W.~Herr, M.~Cornelius, P.~Stromberger, T.~Sternke, C.~Grzeschik
  et~al., \emph{Collective-mode enhanced matter-wave optics},
  \href{https://doi.org/10.1103/PhysRevLett.127.100401}{\emph{Phys. Rev. Lett.}
  {\bfseries 127} (2021) 100401}.

\bibitem{Gaaloul2022}
N.~Gaaloul, M.~Meister, R.~Corgier, A.~Pichery, P.~Boegel, W.~Herr et~al.,
  \emph{A space-based quantum gas laboratory at picokelvin energy scales},
  \href{https://doi.org/10.1038/s41467-022-35274-6}{\emph{Nat. Commun.}
  {\bfseries 13} (2022) 875}.

\bibitem{Roy2016}
R.~Roy, A.~Green, R.~Bowler and S.~Gupta, \emph{Rapid cooling to quantum
  degeneracy in dynamically shaped atom traps},
  \href{https://doi.org/10.1103/PhysRevA.93.043403}{\emph{Phys. Rev. A}
  {\bfseries 93} (2016) 043403}.

\bibitem{Inouye1998}
S.~Inouye, M.R.~Andrews, J.~Stenger, H.-J.~Miesner, D.M.~Stamper-Kurn and
  W.~Ketterle, \emph{Observation of feshbach resonances in a bose-einstein
  condensate}, \href{https://doi.org/10.1038/32354}{\emph{Nature} {\bfseries
  392} (1998) 151}.

\bibitem{Salomon2013}
G.~Salomon, L.~Fouch{\'e}, P.~Wang, A.~Aspect, P.~Bouyer and T.~Bourdel,
  \emph{Gray-molasses cooling of 39 k to a high phase-space density},
  \href{https://doi.org/10.1209/0295-5075/104/63002}{\emph{EPL (Europhysics
  Letters)} {\bfseries 104} (2013) 63002}.

\bibitem{DErrico2007}
C.~D'Errico, M.~Zaccanti, M.~Fattori, G.~Roati, M.~Inguscio, G.~Modugno et~al.,
  \emph{Feshbach resonances in ultracold $^{39}\mathrm{K}$},
  \href{https://doi.org/10.1088/1367-2630/9/7/223}{\emph{New Journal of
  Physics} {\bfseries 9} (2007) 223}.

\bibitem{Enomoto2008}
K.~Enomoto, K.~Kasa, M.~Kitagawa and Y.~Takahashi, \emph{Optical feshbach
  resonance using the intercombination transition},
  \href{https://doi.org/10.1103/PhysRevLett.101.203201}{\emph{Phys. Rev. Lett.}
  {\bfseries 101} (2008) 203201}.

\bibitem{Yan2013}
M.~Yan, B.J.~DeSalvo, B.~Ramachandhran, H.~Pu and T.C.~Killian,
  \emph{Controlling condensate collapse and expansion with an optical feshbach
  resonance}, \href{https://doi.org/10.1103/PhysRevLett.110.123201}{\emph{Phys.
  Rev. Lett.} {\bfseries 110} (2013) 123201}.

\bibitem{Herbst2024a}
A.~Herbst, T.~Estrampes, H.~Albers, V.~Vollenkemper, K.~Stolzenberg, S.~Bode
  et~al., \emph{High-flux source system for matter-wave interferometry
  exploiting tunable interactions},
  \href{https://doi.org/10.1103/PhysRevResearch.6.013139}{\emph{Phys. Rev.
  Res.} {\bfseries 6} (2024) 013139}.

\bibitem{Pandey2021}
S.~Pandey, H.~Mas, G.~Vasilakis and W.~von Klitzing, \emph{Atomtronic
  matter-wave lensing},
  \href{https://doi.org/10.1103/PhysRevLett.126.170402}{\emph{Phys. Rev. Lett.}
  {\bfseries 126} (2021) 170402}.

\bibitem{Albers2022}
H.~Albers, R.~Corgier, A.~Herbst, A.~Rajagopalan, C.~Schubert, C.~Vogt et~al.,
  \emph{All-optical matter-wave lens using time-averaged potentials},
  \href{https://doi.org/10.1038/s42005-022-00825-2}{\emph{Commun. Phys.}
  {\bfseries 5} (2022) 181}.

\bibitem{Herbst2024b}
A.~Herbst, T.~Estrampes, H.~Albers, R.~Corgier, K.~Stolzenberg, S.~Bode et~al.,
  \emph{Matter-wave collimation to picokelvin energies with scattering length
  and potential shape control},
  \href{https://doi.org/10.1038/s42005-024-01621-w}{\emph{Communications
  Physics} {\bfseries 7} (2024) }.

\bibitem{xin2024fast}
M.~Xin, W.S.~Leong, Z.~Chen, Y.~Wang and S.-Y.~Lan, \emph{Fast quantum gas
  formation via electromagnetically induced transparency cooling},
  \href{https://doi.org/10.1038/s41567-024-02677-9}{\emph{Nature Physics}
  {\bfseries 20} (2024) 1}.

\bibitem{Mor00}
G.~Morigi, J.~Eschner and C.H.~Keitel, \emph{Ground state laser cooling using
  electromagnetically induced transparency},
  \href{https://doi.org/10.1103/PhysRevLett.85.4458}{\emph{Phys. Rev. Lett.}
  {\bfseries 85} (2000) 4458}.

\bibitem{Hua21}
C.~Huang, S.~Chai and S.-Y.~Lan, \emph{Dark-state sideband cooling in an atomic
  ensemble}, \href{https://doi.org/10.1103/PhysRevA.103.013305}{\emph{Phys.
  Rev. A} {\bfseries 103} (2021) 013305}.

\bibitem{Kitagawa93}
M.~Kitagawa and M.~Ueda, \emph{Squeezed spin states},
  \href{https://doi.org/10.1103/PhysRevA.47.5138}{\emph{Phys. Rev. A}
  {\bfseries 47} (1993) 5138}.

\bibitem{Hosten2016}
O.~Hosten, N.J.~Engelsen, R.~Krishnakumar and M.A.~Kasevich, \emph{Measurement
  noise 100 times lower than the quantum-projection limit using entangled
  atoms}, \href{https://doi.org/10.1038/nature16176}{\emph{Nature} {\bfseries
  529} (2016) 505}.

\bibitem{Cox2016}
K.C.~Cox, G.P.~Greve, J.M.~Weiner and J.K.~Thompson, \emph{Deterministic
  squeezed states with collective measurements and feedback},
  \href{https://doi.org/10.1103/PhysRevLett.116.093602}{\emph{Phys. Rev. Lett.}
  {\bfseries 116} (2016) 093602}.

\bibitem{huang2023}
M.-Z.~Huang, J.A.~de~la Paz, T.~Mazzoni, K.~Ott, P.~Rosenbusch, A.~Sinatra
  et~al., \emph{Observing spin-squeezed states under spin-exchange collisions
  for a second}, \href{https://doi.org/10.1103/PRXQuantum.4.020322}{\emph{PRX
  Quantum} {\bfseries 4} (2023) 020322}.

\bibitem{szigeti2021}
S.S.~Szigeti, O.~Hosten and S.A.~Haine, \emph{Improving cold-atom sensors with
  quantum entanglement: Prospects and challenges},
  \href{https://doi.org/10.1063/5.0050235}{\emph{Applied Physics Letters}
  {\bfseries 118} (2021) }.

\bibitem{malia2022}
B.K.~Malia, Y.~Wu, J.~{Mart{\'i}nez-Rinc{\'o}n} and M.A.~Kasevich,
  \emph{Distributed quantum sensing with mode-entangled spin-squeezed atomic
  states}, \href{https://doi.org/10.1038/s41586-022-05363-z}{\emph{Nature}
  {\bfseries 612} (2022) 661}.

\bibitem{Greve2022}
G.P.~Greve, C.~Luo, B.~Wu and J.K.~Thompson, \emph{Entanglement-enhanced
  matter-wave interferometry in a high-finesse cavity},
  \href{https://doi.org/10.1038/s41586-022-05197-9}{\emph{Nature} {\bfseries
  610} (2022) 472}.

\bibitem{Cassens24}
C.~Cassens, B.~Meyer-Hoppe, E.~Rasel and C.~Klempt, \emph{An
  entanglement-enhanced atomic gravimeter},  (2024)
  [\href{https://arxiv.org/abs/arXiv:2404.18668}{{\ttfamily
  arXiv:2404.18668}}].

\bibitem{wu2020}
Y.~Wu, R.~Krishnakumar, J.~Mart{\'\i}nez-Rinc{\'o}n, B.K.~Malia, O.~Hosten and
  M.A.~Kasevich, \emph{Retrieval of cavity-generated atomic spin squeezing
  after free-space release},
  \href{https://doi.org/10.1103/PhysRevA.102.012224}{\emph{Physical Review A}
  {\bfseries 102} (2020) 012224}.

\bibitem{Malia2020}
B.K.~Malia, J.~Mart\'{\i}nez-Rinc\'on, Y.~Wu, O.~Hosten and M.A.~Kasevich,
  \emph{Free space {R}amsey spectroscopy in rubidium with noise below the
  {Q}uantum {P}rojection {L}imit},
  \href{https://doi.org/10.1103/PhysRevLett.125.043202}{\emph{Phys. Rev. Lett.}
  {\bfseries 125} (2020) 043202}.

\bibitem{leroux_implementation_2010}
I.D.~Leroux, M.H.~Schleier-Smith and V.~Vuletić, \emph{Implementation of
  {Cavity} {Squeezing} of a {Collective} {Atomic} {Spin}},
  \href{https://doi.org/10.1103/PhysRevLett.104.073602}{\emph{Physical Review
  Letters} {\bfseries 104} (2010) 073602}.

\bibitem{robinson_direct_2022}
J.M.~Robinson, M.~Miklos, Y.M.~Tso, C.J.~Kennedy, T.~Bothwell, D.~Kedar et~al.,
  \emph{Direct comparison of two spin squeezed optical clocks below the quantum
  projection noise limit},  Nov., \textcolor{blue}{arXiv:2211.08621}, 2022.

\bibitem{Braverman2019}
B.~Braverman, A.~Kawasaki, E.~Pedrozo-Pe{\~{n}}afiel, S.~Colombo, C.~Shu, Z.~Li
  et~al., \emph{{Near-Unitary Spin Squeezing in Yb 171}},
  \href{https://doi.org/10.1103/PhysRevLett.122.223203}{\emph{Physical Review
  Letters} {\bfseries 122} (2019) 1}.

\bibitem{Pedrozo-penafiel2020}
E.~Pedrozo-Peñafiel, S.~Colombo, C.~Shu, A.F.~Adiyatullin, Z.~Li, E.~Mendez
  et~al., \emph{Entanglement on an optical atomic-clock transition},
  \href{https://doi.org/10.1038/s41586-020-3006-1}{\emph{Nature} {\bfseries
  588} (2020) 414}.

\bibitem{braverman_impact_2018}
B.~Braverman, A.~Kawasaki and V.~Vuletić, \emph{Impact of non-unitary spin
  squeezing on atomic clock performance},
  \href{https://doi.org/10.1088/1367-2630/aae563}{\emph{New Journal of Physics}
  (2018) 103019}.

\bibitem{hobson_cavity-enhanced_2019}
R.~Hobson, W.~Bowden, A.~Vianello, I.R.~Hill and P.~Gill, \emph{Cavity-enhanced
  non-destructive detection of atoms for an optical lattice clock},
  \href{https://doi.org/10.1364/OE.27.037099}{\emph{Optics Express} {\bfseries
  27} (2019) 37099}.

\bibitem{bowden_improving_2020}
W.~Bowden, A.~Vianello, I.R.~Hill, M.~Schioppo and R.~Hobson, \emph{Improving
  the \${Q}\$ {Factor} of an {Optical} {Atomic} {Clock} {Using} {Quantum}
  {Nondemolition} {Measurement}},
  \href{https://doi.org/10.1103/PhysRevX.10.041052}{\emph{Physical Review X}
  {\bfseries 10} (2020) 041052}.

\bibitem{malia_distributed_2022}
B.K.~Malia, Y.~Wu, J.~Martínez-Rincón and M.A.~Kasevich, \emph{Distributed
  quantum sensing with mode-entangled spin-squeezed atomic states},
  \href{https://doi.org/10.1038/s41586-022-05363-z}{\emph{Nature} {\bfseries
  612} (2022) 661}.

\bibitem{norcia_strong_2016}
M.A.~Norcia and J.K.~Thompson, \emph{Strong coupling on a forbidden transition
  in strontium and nondestructive atom counting},
  \href{https://doi.org/10.1103/PhysRevA.93.023804}{\emph{Phys. Rev. A}
  {\bfseries 93} (2016) 023804}.

\bibitem{pezze_heisenberg-limited_2020}
L.~Pezzè and A.~Smerzi, \emph{Heisenberg-{Limited} {Noisy} {Atomic} {Clock}
  {Using} a {Hybrid} {Coherent} and {Squeezed} {State} {Protocol}},
  \href{https://doi.org/10.1103/PhysRevLett.125.210503}{\emph{Physical Review
  Letters} {\bfseries 125} (2020) 210503}.

\bibitem{Szigeti20}
S.S.~Szigeti, S.P.~Nolan, J.D.~Close and S.A.~Haine, \emph{High-precision
  quantum-enhanced gravimetry with a bose-einstein condensate},
  \href{https://doi.org/10.1103/PhysRevLett.125.100402}{\emph{Phys. Rev. Lett.}
  {\bfseries 125} (2020) 100402}.

\bibitem{Corgier21b}
R.~Corgier, N.~Gaaloul, A.~Smerzi and L.~Pezz\`e, \emph{Delta-kick squeezing},
  \href{https://doi.org/10.1103/PhysRevLett.127.183401}{\emph{Phys. Rev. Lett.}
  {\bfseries 127} (2021) 183401}.

\bibitem{Anders2021}
F.~Anders, A.~Idel, P.~Feldmann, D.~Bondarenko, S.~Loriani, K.~Lange et~al.,
  \emph{Momentum entanglement for atom interferometry},
  \href{https://doi.org/10.1103/PhysRevLett.127.140402}{\emph{Phys. Rev. Lett.}
  {\bfseries 127} (2021) 140402}.

\bibitem{Corgier21a}
R.~Corgier, L.~Pezz\`e and A.~Smerzi, \emph{Nonlinear {B}ragg interferometer
  with a trapped bose-einstein condensate},
  \href{https://doi.org/10.1103/PhysRevA.103.L061301}{\emph{Phys. Rev. A}
  {\bfseries 103} (2021) L061301}.

\bibitem{Hartmann2020}
S.~Hartmann, J.~Jenewein, E.~Giese, S.~Abend, A.~Roura, E.M.~Rasel et~al.,
  \emph{Regimes of atomic diffraction: {R}aman versus {B}ragg diffraction in
  retroreflective geometries},
  \href{https://doi.org/10.1103/PhysRevA.101.053610}{\emph{Phys. Rev. A}
  {\bfseries 101} (2020) 053610}.

\bibitem{Poulsen02}
U.V.~Poulsen and K.~M\o{}lmer, \emph{Quantum beam splitter for atoms},
  \href{https://doi.org/10.1103/PhysRevA.65.033613}{\emph{Phys. Rev. A}
  {\bfseries 65} (2002) 033613}.

\bibitem{Burchianti20}
A.~Burchianti, C.~D'Errico, L.~Marconi, F.~Minardi, C.~Fort and M.~Modugno,
  \emph{Effect of interactions in the interference pattern of bose-einstein
  condensates}, \href{https://doi.org/10.1103/PhysRevA.102.043314}{\emph{Phys.
  Rev. A} {\bfseries 102} (2020) 043314}.

\bibitem{Davis16}
E.~Davis, G.~Bentsen and M.~Schleier-Smith, \emph{Approaching the {H}eisenberg
  limit without single-particle detection},
  \href{https://doi.org/10.1103/PhysRevLett.116.053601}{\emph{Phys. Rev. Lett.}
  {\bfseries 116} (2016) 053601}.

\bibitem{Hosten2016Science}
O.~Hosten, R.~Krishnakumar, N.J.~Engelsen and M.A.~Kasevich, \emph{Quantum
  phase magnification},
  \href{https://doi.org/10.1126/science.aaf3397}{\emph{Science} {\bfseries 352}
  (2016) 1552}.

\bibitem{Corgier23}
R.~Corgier, M.~Malitesta, A.~Smerzi and L.~Pezz\`e, \emph{Quantum-enhanced
  differential atom interferometers and clocks with spin-squeezing swapping},
  \href{https://doi.org/10.22331/q-2023-03-30-965}{\emph{Quantum} {\bfseries 7}
  (2023) 965}.

\bibitem{feldmann2023}
P.~Feldmann, F.~Anders, A.~Idel, C.~Schubert, D.~Schlippert, L.~Santos et~al.,
  \emph{Optimal squeezing for high-precision atom interferometers},
  \textcolor{blue}{arXiv.2311.10241}, 2023.

\bibitem{lezeik2022}
A.~Lezeik, D.~Tell, K.~Zipfel, V.~Gupta, {\'E}.~Wodey, E.~Rasel et~al.,
  \emph{Understanding the gravitational and magnetic environment of a very long
  baseline atom interferometer},
  [\href{https://arxiv.org/abs/arXiv:2209.08886}{{\ttfamily
  arXiv:2209.08886}}].

\bibitem{lotz2017}
C.~Lotz, T.~Frob{\"o}se, A.~Wanner, L.~Overmeyer and W.~Ertmer,
  \emph{{Einstein-Elevator: A New Facility for Research from {$\mu$} to 5}},
  \href{https://doi.org/10.2478/gsr-2017-0007}{\emph{Gravitational and Space
  Research} {\bfseries 5} (2017) 11}.

\bibitem{morel2020}
L.~Morel, Z.~Yao, P.~Clad\'e and S.~Guellati-Kh\'elifa, \emph{Determination of
  the fine-structure constant with an accuracy of 81 parts per trillion},
  \href{https://doi.org/10.1038/s41586-020-2964-7}{\emph{Nature} {\bfseries
  588} (2020) 61}.

\bibitem{weiss1994}
D.S.~Weiss, B.C.~Young and S.~Chu, \emph{Precision measurement of h/m {{Cs}}
  based on photon recoil using laser-cooled atoms and atomic interferometry},
  \href{https://doi.org/10.1007/BF01081393}{\emph{Applied Physics B Lasers and
  Optics} {\bfseries 59} (1994) 217}.

\bibitem{bade2018}
S.~Bade, L.~Djadaojee, M.~Andia, P.~Clad{\'e} and S.~{Guellati-Khelifa},
  \emph{Observation of {{Extra Photon Recoil}} in a {{Distorted Optical
  Field}}},
  \href{https://doi.org/10.1103/PhysRevLett.121.073603}{\emph{Physical Review
  Letters} {\bfseries 121} (2018) 073603}.

\bibitem{carrez2023}
C.~Carrez, \emph{Étude de l’effet des distorsions du front d’onde dans un
  interféromètre atomique avec un condensat de Bose-Einstein}, Ph.D. thesis,
  Sorbonne Universtit\'e, 2023.

\bibitem{Lan2012}
S.-Y.~Lan, P.-C.~Kuan, B.~Estey, P.~Haslinger and H.~M\"uller, \emph{Influence
  of the {C}oriolis force in atom interferometry},
  \href{https://doi.org/10.1103/PhysRevLett.108.090402}{\emph{Phys. Rev. Lett.}
  {\bfseries 108} (2012) 090402}.

\bibitem{Dubetsky2006}
B.~Dubetsky and M.A.~Kasevich, \emph{Atom interferometer as a selective sensor
  of rotation or gravity},
  \href{https://doi.org/10.1103/PhysRevA.74.023615}{\emph{Phys. Rev. A}
  {\bfseries 74} (2006) 531}.

\bibitem{wang2024robust}
Y.~Wang, J.~Glick, T.~Deshpande, K.~DeRose, S.~Saraf, N.~Sachdeva et~al.,
  \emph{Robust quantum control via multipath interference for thousandfold
  phase amplification in a resonant atom interferometer},
  {\emph{\textcolor{blue}{arXiv:2407.11246}} (2024) }.

\bibitem{glick2024coriolis}
J.~Glick, Z.~Chen, T.~Deshpande, Y.~Wang and T.~Kovachy, \emph{{Coriolis force
  compensation and laser beam delivery for 100-m baseline atom
  interferometry}}, \href{https://doi.org/10.1116/5.0180083}{\emph{AVS Quantum
  Science} {\bfseries 6} (2024) 014402}.

\bibitem{Norcia2017}
M.A.~Norcia, J.R.K.~Cline and J.K.~Thompson, \emph{Role of atoms in atomic
  gravitational-wave detectors},
  \href{https://doi.org/10.1103/PhysRevA.96.042118}{\emph{Phys. Rev. A}
  {\bfseries 96} (2017) 042118}.

\bibitem{Chaibi2016}
W.~Chaibi, R.~Geiger, B.~Canuel, A.~Bertoldi, A.~Landragin and P.~Bouyer,
  \emph{Low frequency gravitational wave detection with ground-based atom
  interferometer arrays},
  \href{https://doi.org/10.1103/PhysRevD.93.021101}{\emph{Phys. Rev. D}
  {\bfseries 93} (2016) 173}.

\bibitem{DiPumpo2024}
F.~Di~Pumpo, A.~Friedrich and E.~Giese, \emph{{Optimal baseline exploitation in
  vertical dark-matter detectors based on atom interferometry}},
  \href{https://doi.org/10.1116/5.0175683}{\emph{AVS Quantum Sci.} {\bfseries
  6} (2024) 014404}.

\bibitem{Arvanitaki2015}
A.~Arvanitaki, J.~Huang and K.~Van~Tilburg, \emph{Searching for dilaton dark
  matter with atomic clocks},
  \href{https://doi.org/10.1103/PhysRevD.91.015015}{\emph{Phys. Rev. D}
  {\bfseries 91} (2015) 015015}.

\bibitem{Graham2016}
P.W.~Graham, D.E.~Kaplan, J.~Mardon, S.~Rajendran and W.A.~Terrano, \emph{Dark
  matter direct detection with accelerometers},
  \href{https://doi.org/10.1103/PhysRevD.93.075029}{\emph{Phys. Rev. D}
  {\bfseries 93} (2016) 252}.

\bibitem{Safronova2018}
M.S.~Safronova, D.~Budker, D.~DeMille, D.F.J.~Kimball, A.~Derevianko and
  C.W.~Clark, \emph{Search for new physics with atoms and molecules},
  \href{https://doi.org/10.1103/RevModPhys.90.025008}{\emph{Rev. Mod. Phys.}
  {\bfseries 90} (2018) 025008}.

\bibitem{Derr2023}
D.~Derr and E.~Giese, \emph{{Clock transitions versus Bragg diffraction in
  atom-interferometric dark-matter detection}},
  \href{https://doi.org/10.1116/5.0176666}{\emph{AVS Quantum Sci.} {\bfseries
  5} (2023) 044404}.

\bibitem{DiPumpo2022}
F.~Di~Pumpo, A.~Friedrich, A.~Geyer, C.~Ufrecht and E.~Giese, \emph{Light
  propagation and atom interferometry in gravity and dilaton fields},
  \href{https://doi.org/10.1103/PhysRevD.105.084065}{\emph{Phys. Rev. D}
  {\bfseries 105} (2022) 411}.

\bibitem{Dimopoulos2009}
S.~Dimopoulos, P.W.~Graham, J.M.~Hogan, M.A.~Kasevich and S.~Rajendran,
  \emph{Gravitational wave detection with atom interferometry},
  \href{https://doi.org/10.1016/j.physletb.2009.06.011}{\emph{Phys. Lett. B}
  {\bfseries 678} (2009) 37}.

\bibitem{Graham2016b}
P.W.~Graham, J.M.~Hogan, M.A.~Kasevich and S.~Rajendran, \emph{Resonant mode
  for gravitational wave detectors based on atom interferometry},
  \href{https://doi.org/10.1103/PhysRevD.94.104022}{\emph{Phys. Rev. D}
  {\bfseries 94} (2016) 104022}.

\bibitem{Misner2017}
C.~Misner, K.~Thorne and J.~Wheeler, \emph{Gravitation}, Princeton University
  Press (2017).

\bibitem{Kleinert2015}
S.~Kleinert, E.~Kajari, A.~Roura and W.P.~Schleich, \emph{Representation-free
  description of light-pulse atom interferometry including non-inertial
  effects}, \href{https://doi.org/10.1016/j.physrep.2015.09.004}{\emph{Phys.
  Rep.} {\bfseries 605} (2015) 1}.

\bibitem{DAmico2017}
G.~D'Amico, G.~Rosi, S.~Zhan, L.~Cacciapuoti, M.~Fattori and G.~Tino,
  \emph{Canceling the gravity gradient phase shift in atom interferometry},
  \href{https://doi.org/10.1103/PhysRevLett.119.253201}{\emph{Phys. Rev. Lett.}
  {\bfseries 119} (2017) 253201}.

\bibitem{Roura2017}
A.~Roura, \emph{Circumventing {H}eisenberg's uncertainty principle in atom
  interferometry tests of the {E}quivalence {P}rinciple},
  \href{https://doi.org/10.1103/PhysRevLett.118.160401}{\emph{Phys. Rev. Lett.}
  {\bfseries 118} (2017) 160401}.

\bibitem{Junca2019}
J.~Junca, A.~Bertoldi, D.~Sabulsky, G.~Lef\`evre, X.~Zou, J.-B.~Decitre et~al.,
  \emph{Characterizing {E}arth gravity field fluctuations with the {MIGA}
  antenna for future gravitational wave detectors},
  \href{https://doi.org/10.1103/PhysRevD.99.104026}{\emph{Phys. Rev. D}
  {\bfseries 99} (2019) 3}.

\bibitem{Mitchell2022}
J.~Mitchell, T.~Kovachy, S.~Hahn, P.~Adamson and S.~Chattopadhyay,
  \emph{{MAGIS}-100 environmental characterization and noise analysis},
  \href{https://doi.org/10.1088/1748-0221/17/01/P01007}{\emph{J. Inst.}
  {\bfseries 17} (2022) P01007}.

\bibitem{Schubert2019}
C.~Schubert, D.~Schlippert, S.~Abend, E.~Giese, A.~Roura, W.P.~Schleich et~al.,
  \emph{Scalable, symmetric atom interferometer for infrasound gravitational
  wave detection},
  \href{https://doi.org/10.48550/arXiv.1909.01951}{\emph{arXiv:1909.01951}
  (2019) }.

\bibitem{anderson2017a}
C.A.~Weidner, H.~Yu, R.~Kosloff and D.Z.~Anderson, \emph{Atom interferometry
  using a shaken optical lattice},
  \href{https://doi.org/10.1103/PhysRevA.95.043624}{\emph{Physical Review A}
  {\bfseries 95} (2017) 043624}.

\bibitem{anderson2018}
C.A.~Weidner and D.Z.~Anderson, \emph{Simplified landscapes for optimization of
  shaken lattice interferometry},
  \href{https://doi.org/10.1088/1367-2630/aad36c}{\emph{New Journal of Physics}
  {\bfseries 20} (2018) 075007}.

\bibitem{weidner2018}
C.A.~Weidner, \emph{Shaken {{Lattice Interferometry}}}, Ph.D. thesis,
  University of Colorado Boulder, 2018.

\bibitem{anderson2018a}
C.A.~Weidner and D.Z.~Anderson, \emph{Experimental {{Demonstration}} of
  {{Shaken-Lattice Interferometry}}},
  \href{https://doi.org/PhysRevLett.120.263201}{\emph{Physical Review Letters}
  {\bfseries 120} (2018) 263201}.

\bibitem{gueryodelin}
N.~Dupont, G.~Chatelain, L.~Gabardos, M.~Arnal, J.~Billy, B.~Peaudecerf et~al.,
  \emph{{Quantum State Control of a Bose-Einstein Condensate in an Optical
  Lattice}}, \href{https://doi.org/10.1103/PRXQuantum.2.040303}{\emph{PRX
  Quantum} {\bfseries 2} (2021) 040303}.

\bibitem{Holland2021}
L.-Y.~Chih and M.~Holland, \emph{Reinforcement-learning-based matter-wave
  interferometer in a shaken optical lattice},
  \href{https://doi.org/10.1103/PhysRevResearch.3.033279}{\emph{Phys. Rev.
  Res.} {\bfseries 3} (2021) 033279}.

\bibitem{ledesma}
C.~LeDesma, K.~Mehling, J.~Shao, J.D.~Wilson, P.~Axelrad, M.M.~Nicotra et~al.,
  \emph{A machine-designed optical lattice atom interferometer},
  \textcolor{blue}{arXiv:2305.17603}, 2023.

\bibitem{ledesma2}
C.~LeDesma, K.~Mehling and M.~Holland, \emph{Vector atom accelerometry in an
  optical lattice},  \textcolor{blue}{arXiv:2407.04874}, 2024.

\bibitem{LHCshutdowndates}
``Updated schedule for {CERN}’s accelerators.''
  \url{https://hilumilhc.web.cern.ch/article/updated-schedule-cerns-accelerators}.

\bibitem{Lesko:2015sma}
K.T.~Lesko, \emph{{The Sanford Underground Research Facility at Homestake
  (SURF)}}, \href{https://doi.org/10.1016/j.phpro.2014.12.001}{\emph{Phys.
  Procedia} {\bfseries 61} (2015) 542}.

\bibitem{Gaffet2009}
\emph{{Interdisciplinary And International Deep Underground Science,
  Engineering And Technology Laboratories}}, vol.~All Days of \emph{ISRM
  SINOROCK}, May,
  \textcolor{blue}{https://onepetro.org/ISRMSINOROCK/proceedings-pdf/SINOROCK09/All-SINOROCK09/ISRM-SINOROCK-2009-172/1780316/isrm-sinorock-2009-172.pdf},
  2009.

\bibitem{Joutsenvaara2021}
J.~Joutsenvaara, M.~Holma, O.~Kotavaara and H.J.~Puputti, \emph{Callio {L}ab --
  the deep underground research centre in {F}inland, {E}urope},
  \href{https://doi.org/10.1088/1742-6596/2156/1/012166}{\emph{J. Phys.: Conf.
  Ser.} {\bfseries 2156} (2021) 012166}.

\bibitem{Murphy:2012zz}
A.~Murphy and S.~Paling, \emph{{The Boulby mine underground science facility:
  The search for dark matter, and beyond}},
  \href{https://doi.org/10.1080/10619127.2011.629920}{\emph{Nucl. Phys. News}
  {\bfseries 22N1} (2012) 19}.

\bibitem{Ianni:2016fjt}
A.~Ianni, \emph{{Canfranc Underground Laboratory}},
  \href{https://doi.org/10.1088/1742-6596/718/4/042030}{\emph{J. Phys. Conf.
  Ser.} {\bfseries 718} (2016) 042030}.

\bibitem{Beker_2012}
M.G.~Beker, J.F.J.~van~den Brand, E.~Hennes and D.S.~Rabeling, \emph{Newtonian
  noise and ambient ground motion for gravitational wave detectors},
  \href{https://doi.org/10.1088/1742-6596/363/1/012004}{\emph{Journal of
  Physics: Conference Series} {\bfseries 363} (2012) 012004}.

\bibitem{HawkinsBrown2019}
``Beecroft building university of oxford.''
  \url{https://www.hawkinsbrown.com/projects/beecroft-building-university-of-oxford}.

\bibitem{Carlton2023}
J.~Carlton and C.~McCabe, \emph{{Mitigating anthropogenic and synanthropic
  noise in atom interferometer searches for ultralight dark matter}},
  vol.~108, p.~123004, 2023,
  \href{https://doi.org/10.1103/PhysRevD.108.123004}{DOI}
  [\href{https://arxiv.org/abs/2308.10731}{{\ttfamily 2308.10731}}].

\bibitem{Borde1993}
C.J.~Bord{\'e}, M.~Weitz and T.W.~H{\"a}nsch, \emph{New optical atomic
  interferometers for precise measurements of recoil shifts. application to
  atomic hydrogen}, \href{https://doi.org/10.1063/1.45083}{\emph{{AIP}
  Conference Proceedings} {\bfseries 290} (1993) 76}.

\bibitem{schelfhout2024single}
J.S.~Schelfhout, T.M.~Hird, K.M.~Hughes and C.J.~Foot, \emph{A single-photon
  large-momentum-transfer atom interferometry scheme for {Sr} or {Yb} atoms
  with application to determining the fine-structure constant},
  \href{https://doi.org/10.48550/arXiv.2403.10225}{\emph{arXiv preprint
  arXiv:2403.10225} (2024) }.

\bibitem{Borde1984}
C.J.~Bord{\'{e}}, C.~Salomon, S.~Avrillier, A.~van Lerberghe, C.~Br{\'{e}}ant,
  D.~Bassi et~al., \emph{Optical {Ramsey} fringes with traveling waves},
  \href{https://doi.org/10.1103/physreva.30.1836}{\emph{Physical Review A}
  {\bfseries 30} (1984) 1836}.

\bibitem{Borde1989}
C.~Bord{\'{e}}, \emph{Atomic interferometry with internal state labelling},
  \href{https://doi.org/10.1016/0375-9601(89)90537-9}{\emph{Physics Letters A}
  {\bfseries 140} (1989) 10}.

\bibitem{Parker2018}
R.H.~Parker, C.~Yu, W.~Zhong, B.~Estey and H.~M\"uller, \emph{Measurement of
  the fine-structure constant as a test of the {S}tandard {M}odel},
  \href{https://doi.org/10.1126/science.aap7706}{\emph{Science} {\bfseries 360}
  (2018) 191}.

\bibitem{Zhong2020}
W.~Zhong, R.H.~Parker, Z.~Pagel, C.~Yu and H.~M{\"u}ller, \emph{Offset
  simultaneous conjugate atom interferometers},
  \href{https://doi.org/10.1103/physreva.101.053622}{\emph{Physical Review A}
  {\bfseries 101} (2020) 053622}.

\bibitem{solaro2022}
C.~Solaro, C.~Debavelaere, P.~Clad{\'e} and S.~{Guellati-Khelifa}, \emph{Atom
  {{Interferometer Driven}} by a {{Picosecond Frequency Comb}}},
  \href{https://doi.org/10.1103/PhysRevLett.129.173204}{\emph{Physical Review
  Letters} {\bfseries 129} (2022) 173204}.

\bibitem{debavelaere2024}
C.~Debavelaere, C.~Solaro, S.~{Guellati-Kh{\'e}lifa} and P.~Clad{\'e},
  \emph{Atom interferometer using spatially localized beam splitters},
  \href{https://doi.org/10.1103/PhysRevA.110.013310}{\emph{Physical Review A}
  {\bfseries 110} (2024) 013310}.

\end{thebibliography}\endgroup

\end{document}